%% file: thesis.tex
\documentclass[11pt,a4paper]{book}
\pdfoutput=1

\usepackage{hyperref}
\usepackage{bm}
\usepackage{amsmath,amssymb}
\usepackage{cite}
\usepackage{graphicx}

\numberwithin{figure}{section}
\numberwithin{equation}{section}
\numberwithin{table}{section}

\newcommand{\cutoff}{\,\mathcal{R}_k\,}

\newcommand{\cT}{{\cal T}}
\newcommand{\df}{\,\delta\! f}

\newcommand{\Real}{{\rm Re\,}}

\newcommand{\eps}{\varepsilon}
\newcommand{\vp}{\varphi}
\newcommand{\pb}{\bar \varphi}

\newcommand{\al}{\alpha}
\newcommand{\Vh}{\hat V}
\newcommand{\ph}{\hat \varphi}
\newcommand{\that}{\hat t}
\newcommand{\vh}{\hat v}
\newcommand\tc[1]{\raisebox{.5pt}{\textcircled{\raisebox{-.9pt} {#1}}}}

\newcommand{\kh}{\hat k}
\newcommand{\Kh}{\hat K}
\newcommand{\gmn}{g_{\mu\nu}}
\newcommand{\bgmn}{\bar g_{\mu\nu}}
\newcommand{\hmn}{h_{\mu\nu}}

\newcommand{\hess}{\Gamma^{(2)}}
\newcommand{\eipx}{e^{i p\cdot x}}
\newcommand{\emipx}{e^{-ip \cdot x}}
\newcommand{\dclnf}{\,\partial_\chi\! \ln\! f \,}
\newcommand{\cb}{\bar \chi}
\newcommand{\phih}{\hat \phi}
\newcommand{\bp}{\bar \varphi}

\newcommand{\be}{\begin{equation}}
\newcommand{\ee}{\end{equation}}
\newcommand{\bea}{\begin{eqnarray}}
\newcommand{\eea}{\end{eqnarray}}

\begin{document}
\frontmatter

\thispagestyle{empty}

\begin{center}
\mbox{}


{\Huge University of Southampton}	\\
\vspace{0.5cm}
{\huge Faculty of Physical Sciences and Engineering}

\vfill

\vspace{2cm}

\textbf{\huge{Functional truncations \\ \vspace{0.5cm} in asymptotic safety for quantum gravity}}\\

\vspace{4cm}

{\huge
J\"urgen Albert Dietz
}
 \vspace{3cm}

{\Large Submitted for the degree of}\\
\vspace{0.5cm}
\begin{Large}\MakeUppercase{Doctor of Philosophy} \end{Large}\\

\vspace{1cm}
\vfill

{\Large
November 2014
}
\vfill
\end{center}


\newpage
\thispagestyle{empty}
\mbox{}


\newpage
\thispagestyle{empty}

\begin{center}

\MakeUppercase{University of Southampton}
\vfill
\underline{\MakeUppercase{{\Large Abstract}}}
\vfill
\MakeUppercase{Faculty of Physical Sciences and Engineering}

Physics and Astronomy
\vfill
\underline{Doctor of Philosophy}
\vfill
\MakeUppercase{{\Large Functional truncations in asymptotic safety for quantum gravity}}
\vfill
J\"urgen Albert Dietz
\vfill 
\end{center}

\noindent
Finite dimensional truncations and the single field approximation have thus far played dominant roles in investigations of asymptotic safety for quantum gravity. This thesis is devoted to exploring asymptotic safety in infinite dimensional, or functional, truncations of the effective action as well as the effects that can be caused by the single field approximation in this context. It begins with a comprehensive analysis of the three existing flow equations of the single field $f(R)$ truncation by determining their spaces of global fixed point solutions and, where applicable, of corresponding eigenoperator solutions. This becomes possible through the use of the powerful parameter counting method and a combination of analytical and numerical methods, all of which can be applied much more generally when faced with similar differential equations. As a second result, it is then shown that one incarnation of the single field $f(R)$ approximation actually breaks down in the sense that there is no physical content left to explore. In order to clarify whether such drastic findings can be caused by the approximations used in setting up the renormalisation group flow, we identify the single field approximation as a prime candidate and show in the more familiar context of scalar field theory that it can indeed lead to many types of non-physical results. As a way to avoid such non-physical behaviour we highlight the importance of the previously known split Ward identity and exemplify its usefulness by fully restoring the correct physical picture in scalar field theory. Taking this result as evidence that the split Ward identity may lead to well behaved functional truncations also in gravity, we derive the flow equations of conformal gravity in a bi-field truncation of the effective action that goes beyond the local potential approximation in the fluctuation field. It is found that the split Ward identity leads to a simplified set of renormalisation group equations for the conformal factor that, while differing at crucial points, bear close resemblance to flow equations obtained in scalar field theory.
\vfill

\newpage
\thispagestyle{empty}

\tableofcontents

\listoffigures

\newpage
\thispagestyle{empty}

\listoftables

\newpage

\thispagestyle{empty}
\noindent
\textbf{{\LARGE Declaration of Authorship}}
\vspace{1cm}

\noindent
I declare that this thesis and the work presented in it are my own and has been generated by me as the result of my own original research.

\mbox{}

\noindent
I confirm that:
\begin{itemize}
\item This work was done wholly or mainly while in candidature for a research degree at this University;
\item No part of this thesis has previously been submitted for a degree or any other qualification at this University or any other institution;
\item Where I have consulted the published work of others, this is always clearly attributed;
\item Where I have quoted from the work of others, the source is always given. With the exception of such quotations, this thesis is entirely my own work;
\item I have acknowledged all main sources of help;
\item The parts of this thesis based on work done mainly by the author are sec. \ref{numerics}, chapter \ref{ch:LPA} with the exception of sec. \ref{sec:LPALitim}, and chapter \ref{sec:conformal}. Research underlying sec. \ref{sec:fpanaquali}, \ref{sec:asy-fp}, \ref{sec:eopsasy} and chapter \ref{sec:red-ops} was done in large part by the author's supervisor Prof. Tim Morris while the main part of the work for sec. \ref{sec:LPALitim} was carried out by Hamzaan Bridle;
\item Results of chapters \ref{sec:f-of-R}, \ref{sec:red-ops} and \ref{ch:LPA} of this thesis have respectively been published in \cite{DietzMorris:2013-1}, \cite{DietzMorris:2013-2} and \cite{Bridle:2013sra};
\end{itemize}

\mbox{}

Signed:	.............................. \hspace{2cm} Date: ................................

\vfill

\newpage
\thispagestyle{empty}
\noindent
\textbf{{\LARGE Acknowledgements}}
\vspace{1cm}

\noindent
First and foremost, I would like to express my sincere gratitude for the guidance provided by my supervisor Tim Morris over the whole period of my PhD. His clear scientific insight and continuous support have contributed enormously to making this thesis possible. Special thanks go to my fellow researchers at the postgraduate offices, Jason Hammett, Marc Thomas, Shane Drury, Daniele Barducci, Maria Dimou, Hamzaan Bridle, Anthony Preston, Miguel Rom\~ao and Juri Fiaschi for an uncountable number of discussions and an excellent office atmosphere. In the same way I would like to extend my gratitude to the whole Southampton Particle Physics Group for creating a thoroughly stimulating and enjoyable work environment throughout.

\newpage 
\thispagestyle{empty}
\mainmatter
\input{intro/intro.tex}
\input{f-of-R/f-of-R.tex}

\input{red-ops/red-ops.tex}

\input{LPA/LPA.tex}

\input{conformal/conformal.tex}
\input{conclusions/conclusions.tex} 

\addcontentsline{toc}{chapter}{Bibliography} 
\bibliographystyle{utphys}
\bibliography{refs}

\end{document}

%% file: intro/intro.tex
\chapter{Introduction and Fundamentals}
One of the most prominent and long standing open questions in theoretical physics concerns the unification of general relativity with the principles of quantum mechanics \cite{kiefer2012quantum, hamber2009}. The desire to modify general relativity to incorporate quantum mechanical effects was expressed by Einstein in the context of gravitational waves already in 1916 \cite{einstein1916}:
\begin{quote}
\emph{
''Nevertheless, due to the inner-atomic movement of electrons, atoms would have to radiate not only electro-magnetic but also gravitational energy, if only in tiny amounts. As this is hardly true in Nature, it appears that quantum theory would have to modify not only Maxwellian electrodynamics, but also the new theory of gravitation.''}
\end{quote}
At the time of this statement Einstein likely based his conclusion on the idea of an eternal, static universe. Nowadays, atomic gravitational radiation cannot be taken as experimental evidence for the necessity of unifying gravity with quantum mechanics in the same way as this was the case for electrodynamics since the time scales associated with the collapse of atoms due to gravitational radiation are far longer than the age of the universe \cite{gorelik}. Nevertheless, it already expresses one of the main reasons that provides the driving force for many physicists of today on the quest for a quantum theory of gravity, which is the belief that the principles of quantum field theory (QFT) or generalisations thereof have to apply to all forces in Nature. That this is the case is only an apparently slight generalisation of experimental evidence as provided by the thorough success of QFT in the form of the Standard Model of particle physics for modelling all forces known in Nature apart from gravity.

The need for accommodating quantum mechanical principles in a generalised theory of gravity is also more directly motivated by the simple observation that the nature of the matter content of spacetime on the right hand side of Einstein's equations,
\begin{equation}
 R_{\mu\nu} - \frac{1}{2}g_{\mu\nu}R = 8\pi G \, T_{\mu\nu},
\end{equation}
as expressed by the energy-momentum tensor is intrinsically quantum mechanical. At this point one could in principle try to proceed by turning the right hand side into a classical quantity to obtain a so-called semi-classical theory for gravity, but so far this has not led to a fully consistent result, see \cite{kiefer2012quantum}. The alternative route is to advocate the opposite proposal that requires to generalise Einstein's equations to a fully quantum mechanical theory that contains general relativity in the classical limit. This is the point of view supported in this work.

When presented with a classical field theory, its quantisation is usually carried out in the framework of perturbative quantisation. This has been an extremely successful approach as is testified by the Standard Model, but it is well known that if it is applied to general relativity, the inevitable result is a perturbatively non-renormalisable QFT \cite{'tHooft:1974bx, Goroff:1985sz, vandeVen:1991gw}. It entails the loss of predictivity of the theory at a fundamental level as an infinite number of parameters have to be determined experimentally.

With this result in mind, it might seem that the very diverse field of quantum gravity research sometimes conveys the impression that gravity and quantum mechanics are entirely incompatible, and that any attempt of quantising gravity results in unacceptable complications. Despite perturbative non-renormalisability however, and at energies well below the Planck scale, gravity can be successfully quantised in the form of an effective field theory \cite{Burgess:2003jk, Donoghue:2012zc}, leading to universally valid corrections for the Newtonian potential, for example. As infrared predictions of quantum gravity, these corrections are universally valid in the sense that they are independent of the ultimate ultraviolet completion of quantum gravity. Unfortunately, this modification of the Newtonian potential and other predictions are too small to be verified experimentally for the time being. Nevertheless, effective field theories have been used successfully in other areas of particle physics, such as chiral perturbation theory in the low energy regime of QCD, e.g. \cite{Scherer:2002tk}, where experimental confirmation is available and we would therefore tend to trust the results of the effective field theory also in the case of gravity.

Thus, when we refer to the unification of quantum mechanics and gravity, the real goal is to find a viable ultraviolet completion of the effective field theory of quantum gravity that takes over once the latter fails at energies comparable to the Planck scale. Of course, there is the possibility that the ultraviolet regime of gravity is simply not amenable to the quantum field theoretical process of quantisation, and instead there is a different framework that correctly describes gravity at very high energies which then reproduces the quantum field theoretical description at low energies (assuming the effective field theory treatment described above is correct). These theories leave the realm of conventional QFT by introducing novel structures as is the case in string theory or loop quantum gravity.

Such fundamentally new frameworks are however by no means the only way to attempt a successful quantisation of gravity. One can also explore a more conservative route that does not go beyond the framework of QFT. Even though there is no doubt about the perturbative non-renormalisability mentioned before, which at first sight might seem to question the usefulness of QFT as the correct framework, we cannot a priori exclude the possibility that gravity may be renormalisable in a non-perturbative way. In other words, treating quantum gravity in a perturbative fashion may not be a powerful enough approach to investigate its ultraviolet behaviour, where crucial non-perturbative dynamics could lead to a well defined QFT of gravity. This is the idea that lies at the heart of what is known as asymptotic safety for quantum gravity, which we will pick up in sec. \ref{sec:asymptotic-safety} after having introduced the necessary tools for its discussion.

\section{Non-perturbative renormalisation}
\label{sec:non-pert-renorm}
An appropriate framework in which non-perturbative renormalisation can be investigated is provided by the functional renormalisation group. It is based on the idea of Wilsonian renormalisation \cite{Wilson:1973,Wegner:1972ih} and is also known as the exact or continuous renormalisation group. The philosophy behind this technique is that it can prove useful for practical purposes to integrate out only those modes in the path integral that possess momenta larger than some infrared cutoff scale $k$. Varying $k$, a non-perturbative renormalisation group flow is generated with the property that upon sending $k \rightarrow 0$ we recover the information contained in the full path integral. In order to illustrate the concepts associated with the functional renormalisation group, it is useful to look at its simplest incarnation, given by scalar field theory, and then elucidate the additional complexities arising from the gravitational context in the next section. For introductions and reviews on the functional renormalisation group for  non-gravitational theories, see \cite{Berges:2000,Morris:1998,Bagnuls:2000,Rosten:2010,Gies:2006wv,Pawlowski:2005xe}.

For setting up the non-perturbative renormalisation group flow of the functional renormalisation group, the essential ingredient is the cutoff action $S_k$ which depends on the renormalisation group (RG) scale $k$ and is added to the bare action, leading to the desired modification of the path integral. For a single scalar field the starting point is therefore the partition function, 
\begin{equation} \label{partfunc}
 Z_k[J]=\int \mathcal{D}\phi\, \exp\left(-S[\phi]-S_k[\phi]+ J\cdot \phi\right)
\end{equation}
where with the source term we have introduced the notation
\begin{equation} \label{dotconvention}
 J\cdot\phi \equiv J^a\phi_a \equiv \int_x J(x)\phi(x) \equiv \int d^dx \, J(x)\phi(x),
\end{equation}
with the index $a$ replacing the coordinates $x$, and $d$ being the dimension of space. These notationally different ways of expressing an integral will also be used where convenient at later stages.  Note that the path integral is written in its Wick rotated formulation as it is easier to work in Euclidean signature. 

\begin{figure}[h]
\centering
\includegraphics[scale=0.6]{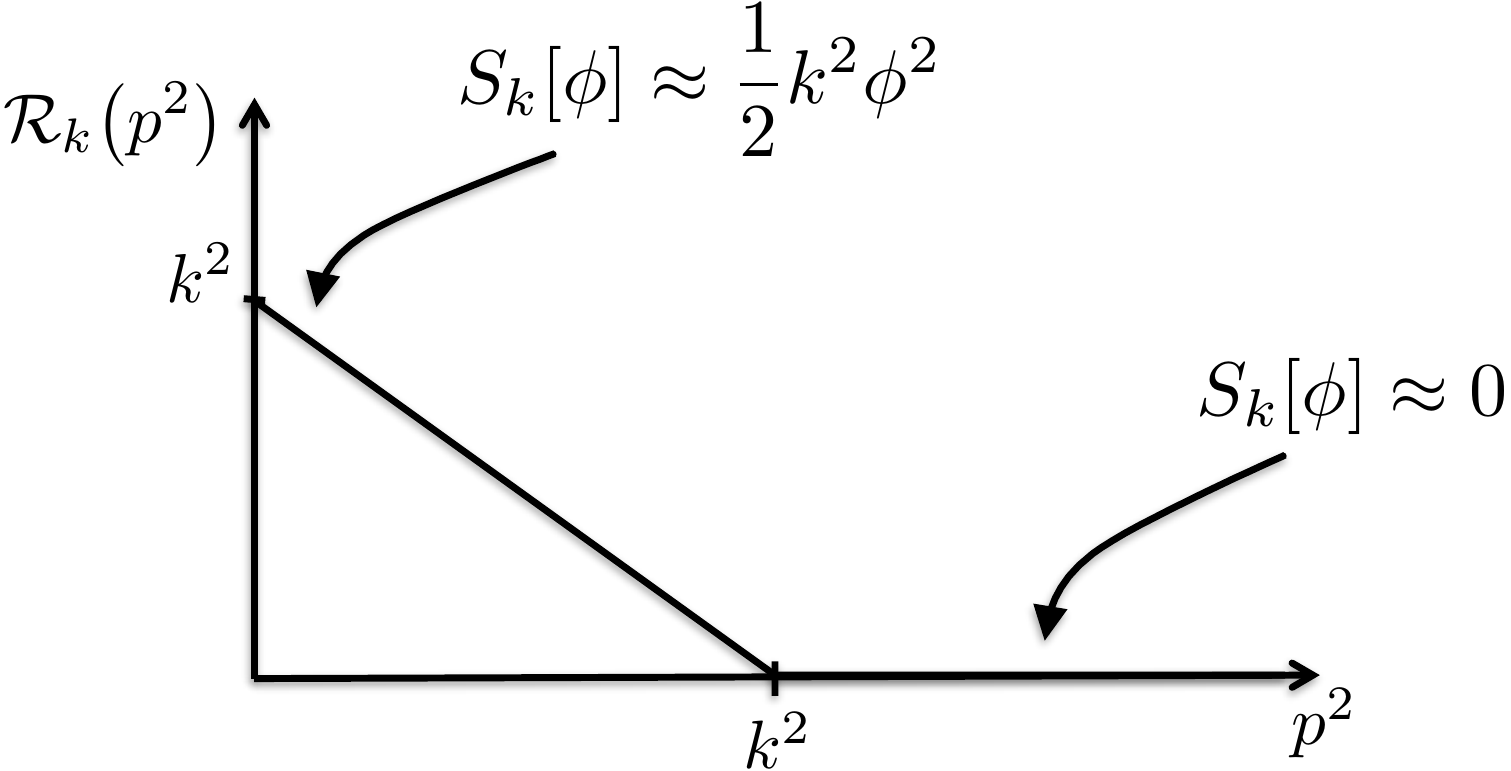}
\caption[The optimised cutoff]{A typical choice of cutoff operator used in many applications of the functional renormalisation group. Here, the cutoff action is evaluated for momentum eigenmodes of the Laplacian.}
\label{fig:optimised-cutoff}
\end{figure}

The cutoff action is taken to be of the following form,
\begin{equation} \label{equ:cutoffaction}
 S_k[\phi] = \frac{1}{2} \int_x \phi \cutoff \phi,
\end{equation}
where the cutoff operator $\cutoff = \mathcal{R}_k\!\left(-\nabla^2\right)$ is a function of the Laplacian and the RG scale $k$ and turns $S_k$ into a momentum dependent mass term by acting on the field $\phi$. The precise form of $\cutoff$ is irrelevant as long as it satisfies the two limits
\begin{equation} \label{cutoff-props}
\lim_{k^2/p^2 \to 0} \cutoff\!\big(p^2\big) = 0 \qquad \text{and} \qquad \lim_{k^2/p^2 \to \infty} \cutoff\!\big(p^2\big) >0,
\end{equation}
ensuring, loosely speaking, that modes with momenta above the cutoff scale $k$ are integrated out and modes with momenta below the cutoff scale are suppressed in the path integral \eqref{partfunc}. An example of a cutoff operator that has been widely used in many instances of the functional renormalisation group is the so-called optimised cutoff \cite{Litim:2001} which is depicted in fig. \ref{fig:optimised-cutoff}.

The partition function \ref{partfunc} can then be rewritten as a functional differential equation, known as the functional renormalisation group equation \cite{Wetterich:1992,Morris:1993},
\begin{equation} \label{equ:FRGE}
\frac{\partial}{\partial t}\Gamma_k = \frac{1}{2} \, \mathrm{Tr} \left[ \left(\Gamma^{(2)}_k + \cutoff\right)^{-1} \frac{\partial}{\partial t} \cutoff \right].
\end{equation}
The central object in this equation is the effective average action $\Gamma_k[\phi^c]$ which is a functional of the classical field $\phi^c=\langle \phi \rangle$ and, with $W_k = \ln Z_k$, is related to the path integral through a modified Legendre transformation with respect to the source $J$,
\begin{equation*}
\Gamma_k[\phi^c] = J \cdot \phi^c - W_k[J] - S_k[\phi^c].
\end{equation*}
The Hessian $\Gamma_k^{(2)}$, i.e. the second functional derivative with respect to the classical field, of the effective average action appears on the right hand side of \eqref{equ:FRGE}, where the effect of the cutoff operator as an effective mass term added to the inverse propagator $\Gamma_k^{(2)}$ becomes apparent. We have also introduced the so-called renormalisation group time, $t = \ln(k/k_0)$, where $k_0$ is an arbitrary reference scale. The trace in the functional renormalisation group equation is taken over space (or momentum) coordinates as well as over possible field indices, if multiple fields are present.

The result \eqref{equ:FRGE} is also sometimes referred to as the exact renormalisation group equation, but in the following we will simply refer to it as the flow equation. Similarly, the effective average action is also known as the Legendre effective action, although we will use an abbreviated terminology by referring to it as the effective action. It differs from the standard effective action in QFT only by the presence of the cutoff action in the path integral \eqref{partfunc}. For $k \to 0$ the cutoff drops out because of the first property in \eqref{cutoff-props} and the effective action of the present context coincides with the standard effective action.

The flow equation describes the non-perturbative renormalisation group flow of the effective action in the space of all action functionals $\Gamma_k$. This so-called theory space is only constrained by the symmetries present in the theory. In the case of single component scalar field theory, the only symmetry the effective action has to respect is Lorentz symmetry. For a scalar field with $n$ components, we could additionally require invariance under rotations amongst the individual components, for a gauge theory we would impose gauge invariance and in the case of gravity the additional requirement is coordinate invariance, as we will see later. If we think of the effective action as expanded in a set of basis operators of theory space,
\begin{equation}
\Gamma_k[\phi^c] = \sum_n g_n(k)\, \mathcal{O}_n[\phi^c]
\end{equation}
with $k$-dependent couplings $g_n(k)$, substitute this into the flow equation \eqref{equ:FRGE} and organise the right hand side as the same type of expansion, the beta functions of each coupling $g_n(k)$ can be read off as the coefficient of the corresponding operator: $\dot g_n = \frac{d}{dt} g_n=\beta_n(g_1,g_2,\dots)$.

For every value of the sliding energy scale $k$, or equivalently for any renormalisation group time $t$, the effective action $\Gamma_k$ is represented by a point in theory space. Starting at an initial time with an initial effective action, a solution to the flow equation \eqref{equ:FRGE} takes the form of a curve $k \mapsto \Gamma_k$ in theory space, with the right hand side of the flow equation describing the tangent vector along this curve. The initial point of such a curve can be taken to be at a bare or ultraviolet cutoff scale $k=\Lambda$, where it is convenient to regularise the path integral \eqref{partfunc} in the ultraviolet by imposing this ultraviolet cutoff from the outset. Noting that adding the cutoff action in \eqref{partfunc} provides infrared regularisation, this implies that the theory we start out with is fully regularised. Of course, the goal then is to remove the ultraviolet cutoff to obtain a renormalisation group trajectory $t \mapsto \Gamma_k$ defined for all $-\infty<t<\infty$. This can be done if the trajectory tends to a fixed point, as we discuss now.

In order to exhibit the correct RG flow, and fixed points in particular, all quantities in the flow equation \eqref{equ:FRGE} have to be expressed in dimensionless variables with the help of the RG scale $k$. We apply this change of variables to the flow equation and rewrite the result in the general form
\begin{equation}\label{flow-gen}
\partial_t \Gamma_k[\phi] = \mathcal{F}[\Gamma_k],
\end{equation}
where we have dropped the superscript of the classical field and kept the same notation for the dimensionless quantities. The term $\mathcal{F}$ contains the right hand side of \eqref{equ:FRGE} but also additional terms arising from the change of variables on the left hand side. The $t$-derivative on the left of \eqref{flow-gen} can be thought of as acting only on the dimensionless couplings contained in the effective action to its right.

Fixed points are stationary solutions of \eqref{flow-gen}, i.e. solutions independent of time $t$, for which we reserve the notation $\Gamma_*$. A fixed point can be non-interacting, in which case it is called a Gaussian fixed point, or it can include interactions and is referred to as a non-Gaussian fixed point or often as a non-perturbative fixed point, assuming its couplings are non-perturbative.
\begin{figure}[ht]
\centering
\includegraphics[scale=0.4]{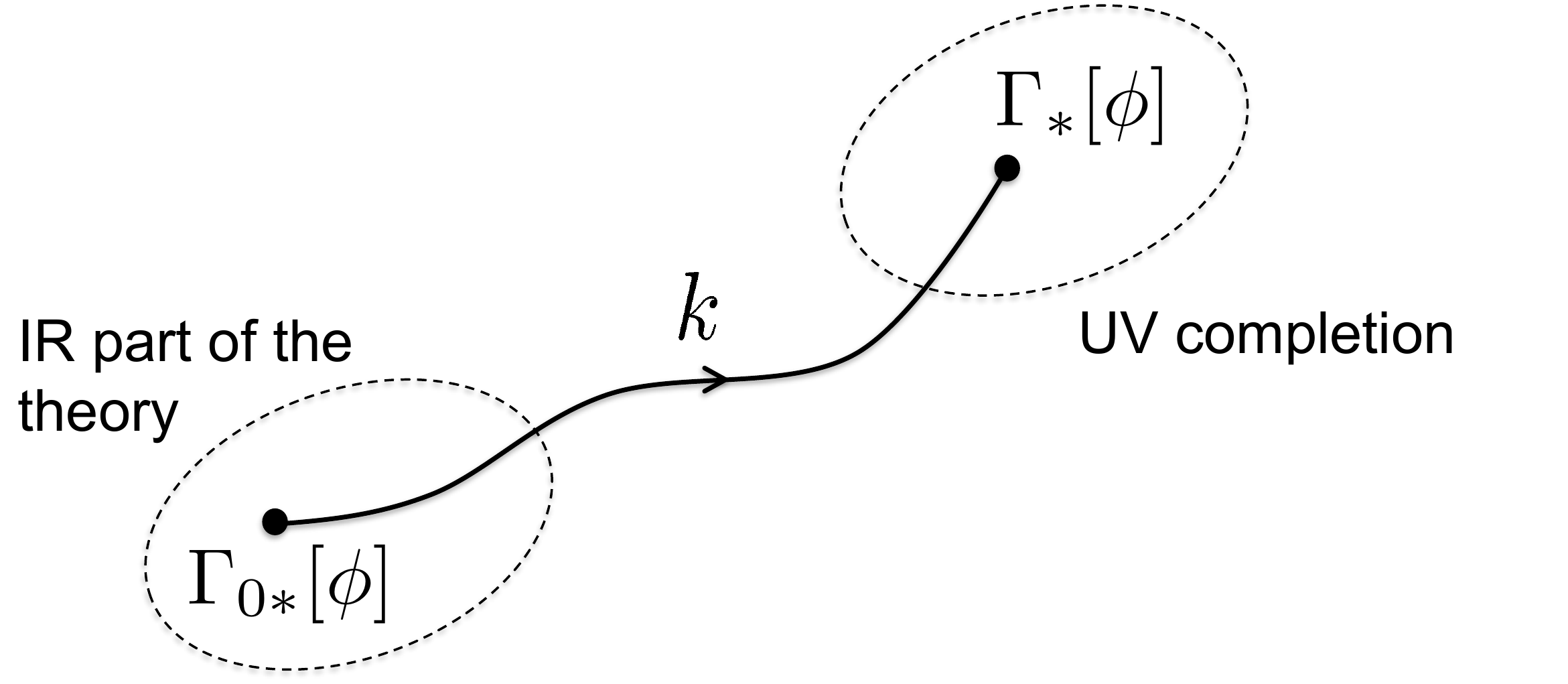}
\caption[A global renormalised trajectory]{A renormalisation group trajectory connecting a Gaussian fixed point denoted $\Gamma_{0*}$ in the infrared with an
interacting non-perturbative fixed point $\Gamma_*$ in the ultraviolet.}
\label{fig:renQFT}
\end{figure}
Global renormalisation group trajectories that connect a fixed point in the infrared, $k \to 0$, to a fixed point in the ultraviolet, $k \to \infty$, represent QFTs defined on all scales since the bare cutoff $\Lambda$ on the theory can be removed. The fact that $\Gamma_k$ tends to an infrared fixed point for $k \to 0$ guarantees infrared finiteness. An example of such a configuration in theory space is illustrated in fig. \ref{fig:renQFT}. It is expected in general that trajectories ending at fixed points in this fashion are the only global solutions of the flow equation that describe fully defined QFTs.

It is worth noting that in the vicinity of the Gaussian fixed point, where couplings are small, an expansion of the flow equation \eqref{equ:FRGE} in the couplings results in the usual perturbative beta functions. As mentioned earlier, the real power of the flow equation lies in the fact that it also describes renormalisation in quantum field theory when there are no small parameters to expand in.

The renormalisation group flow in the vicinity of any fixed point $\Gamma_*$ can be determined by linearising the flow equation around the fixed point. This is achieved by writing
\begin{equation} \label{perturbations}
\Gamma_k[\phi] = \Gamma_*[\phi] + \eps\, \mathcal{O}[\phi]\exp(-\lambda t),
\end{equation}
where the dimensionless $\eps$ is introduced since we consider infinitesimal perturbations around the fixed point action $\Gamma_*$. We are interested in perturbations whose $t$-dependence is separable, in which case they take the form given here. The integrated operator $\mathcal{O}$ satisfies the eigenoperator equation
\begin{equation}\label{eigenopequ}
-\lambda\,\mathcal{O} = \left. \frac{\delta \mathcal{F}}{\delta \Gamma_k}\right|_*\cdot \mathcal{O},
\end{equation}
with the right hand side being the variational derivative of the right hand side of the flow equation \eqref{flow-gen} in the direction of $\mathcal{O}$ taken at the fixed point,
\begin{equation}
\left. \frac{\delta \mathcal{F}}{\delta \Gamma_k}\right|_*\cdot \mathcal{O} = \lim_{\eps\to 0}\frac{1}{\eps}\left\{\mathcal{F}[\Gamma_*+\eps\,\mathcal{O}]-\mathcal{F}[\Gamma_*]\right\}.
\end{equation}
Corresponding to the fact that all variables are expressed in terms of dimensionless quantities, we see after using the definition of $t$ that the eigenoperator $\mathcal{O}$ in \eqref{perturbations} has the infinitesimal dimensionless coupling $\eps\left(k_0/k\right)^\lambda$ corresponding to the physical coupling $\eps k_0^\lambda$ of mass dimension $\lambda$. Eigenoperators in \eqref{perturbations} with dimensionless couplings that grow (decay) when flowing towards the infrared, i.e. decreasing $t$, are called relevant (irrelevant). If they have a non-vanishing RG eigenvalue $\lambda$ it will be positive (negative). Eigenoperators with vanishing RG eigenvalue at the linearised level \eqref{eigenopequ} may still be exposed as relevant or irrelevant beyond the linearisation around the fixed point, in which case they are marginally relevant (irrelevant). An eigenoperator can also be exactly marginal, if it belongs to neither of these categories. 

A QFT is fully specified by determining the values of couplings of all (marginally) relevant infinitesimal perturbations as in \eqref{perturbations} in the vicinity of a fixed point and setting all (marginally) irrelevant couplings to zero. In this way, the set of (marginally) relevant perturbations describe how the RG trajectory leaves the fixed point and are sufficient to determine its global evolution uniquely as obtained from the flow equation. The relevant eigenoperators span the so-called critical surface around a fixed point and since they are ultimately obtained from experiment, the result of this process is a fully predictive QFT if the critical surface is of finite dimensionality.\footnote{Wilson defined the critical surface in the vicinity of a fixed point as consisting of all irrelevant perturbations, i.e. perturbations that are drawn into the fixed point for decreasing $k$ \cite{Wilson:1973}. In the context of gravity the critical surface is often defined as described here, cf. \cite{Reuter:2012}.}

Thus, a QFT that is non-renormalisable when treated perturbatively may still possess a fixed point in the vast non-perturbative region of theory space with a finite number of relevant eigenoperators. It can therefore still turn out to be a valid, fully predictive theory, provided other issues such as unitarity, that have to be investigated separately and will be left aside in the following, can be resolved.

\section{Adaptations for gravity}
\label{sub:adaptations-gravity}
In principle, the approach of non-perturbative renormalisation discussed in the previous section can be applied to any QFT. However, depending on the specific theory additional technical difficulties have to be overcome compared to the scalar field theory case treated before. This is also true for gravity, as we review now. The purpose here is only to highlight the important extensions necessary in order to deal with gravity and to draw attention to aspects of the formalism that will become particularly relevant later on. For the original derivation of the flow equation for quantum gravity we refer to \cite{Reuter:1996}.

The quantisation of any QFT other than gravity always makes use of some classical background spacetime. In gravity, the fundamental field we would like to quantise is the metric field $\gmn$ itself, and a priori there is no ambient background spacetime this can be done on. This intrinsic problem is circumvented by the use of the background field method, the first step of which is to linearly split the total metric into a background metric $\bgmn$ and a fluctuation field,
\begin{equation}\label{background-split}
\gmn = \bgmn + \hmn.
\end{equation}
This split is arbitrary, and in particular does not imply that the fluctuation field, which need not be a metric field, is perturbative. The analogous path integral to \eqref{partfunc} over the metric $\gmn$ is then shifted to an integration over $\hmn$, i.e. it is the fluctuation field that is quantised.

The bare action $S[\gmn]$ in the path integral is required to be diffeomorphism invariant, with the infinitesimal action of a diffeomorphism generated by the vector field $v$ on the metric being given by the Lie derivative
\begin{equation}
\delta \gmn =\mathcal{L}_v \gmn = v^\kappa\partial_\kappa \gmn +\partial_\mu v^\kappa  g_{\kappa\nu}+\partial_\nu v^\kappa g_{\mu\kappa}.
\end{equation}
The gauge transformations
\begin{equation} \label{quantgaugetrafo}
\delta \bgmn = 0 \qquad \text{and} \qquad \delta \hmn = \delta \gmn
\end{equation}
need to be gauge-fixed by an appropriate gauge condition $F_\mu[\bar g, h] = 0$, which is taken to be linear in the quantum field $\hmn$. It also depends on the background field and is chosen in such a way that background field covariance is implemented. This means that although the gauge condition $F_\mu$ breaks the invariance under the quantum gauge transformations \eqref{quantgaugetrafo} as required, it is chosen such that it is invariant under background gauge transformations as defined in
\begin{equation} \label{background-gauge-trafo}
\delta \bgmn = \mathcal{L}_v \bgmn, \qquad \text{and} \qquad \delta \hmn = \mathcal{L}_v \hmn.
\end{equation}
The functional determinant that appears as part of the Faddeev-Popov procedure applied in the present context is exponentiated and thus leads to a ghost action that also needs to be included in the path integral. Furthermore, the cutoff action \eqref{equ:cutoffaction} now not only contains a term for the fluctuation field $\hmn$ but also for the ghost fields. 

An important point concerning the cutoff action is that the cutoff operator is now a function of the covariant background field Laplacian $\cutoff = \cutoff\!\!\left(-\bar \nabla^2\right)$. It cannot be taken to depend on the Laplacian of the total metric instead as the cutoff action \eqref{equ:cutoffaction} has to stay quadratic in the fluctuation field to preserve the structure of the flow equation \eqref{equ:FRGE}. Thus, it is the spectrum of the Laplacian associated with the background metric that provides the momentum scales of fluctuation field modes $\hmn$ that are compared to the cutoff scale $k$ to distinguish between low and high momentum modes that are consequently suppressed or integrated out as described in the previous section. One of the main reasons for the use of the background field method is to make this construction possible, and it shows the crucial role played by the background field in this setup.

Once the gauge fixing term, the ghost action and the extended cutoff sector have been included besides the bare action in the path integral, one proceeds in the same way as for the path integral \eqref{partfunc} to obtain the analogous version of the flow equation \eqref{equ:FRGE} for the gravitational effective action 
\begin{equation}\label{effacgrav}
\Gamma_k = \Gamma_k[h,\bar g, \tau, \bar \tau].
\end{equation}
It depends on the classical fluctuation field $\hmn$, for which we have taken the liberty of using the same notation, and the background field, as well as the two classical ghost fields $\tau^\mu$ and $\bar \tau_\mu$. For each field $\Phi \in \{h,\tau\}$ the right hand side of the flow equation now has a trace as in \eqref{equ:FRGE} with an additional minus sign for the Grassmannian ghost fields.

Invariance under background gauge transformations as in \eqref{background-gauge-trafo} implies that the effective action $\Gamma_k[h,\bar g, \tau, \bar \tau]$ remains unchanged when all its arguments transform according to the Lie derivative acting on them, implying that it is a diffeomorphism invariant functional of its arguments. It is another virtue of the background field formalism with the metric split \eqref{background-split} that this becomes possible. 

We already remark here, that any viable theory of quantum gravity has to satisfy background independence. In the present case this translates into the requirement that even though the effective action in general depends on the background field $\bar g$, this dependence has to drop out at the level of physical observables. The condition of background independence will be a recurring theme throughout this work, and different aspects of it will be investigated in later sections.

While this is the basic setup, there are many additional aspects to consider for the actual evaluation of the flow equation. We will comment on them in more detail where they are needed in chapter \ref{sec:f-of-R}. 

Solving the full flow equation \eqref{equ:FRGE} is equivalent to solving the path integral it is derived from. It therefore comes as no surprise that it is necessary to apply some approximation scheme in order to make progress with the flow equation, which at the same time is a crucial advantage over the path integral itself as the latter does not lend itself to equally powerful approximation techniques in a comparable way. The general strategy is to truncate the effective action $\Gamma_k$ to only contain operators of a certain specified class, i.e. to project the full theory space onto the subspace spanned by this set of operators, and to evaluate the flow equation in this subspace. Note that this requires projecting the right hand side of the flow equation \eqref{equ:FRGE} onto this subspace even after a truncated effective action has been used in the calculation.

A very general truncation ansatz for gravity is given by
\begin{equation}
\label{truncation-effaction}
\Gamma_k[h,\bar g,\tau, \bar \tau] = \Gamma_k[g] + \hat \Gamma_k[h,\bar g] + S_{\mathrm{gf}}[h, \bar g] + S_{\mathrm{gh}}[h,\bar g, \tau, \bar\tau],
\end{equation}
where $\gmn=\bgmn +\hmn$ is now the total classical metric analogous to its quantum version in \eqref{background-split}, and the classical, $t$-independent gauge fixing and ghost actions have been separated out. We have also defined $\Gamma_k[g] := \Gamma_k[0,g,0,0]$ by substituting the total field for the background field in the effective action and replacing all its other arguments by zero (note that this does not imply that $\hmn$ in the argument $\gmn$ is also set to zero). We have further allowed for a remainder term $\hat \Gamma_k[h,\bar g]$ that depends on the fluctuation field and the background field separately but further truncate by assuming it does not contain any additional dependence on the ghost fields. Therefore, this ansatz neglects any RG evolution in the gauge-fixing or ghost sectors. By its definition, we can interpret $\hat \Gamma_k[h,\bar g]$ as capturing the difference between the total metric being equal to the background metric $g=\bar g$, modulo ghost contributions.

The truncation \eqref{truncation-effaction} is very general with a particular difficulty residing in the fact that $\hat\Gamma_k$ depends on both fields. Further truncating by setting $\hat \Gamma_k =0$ leads to the still very large class of so-called single metric truncations. The result is an ansatz for the effective action in which one has to effectively deal with the functional $\Gamma_k[g]$ only. The majority of studies in asymptotic safety so far have made use of the single field approximation and it has also been employed for the three $f(R)$ flow equations that are the topic of chapter \ref{sec:f-of-R}. After setting $\hat \Gamma_k=0$ in \eqref{truncation-effaction} the implementation of the single field approximation proceeds by evaluating the flow equation \eqref{equ:FRGE} for the remaining three terms in \eqref{truncation-effaction} (which generally involves projecting onto the subspace of the chosen truncation for $\Gamma_k[g]$) and finally setting the fluctuation field $h=0$, or equally $g=\bar g$. The last step is necessary to avoid a parametric dependence of the flow equation on the background field which would make the search for solutions much more challenging. On the other hand, no longer distinguishing between the total metric and the background metric leads to this parametric dependence on the background field to become additional dependence on the total field. Such terms originate from the gauge fixing and the ghost term in \eqref{truncation-effaction} but importantly also from the cutoff which, as discussed above, brings background field dependence in through the corresponding Laplacian, $\cutoff=\cutoff\!\!\left(-\bar \nabla^2\right)$. Since the flow equation takes the form of a differential equation these additional terms are now on equal footing with all $g$-dependent terms present beforehand and can lead to significant alterations. This problem is central to chapter \ref{ch:LPA}.

Another issue in this context is that since any background field dependence of the effective action becomes invisible in the single field approximation it becomes impossible to investigate background independence. Instead, this central requirement can only be analysed in bi-metric truncations of the effective action, where $\hat \Gamma_k$ in \eqref{truncation-effaction} is no longer neglected.

\section{Asymptotic safety}
\label{sec:asymptotic-safety}
The formalism described in the previous section opens the door to an investigation of non-perturbative renormalisation in quantum gravity. As alluded to previously, it is centred around the idea that non-perturbative dynamics may lead to a well defined ultraviolet completion of quantum gravity, despite perturbative non-renormalisability.

The first ingredient needed for this scenario to be viable is a non-Gaussian fixed point in the theory space of quantum gravity, i.e. a fixed point effective action $\Gamma_*[h,\bar g,\tau, \bar\tau]$ that can be used to remove the bare cutoff and thereby take the continuum limit. For gravity this would be a non-perturbative ultraviolet fixed point, reached in the limit $k\to \infty$. The second requirement for asymptotic safety is that any viable non-perturbative fixed point has to exhibit only a finite number of relevant eigenoperators in order to retain predictivity, as discussed at the end of sec. \ref{sec:non-pert-renorm}. Once these two properties are satisfied, it also has to be possible to find a renormalised trajectory emanating from the non-perturbative fixed point that reproduces the classical behaviour described by the Einstein-Hilbert action in the infrared. Reassuringly, all of these requirements can in principle be investigated using the flow equation for the effective action of the previous section.

\begin{figure}[h]
\centering
\includegraphics[scale=0.5]{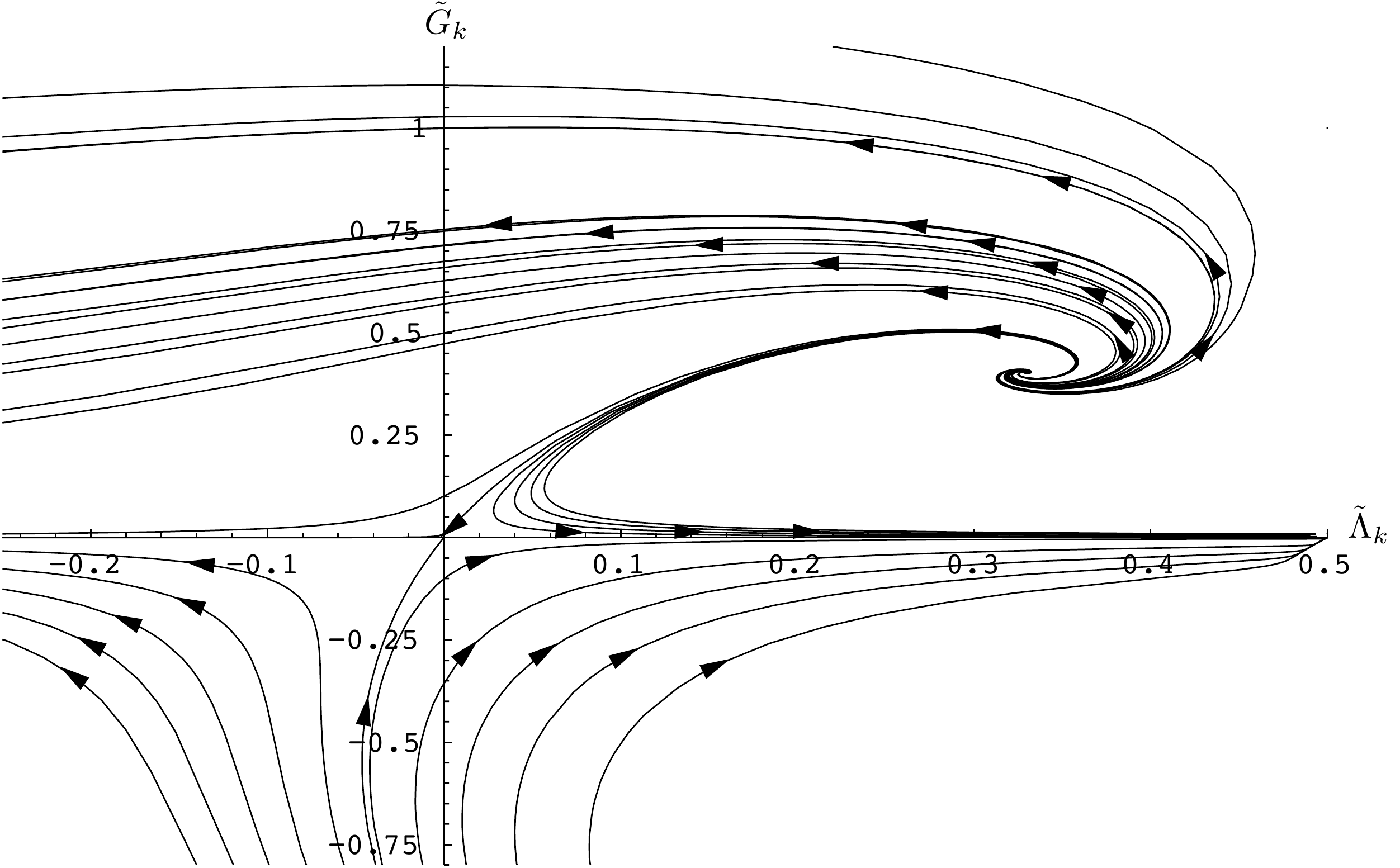}
\caption[Phase portrait of the Einstein-Hilbert truncation]{The RG flow of Newton's constant $\tilde G_k$ and the cosmological constant $\tilde \Lambda_k$ in the Einstein-Hilbert truncation \cite{Reuter:2001ag}.}
\label{fig:flow-EH}
\end{figure}

The concept of asymptotic safety was suggested in \cite{Weinberg:1980} and ever since the flow equation was first formulated in \cite{Reuter:1996} a large number of studies have contributed to growing confidence in its validity. For reviews and introductions we refer to \cite{Ashtekar:2014kba,Reuter:2012,Litim:2011cp,Niedermaier:2006wt,Percacci:2007sz,Nagy:2012ef}. An early example is given by the RG flow of the Einstein-Hilbert truncation. It is a single field truncation, i.e. $\hat \Gamma_k =0$ in \eqref{truncation-effaction}, and 
\begin{equation}\label{EH}
\Gamma_k[g] = \frac{1}{16\pi G_k}\int d^4x \,\sqrt{g}\left(-R+2\Lambda_k\right).
\end{equation}
The resulting phase diagram of the dimensionless Newton's constant $\tilde G_k = k^2 G_k$ and cosmological constant $\tilde \Lambda_k = \Lambda_k/k^2$ as calculated in \cite{Reuter:2001ag} is shown in fig. \ref{fig:flow-EH}.
It displays the Gaussian fixed point at the origin and a non-perturbative fixed point in the quadrant of positive $\tilde G_k$ and $\tilde \Lambda_k$, as well as RG trajectories emanating from the non-perturbative fixed point that approach the Gaussian fixed point arbitrarily closely in the infrared. Linearising around the non-perturbative fixed point, one finds a complex conjugate pair of relevant eigen-directions. After the original study \cite{Reuter:1996} the Einstein-Hilbert truncation has been re-considered in various different contexts in studies including \cite{Lauscher:2001ya,Souma:1999at,Litim:2003vp,Fischer:2006fz,Codello:2008,Benedetti:2010nr,Harst:2012ni} and most recently \cite{Falls:2014zba}.

\begin{figure}[ht]
\centering
\includegraphics[scale=0.6]{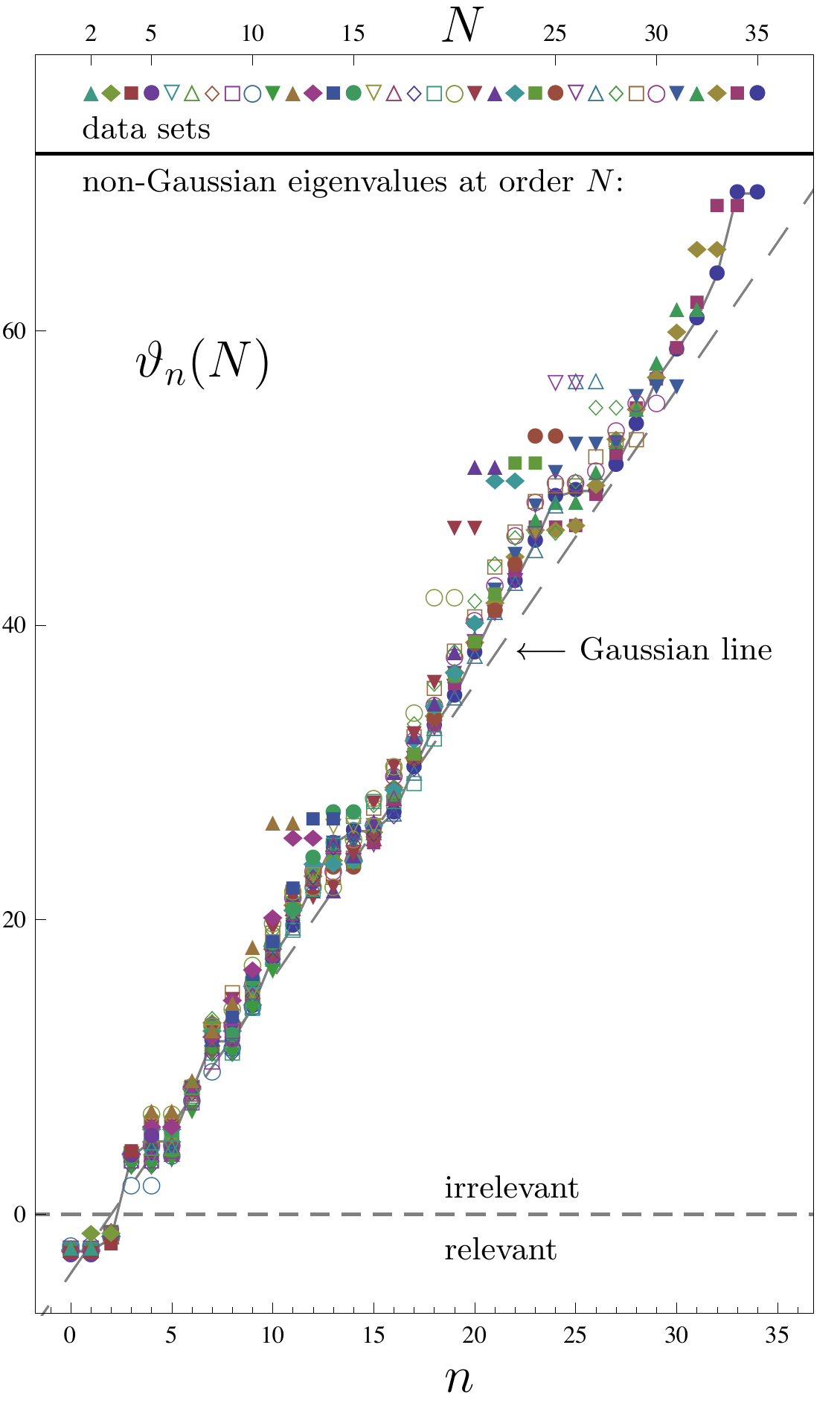}
\caption[Eigenvalues for highest polynomial truncation in $R$ to date]{Eigenvalues calculated for polynomial truncations of $\Gamma_k[g]$ in $R$ at the non-perturbative fixed point \cite{Falls:2013}. With $N$ referring to the truncation order in $R$, the eigenvalues are counted by $n$ and displayed by magnitude with a different symbol for each $N$ as indicated in the top panel.}
\label{fig:poly-trunc}
\end{figure}

The ansatz \eqref{EH} has been extended to include higher curvature terms to test if the non-perturbative fixed point found in the Einstein-Hilbert truncation persists and to find an upper bound on the number of relevant eigenoperators. After taking into account an $R^2$-term in \cite{Lauscher:2002sq}, additional curvature-squared terms have been investigated in \cite{Codello:2006in,Benedetti:2009rx}, while higher polynomial truncations in the Ricci scalar $R$ have been studied in \cite{Codello:2008,Machado:2007} and have been pushed to very high order in \cite{Falls:2013,Falls:2014tra}, see fig. \ref{fig:poly-trunc}.
Collectively these works show that the existence of a non-perturbative fixed point seems to be robust under inclusion of additional operators in the effective action and the dimensionality of the critical surface seems to stabilise at three.

Moreover, promising results have also been obtained in conformally reduced gravity in $d=3$ and $d=4$ dimensions, see \cite{Reuter:2008qx,Reuter:2008wj,Machado:2009ph,Demmel:2014sga,Demmel:2013myx,Demmel:2012ub}, and under the inclusion of different types of matter in different truncations for the gravitational part of the effective action, see e.g. \cite{Dou:1997fg,Benedetti:2009gn, Dona:2013qba, Dona2014}, although with indications that asymptotic safety may be lost for many matter configurations that go significantly beyond the Standard Model \cite{Dona:2013qba, Dona2014}.

With the exception of the last two, all studies mentioned so far have used the single field approximation with $\hat \Gamma_k =0$ in \eqref{truncation-effaction}. More recently, asymptotic safety has also been tested in several studies that go beyond the single field approximation in different ways, such as \cite{Becker:2014qya,Manrique:2010am,Manrique:2009uh,Manrique:2010mq,Codello:2013fpa,Christiansen:2014raa,Groh:2010ta,Eichhorn:2010tb,Becker:2014jua}, also presenting evidence for the presence of a non-perturbative fixed point with indications from some works that the number of relevant eigen-perturbations may increase to four.

It should be mentioned that the references to literature in this section are illustrative and by no means comprehensive. A good resource with a continuously maintained and annotated list of literature on asymptotic safety and other topics related to it can be found under \cite{Percacci-lit}.

\section{Thesis outline and scope}
We will begin our investigations surrounding asymptotic safety in chapter \ref{sec:f-of-R} with a comprehensive analysis of three previously derived flow equations of the so-called $f(R)$ approximation. This approximation is given by 
\begin{equation}\label{ansatz-fofR}
\Gamma_k[g] = \int d^4x\sqrt{g}\, f_k(R)
\end{equation}
and $\hat \Gamma_k =0$ in \eqref{truncation-effaction}. Hence, it is a single field approximation since $R=R(g)$ is the Ricci scalar of the total metric, but it goes beyond previous truncations in asymtptotic safety as $f_k$ is an arbitrary function of the Ricci scalar which will not be truncated to a polynomial. In this sense it is similar to the local potential approximation (LPA) in scalar field theory to which we will compare in several places later on. Three versions of the infinite dimensional flow for the function $f_k$ have been derived in \cite{Benedetti:2012dx,Codello:2008,Machado:2007}. The purpose of chapter \ref{sec:f-of-R} is to subject these equations to an analysis with respect to the existence of global fixed point solutions as well as their eigenoperator solutions, if applicable.

The main result will be that the two fixed point equations of refs. \cite{Codello:2008,Machado:2007} do not admit global fixed point solutions due to conditions imposed by a number fixed singularities inherent in the equations. Due to a different approach for deriving the $f(R)$ flow equation, this is not the case for ref. \cite{Benedetti:2012dx} for which several continuous sets of global fixed point solutions are found. An analysis of their eigenspectra further reveals that each of them supports a continuous set of relevant eigenoperators.

Of course, a continuous set of relevant eigenvalues would represent an untenable situation for asymptotic safety. Given the large amount of favourable evidence for asymptotic safety obtained with truncations of the effective action to a finite number of operators as discussed in the previous section, a close look at the reasons behind obtaining either an empty space of solutions or such a large set of solutions is in order. This will be the subject of chapters \ref{sec:red-ops} and \ref{ch:LPA}. The goal in chapter \ref{sec:red-ops} will be to show that all eigenoperators for the flow equation that leads to continuous sets of fixed point and eigenoperator solutions are actually redundant operators, i.e. eigenoperators that are generated by redefinitions of the metric field and are thus unphysical. This happens because the equations of motion of the $f(R)$ truncation at a fixed point, on spaces of constant Ricci-curvature
\begin{equation*}
E_*(R) = 2 f_*(R) - Rf_*'(R)
\end{equation*}
do not have a solution for any admissible $R$. The physical space of eigenoperators, obtained by factoring out any redundant operators, therefore turns out to be empty. We will refer to this as the breakdown of the $f(R)$ approximation.

Such a drastic result points towards the possibility that infinite dimensional truncations such as the $f(R)$ approximation may require particular care. This is further supported by the fact that, as mentioned above, the other two fixed point equations for $f_*(R)$ behave completely differently in that they do not admit global solutions in the first place, instead of continuous sets of solutions, albeit unphysical. The task has to be to investigate to what extent the approximations that are made in the $f(R)$ truncation have an effect on the observed difficulties. Again judging from the success of finite dimensional truncations, the expectation is that it may be possible to attribute the breakdown of the $f(R)$ truncation to one or several of these approximations.

In particular, the flow equations of the $f(R)$ truncation have been derived in the single field approximation by setting $\hat \Gamma_k =0$ in \eqref{truncation-effaction}. This approximation allows one to identify the total metric with the background metric once the Hessian on the right hand side of the flow equation \eqref{equ:FRGE} has been evaluated, as discussed at the end of sec. \ref{sub:adaptations-gravity}. As a consequence, all dependence of the effective action on the background field is converted into a dependence on the total field. In order to understand the possible implications of this procedure, we perform the same single field approximation in a background field formulation of single component scalar field theory in chapter \ref{ch:LPA}. Unlike in gravity, non-perturbative renormalisation has been studied extensively for scalar fields and we are therefore able to compare the effects of the single field approximation to established results. The conclusion of chapter \ref{ch:LPA} will be that even at the level of the simplest infinite dimensional truncation of scalar field theory, the so-called local potential approximation, adopting the single field approximation can lead to drastically different and unphysical behaviour.

The second part of chapter \ref{ch:LPA} is then devoted to a way of avoiding the problems of the single field approximation by performing the appropriate calculations in the corresponding bi-field truncation of the effective action, where now $\hat \Gamma_k \neq 0$ in the analogue of \eqref{truncation-effaction} for scalar fields. These calculations become possible by supplementing the flow equation \eqref{equ:FRGE} by a second, similar identity that expresses the arbitrariness of splitting the total field into a background field and a fluctuation field. This split Ward identity is what allows us to recover the correct physical results of single component scalar field theory in three dimensions in the local potential approximation, in the presence of a background field.

The following chapter \ref{sec:conformal} is then aimed at the first step of implementing the same ideas that proved successful in scalar field theory in quantum gravity. Instead of generalising the $f(R)$ truncation to an $f(R,\bar R)$ truncation, where $\bar R$ is the Ricci scalar of the background metric, which would require to overcome so far unresolved issues in evaluating the flow equation, this will be carried out in the context of conformal gravity. In doing so, we will be careful to adapt the flow equation \eqref{equ:FRGE} and the split Ward identity to reflect the fact that truncating to the conformal degree of freedom is a truncation of the full quantum gravitational theory space, in spite of the conformal field being a scalar field. Since the single field approximation is no longer used, the effective action now depends on the conformal fluctuation field and the conformal background field. The analysis of chapter \ref{sec:conformal} goes beyond the bi-field LPA by not just keeping a generic potential depending on both fields but also including a general kinetic coefficient function at second order in the derivative expansion for the fluctuation field. The outcome will be that the flow equation can indeed be combined with the split Ward identity for the conformal factor, leading to a simplified reduced renormalisation group flow in which all explicit dependence on the background field has been eliminated.

Finally, in chapter \ref{sec:conclusions} we discuss the results of the individual chapters in light of the outcomes of the whole thesis, with a view on possible lessons for, and future developments in, asymptotic safety.

%% file: f-of-R/f-of-R.tex
\chapter{Asymptotic safety in the $f(R)$ approximation}
\label{sec:f-of-R}

There is no doubt that it would be desirable to confirm the asymptotic safety scenario in an infinite dimensional truncation of the effective action retaining all metric degrees of freedom. The motivation for this is laid out in more detail in sec. \ref{sec:beyondpoly} of this chapter. We then recapitulate the main aspects of the derivation of the $f(R)$ truncation in sec. \ref{sub:flows-f(R)} to prepare the ground for the subsequent qualitative and quantitative analysis of the resulting flow equations. As a corner stone in the qualitative treatment of fixed point and eigenoperator equations we describe and employ the parameter counting method in sec. \ref{sec:fpanaquali} and \ref{sec:eopsparcount}. It is a methodology that has previously proved very successful in the context of functional renormalisation, notably for scalar field theory, see e.g. \cite{Morris:1998}. For the case of ref. \cite{Benedetti:2012dx}, where parameter counting at finite $R$ does not exclude the existence of global fixed point solutions, a detailed analysis of the asymptotic behaviour of both the fixed point solutions themselves, sec. \ref{sec:asy-fp}, and the associated eigenoperator solutions, sec. \ref{sec:eopsasy}, has to be carried out. Once we have thus obtained the complete results of parameter counting on the structure of solution space, a detailed numerical analysis of the fixed point equation of \cite{Benedetti:2012dx} is shown to confirm these predictions in sec. \ref{numerics}. Finally, a brief look at the asymptotic properties of finite eigen-perturbations is contained in sec. \ref{sec:finiteperts}.

\section{The need to go beyond polynomial truncations}
\label{sec:beyondpoly}
As mentioned in sec. \ref{sec:asymptotic-safety}, the vast majority of truncations studied thus far in asymptotic safety are given by polynomial truncations of the effective action. Such truncations retain a finite number of invariants of one or several types up to a certain maximum polynomial degree. A prominent example is given by taking $f_k$ in \eqref{ansatz-fofR} to be a polynomial in $R$ of some degree $N$, for others see \cite{Reuter:2012}.

The computational advantage of polynomial truncations is that the fixed point equations reduce to a finite number of algebraic relations for the couplings of the operators retained in the effective action and the full flow equation is an ordinary differential equation in RG time $t$. By contrast, as soon as general functions of field invariants are kept in the effective action, the fixed point equation itself becomes an ordinary differential equation, while the flow equation belongs to the difficult arena of partial differential equations.

It is well known however, that polynomial truncations can give rise to spurious fixed point solutions. In single component scalar field theory in the LPA it has been shown that unphysical fixed point solutions persist to high degrees in a polynomial truncation of the potential \cite{Morris:1994ki}, while the correct results are reproduced when a general potential is retained in the effective action.

A second example is provided by the RG study \cite{Morris:1995he} of a $U(1)$ vector field in $d=3$ dimensions, where the effective action was truncated to a general function $f\!\left(F_{\mu\nu}^2\right)$. Here too, one finds spurious non-Gaussian fixed points when the function $f$ is restricted to polynomial form, but concludes that no viable, non-trivial fixed points exist from an analysis of the fixed point equation for the full function $f$.

The general strategy adopted to discard spurious singularities in polynomial truncations is to retain only those solutions at each degree of truncation that display sufficiently fast convergence as the polynomial degree is increased. It was argued in \cite{Litim:2002cf} that this leads to satisfactory results in scalar field theory provided one uses a particular type of cutoff operator. The same approach has been used in gravity, where fixed points have been identified according to their convergence properties, e.g. \cite{Codello:2008,Falls:2013,Falls:2014tra}.

Despite the success of polynomial truncations in gravity, the above examples illustrate that polynomial truncations are not without pitfalls. Independently and as alluded to before, it would certainly be a crucial step for asymptotic safety to confirm the observed evidence for a non-perturbative fixed point from polynomial truncations in an infinite dimensional truncation such as the $f(R)$ truncation.

One can also argue that apart from computational issues there is a compelling qualitative reason for going beyond finite dimensional truncations. By construction, they only explore the small curvature properties of quantum gravity and are insensitive to effects that may appear at curvatures $R \gtrsim \mathcal{O}(1)$. For example, the record polynomial truncation studies \cite{Falls:2013,Falls:2014tra} are based on the $f(R)$ fixed point equation of \cite{Codello:2008} that, as we will see below, does not admit a global solution valid for $R$ arbitrarily large. In fact, there is a fixed singularity already at $R = R_c =2.0065$ that any viable solution would have to cross. Truncating $f(R)$ to a polynomial supplies the first terms of a series expansion of $f(R)$ around $R=0$ whose radius of convergence for the fixed point solution of \cite{Falls:2013} was estimated at $R_\mathrm{max} \approx 0.8$ in \cite{Falls:2013bma}. Thus, even if this expansion were taken to infinite order, the fixed singularity at $R=R_c$ would still be invisible and indeed, since there are no global solutions to the underlying fixed point equation, this Taylor series is only a partial solution valid within its radius of convergence and any attempt of extending it will eventually end in singular behaviour. 

A further explicit example of differences that become visible only for large curvatures is provided by the study \cite{Demmel:2014sga} of conformal gravity in $d=3$ dimensions, where particular modes on the sphere are shown to contribute to the RG flow that are not present in the local heat kernel expansion. This can be seen as a topological effect and is invisible in polynomial truncations.

A related aspect of the non-perturbative renormalisation group approach for gravity that may require investigations considering large curvatures is background independence. In order to verify background independence it may not be enough to consider only small curvatures as the Minkowski space limit $R\to 0$ could instead preclude any effects the background configuration might have on the RG flow.

We note here that there are several studies that have indeed been carried out on infinite dimensional subspaces of theory space. Amongst them are those that are the subject of the analysis following this section, \cite{Benedetti:2012dx,Codello:2008,Machado:2007}, as well as the rather general treatment \cite{Benedetti:2013jk} of the $f(R)$ truncation and the conformal truncation studies \cite{Demmel:2012ub,Demmel:2013myx,Demmel:2014sga,Reuter:2008qx}.

\section{Flow equations of the $f(R)$ truncation}
\label{sub:flows-f(R)}
Building on the approach outlined in sec. \ref{sub:adaptations-gravity}, the goal of the present section is to review additional aspects and assumptions that go into the derivation of the $f(R)$ flow equations of \cite{Benedetti:2012dx,Codello:2008,Machado:2007}. Since our analysis in later sections will be carried out in four dimensions, we set $d=4$ in the following.

After the background field split \eqref{background-split}, the fluctuation field (both classical and quantum) is decomposed according to a transverse traceless decomposition \cite{York:1973ia,Dou:1997fg,Lauscher:2001ya}:
\be \label{equ:TT}
h_{\mu\nu} = h_{\mu\nu}^{T} + {\bar\nabla}_\mu \xi_\nu + {\bar\nabla}_\nu \xi_\mu + {\bar\nabla}_\mu {\bar\nabla}_\nu \sigma + \frac{1}{4} { \bar g}_{\mu\nu} {\bar h}\,.
\ee
The advantage of switching to this new set of fluctuation fields is that it helps to diagonalise the Hessian of \eqref{equ:FRGE} in field space. Here, $\bar \nabla_\mu$ is the covariant derivative associated to the background metric $\bar g_{\mu\nu}$
and the component fields satisfy
\be \label{TT-const}
h_{\mu}^{T\,\mu} = 0 \, , \quad {\bar\nabla}^\mu h_{\mu\nu}^{T} = 0
\, , \quad {\bar\nabla}^\mu \xi_\mu = 0 \, , \quad {\bar h} = h -{\bar\nabla}^2 \sigma \, ,
\ee
where $h=h^\mu_{\phantom{\mu}\mu}$ is the trace of the fluctuation field and indices are lowered and raised with the background metric $\bar g_{\mu \nu}$. The fields $\xi_\mu$ and $\sigma$ in this decomposition are gauge degrees of freedom
for the gauge choice
\be \label{gauge-cond}
{\bar\nabla}^\mu h_{\mu\nu} ={1\over 4}{\bar\nabla}_\nu h\,,
\ee
whereas the transverse traceless component $h_{\mu\nu}^{T}$ and the scalar field $h$ are physical. The ensuing ghost fields are also decomposed into their transverse and longitudinal components. The transverse traceless decomposition and the decomposition of ghost fields both lead to functional Jacobians that are accounted for by introducing a number of auxiliary fields with exponentiated kinetic terms in the path integral. Correspondingly, the cutoff action is enlarged by adding cutoff terms to suppress the low momentum modes of these auxiliary fields.

The ansatz \eqref{truncation-effaction} for the effective action in the $f(R)$ truncation is slightly modified to
\begin{equation}
\label{truncation-ansatz-f(R)}
\Gamma_k[h,\bar g,\xi, \bar \xi] = \Gamma_k[g] + S_{\mathrm{gf}}[h, \bar g] + S_{\mathrm{gh}}[h,\bar g, \tau, \bar\tau] +S_\mathrm{aux},
\end{equation}
where the fluctuation field $\hmn$ and the ghost fields have to be expressed in terms of their component fields. Apart from the classical gauge-fixing and ghost actions the authors of \cite{Benedetti:2012dx,Codello:2008,Machado:2007} also extract the classical action of the auxiliary fields. The approximations associated with this ansatz are to neglect the running of the gauge-fixing, ghost and auxiliary sectors, adopting the single field approximation by setting $\hat \Gamma_k =0$ in \eqref{truncation-effaction} and finally truncating $\Gamma_k[g]$ to take the form \eqref{ansatz-fofR}.

Furthermore, important simplifications for evaluating the trace in \eqref{equ:FRGE} are achieved by fixing the background metric $\bgmn$ to that of a four sphere, parametrised by the Ricci scalar $R$. The final common feature of all three studies is the use of the optimised profile function \cite{Litim:2001}
\begin{equation}\label{optimised-profile}
r(z) = (1-z)\,\theta(1-z)
\end{equation} 
in the cutoff operator $\cutoff$ as this leads to further simplifications that we will discuss below.

We first focus on the flow equation of \cite{Machado:2007}. For each of the metric fluctuation fields $h_{\mu\nu}^{T}, \xi^\mu, \sigma, h$, the components of the ghost fields and the auxiliary fields, a cutoff term as in \eqref{equ:cutoffaction} is implemented such that the Laplacian $\Delta = -\bar \nabla^2$ in the two point function of the flow equation \eqref{equ:FRGE} is modified according to the replacement rule
\be
\label{typeI}
\Delta \mapsto \Delta + k^2 r(\Delta/k^2)\,.
\ee
This implements the suppression of covariant background momentum modes below $k$. The simplification achieved by \eqref{optimised-profile} is that it causes $\partial_t \cutoff$ to vanish for eigenmodes of $\Delta$ with eigenvalues above $k^2$ and thereby restricts the trace to a sum over eigenvalues below $k^2$, in which region $\Delta$ is simply replaced by $k^2$.

The evaluation of the traces is then performed with an asymptotic heat kernel expansion based on the background field Laplacian $\Delta$. In the process a number of modes for several fields are excluded from the trace as a consequence of the change of variables to component fields. Based on \eqref{equ:TT}, Killing vectors $\xi_\mu$, constant scalars $\sigma$ as well as scalars satisfying
\begin{equation} \label{conf-Killing}
\bar\nabla_\mu \bar\nabla_\nu \sigma= \frac{1}{4}\bgmn \bar \nabla^2 \sigma
\end{equation}
do not contribute to $\hmn$. Excluding the modes $\sigma$ of this last equation is justified by regarding $h$ as the physical scalar mode instead of $\bar h$. Similar exclusions are made for the ghost and auxiliary fields.

As discussed in sec. \ref{sec:non-pert-renorm}, the final step is passing to dimensionless variables. If we change the notation of the variables used so far by adding a tilde, the dimensionless variables are given by
\be \label{scaled-vars}
R =  \frac{\tilde R}{k^2} \qquad  \text{and} \qquad f_k(R) = \frac{\tilde f_k(\tilde R)}{k^4}.
\ee
From here on, dimensionless variables will be used. The flow equation of ref. \cite{Machado:2007} then becomes
\bea\label{flow-Machado} 
&& 384 \pi^2   \left( \partial_t f + 4 f - 2 R f^{\prime} \right) = 
\\ \nonumber
&& \quad \Big[ 5 R^2 \theta\!\left(1-\tfrac{R}{3}\right) 
-  \left( 12 + 4 \, R - \tfrac{61}{90} \, R^2 \right)\Big]
\Big[1 - \tfrac{R}{3} \Big]^{-1}  
+ \Sigma \\ \nonumber
&& + \Big[ 10 \, R^2 \, \theta\!\left(1-\tfrac{R}{4}\right) - R^2 \, \theta\!\left(1+\tfrac{R}{4}\right) 
-   \left( 36 + 6 \, R - \tfrac{67}{60} \, R^2 \right) \Big] 
\Big[ 1 - \tfrac{R}{4}\Big]^{-1} \\ \nonumber
&& +  \Big[ (\partial_t f' +2f'-2Rf'') \, \left( 10 - 5  R - \tfrac{271}{36}  R^2 + \tfrac{7249}{4536}  R^3 \right) 
 \Big. \\ \notag
&& \quad \Big. + f'\left( 60 - 20  R - \tfrac{271}{18}  R^2 \right)
\Big] \left[ f+f'\left(1  - \tfrac{R}{3}\right) \right]^{-1}
 + \tfrac{5R^2}{2} \, \Big[ 
 (\partial_t f' +2f'-2Rf'')  \Big. \\ \notag  
&& \Big. \quad  \left\{r\!\left(-\tfrac{R}{3}\right)+ \,2r\!\left(-\tfrac{R}{6}\right) \right\} 
 + 2 f'\theta\!\left(1+\tfrac{R}{3}\right) + 4 f'\theta\!\left(1+\tfrac{R}{6}\right) 
\Big]  
 \left[ f+f'\left(1  - \tfrac{R}{3}\right) \right]^{-1} \\ \nonumber
&& + 
\Big[
(\partial_t f' +2f'-2Rf'')
\left(6 + 3 R + \tfrac{29}{60} R^2 + \tfrac{37}{1512} \, R^3 \right)\\ \nonumber
&& \quad+ \left( \partial_t f^{\prime \prime}  - 2 R f^{\prime \prime \prime} \right) 
\left( 27 - \tfrac{91}{20} R^2  - \tfrac{29}{30} \, R^3 - \tfrac{181}{3360} \, R^4 \right) 
+ f^{\prime \prime} \left( 216 - \tfrac{91}{5}  R^2 - \tfrac{29}{15}  R^3 \right)
 \\ \nonumber
&&  \quad   
+  f^\prime \left( 36 + 12  R + \tfrac{29}{30}  R^2 \right)
\Big] \Big[ 2 f + 3 f^\prime \left(1-\tfrac{2}{3}R\right) + 9 f^{\prime \prime} \left(1-\tfrac{R}{3}\right)^2 \Big]^{-1} \, ,
\eea
where
\be
\label{Sigma1}
\Sigma = 10 \, R^2 \, \theta\left(1 -\tfrac{R}{3}\right)\,.
\ee
In order to lighten the notation somewhat, we have suppressed the dependence of $f_k$ on the RG scale. Primes denote differentiation with respect to $R$, and $r$ refers to the profile function \eqref{optimised-profile}. The first two lines originate from ghost and auxiliary field contributions as well as the gauge degrees of freedom $\xi_\mu$ and $\sigma$ of the metric sector. The next two ratios stem from the transverse traceless metric component $\hmn^T$ and the last ratio in turn belongs to the physical scalar $h$ in \eqref{equ:TT}.

We now proceed to the flow equation of \cite{Codello:2008} whose approach is in many ways very similar. The resulting flow equation can be obtained from \eqref{flow-Machado} with the following small modifications. Since the authors are interested in polynomial truncations for $f$ from the outset, all $\theta$ functions are set to one. Their gauge choices are two different limits in which \eqref{gauge-cond} is realised, each of them leading to a different treatment of modes excluded for \eqref{flow-Machado}. They also consider a third possibility for excluding modes that was previously used in \cite{Codello:2008}. The three different approaches only change $\Sigma$ in \eqref{flow-Machado} to one of the following choices,\footnote{Comparing the flow equations of \cite{Machado:2007} and \cite{Codello:2008} results in a difference given by the overall factor of $384 \pi^2$ and the coefficient of the $R^3(\partial_t f' +2f'-2Rf'')/\left[ f+f'\left(1  - \tfrac{R}{3}\right) \right]$ term in \eqref{flow-Machado}. However, they will not affect our results.}
\be
\label{Sigma2}
\Sigma = 0\,,\qquad - \frac{10 R^2 (R^2-20R+54)}{(R-3)(R-4)}\qquad{\rm or}\qquad   \frac{10(11R-36)}{(R-3)(R-4)}\,,
\ee
but the conclusions we will come to in sec. \eqref{sec:fpanaMachado} will actually be independent thereof.

We remark here that the use of the asymptotic heat kernel expansion to evaluate the traces of \eqref{equ:FRGE} simplifies from an infinite series to a finite sum when the cutoff profile \eqref{optimised-profile} is chosen. That this is actually true only once infinitely many terms proportional to $\delta(k^2)$ and its derivatives are neglected has been stated in \cite{Machado:2007}. It was noted in \cite{Benedetti:2012dx} that using heat kernel methods also introduces an implicit smoothing of the traces in the flow equation whose exact evaluation from summing eigenspectra leads to staircase functions.

This leads us naturally to the latest version of the $f(R)$ truncation \cite{Benedetti:2012dx} that was obtained with an approach differing in crucial places to the previous two. Firstly in the transverse traceless decomposition \eqref{equ:TT} with \eqref{TT-const} it is not $h$ that is regarded as the physical scalar but $\bar h$. As a consequence the $\sigma$-modes satisfying \eqref{conf-Killing} are not excluded from the trace.

The ghosts have been implemented according to \cite{Benedetti:2011ct} which has the advantage of leading to stronger simplifications between gauge and ghost sectors than for the previous two versions, at least in the particular limit of the gauge \eqref{gauge-cond} implemented in \cite{Benedetti:2012dx}.

A further important difference is the cutoff implementation. While the same profile function \eqref{optimised-profile} is used, it is not the replacement rule \eqref{typeI} that is realised. Instead it is noticed that the Hessian of $\Gamma_k$ in \eqref{equ:FRGE} for the ansatz \eqref{ansatz-fofR} depends on the particular combinations 
\begin{equation} \label{Deltas}
\Delta_0=\Delta - \frac{R}{3}, \qquad \Delta_1 = \Delta - \frac{R}{4}, \qquad \Delta_2 = \Delta-\frac{R}{6}
\end{equation}
for the scalar, vector and tensor contributions of \eqref{equ:TT}. For example, if the replacement \eqref{typeI} is used $\Delta_0$ becomes $1-R/3$ in scaled variables, which leads to the corresponding denominator in the first line of \eqref{flow-Machado}. In order to avoid such denominators vanishing at specific values of $R$, the replacement rule \eqref{typeI} is implemented for the individual operators \eqref{Deltas} instead. This will play a crucial role in our later analysis.

In contrast to \cite{Machado:2007, Codello:2008}, the traces of the flow equation in \cite{Benedetti:2012dx} are evaluated using spectral sums over the eigenspectra of the operators \eqref{Deltas}. Since spheres are compact spaces, these eigenspectra are discrete and thus give rise to a staircase behaviour of the evaluated traces. To obtain a smooth flow equation these discontinuities are eliminated by approximating the traces with their leading behaviours for $R\to 0$ and $R\to \infty$ added together. The flow equation then becomes:
\be \label{flow-Dario}
{384\pi^2} \left( \partial_t f + 4 f - 2 R f'    \right) = \cT_2 + \cT_1 +\cT^{\text{np}}_0+\cT^{{\bar h}}_0 \, ,
\ee
where the right hand side is split up into the contribution from $\hmn^T$,
\be
\label{T2}
\cT_2 = -\frac{20 \left(\partial_tf'-2 R f''+8
   f'\right)}{(R-2) f'-2
   f} \, ,    
\ee
from the vector modes,
\be
\label{t1}
\cT_1 = -36 \, ,
\ee
from the non-physical scalar modes,
\be
\label{tnp}
\cT^{\text{np}}_0 = - 12-5 R^2 \, ,
\ee
and finally from the physical scalar mode ${\bar h}$:
\begin{multline}
\label{Tbh}
\cT^{{\bar h}}_0 = \left[(R^4-54 R^2-54)( \partial_tf'' -2R  f''') 
						- (R^3+18 R^2+12)(\partial_tf' -2R f'' + 2f') \right. \\
				\left.  -36(R^2+2)(f'+6f'')\right] \left[ 2 (-9 f''+(R-3) f'-2f)\right]^{-1}\,.
\end{multline}
As before, $f$ is implicitly $t$-dependent and primes denote differentiation with respect to $R$.

An additional important aspect of this flow equation that will become important later on is the coefficient of the highest derivative with respect to the Ricci scalar $f'''$, appearing in \eqref{Tbh}. Its zeros will have a great influence on the structure of the space of fixed point solutions. To begin with, even though the flow equation \eqref{equ:FRGE} contains only the second functional derivative of the effective action, a third derivative of $f$ appears as a consequence of the replacement rule \eqref{typeI} for the operators \eqref{Deltas}. This leads to a cutoff operator $\cutoff$ that depends on $f''$ which gets differentiated once more when the $t$-derivative in \eqref{equ:FRGE} is taken in scaled variables \eqref{scaled-vars}. The same process also produces the factor of $R$ in front of $f'''$. Note that this last comment is not restricted to the flow \eqref{flow-Dario} but also applies to \eqref{flow-Machado} and therefore leads to a zero at $R=0$ of the coefficient of $f'''$ in both equations. The additional factor $R^4-54 R^2-54$ in \eqref{Tbh} has further zeros at
\be
\label{Rplus}
R_+=-R_-= \sqrt{27+3\sqrt{87}}\,.
\ee
As recognised in \cite{Benedetti:2012dx} they originate from the fact that the lowest mode of the scalar operator $\Delta_0$ in \eqref{Deltas} is $\lambda_0=-R/3$ and therefore negative on spheres. When the scalar trace is evaluated by a spectral sum, $f'''$  is multiplied by $\sum_{\lambda_n<1}(1-\lambda_n^2)$, where $\lambda_n = 1/12(n(n+3)-4)R$ are the scaled eigenvalues of $\Delta_0$. Due to the restriction $\lambda_n<1$ coming from the cutoff profile \eqref{optimised-profile} this sum is a finite sum for all $R\geq0$ and becomes shorter as $R$ is increased. Eventually all but the $n=0$ term drop out. This last term however remains and thus causes the above sum to vanish at some large enough $R$. After smoothing this sum is converted into the polynomial $R^4-54 R^2-54$ we find in \eqref{Tbh} and the zero is given by $R_+$ of \eqref{Rplus}.

\section{Qualitative fixed point analysis}
\label{sec:fpanaquali}
We now specialise the flow equations to fixed points by requiring $t$-independence through $\partial_t f =0$. Since we will exclusively focus on fixed points in this section we will again use $f(R)$ instead of $f_*(R)$ to refer to them. With this, the fixed point equation of ref. \cite{Machado:2007}, obtained from \eqref{flow-Machado}, reads
\bea\label{fp-Machado} 
&& 384 \pi^2   \left( 4 f - 2 R f^{\prime} \right) = 
\\ \nonumber
&& \quad \Big[ 5 R^2 \theta\!\left(1-\tfrac{R}{3}\right) 
-  \left( 12 + 4 \, R - \tfrac{61}{90} \, R^2 \right)\Big]
\Big[1 - \tfrac{R}{3} \Big]^{-1}  
+ \Sigma \\ \nonumber
&& + \Big[ 10 \, R^2 \, \theta\!\left(1-\tfrac{R}{4}\right) - R^2 \, \theta\!\left(1+\tfrac{R}{4}\right) 
-   \left( 36 + 6 \, R - \tfrac{67}{60} \, R^2 \right) \Big] 
\Big[ 1 - \tfrac{R}{4}\Big]^{-1} \\ \nonumber
&& +  \Big[ (2f'-2Rf'') \, \left( 10 - 5  R - \tfrac{271}{36}  R^2 + \tfrac{7249}{4536}  R^3 \right) 
 \Big. \\ \notag
&& \quad \Big. + f'\left( 60 - 20  R - \tfrac{271}{18}  R^2 \right)
\Big] \left[ f+f'\left(1  - \tfrac{R}{3}\right) \right]^{-1}
 + \tfrac{5R^2}{2} \, \Big[ 
 ( 2f'-2Rf'') \left\{r\!\left(-\tfrac{R}{3}\right) \right. \Big. \\ \notag  
&& \quad \left. + \,2r\!\left(-\tfrac{R}{6}\right) \right\} 
\Big. + 2 f'\theta\!\left(1+\tfrac{R}{3}\right) + 4 f'\theta\!\left(1+\tfrac{R}{6}\right) 
\Big]  
 \left[ f+f'\left(1  - \tfrac{R}{3}\right) \right]^{-1} \\ \nonumber
&& + 
\Big[
(2f'-2Rf'') 
\left(6 + 3 R + \tfrac{29}{60} R^2 + \tfrac{37}{1512} \, R^3 \right)\\ \nonumber
&& \quad - 2 R f^{\prime \prime \prime}
\left( 27 - \tfrac{91}{20} R^2  - \tfrac{29}{30} \, R^3 - \tfrac{181}{3360} \, R^4 \right) 
+ f^{\prime \prime} \left( 216 - \tfrac{91}{5}  R^2 - \tfrac{29}{15}  R^3 \right)
 \\ \nonumber
&&  \quad   
+  f^\prime \left( 36 + 12  R + \tfrac{29}{30}  R^2 \right)
\Big] \Big[ 2 f + 3 f^\prime \left(1-\tfrac{2}{3}R\right) + 9 f^{\prime \prime} \left(1-\tfrac{R}{3}\right)^2 \Big]^{-1} \, ,
\eea
where $\Sigma$ is given by \eqref{Sigma1}. The fixed point equation of \cite{Codello:2008} is the same up to $\Sigma$ which now takes one of the forms in \eqref{Sigma2}. Benedetti and Caravelli's \cite{Benedetti:2012dx} fixed point equation becomes
\be \label{fp-Dario}
{768\pi^2} \left( 2 f - R f'    \right) = {\tilde \cT}_2 + \cT_1 +\cT^{\text{np}}_0+{\tilde \cT}^{{\bar h}}_0 \, ,
\ee
where
\be
\label{t2}
{\tilde \cT}_2 = \frac{40 \left(R f''-4
   f'\right)}{(R-2) f'-2
   f} \, ,
\ee
and
\be
\label{th}
{\tilde \cT}^{{\bar h}}_0 = {R\left(R^4-54 R^2-54\right)  f''' - \left(R^3+18 R^2+12\right) \left(  R f'' -     f' \right) +18\left(R^2+2\right)\left(f'+6f''\right)\over  9 f''-(R-3) f'+2f}\,.
\ee
The terms $\cT_1$ and $\cT^{\text{np}}_0$ are given by \eqref{t1} and \eqref{tnp}, respectively. We are interested in finding global solutions to these fixed point equations. Since these equations have been derived on background spheres, global in this context refers to $0\leq R < \infty$. The value $R=0$ is included by requiring continuity of $f$ and all its derivatives in the limit $R \to 0$.

The fundamental property that any effective fixed point Lagrangian $f(R)$ has to satisfy is that it be smooth for all non-negative $R$. In particular it is required to be well defined for any finite $R\geq 0$ and is not allowed to develop singularities at any of these values. This is a physical requirement as it is unclear how an effective Lagrangian could be used or even interpreted if it diverged at some finite value of its argument.

This strategy has by now been applied in other studies in asymptotic safety such as the $f(R)$ truncation in $d=3$ based on conformal gravity \cite{Demmel:2014sga,Demmel:2013myx,Demmel:2012ub} and led to promising and motivating results.

It is clear however, that smoothness of $f$ can be destroyed by unfortunate choices of cutoff. In principle this is the case for the optimised cutoff profile \eqref{optimised-profile} that gives rise to the $\theta$-functions in \eqref{fp-Machado} and also in \eqref{fp-Dario} before the smoothing process discussed earlier. For \eqref{fp-Machado} there is no chance that $f'''$ is smooth at the points where the $\theta$-functions strike. Nevertheless, the optimised profile can be accommodated by requiring smoothness only on each interval between two step functions and allowing $f'''$ and all higher derivatives to change discontinuously at the endpoints. Even though smoothness is lost in this way, it is due to a technical aspect of the setup and does not represent a fundamental issue of the RG flow.

\subsection{The concept of parameter counting}
\label{par-counting}
The fundamental requirements on any fixed point solution discussed at the end of the last section can be exploited to gain a qualitative understanding of the space of global solutions to the equations \eqref{fp-Machado}, \eqref{fp-Dario}. For this, it is convenient to write the fixed point equations in normal form,
\be
\label{normal}
f'''(R)=F \!\left(f,f',f'',R\right)\,,
\ee
such that $F$ does not depend on the highest derivative $f'''$. The solution space to such an equation is then restricted by two different types of singularities.

Let us first assume that $F$ is well defined at some generic point $R_p$. The standard result of local solutions for ordinary differential equations then guarantees that there is some interval $\mathcal{D}=(R_p-\rho,R_p+\rho)$ around $R_p$ on which there exists a three parameter set of local solutions. These solutions can be taken to be parametrised by the first three coefficients of the Taylor expansion
\be
\label{Taylor}
f(R)=f(R_p)+f'(R_p)(R-R_p)+\frac{1}{2}f''(R_p)(R-R_p)^2+\frac{1}{6}f'''_\pm(R_p)(R-R_p)^3+\dots
\ee
around $R_p$. In case $R_p$ coincides with the location where one of the $\theta$-functions in \eqref{fp-Machado} changes discontinuously, we have allowed for differing values of the third derivative corresponding to the left and right of $R_p$. The reason why it may well happen that $\rho$ cannot be chosen arbitrarily large is that the right hand side $F$ in \eqref{normal} diverges as $R \to R_c$ for some $R_c>R_p$. If the value of $R_c$ depends on the initial conditions
\be
\label{generic-ICs}
f(R_p)\,,\quad f'(R_p)\quad{\rm and}\quad f''(R_p)\,,
\ee
this phenomenon is referred to as a moveable singularity. With the requirement of a well-defined fixed point solution, this would disqualify the initial conditions \eqref{generic-ICs} that lead to a singularity at $R=R_c$.

On the other hand, it is also possible that $F$ diverges at $R=R_c$ due to an algebraic pole. In this case $R_c$ is a fixed singularity as its location is independent of the initial conditions \eqref{generic-ICs} and we may rewrite
\begin{equation} \label{pole}
F \!\left(f,f',f'',R\right) = \frac{G \!\left(f,f',f'',R\right)}{R-R_c}.
\end{equation}
We assume for the purposes of the discussion here that $G$ no longer diverges at $R=R_c$, i.e. $R_c$ is a single pole of $F$. Since we require $f$ to be smooth everywhere, we can substitute an analogous Taylor expansion to \eqref{Taylor} around $R=R_c$ into \eqref{normal} with \eqref{pole}. The result will take the form
\begin{equation*}
\text{regular in }R = \frac{u\!\left(f(R_c),f'(R_c),f''(R_c)\right)}{R-R_c} + \text{regular in } R
\end{equation*}
leaving us with the constraint
\begin{equation}
\label{pole-const}
u\!\left(f(R_c),f'(R_c),f''(R_c)\right)=0
\end{equation}
on the parameter space at $R=R_c$. The effect of this condition on parameter space is to eliminate one of the three parameters in favour of the other two, thereby reducing its dimension by one.

If the right hand side $F$ in the normal form \eqref{normal} has several such simple poles, each will act as a fixed singularity. Unless there is some intrinsic symmetry or simplification hidden in $F$, different fixed singularities will result in independent constraints on solution space, each reducing its dimension. An investigation as to the presence of such structures in $F$ has to be carried out on a case by case basis to ensure parameter counting based on fixed singularities remains valid.

\subsection{Parameter counting for the fixed point equations of \cite{Machado:2007, Codello:2008}}
\label{sec:fpanaMachado}
We now apply the parameter counting strategy to the fixed point equation \eqref{fp-Machado}. The first two lines feature the two fixed singular points $R_c=3,4$ and it is straightforward to check that these linear factors are not cancelled by similar factors contained in the numerators on either side of the $\theta$-functions. As a result, the first line  diverges for both $R\nearrow 3$ and $R\searrow 3$ and the second line for the same two limits at $R=4$. It is also clear that these two single poles remain once the fixed point equation is cast into normal form \eqref{normal}. In normal form however, the right hand side $F$ of \eqref{normal} will contain the term
\begin{equation*}
R \left( 27 - \frac{91}{20} R^2  - \frac{29}{30} \, R^3 - \frac{181}{3360} \, R^4 \right)
\end{equation*}
in its denominator. This polynomial has two positive real roots given by $R=0, 2.0065$ that amount to single poles of $F$. All in all, the normal form of \eqref{fp-Machado} therefore suffers from the four fixed singularities $R=0,2.0065,3,4$ in the domain $R\geq 0$.

The constraints \eqref{pole-const} are readily found using Taylor expansions \eqref{Taylor} for each of these singular points in the normal form. They are
\begin{equation*}
\label{34}
f'(3)=\frac{2}{3}f(3)\qquad{\rm and}\qquad f''(4)=5f'(4)-2f(4)\,,
\end{equation*}
for $R=3,4$ and the constraint at the origin is:
\begin{equation*}
f''(0)=-\frac{2}{9}{\frac{192 \pi^{2}f(0)^{3}+ 6f(0)^{2}[1+ 80 \pi^2 f'(0)]
	+2 f(0)f'(0)[ 1+ 144 \pi^{2}f'(0)] -9 {f'(0)}^2}
	{192{\pi }^{2} f(0)^{2} +3{ f(0)}+192{\pi }^{2}{ f(0)}{ f'(0)}-7{ f'(0)}}}\,.
\end{equation*}
The fourth constraint at $R=2.0065$ is also non-linear and even longer. Taken together, they represent four conditions on the three-dimensional parameter space of the fixed point equation \eqref{flow-Machado}. It is certainly not obvious that there is a hidden structure in \eqref{flow-Machado} that would destroy independence of these conditions of each other. If there were it would certainly point to an underlying symmetry of the flow equations of quantum gravity that would have to survive all the approximations that have been made to obtain \eqref{flow-Machado}. Since this is an extremely unlikely scenario we can conclude that the above conditions over-constrain the parameter space of global solutions leading to no fixed point solutions of \eqref{flow-Machado} valid for all $R\geq 0$.

Since this conclusion is not affected by the different choices \eqref{Sigma2} for $\Sigma$ in \eqref{flow-Machado}, also the three different variants of the fixed point equation \eqref{flow-Machado} of ref. \cite{Codello:2008} do not possess any global solutions.

\subsection{Parameter counting for the fixed point equations of \cite{Benedetti:2012dx}}
\label{par-count-Benedetti}
Inspecting the pertaining fixed point equation \eqref{fp-Dario}, we see that the denominators of the terms on the right hand side can lead at most to moveable singularities. The fixed singularities relevant for the range $R\geq 0$ are confined to the coefficient of $f'''$ in \eqref{th} and are given by $R=0$ and $R_+$ defined in \eqref{Rplus}. Defining $a_n = f^{(n)}(0)$ the constraint \eqref{pole-const} at $R=0$ becomes
\be
\label{expra2}
a_2 = -{2\over9}\,{\frac {6{a_0}^{2}-9{a_1}^{2}+480{\pi }^{2}{{
 a_0}}^{2}a_1+2a_0a_1+192{\pi }^{2}{a_0}^{
3}+288{\pi }^{2}a_0{a_1}^{2}}{3a_0-7a_1+192{\pi }^{
2}{a_0}^{2}+192{\pi }^{2}a_0a_1}}\,.
\ee
A second constraint is obtained for the singular point $R=R_+$. It is an expression quadratic in $f''(R_+)$ with dependence on both $f'(R_+)$ and $f(R_+)$ that has already been reported in \cite{Benedetti:2012dx}.

As before, we can be confident that these two conditions act independently in restricting the parameter space of global solutions. Based on this we would expect possibly disconnected one-dimensional sets of solutions to the fixed point equation \eqref{fp-Dario}. In principle, additional constraints could arise from the behaviour of solutions as $R\to \infty$. As we will show in the next section, the asymptotic solution of \eqref{fp-Dario} takes the form
\begin{align}
\label{asymp}
f(R) &= A \,R^2 + R\left\{\frac{3}{2}A+B\cos\ln R^2 + C\sin\ln R^2\right\}\notag \\ \notag
 &-\frac {3}{68}\left\{ (103B-149C)+ 6912 A (11C+7B ) {\pi }^{2}\right\}\cos \ln R^2\\ \notag
 &-\frac {3}{68}\left\{ (149B+103C ) -6912 A (-7C+11B ) {\pi }^{2}\right\}\sin \ln R^2
  \\ \notag
 &-\frac{192\pi^2}{37}\left\{({B}^{2}+12BC-{C}^{2} )\cos \ln R^4 +2 (3{B}^{2}-BC-3{C}^{2})\sin \ln R^4\right\}
 \\ 
 &+{\frac {63}{4}}A-96 (9{A}^{2}+2{B}^{2}+2{C}^{2} ) {\pi }^{2}+O(1/R)\,,
\end{align}
where $A,B$ and $C$ are real parameters that are only constrained by the inequality
\begin{equation*}
\frac{121}{20}A^2 > B^2 + C^2\,.
\end{equation*}
Somewhat surprisingly perhaps, the asymptotic regime therefore does not restrict the dimensionality of solution space any further and we still expect lines of fixed point solutions. A careful and detailed numerical analysis in sec. \ref{numerics} will confirm this prediction of parameter counting for the fixed point solutions.

A continuous set of fixed point solutions is a result that, albeit unexpected, can in principle be accommodated in the asymptotic safety scenario since the fixed point solution realised in Nature may be determined by just a few experimental measurements. However, as we will see in sec. \ref{sec:eops}, the parameter counting associated to the eigenoperator equation inherits its main features from the parameter counting for the fixed point equation and thus leads to a continuous set of relevant eigen-perturbations. This corresponds to a situation worse than perturbative non-renormalisability and asymptotic safety would be no longer viable.

A possible way out of this dilemma is suggested by the parameter counting of this section itself. If we consider the fixed point equation \eqref{fp-Dario} on the whole real line $-\infty<R<\infty$, despite its derivation on four spheres, we pick up an additional fixed singularity at $R=R_-$ as given in \eqref{Rplus} whose associated constraint has to be imposed on parameter space. From the perspective of fixed singularities, the continuous lines of solutions are therefore reduced to a discrete set, and we will see in sec. \ref{sec:eopsparcount} that eigenspectra would now also be quantised, opening the door to the possibility of only finitely many relevant eigenoperators. The derivation of the asymptotic behaviour \eqref{asymp} in the next section is also valid for the limit $R\to -\infty$. Thus, since the fixed point equation \eqref{fp-Dario} is not symmetric under $R \mapsto -R$, such a truly global fixed point solution would have to obey \eqref{asymp} at $R\to \pm \infty$ with corresponding parameters $A_\pm,B_\pm,C_\pm$ on both sides, which in turn have to satisfy the cone condition \eqref{safedisc}. By continuing the asymptotic solution \eqref{asymp} into the complex plane, we see that the sub-leading and subsequent terms contain an obstruction to this process due the logarithms they contain, but the leading term becomes an entire function. Hence, the asymptotic parameters of this term for $R\to \pm \infty$ have to coincide for any fully global solution: $A_-=A_+=A$. This is a consequence of the asymptotic structure of the fixed point equation \eqref{fp-Dario} as we comment below \eqref{f0-equ} and does not represent a further constraint on parameter space.

Thus we still expect a discrete set of fixed point solutions as described above. Promising as this may seem, it is a picture that has to withstand the constraining powers of moveable singularities to survive. While there is indeed a chance of this happening, as is the case in scalar field theory \cite{Morris:1996xq}, only a numerical study of the fixed point equation can decide this point. Unfortunately, as is detailed in sec. \ref{numerics},  moveable singularities eliminate the solutions on $-\infty<R<\infty$ allowed in principle from parameter counting based on fixed singularities alone for the fixed point equation \eqref{fp-Dario}.

\section{Asymptotic analysis of the fixed point solutions}
\label{sec:asy-fp}
The aim of this section is to derive the asymptotic solution \eqref{asymp} to the fixed point equation \eqref{fp-Dario}. It is generally a restrictive requirement to insist for a solution of a non-linear ordinary differential equation to exist for arbitrarily large values of the independent variable. As discussed in sec. \ref{par-counting} this is the case since any such solution has to avoid moveable singularities intrinsic to the differential equation. The result \eqref{asymp} shows that in the present case this does not lead to a reduction of the dimension of parameter space but it is sufficient to allow the fixed point equation to be solved analytically in an asymptotic expansion.

We note that the derivation in this section of the asymptotic expansion \eqref{asymp} also holds in the limit $R\to -\infty$.

A fruitful strategy to determine the leading behaviour of the asymptotic solution is to neglect the right hand side of \eqref{fp-Dario} and to solve the left hand side separately. At a mathematical level, the motivation for this approach comes from the fact that the right hand side contains all sources for moveable singularities that may stand in the way for reaching $R\to \infty$. The result is $f(R)=AR^2$ for a free real parameter $A$. But even without this assumption, upon substitution of a power-law ansatz $f(R) \propto R^p$ into the fixed point equation \eqref{fp-Dario} one confirms that for $p>2$ the left hand side of \eqref{fp-Dario} has to be satisfied on its own, thus excluding this ansatz. This holds unless the denominator of \eqref{t2} or \eqref{th} vanishes to leading order. However, the leading order solutions of these denominators is again $\propto R^2$ which is incompatible with $p>2$. Moreover quadratic growth is the only behaviour that was observed in an extensive numerical investigation of the fixed point equation, cf. sec. \ref{numerics}.

For a systematic derivation of the sub-leading terms of the asymptotic expansion, we will make use of the following scaling techniques. We first introduce a small parameter $\eps$ according to 
\be \label{feps}
f_\eps(R) = \eps^2 f(R/\eps) \, ,
\ee
so that the limit $R\to \infty$ translates into $\eps \to 0$. This definition of $f_\eps$ is motivated by the leading behaviour $AR^2$ and would have to be adapted for other cases. We can rewrite the fixed-point equation \eqref{fp-Dario} in terms of $f_\eps(R)$ by making the substitutions 
\begin{equation} \label{change-R-Reps}
R \mapsto R/\eps \qquad \text{and} \qquad  f^{(n)}(R) \mapsto \eps^{n-2}f^{(n)}_\eps(R) \,.
\end{equation}
If the result is then expanded as a series in $\eps$ the term of lowest order in $\eps$ reproduces the left hand side of \eqref{fp-Dario}, thus confirming the leading term of the asymptotic expansion.

To access the sub-leading terms it will be convenient to write
\be \label{fepsexp}
f_\eps(R) = f_0(R) + \eps f_1(R) +\eps^2 f_2(R) + \dots \, ,
\ee
in order to divide $f_\eps$ into contributions $f_n$ order by order in $\eps$. We note that these individual contributions still depend on the scaling parameter $\eps$ but we require them to not vanish faster than $\eps$ or diverge faster than $1/\eps$ so that the separation into orders above is meaningful. We also require this dependence to be such that the limit $\eps\to 0$ can be taken at each order of the expansion, as is necessary to achieve the corresponding limit $R\to \infty$. We will see in the following that this is always possible. To achieve algebraic simplifications of later expressions, it is furthermore useful to define
\be
\label{defgn}
f_n(R)=R^{2-n}g_n(R)\,. 
\ee
Consequently, $g_n$ inherits an $\eps$-dependence from $f_n$ but it is invariant under the combined transformation $R\mapsto sR$ and $\eps \mapsto s\,\eps$ as follows from the definitions \eqref{feps} and \eqref{fepsexp} and the fact that $f$ is independent of $\eps$. In accordance with the ansatz \eqref{fepsexp} and the scaling $R \mapsto R/\eps$ in \eqref{feps}, a solution for $g_n(R)$ is admissible if it satisfies
\be \label{gnbounds}
\frac{1}{R} < g_n(R) < R
\ee
in the limit $R \to \infty$. After expressing the fixed point equation \eqref{fp-Dario} in terms of $f_\eps$ as described in \eqref{change-R-Reps} and substituting \eqref{fepsexp}, we expand the result in powers of $\eps$. At each order in $\eps$ this leads to an ordinary differential equation for the individual contributions in \eqref{fepsexp}.

If this process is carried out with a general $f_0(R)$ in \eqref{fepsexp}, the result at $\mathcal{O}\!\left(\eps^{-2}\right)$ is
\begin{equation} \label{f0-equ}
768\pi^2 \left( 2f_0-R f_0'\right) + 5R^2 -R^2 {R^3  f_0''' - R \left(  R f_0'' -     f_0' \right) \over  2f_0-R f_0'}=0 \, ,
\end{equation}
where the first term originates from the left hand side of \eqref{fp-Dario}, the middle term from \eqref{tnp} and the last term from \eqref{th}. Using 
\begin{equation} \label{f0}
f_0(R) = AR^2
\end{equation}
in this equation as the leading order of the asymptotic expansion leads to an ill-defined ratio $0/0$ for the third term. The significance of this is that the expansion in $\eps$ of the fixed point equation \eqref{fp-Dario} has to be performed directly around the explicit solution $f_0(R) = AR^2$, with a non-vanishing sub-leading contribution $f_1$ in \eqref{fepsexp}, in order to obtain a well-defined series in $\eps$.

We can nevertheless see that the leading asymptotic behaviour results from balancing the last two terms in \eqref{f0-equ} which represent the non-physical scalar contribution and the physical scalar contribution, respectively. At the same time this leading term is a solution of the left hand side of the fixed point equation and thus follows from classical scaling. This situation is markedly different from scalar field theory in the LPA, where the leading asymptotic solution solves the left hand side of the fixed point equation and the quantum corrections encoded in the right hand side are sub-leading, see e.g. \cite{Morris:1998}.

The leading order relation \eqref{f0-equ} is symmetric under $R\mapsto -R$ which implies that the fixed point equation \eqref{fp-Dario} enjoys the same symmetry to leading order as $R\to \infty$. This leads again to the result obtained at the end of the previous section that the two asymptotic coefficients $A_\pm$ for the two limits $R\to \pm\infty$ have to coincide.

Following this insight and expanding in $\eps$ with $f_0(R) = AR^2$ in \eqref{fepsexp} leads to contributions again arising at order $\mathcal{O}\!\left(\eps^{-2}\right)$ which in turn reflects the singular ratio in \eqref{f0-equ}. Expressing the result in terms of \eqref{defgn} leads to the differential equation
\be
\label{g1eq}
R^3 g'''_1+2R^2g''_1+4Rg'_1-4g_1=-6A \,.
\ee
One solution of this equation is $\propto R$ and with \eqref{defgn} just reproduces the leading term $f_0$. It is excluded by the condition \eqref{gnbounds}. The other two solutions give rise to
\be
\label{f1}
f_1(R) = R\left\{ \frac{3}{2} A + B  \cos\ln R^2 + C\sin\ln R^2 \right\}\,,
\ee
with two additional real parameters $B,C$. The aforementioned implicit dependence on $\eps$ is now contained in the parameters $B$ and $C$ and is such that the term in brackets is invariant under $R\mapsto sR$ and $\eps \mapsto s\,\eps$. 

We therefore recover the result announced with \eqref{asymp} that the asymptotic solutions depend on three parameters in total and therefore do not restrict the dimension of parameter space. This is in contrast to the asymptotic expansion
\be
\label{Aexp}
f(R) = A R^2 + \frac{3}{2} A R +{\frac {63}{4}}A-864{\pi }^{2}{A}^{2}
+O(1/R)\,,
\ee
obtained in \cite{Benedetti:2012dx} which displays the same leading behaviour but depends on only one parameter $A$. Since the fixed point equation \eqref{fp-Dario} is a third order ordinary differential equation one would however expect a three parameter set of solutions. In principle, some of these parameters may be eliminated by the requirement that an asymptotic solution has to be valid for all $R\to \infty$ via the restrictions imposed by moveable singularities. Our analysis shows that this is not the case for the present fixed point equation. We remark that once a first asymptotic solution such as \eqref{Aexp} has been found, a convenient way of searching for additional parameters is to perturb that solution according to $f(R) \mapsto f(R)+\delta\!f(R)$ and to expand the fixed point equation \eqref{fp-Dario} to linear order in $\delta\!f(R)$. If this is carried out with $f(R)$ given by the first two terms in \eqref{Aexp} for the present case one recovers the additional two solutions parametrised by $B$ and $C$ in \eqref{f1}.

The next order in the expansion \eqref{fepsexp} is accessed at $\mathcal{O}\!\left(\eps^{-1}\right)$, where now the left hand side of \eqref{fp-Dario} and the physical scalar sector \eqref{th} contribute. It is useful for this to change variables according to $|R|=\exp u$. By taking into account the modulus of $R$ we ensure that the derivation of the asymptotic series is also applicable to the range $R<0$, cf. the end of this section. The differential equation for $g_2(u)$ then becomes
\begin{multline}
\left\{ \partial^3_u -4\partial^2_u+9\partial_u -10 \right\} g_2(u)= 
{-{315}A/{2}}+960( 9{A}^{2}+2{B}^{2}+2{C}^{2} ) {\pi }^{2} \\
-3 ( 2C+31B ) \cos2u
+6912A ( B-2C ) {\pi }^{2}\cos2u 
 -384 ( B+3C )  ( 3B-C ) {\pi }^{2}\cos 4u  \\
+3 ( 2B-31C ) \sin 2u 
+6912A ( 2B+C ) {\pi }^{2}\sin 2u 
+768 ( 2B+C )( B-2C ) {\pi }^{2}\sin 4u\,.
 \end{multline}
The general solution of this equation, given by solving its left hand side, reproduces the three solutions of the previous orders given in \eqref{f0} and the two terms parametrised by $B,C$ in \eqref{f1}. They are ruled out by the constraint \eqref{gnbounds}. It is the special solution that leads to the next term in the asymptotic series, and expressed in terms of $R$ it reads
\begin{align}
\label{f2}
f_2(R) = g_2(R)&= {\frac {63}{4}}A-96 (9{A}^{2}+2{B} ^{2}+2{C}^{2} ) {\pi }^{2} \\
&-\frac {3}{68}\left\{ (103B-149C)+ 6912 A (11C+7B ) {\pi }^{2}\right\}\cos \ln R^2 \notag \\
&-\frac {3}{68}\left\{ (149B+103C ) -6912 A (-7C+11B ) {\pi }^{2}\right\}\sin \ln R^2 \notag \\
&-\frac{192\pi^2}{37}\left\{({B}^{2}+12BC-{C}^{2} )\cos \ln R^4 +2 (3{B}^{2}-BC-3{C}^{2})\sin \ln R^4\right\}  . \notag
\end{align}
This pattern is repeated at higher orders of the asymptotic series. The general solution always reproduces the known three leading solutions parametrised by $A,B,C$ and has to be discarded because of \eqref{gnbounds}, whereas the special solution provides the corresponding term in \eqref{fepsexp}. At the next order $\mathcal{O}\!\left(\eps^{0}\right)$ all terms in \eqref{fp-Dario} contribute to the differential equation for $g_3(u)$, making it correspondingly longer. It can be expressed in the form
\be
\label{g3eq}
\left\{ \partial^3_u -7\partial^2_u+20\partial_u-24\right\} g_3(u) = {\cal N}(u)/d(u)\,,
\ee
where the numerator $\cal N$ is a lengthy expression, quartic in the parameters $A,B,C$ and containing sines and cosines with arguments $2nu, \, n=1,\dots,4$. Re-expressed in terms of $R$ the denominator is
\be
\label{singpoint}
d(R)=11A/2+(B-2C)\cos\ln R^2 + (2B+C)\sin\ln R^2 \,.
\ee
The special solution to \eqref{g3eq} is again a rather long expression of sines and cosines also containing integrals of ratios of them. It provides $f_3$ in \eqref{fepsexp} once it is written in terms of $R$. An important observation at this point of the asymptotic series however is that \eqref{g3eq} allows for moveable singularities whenever the denominator \eqref{singpoint} on its right hand side vanishes. These moveable singularities are avoided, provided the parameters $A,B,C$ satisfy the following cone condition
\be
\label{safedisc}
\frac{121}{20}A^2 > B^2 + C^2\,.
\ee
If this inequality is violated $g_3(R)$ gives rise to an infinite series of singularities at locations $R_c$ with multiplicative periodicity $R_c \mapsto R_c \mathrm{e}^\pi$ in the asymptotic regime $R\to \infty$.

The asymptotic series can be continued to orders $\mathcal{O}\!\left(\eps^{n}\right)$ with $n>0$ or equivalently $\mathcal{O}\!\left(R^{-n-1}\right)$. The resulting differential equations for the corresponding terms in \eqref{fepsexp} take a form similar to \eqref{g3eq} with higher powers of the factor \eqref{singpoint} in the denominators on their right hand sides. One also finds a second factor appearing in the denominators,
\be
15A/2+(B-2C)\cos\ln R^2 + (2B+C)\sin\ln R^2\,,
\ee
which however does not lead to additional moveable singularities as long as the constraint \eqref{singpoint} is satisfied.

For our numerical investigation in sec. \ref{numerics} it will be sufficient to work with the first three terms of the asymptotic expansion. These are obtained by substituting \eqref{f2}, \eqref{f1} and the leading solution \eqref{f0} into \eqref{fepsexp} using \eqref{feps} with the book-keeping parameter $\eps$ set to one, thus recovering \eqref{asymp}.

To see that the derivation of the asymptotic series here stays valid for $R<0$ one only needs to replace the limit $R\to \infty$ with $R\to -\infty$ throughout.

\section{Numerical solutions}
\label{numerics}
In this section we will follow up the asymptotic analysis of the previous section with a detailed numerical analysis of the fixed point equation \eqref{fp-Dario}. This will confirm the expectation gained from the qualitative result of sec. \ref{par-count-Benedetti} that we should find one-dimensional parameter sets of solutions valid in the range $0\leq R < \infty$. However, we take into account from the beginning that \eqref{fp-Dario} may be considered also for $R<0$ which will lead us to the conclusions announced at the end of sec. \ref{par-count-Benedetti}.
\subsection{General approach}\label{approach}
The numerical analysis of the fixed point equation \eqref{fp-Dario} is dominated by the presence of the three fixed singularities $R_+, 0, R_-$. It is therefore sensible to divide the real axis into the following intervals
\be
I_{-\infty} = (-\infty,R_-], \quad I_- = [R_-,0], \quad I_+ = [0,R_+], \quad I_\infty = [R_+,\infty).
\ee
A careful numerical search for solutions has to be carried out separately on each interval and any possible solutions for $R \leq0$ have to be regarded as tentative since \eqref{fp-Dario} was derived under the assumption $R >0$. In order to bridge the fixed singularities we Taylor expand any potential solution around them. In the case of $R_+$ we have
\be \label{tayexpRp}
f(R) =b_0 + b_1(R-R_+)+\sum_{n=2}^5 \frac{b_n(b_0,b_1)}{n!}(R-R_+)^n.
\ee
Similarly,
\be\label{tayexp0}
f(R) =a_0 + a_1 R + \sum_{n=2}^5 \frac{a_n(a_0,a_1)}{n!}R^n
\ee
is the Taylor expansion around $R=0$ and finally
\be\label{tayexpRm}
f(R) =\beta_0 + \beta_1(R-R_-)+\sum_{n=2}^5 \frac{\beta_n(\beta_0,\beta_1)}{n!}(R-R_-)^n.
\ee
bridges across $R_-$. In each case, the requirement that the solutions $f(R)$ be regular at the fixed singularity (cf. sec. \ref{par-counting}) translates into the fact that only the first two coefficients are independent, i.e. we have implemented the condition \eqref{pole-const}. All higher coefficients, starting with the second, can then be expressed as functions of the first two, e.g. $a_2(a_0,a_1)$ is given by \eqref{expra2}. It should be noted that these expressions are unique for all coefficients except $b_2$ and $\beta_2$ for which we find quadratic constraints. Their solutions can be uniquely written as
\be\label{b2equ}
b_2^\pm(b_0,b_1)=\frac{1}{720 R_+}\left(\, p(b_0,b_1) \pm \sqrt{q(b_0,b_1)}\,\right)
\ee
and
\be\label{beta2equ}
\beta_2^\pm(\beta_0,\beta_1)=-\frac{1}{720 R_+}\left(\, \tilde p(\beta_0,\beta_1) \pm \sqrt{\tilde q(\beta_0,\beta_1)}\,\right),
\ee
where $p, \, q$ and $\tilde p, \, \tilde q$ are polynomials in $b_0, \, b_1$ and $\beta_0,\, \beta_1$, respectively. As we will see, both possibilities for $b_2$ and $\beta_2$ lead to fixed point solutions. As displayed above, the Taylor expansions we have used for our computations are of order five with the higher coefficients quickly becoming complicated expressions of the first two.\footnote{In fact we extend \eqref{tayexp0} to sixth order for part of the analysis for reasons discussed at the end of sec. \ref{sec:aplane}.} Especially for the expansions around $R_\pm$ it would be a challenging task to handle them for even higher coefficients.

Our search for numerical solutions was performed within the Maple(TM) package \cite{Maple} employing the shooting method. We picked the numerical integration method dverk78 which is a seventh-eighth order Runge-Kutta integrator and is able to operate at arbitrarily high precision.

We now illustrate the individual steps that led to the discovery of solutions in the region $R \ge 0$. If we started shooting down from some large $R_\infty$ to $R=R_+$ we would face the problem of having to search through the $3$-dimensional parameter space of the asymptotic expansion \eqref{asymp}. It is therefore more reasonable to shoot out from $R=R_+$ or $R=0$ where the parameter space is only two-dimensional. Moreover, we start looking for solutions on $I_+$ as opposed to shooting from $R_+$ towards large $R$ as we expect the additional constraint at $R=0$ to reduce our parameter space by one dimension which is a more severe condition than \eqref{safedisc} for the asymptotic series. Most solutions we were able to find can be obtained by both shooting out from zero towards $R_+$ and from $R_+$ towards zero, in particular those which are valid for all $R\geq0$. The exceptions were solutions valid only on $I_+$ with a moveable singularity appearing shortly after $R_+$, e.g. at $R_c \approx 8$; in this case integrating from $R_+$ to zero turned out to be difficult.\footnote{Similarly, integrating from $R_-$ to zero was ill behaved for partial solutions with a moveable singularity at $R_c \approx -8$.} For the purposes of illustration however, let us focus on shooting out from zero towards $R_+$ in the following.

\begin{figure}[h]
\begin{center}
 \includegraphics[scale=0.4]{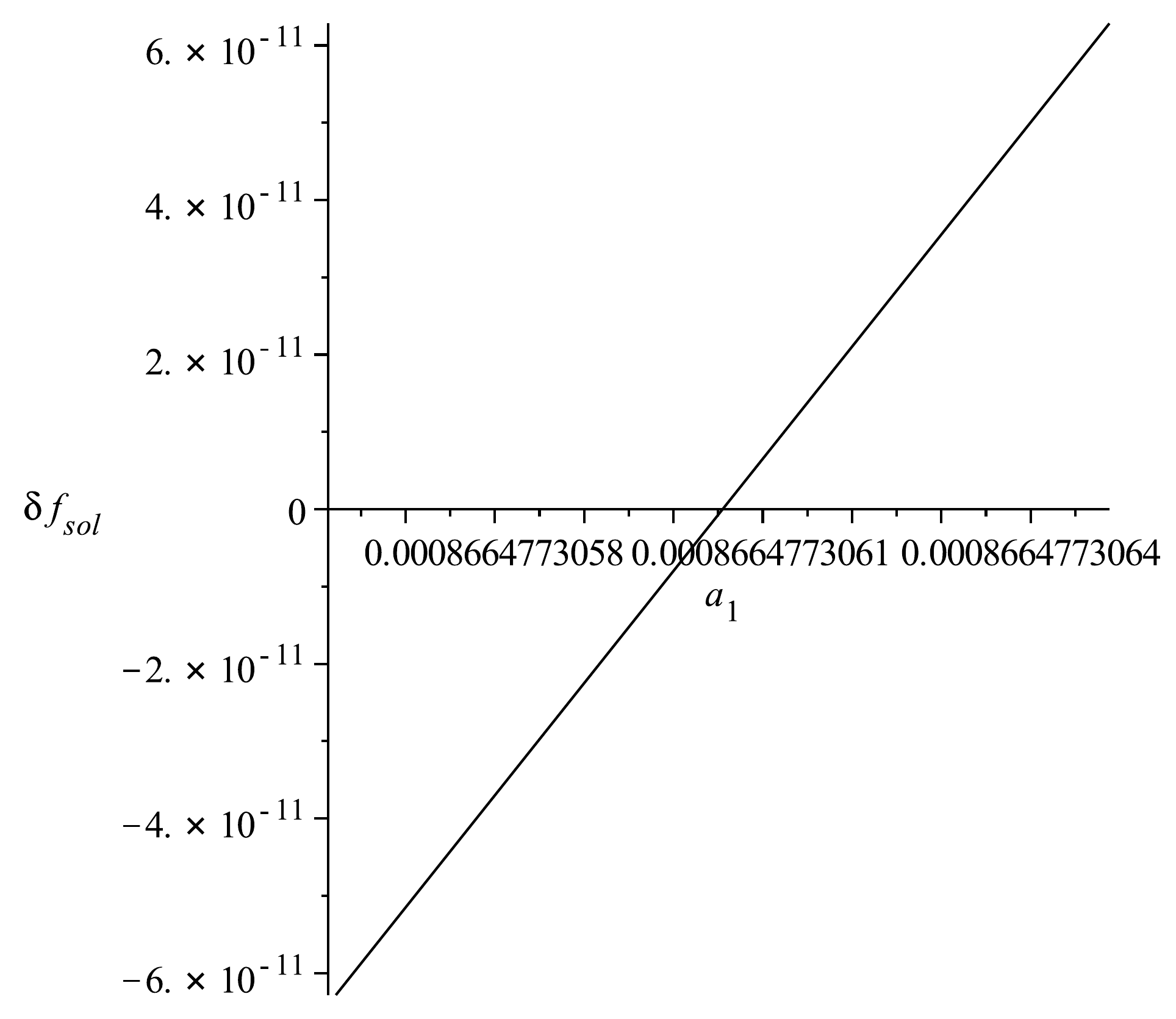}
 \caption[Solution criterion for $\delta f_\mathrm{sol}$]{The quantity $\delta f_\text{sol}$ varies linearly across zero as a function of $a_1$ keeping $a_0$ fixed. If this is the case we recognise the pair $(a_0,a_1)$ as a solution point for $I_+$. }
 \end{center}
 \label{fig:solcrit}
\end{figure}

Given any initial pair $(a_0,a_1)$ we use \eqref{tayexp0} to compute the three initial values 
\be \label{initconds0}
f(\eps_0), \quad f'(\eps_0), \quad f''(\eps_0)
\ee
at distance $\eps_0$ to the right of $R=0$. Shooting out from zero towards $R_+$ we will find that most pairs $(a_0,a_1)$ end at a moveable singularity. For some however, the numerical integrator supplies us with the solution $f_n$ valid close enough to $R_+$ so that by using \eqref{tayexpRp} 
we can solve the system
\be \label{bsystem}
f(R_+-\eps_1)=f_n(R_+-\eps_1), \quad f'(R_+-\eps_1)=f'_n(R_+-\eps_1)
\ee
at $\eps_1$ away from $R_+$ for $b_0$ and $b_1$. Using \eqref{tayexpRp} again, we are then in a position to compute $f''(R_+-\eps_1)$ and compare it to the actual value $f''_n(R_+-\eps_1)$ via the quantity
\be \label{solcrit}
\delta f_\text{sol} = \frac{f_n''(R_+-\eps_1) - f''(R_+-\eps_1)}{f''(R_+-\eps_1)}.
\ee
This represents the additional matching condition at $R_+$ and implements the constraint coming from the singularity at $R=R_+$. A solution is found if the two second derivatives agree to sufficient accuracy. In fact, our criterion for a valid solution on $I_+$ will be that $\delta f_\text{sol}$ varies smoothly across zero upon variation of $a_1$ while keeping $a_0$ fixed. This condition on $\delta f_\text{sol}$ is illustrated in fig. \ref{fig:solcrit}.
Small variations of $a_1$ lead to the expected linear behaviour of $\delta f_\text{sol}$. Varying $a_1$ by a larger amount brings the non-linear form of $\delta f_\text{sol} (a_1)$ to light whereas if we go to very small variations of $a_1$ we find numerical fluctuations superimposed on the linear behaviour. There are good reasons for this solution criterion and why we should not just require $|\delta f_\text{sol}|$ to be small, as we will explain in detail in sec. \ref{sec:aplane}.

A priori there is no special region in the $a$-plane of points $(a_0,a_1)$ where solutions are to be expected. However, there is an important structure to the $a$-plane which is connected to the Taylor expansion \eqref{tayexp0}. The coefficients $a_2(a_0,a_1),\, a_3(a_0,a_1), \dots$ become singular along the dashed lines $\gamma_i$ in fig. \ref{fig:as} where the expansion \eqref{tayexp0} is no longer valid. For reasons that will be explained in sec. \ref{sec:aplane}, along these lines there are no solutions except at finitely many points. It is therefore reasonable to search for the first solution between these singular lines. We can do this by using similar graphs as in fig. \ref{fig:solcrit} where we plot $\delta f_\text{sol} (a_1)$ in the range between two singular lines while keeping $a_0$ fixed or, if moveable singularities prevent the solver from reaching $R_+$, instead of $\delta f_\text{sol} (a_1)$ we similarly plot the maximum $R$ reached. In this way it is possible to find the first solution point in the $a$-plane. Once we have a solution $(a_0,a_1)$, we create two new $a$-vectors $(a_0+\delta a_0,a_1+\delta a_1)$ and $(a_0+\widetilde{\delta a_0}, a_1+\widetilde{\delta a_1})$, where the two variations $(\delta a_0,\delta a_1)$ and $(\widetilde{\delta a_0},\widetilde{\delta a_1})$ are orthogonal to each other and small enough so that we are still able to integrate from zero to $R_+$. Generically $\delta f_\text{sol}$ will be large at these new points and so we repeatedly apply the secant method to $\delta f_\text{sol}(a_0,a_1)$ until we find a new pair $(a_0,a_1)$ with the behaviour shown in fig. \ref{fig:solcrit}. Given at least two solution points $(a_0,a_1)$ and $(\tilde a_0,\tilde a_1)$ we can optimise the procedure we have just described by using their difference as a guide for the direction in which to look for the next solution point, i.e. we define
\be
(\delta a_0,\delta a_1) \propto  (\tilde a_0 -a_0,\tilde a_1-a_1).
\ee
These two methods constitute the way by which we extended into lines of solutions starting with only one solution point.

Clearly, the numerical approach to the search for solutions on $I_-$ proceeds in a similar way. We start off with a pair $(a_0,a_1)$ and use \eqref{tayexp0} at $\eps_0$ to the left of zero, to determine the initial conditions which are needed for shooting towards $R_-$. If the (partial) solution makes it to $R_-+\eps_1$, we solve the corresponding version of \eqref{bsystem} and adapt \eqref{solcrit} to judge whether we found a solution. The same methods as for $I_+$ are also employed to find the very first solution on $I_-$ and to extend into solution lines.

When it comes to finding solutions on $I_\infty$ the degree of difficulty of the problem at hand has already been lowered by the fact that we can restrict ourselves to check if any of the solutions valid on $I_+$ will make it out to infinity, i.e. we are left with a one dimensional parameter space. The strategy therefore is to select a possible candidate pair $(b_0,b_1)$ and to exploit \eqref{tayexpRp} at $R_+ + \eps_1$ to find the initial values. We then integrate up to some large $R_\infty$ where we finally match to the asymptotic expansion \eqref{asymp}. The accuracy of the three asymptotic parameters $A,B,C$ will depend on the value of $R_\infty$ which is why it should be taken reasonably large.\footnote{We defer a discussion of this important point to sec. \ref{errorana}.} If the constraint \eqref{safedisc} is satisfied the existence of the solution is guaranteed for all $R>R_\infty$. On the other hand, if the asymptotic parameters $A,B,C$ are such that \eqref{safedisc} is not fulfilled, we have found only a partial solution since a moveable singularity is bound to appear at some $R>R_\infty$, cf. sec. \ref{sec:asy-fp}. In principle this same method can be used to find solutions on $I_{-\infty}$ but as we will see, none of the solutions we found on $I_-$ extend to $I_- \cup I_{-\infty}$.

\subsection{The solutions} \label{sec:sols}
After extensive numerical searches using the approach outlined in the previous section we were able to obtain a total of five solution lines valid on the range $R\geq0$, four additional solution lines representing solutions on $I_-$, and we found evidence for more solution lines. For clarity of presentation we devote a separate section to each set of solution lines.

\subsubsection{Solutions for $R\geq0$:}
\begin{figure}[ht]
\begin{center}
\includegraphics[width=0.6\textwidth,height=250pt]{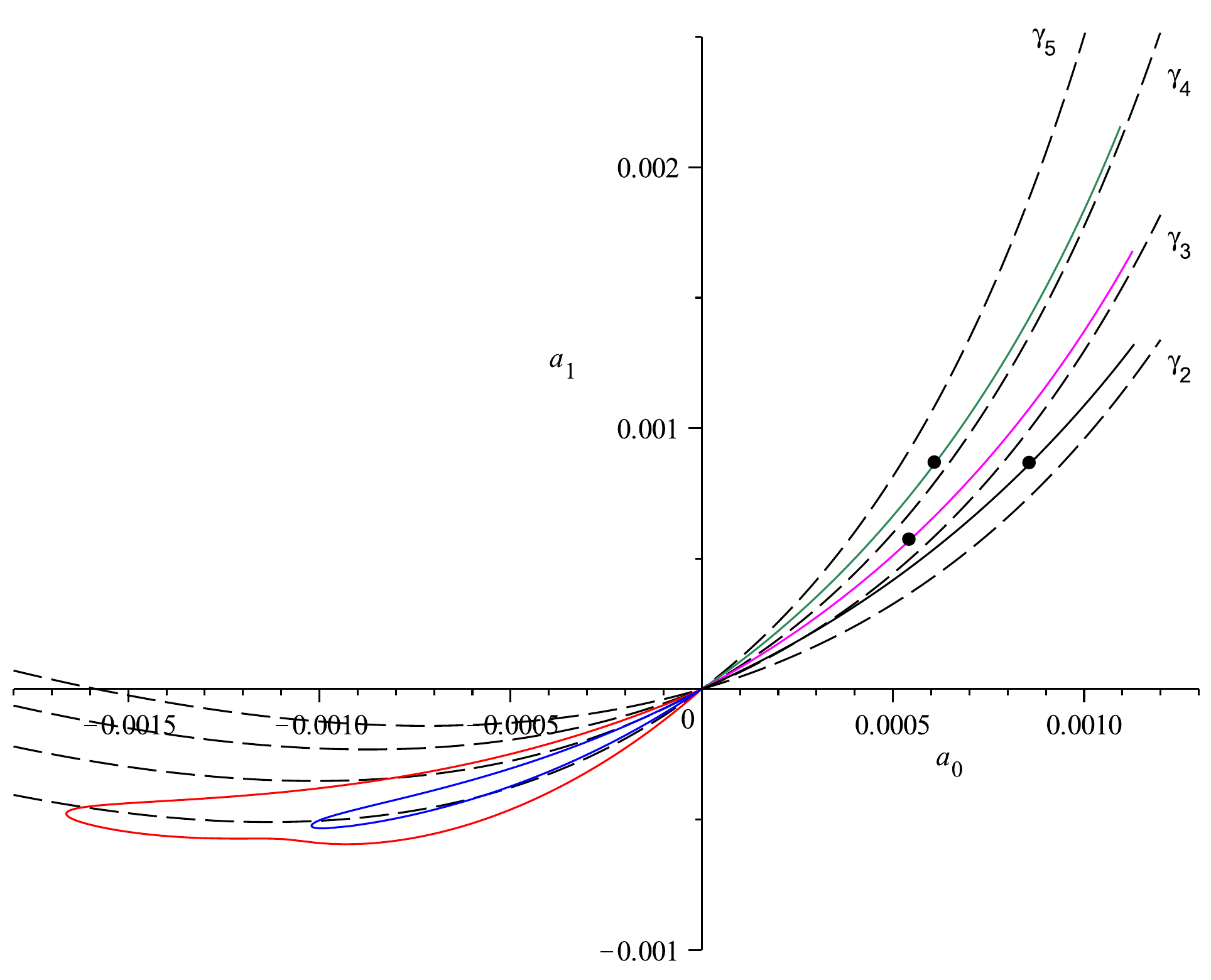}
\end{center}
\caption[Solution lines for $R\geq 0$ in the $a$-plane]{Five lines of solutions in the $a$-plane corresponding to the $b$-lines in fig. \ref{fig:bs}. Different solution lines are represented consistently in  different colours (these are red and blue in the range $a_0<0$ and, from bottom to top, black, magenta and green in the range $a_0>0$). The black points represent example solutions which are plotted in fig. \ref{fig:expls}. Along each of the black dashed lines a coefficient in \eqref{tayexp0} becomes singular (cf. sec. \ref{sec:aplane}).}
\label{fig:as}
\end{figure}

\begin{figure}[ht]$
\begin{array}{cc}
\includegraphics[width=0.4\textwidth,height=150pt]{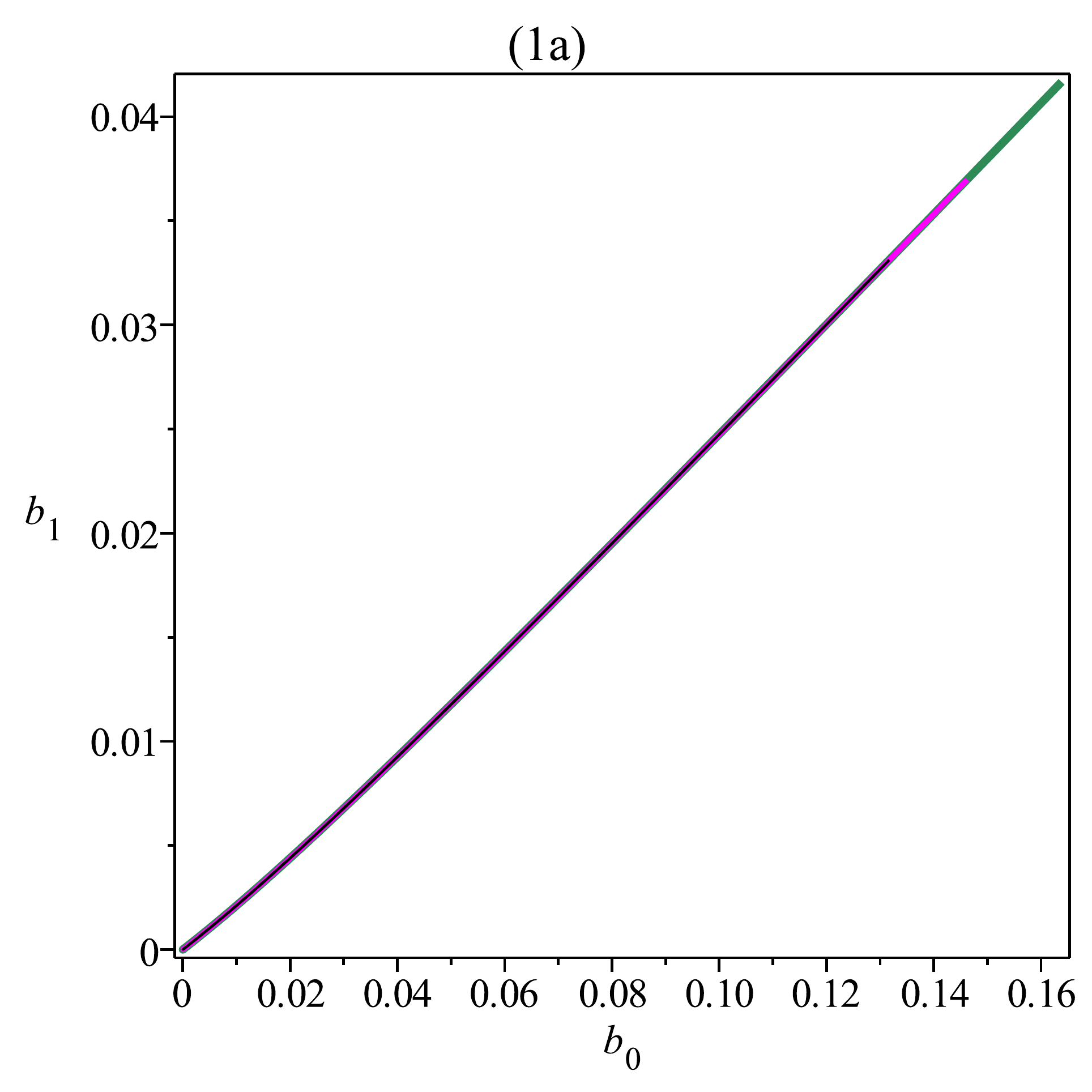} &
\includegraphics[width=0.55\textwidth,height=150pt]{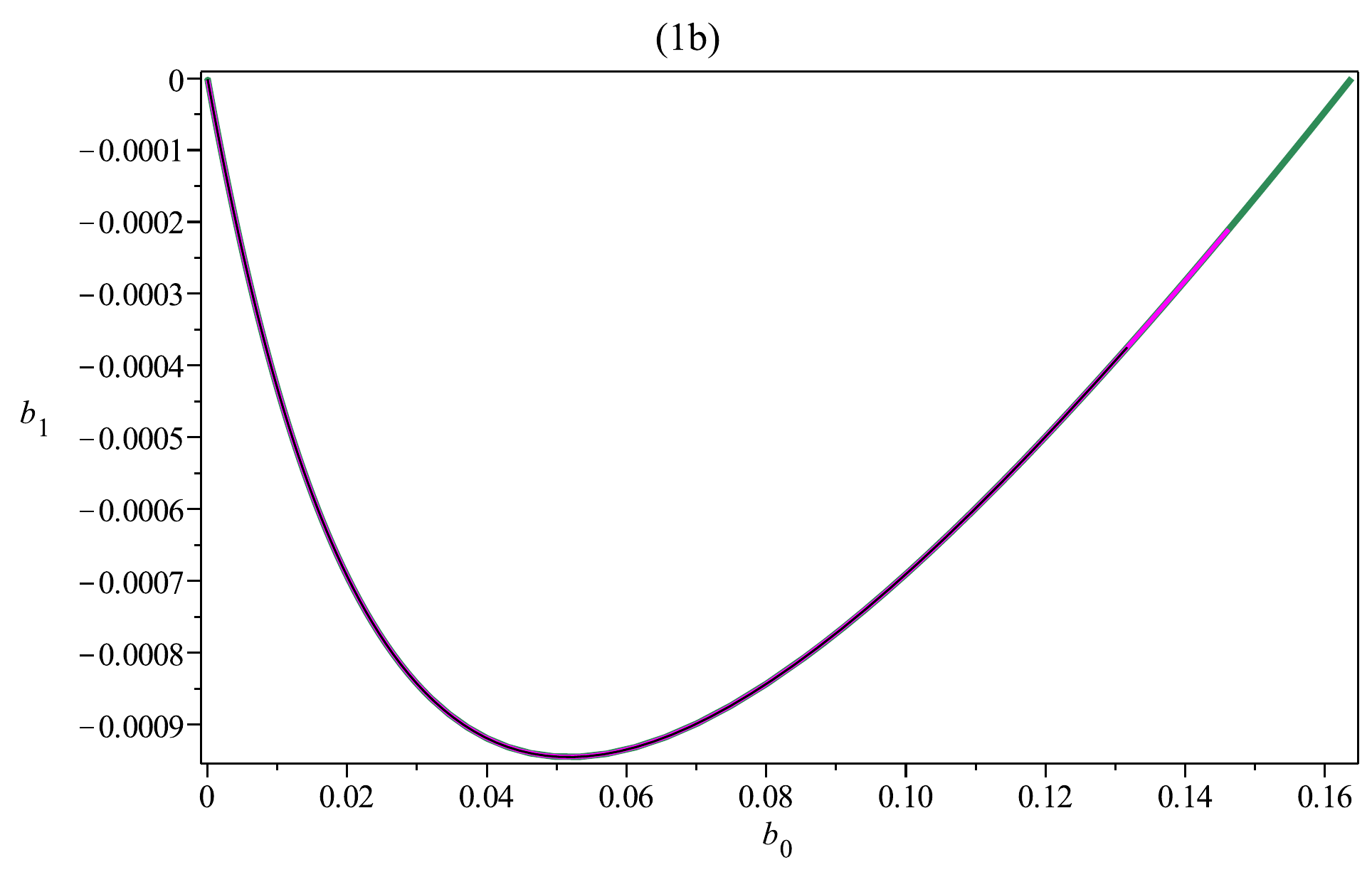} \\
\includegraphics[width=0.4\textwidth,height=150pt]{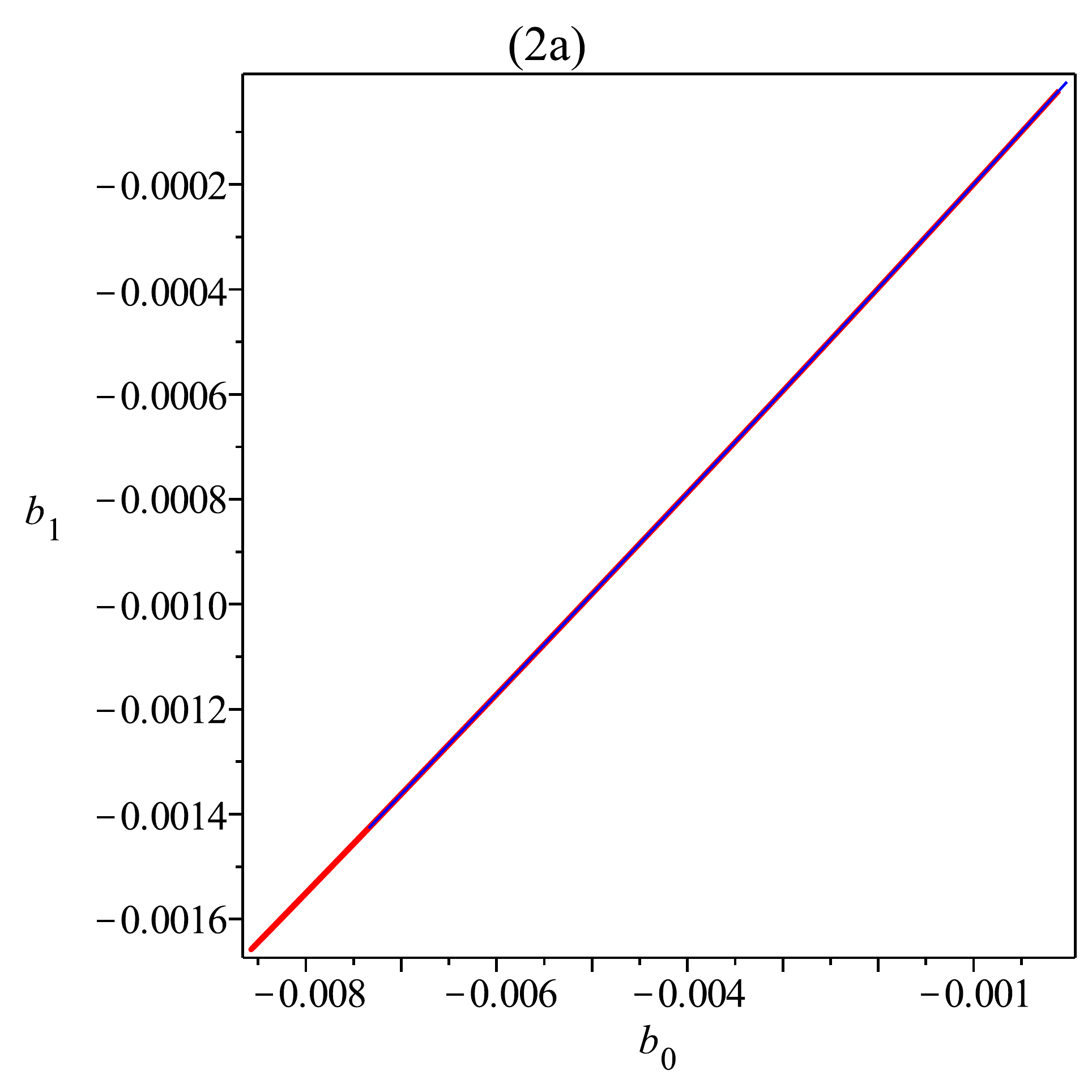} &
\includegraphics[width=0.55\textwidth,height=150pt]{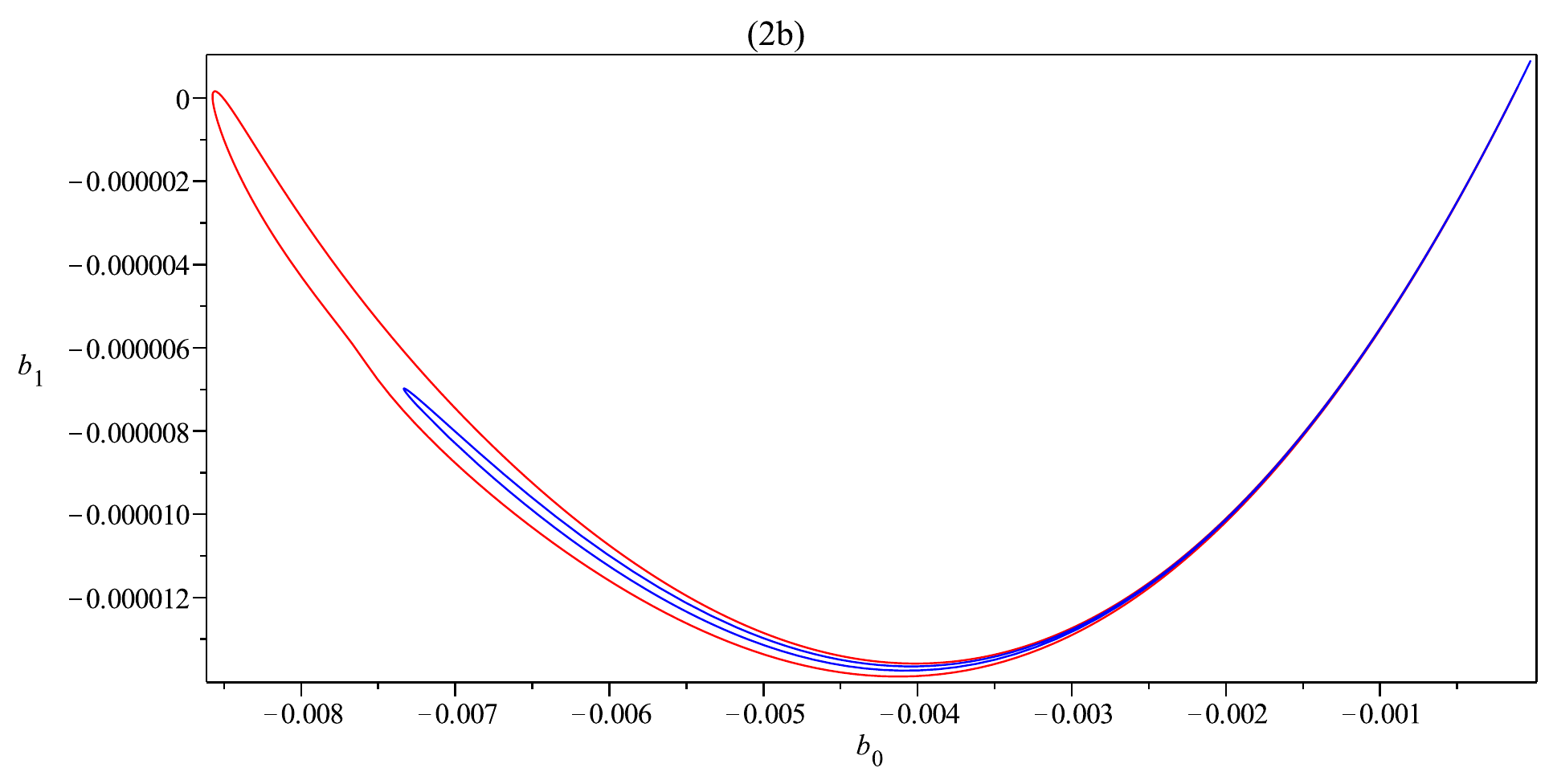} 
\end{array}$
\caption[Solution lines in the $b$-plane for $R\geq0$]{Three lines of solutions on $I_+ \cup I_\infty$ in the range $b_0>0$ in (1a) and two additional solution lines in the range $b_0<0$ in (2a). In both cases we have subtracted the straight line given by  $b_1=0.254352\cdot b_0 -2.5\cdot 10^{-7}$ and plotted the result on the right.}
\label{fig:bs}
\end{figure}

The lines of solutions defined on the original domain of validity $I_+ \cup I_\infty$ of the fixed point equation \eqref{fp-Dario} are shown in fig. \ref{fig:as}. We have three lines of fixed point solutions in the range $a_0>0$, all starting at the origin and moving out between two singular lines of the expansion \eqref{tayexp0}.\footnote{Further such solution lines are expected between singular lines $\gamma_n$ with $n\ge5$, as we discuss in sec. \ref{sec:aplane}.} In fact, the bottom two solution lines are both situated between the two singular lines $\gamma_3$ and $\gamma_4$ in the region close to the origin, cf. fig. \ref{fig:als}, but the black solution line crosses the singular line $\gamma_3$ in its early stages and moves out between $\gamma_2$ and $\gamma_3$. In order to find these fixed point solutions in the region of positive $a_0$ it was crucial to use the $b^+_2(b_0,b_1)$ solution in \eqref{b2equ}. Furthermore, all these solutions are valid for all $R\geq0$, i.e. satisfy the asymptotic constraint \eqref{safedisc}. The corresponding solution lines in the $b$-plane are shown in fig. \ref{fig:bs} (1a), (1b). All three start at the origin and approximately follow a straight line towards larger values of $b_0$. Even after subtraction of this linear component, the result being shown in fig. \ref{fig:bs} (1b), it is impossible to distinguish them by eye. The distance between these solution lines varies but can be as small as $\Delta b_1/b_1= 10^{-10}$.

In the range $a_0<0$ there are two more solution lines but they behave differently, see fig. \ref{fig:as}. Leaving the origin they run towards the left but eventually both turn around to approach the origin again. In order to satisfy the constraint arising from the fixed singularity at $R_+$ it was necessary to employ the $b_2^-(b_0,b_1)$ solution in \eqref{b2equ} in this region of the $a$-plane. If we look at the corresponding solution lines in the $b$-plane shown in fig. \ref{fig:bs} (2a), (2b) we again observe a linear behaviour. Unlike in the case for $b_0 >0$ however, we can distinguish the two solution lines clearly after subtracting their straight line approximation. It then also becomes clear that, just like the solution lines in the $a$-plane, after moving away from the origin they turn around and approach it again. In both the $a$- and the $b$-plane the blue solution line is contained inside the region bounded by the red solution line.

The fact that the solution lines in the $b$-plane can be approximately described by a straight line can be attributed to the following observation. If we rescale according to
\begin{equation*}
f(R)=b_1 \bar f(R) =b_1 \left(\frac{b_0}{b_1} + (R-R_+) + \frac{b_2}{b_1}(R-R_+)^2 + \dots \right)
\end{equation*}
and consider the limit $b_1 \to 0$, the fixed-point equation \eqref{fp-Dario} becomes  
\begin{equation*}
\left.{\tilde \cT}_2 + \cT_1 +\cT^{\text{np}}_0+{\tilde \cT}^{{\bar h}}_0 \right|_{f= \bar f} =0 \,.
\end{equation*}
This follows from the fact that the right hand side in \eqref{fp-Dario} is invariant under rescaling of $f$ but the left hand side is linear. Hence we would expect that, close to the origin of the $b$-plane, any solution line should show predominantly linear behaviour. The same conclusions can be drawn for solution lines in the $a$-plane. However, we find that the slopes vary linearly, implying quadratic behaviour of the solution lines close to the origin. Given the huge magnification factor implied above, which occurs as we integrate from $R_+$ to zero, we would expect to recover linear behaviour of the solution lines at distances much closer to the origin of the $a$-plane than we have probed here.

The non-linear components of the solution lines in the $b$-plane in fig. \ref{fig:bs} were obtained by subtracting the same straight line from both the solution lines with positive and negative $b_0$. This is a first indication that it could be possible to match these solution lines across the origin in the $b$-plane. If one computes the second coefficient $b_2$ along the solution lines the two possibilites \eqref{b2equ} also seem to match across the origin. However, since the $b$-lines are so close to each other it is difficult to give a definite answer to this question in the $b$-plane. The solution lines in the $a$-plane can be separated clearly and computing their slopes close to the origin it seems to be possible to match the lines with positive $a_0$ uniquely to a solution line with negative $a_0$, e.g. the black solution line matches to the upper part of the red solution line and the solution line in magenta can be matched to the upper part of the blue solution line. The lower part of the blue solution line can be continued by the green solution line whereas the lower part of the red solution line seems to belong to an as yet undeveloped line on the other side of the origin. Although these pairings are confirmed by also matching the other coefficients of the $a$-series \eqref{tayexp0} across the origin, none of these methods would be able to distinguish solution lines that become tangential as they approach the origin.

The solution lines on the right hand side in fig. \ref{fig:as} can be continued towards larger values of $a_0$ but the numerical integration becomes ever more time consuming. We have also found a first solution point on four additional solution lines between $\gamma_4$ and $\gamma_5$, on both sides of the origin. However, these first solutions are only partial solutions, not reaching $R\to \infty$.

The fixed point solutions in the top right quadrant of fig. \ref{fig:as} are bounded below since they all have a positive asymptotic coefficient $A$. Since the limit $k\to 0$ corresponds $R\to \infty$ for fixed physical curvature $\tilde R$, cf. \eqref{scaled-vars}, the functional integral can be completed at these fixed points using \eqref{asymp} with the full effective action taking the form $\Gamma_{*, k=0}=\int d^4x \sqrt{g}\,A\tilde R^2$ in dimensionful variables \cite{Benedetti:2012dx}. By contrast the solutions in the bottom left quadrant cannot be physically sensible as they have $A<0$ and are thus unbounded below. Furthermore, one may interpret the sign of $-a_1$ (and $-a_0/a_1$) as that of Newton's constant (and the cosmological constant) at the ultraviolet fixed point by comparison with the Einstein-Hilbert truncation \eqref{EH}. But these should not be directly compared to the signs of Newton's constant and the cosmological constant as measured by experiment. To do this, one would have to follow the RG trajectory spawned by all (marginally) relevant eigen-perturbations from the ultraviolet fixed point to the infrared and extract the corresponding effective Newton's constant and cosmological constant from the result. On the other hand, in all previous investigations in asymptotic safety where RG trajectories have been computed, Newton's coupling never changes sign along the flow. It would be interesting to study this point in the present case.

\begin{figure}[ht]
\begin{center}$
\begin{array}{ccc}
\includegraphics[width=0.4\textwidth,height=115pt]{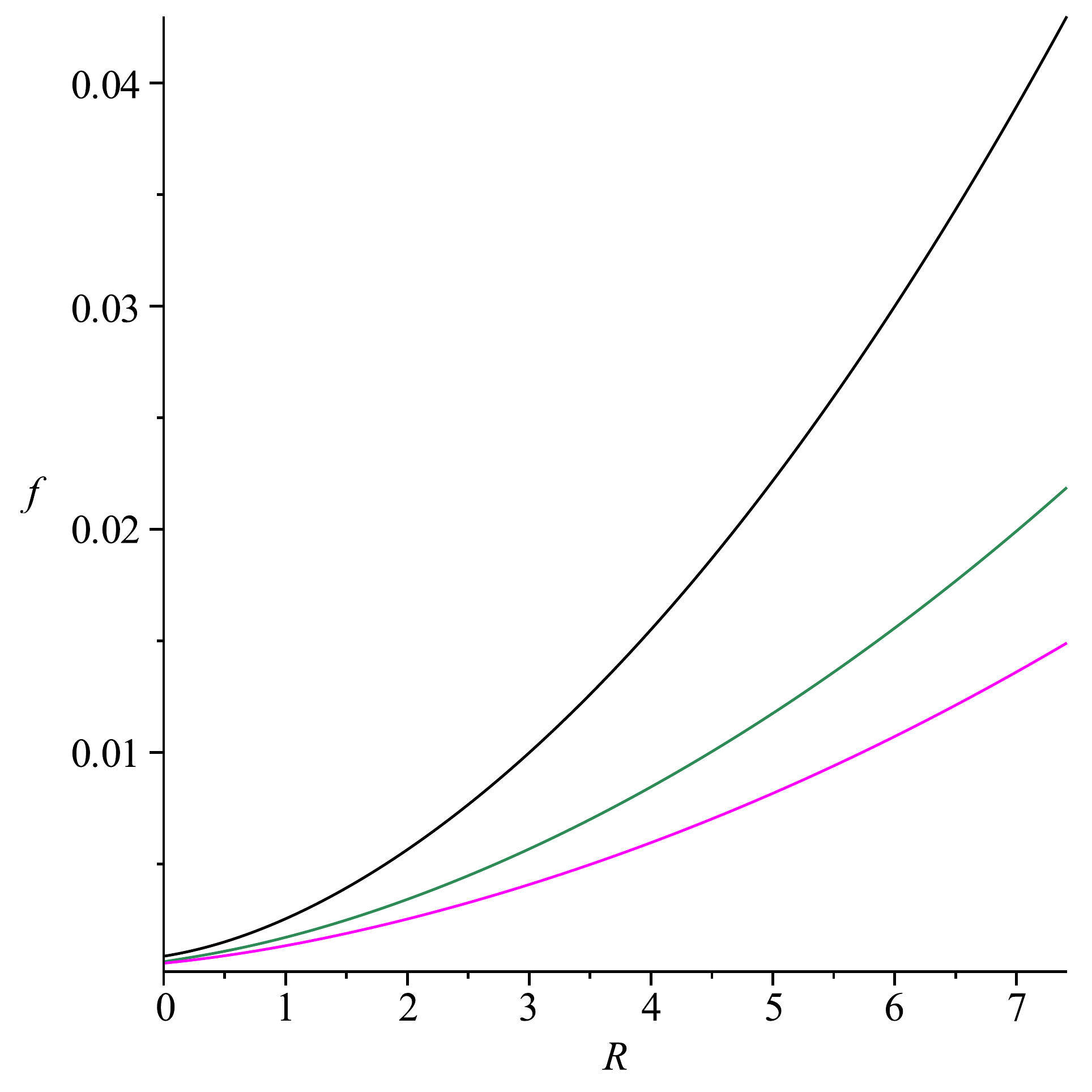} & & \includegraphics[width=0.4\textwidth,height=115pt]{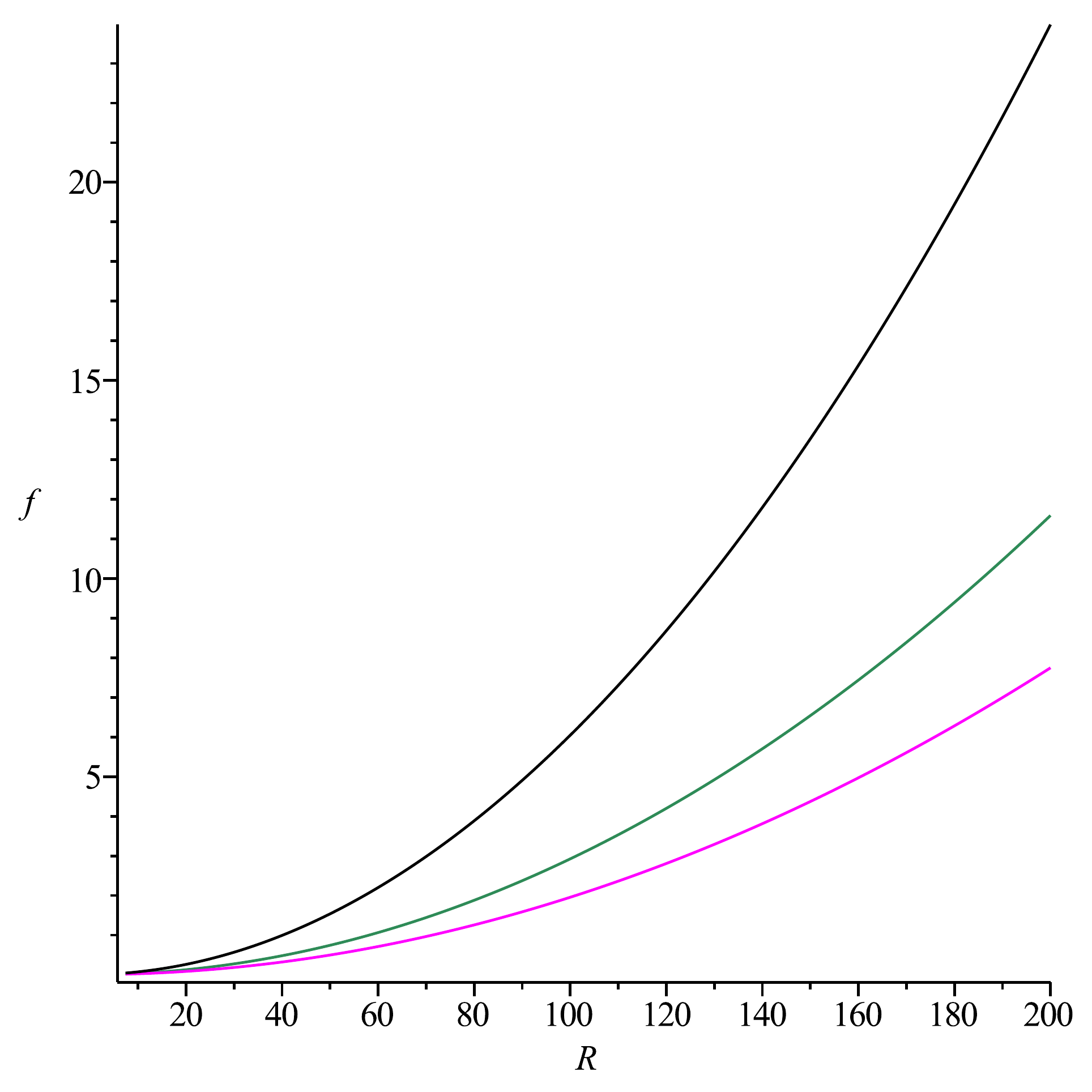} \\
 \includegraphics[width=0.4\textwidth,height=115pt]{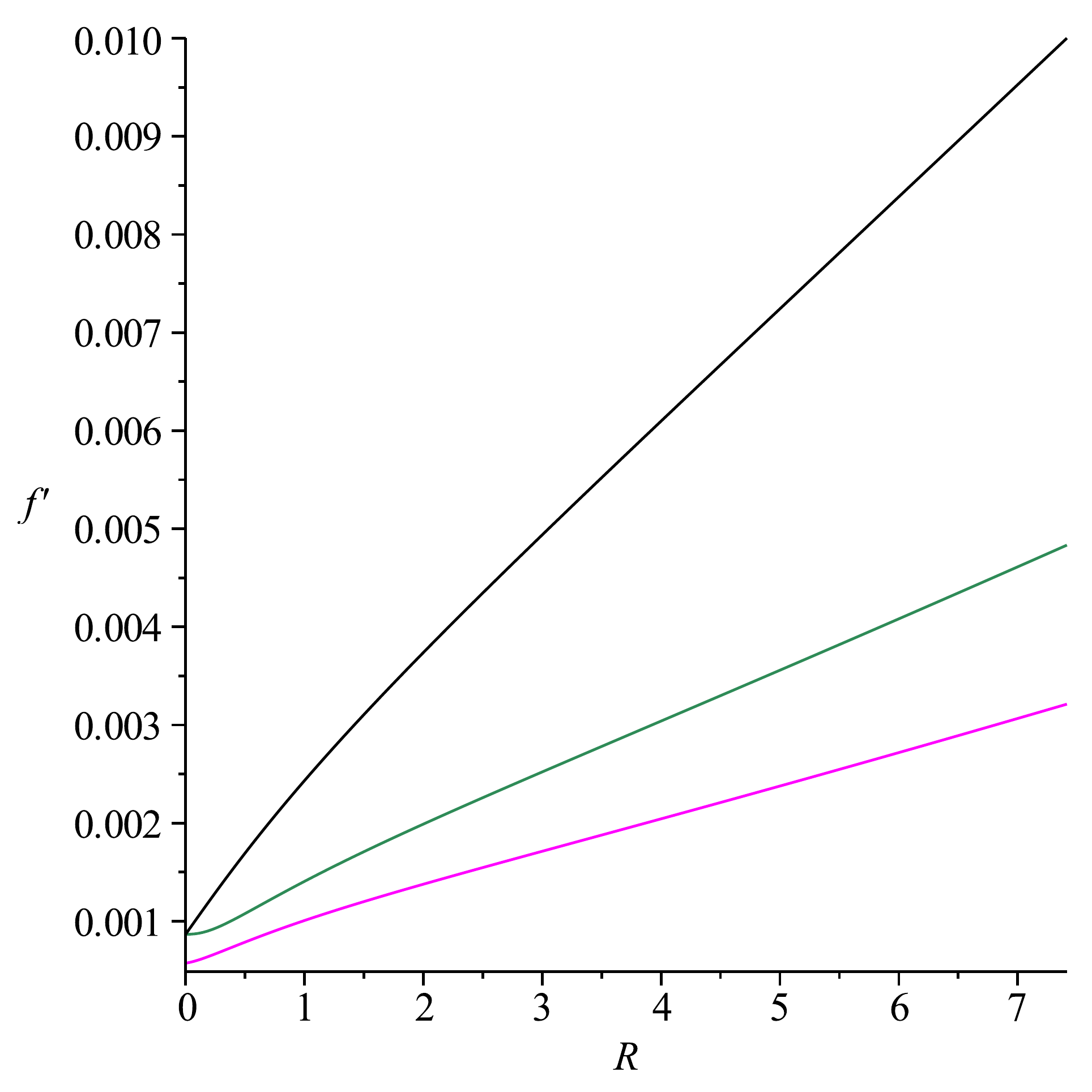} & & \includegraphics[width=0.4\textwidth,height=115pt]{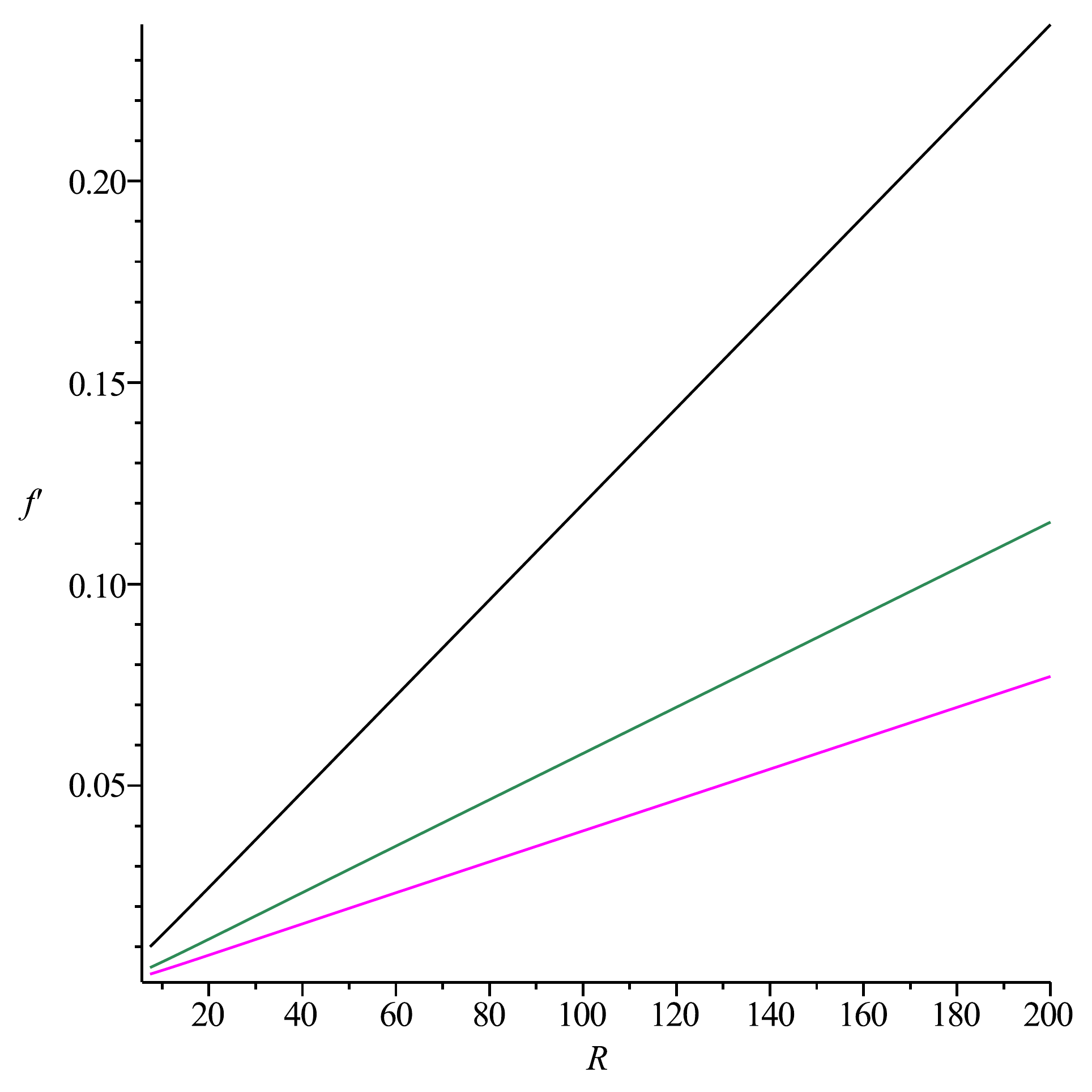} \\
\includegraphics[width=0.4\textwidth,height=115pt]{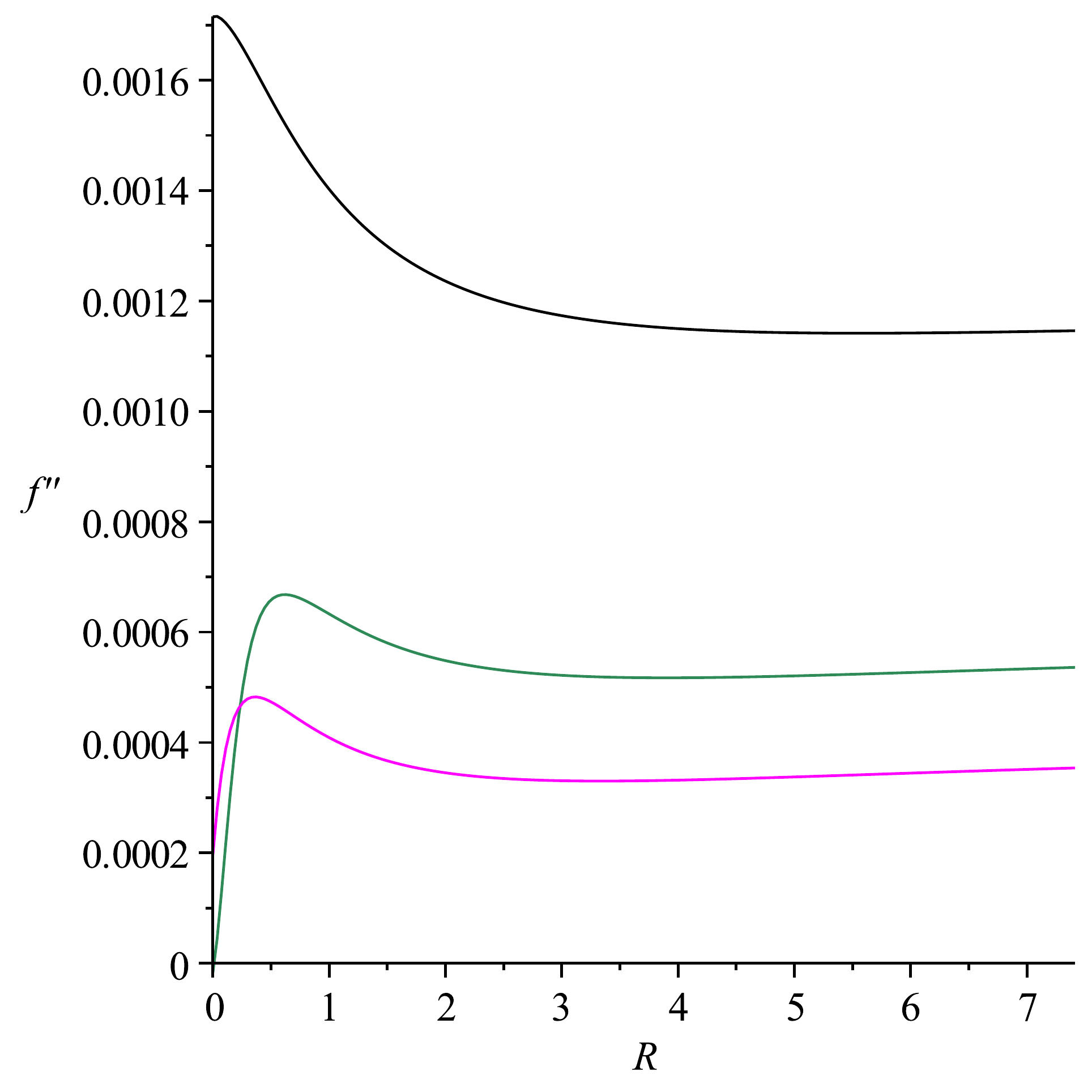} & &
\includegraphics[width=0.4\textwidth,height=115pt]{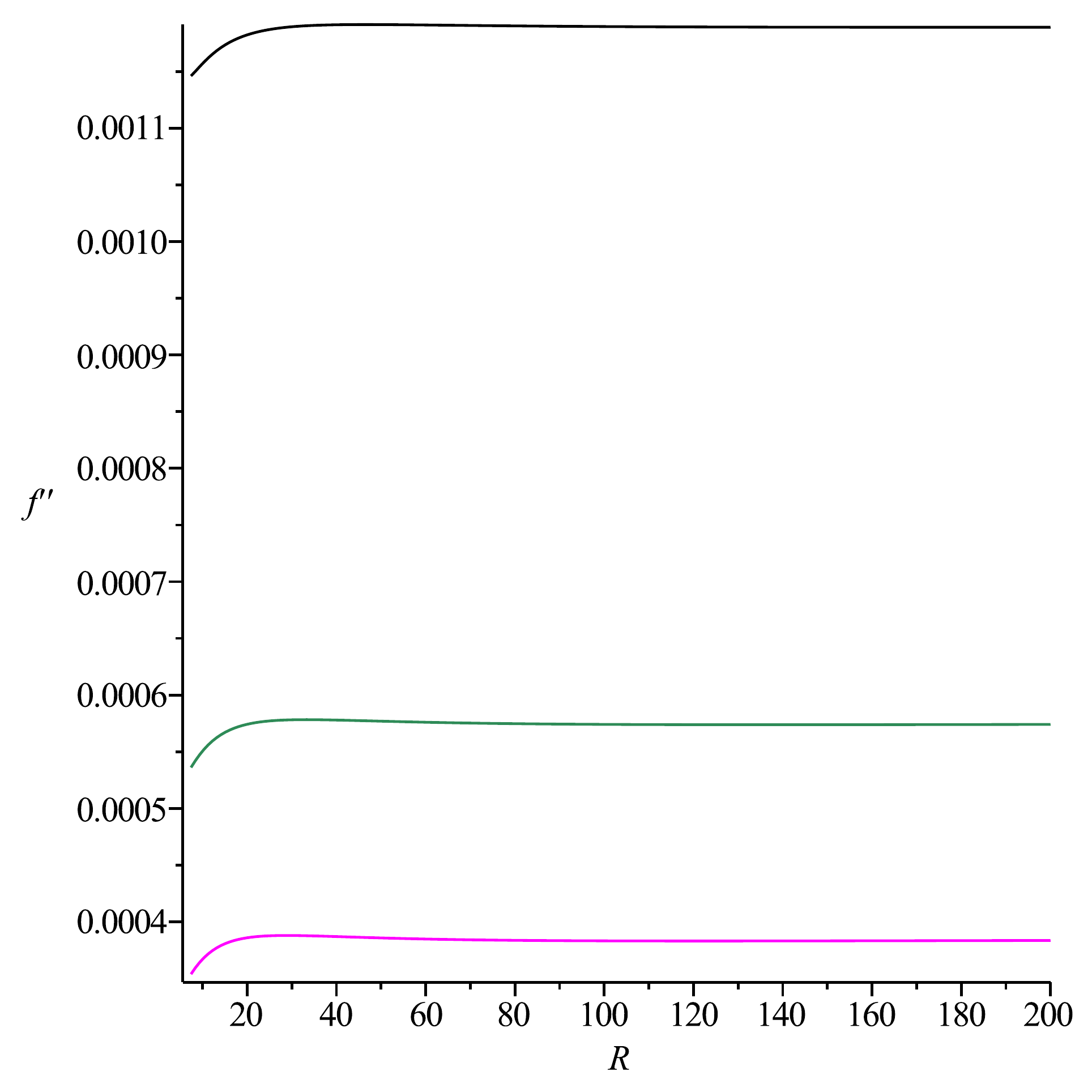}
\end{array}$
\end{center}
\caption[Plot of $f(R)$ example fixed point solutions]{Fixed point functions and their first and second derivatives represented by the points shown in fig. \ref{fig:as} in corresponding colours (from bottom to top these are magenta, green, black) plotted over the interval $I_+$ in the left column and over the range $R_+ \leq R \leq R_\infty$ on the right.}
\label{fig:expls}
\end{figure}

As indicated by the points in fig. \ref{fig:as}, we have selected one example solution on each of the solution lines for $R \geq 0$ with $A>0$ and plotted the fixed point function $f(R)$ and its first and second derivatives in fig. \ref{fig:expls}. All three fixed point functions assume positive values only. The first and second derivatives already show that we approach quadratic behaviour close to $R_+$ as the plots on the right hand side in fig. \ref{fig:expls} confirm. Nevertheless, in agreement with what is already built into the asymptotic expansion \eqref{asymp}, we can also discern the oscillatory pattern from the plots of the second derivative. The parameter values for these solutions are listed in table \ref{tab:par}. 

\begin{table}[h]
\begin{center}
\begin{tabular}{|c|c|c|c|c|c|c|c|}
 \hline
  FP & $a_0 \, [10^{-4}]$ & $a_1\, [10^{-4}]$ & $b_0 \,[10^{-3}]$ & $b_1 \,[10^{-3}]$ & $A\, [10^{-5}]$ & $B\, [10^{-5}]$ & $C \,[10^{-5}]$ \\
 \hline
  green & 6.1053 & 8.6648 & 21.869 & 4.8337 & 28.741 & -3.2321 & -1.3074 \\
 \hline
  magenta & 5.4421 & 5.7113 & 14.899 & 3.2125 & 19.207 & -3.3397 & -0.4781 \\
 \hline
 black & 8.5841 & 8.6461 & 42.986 & 10.002 & 59.479 & -1.5161 & -2.1547 \\
 \hline
\end{tabular}
\end{center}
\caption[Parameter values of $f(R)$ example fixed point solutions]{Parameter values for the fixed point functions plotted in fig. \ref{fig:expls}. Note that many of these parameters are known to higher accuracy than given here, cf. sec. \ref{errorana} and table \ref{tab:errs}.}
\label{tab:par}
\end{table}

\subsubsection{Solutions for $R_-\leq R \leq 0$:}
The fact that the fixed point equation \eqref{fp-Dario} admits the uncountably infinite number of solutions presented in the previous section can be expected from the parameter counting method. As mentioned before, although the derivation of \eqref{fp-Dario} was carried out on spaces of positive scalar curvature $R$ we have nevertheless extended our analysis to the range $R\leq0$ as this will constrain the solution space to at most a finite number of physically sensible fixed point functions valid for $-\infty < R < \infty$. Since in this regard we are interested in fixed point solutions with positive quadratic asymptotic component, i.e. $A>0$, we have concentrated our efforts on finding solutions valid on $I_-$ which are situated in the $a_0>0$ half-plane.

\begin{figure}[h]
\begin{center}
\includegraphics[width=0.49\textwidth,height=200pt]{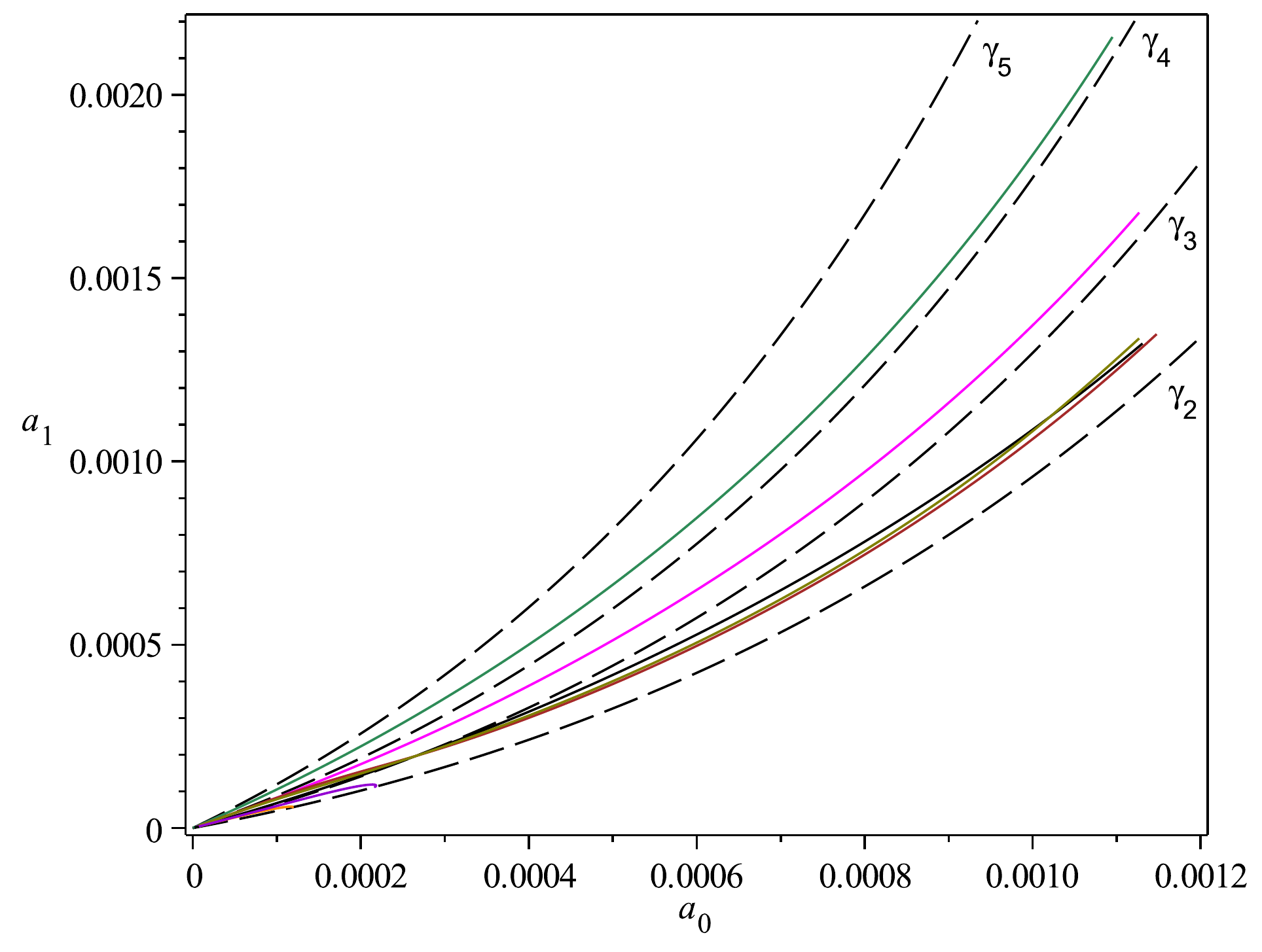}
\includegraphics[width=0.5\textwidth,height=200pt]{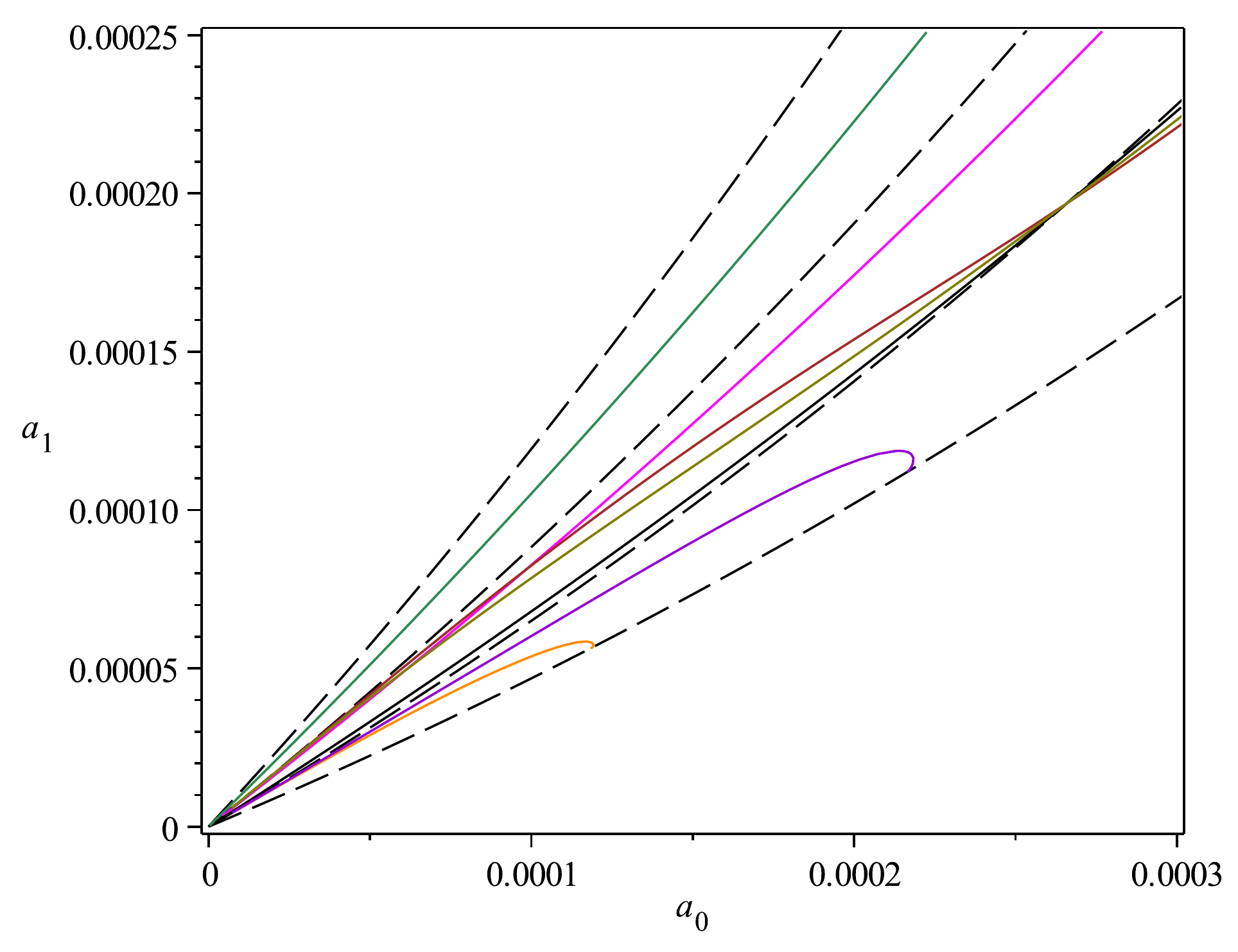}
\end{center}
\caption[Solution lines for $I_-$ in the $a$-plane]{Four solution lines for $I_-$ in the $a$-plane together with three of the solution lines displayed in fig. \ref{fig:as} on the left and a magnification of the region close to the origin on the right. In the magnification the colouring from bottom to top for the solution lines on $I_-$ is orange, violet, olive and brown. The black dashed lines are the same singular lines $\gamma_i$ as in fig. \ref{fig:as}.}
\label{fig:als}
\end{figure}

We have been able to find four solution lines in this range whose form in the $a$-plane is displayed in fig. \ref{fig:als}. Two of these solution lines extend as far as the previous solution lines for $R\geq0$ whereas the other two turn around while still close to the origin running alongside the bottom singular line $\gamma_2$. Because the latter two lines are so close to $\gamma_2$ we have refrained from developing them all the way back to the origin but have checked that they are still present at $a_0=10^{-4}$ where they are approximately $2\cdot 10^{-9}$ away from $\gamma_2$.

The situation is different for the brown and olive solution line in fig. \ref{fig:als}. They start out between the same singular lines $\gamma_3$ and $\gamma_4$ as the two solution lines for positive $R$ shown in black and magenta and both cross $\gamma_3$ where the black solution line does but approach the black solution line for large values of $a_0$. The olive solution line then intersects the black solution line and thus gives rise to a solution valid on the interval $R_- \leq R < \infty$. Although the brown solution line does not cross the black solution line again in the range we have developed it in, it is to be expected that it will do so if the two lines are continued further. Close to the origin we find two more solutions valid on $R_- \leq R < \infty$: the solution lines in brown and olive both intersect the solution line in magenta for $I_+$ exactly once (this is barely discernible in fig. \ref{fig:als}). The former at $(a_0,a_1) = (9.8\cdot 10^{-5},8.1\cdot 10^{-5})$ and the latter at $(a_0,a_1)= (6.07 \cdot 10^{-5},4.9\cdot 10^{-5})$. 

\begin{figure}[h]
\begin{center}
\includegraphics[width=0.49\textwidth,height=200pt]{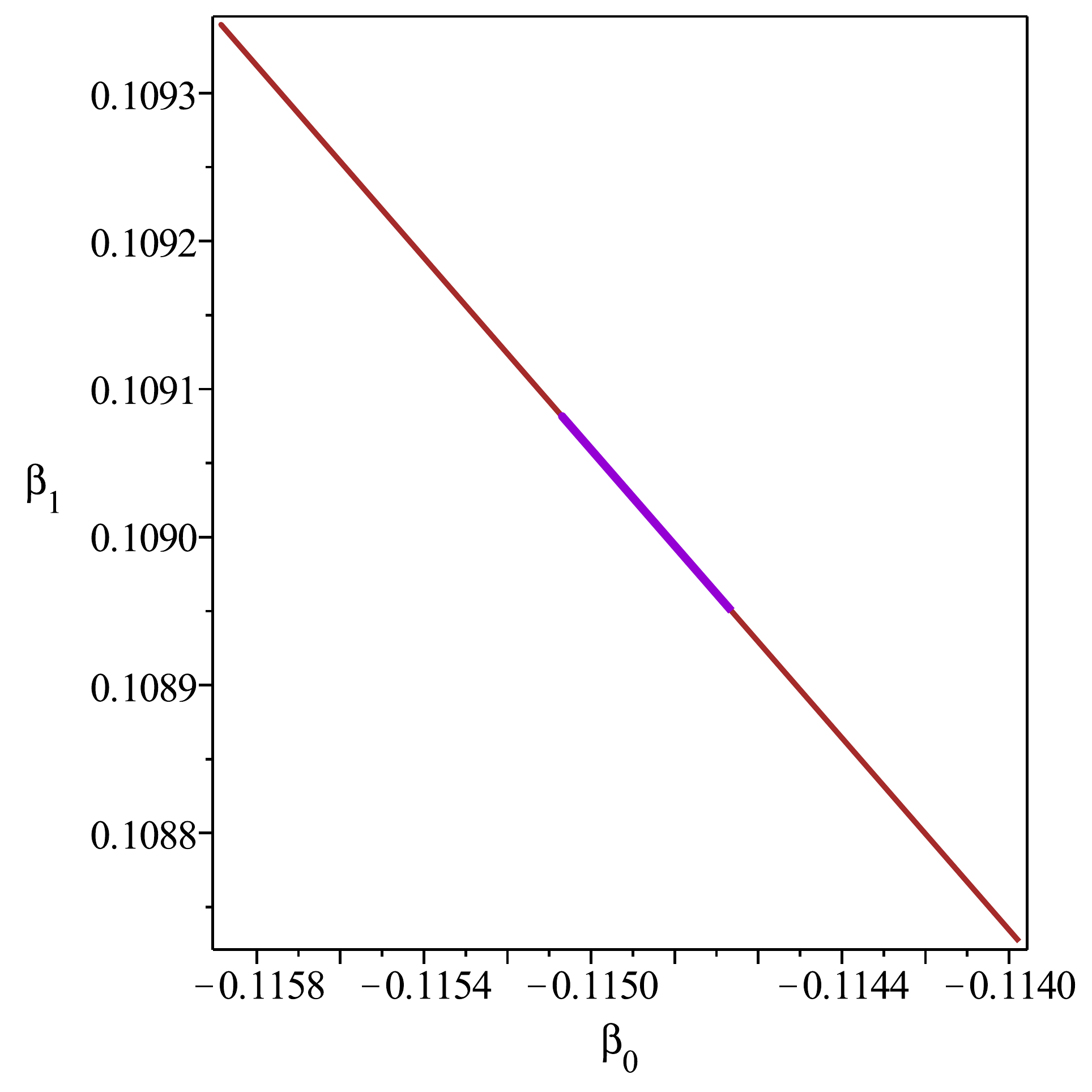}
\includegraphics[width=0.5\textwidth,height=200pt]{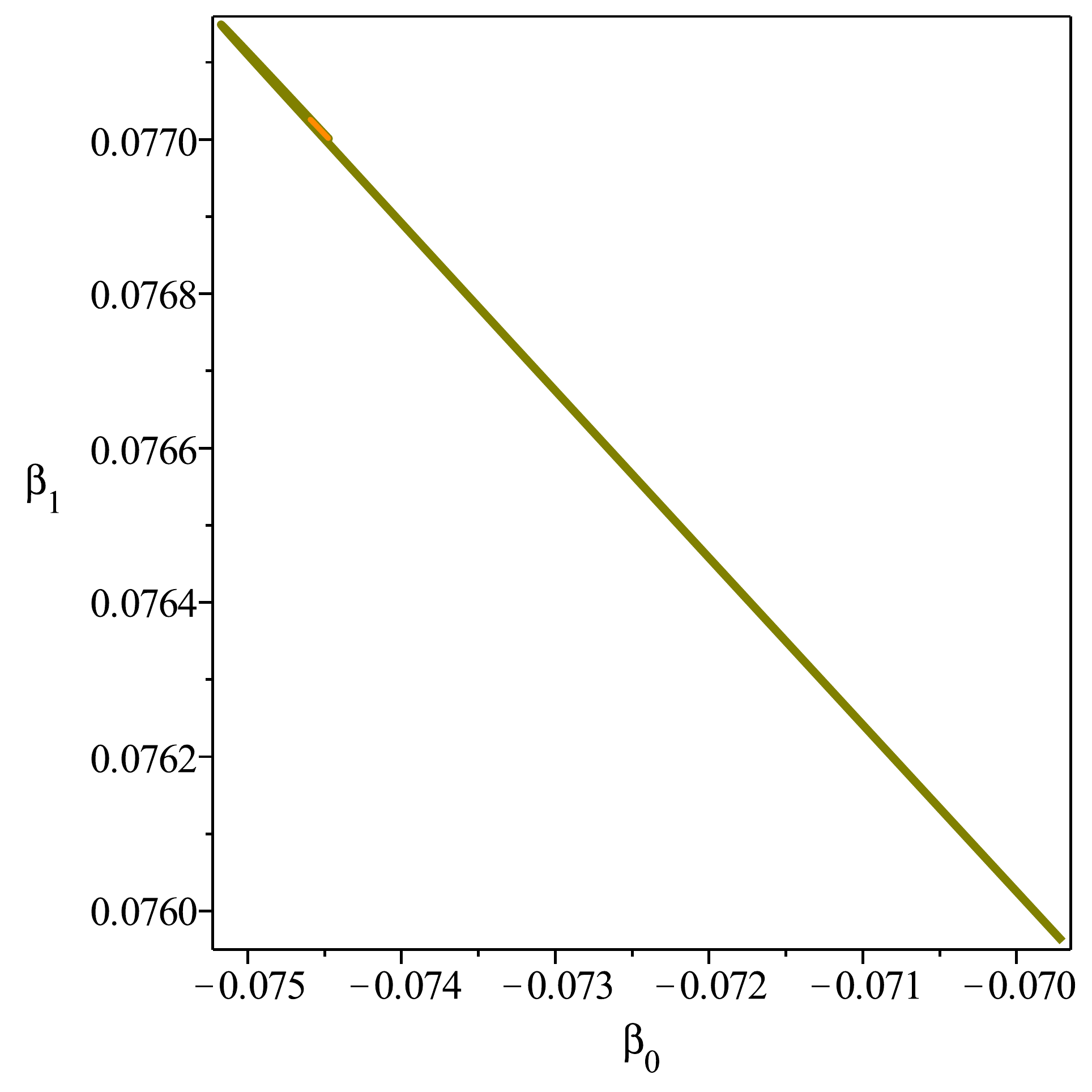}
\end{center}
\caption[Solution lines for $I_-$ in the $\beta$-plane]{The four solution lines for $I_-$ in the $\beta$-plane in the same colours as the corresponding lines shown in fig. \ref{fig:als}.}
\label{fig:bts}
\end{figure}

The corresponding solution lines in the $\beta$-plane are shown in fig. \ref{fig:bts}. Like the solution lines
in the $b$-plane they are governed by an underlying linear behaviour with superimposed non-linear variations
which in the case of the left plot are of order $10^{-8}$ and for the plot on the right they can be as large
as $3\cdot10^{-5}$. Note the two different ways in which these four lines are paired: in the $a$-plane
the two long and the two short solution lines are paired together whereas in the $\beta$-plane it is always
one $\beta$-line corresponding to a long solution line in the $a$-plane that is paired with a $\beta$-line
corresponding to a short solution line in the $a$-plane. As we would expect from continuity, varying the initial pair $(a_0,a_1)$ along a smooth curve from one of the short solution lines in the $a$-plane to the other leads to a smooth curve of  points $(\beta_0,\beta_1)$ in the $\beta$-plane connecting the two corresponding solution lines in the two different regions of the $\beta$-plane. However, this connecting curve does not consist of solutions since the constraint at $R_-$ is not satisfied.
Curiously, despite the fact that the solution lines in the $a$-plane approach the origin, the corresponding lines in the $\beta$-plane do not. Another difference to the situation for $R\geq0$ is that we have not found solution lines between the upper two singular lines $\gamma_4$ and $\gamma_5$.

We have also found evidence for a line of solutions valid on $I_-$ between the middle two singular lines $\gamma_3$ and $\gamma_4$ in the range of negative $a_0$. As we would have to match any such solution at $R=0$ to a solution with negative asymptotic parameter $A$, these solutions are physically not relevant.

With regard to extending these solutions on $I_-$ to the range of negative scalar curvatures $I_{-\infty} \cup I_-$ one finds that they all end at a moveable singularity at $R_c \approx -8 <R_-$ to the left of the singular point $R=R_-$. In particular, the three solutions valid for all $R\geq R_-$ found above are thus not extendible to fully global solutions valid on the whole real line.

In this context we remark that for certain setups of the RG flow on the sphere there are arguments implying that the function $f(R)$ is not required to be well-defined for arbitrarily large values of $R$, see the recent work \cite{Demmel:2014sga}. The possibility of realising this proposal for the flow equations of this chapter would have to be investigated separately.

\subsection{Error analysis}
\label{errorana}

The actual numerical computations have been carried out with $25$ significant digits in Maple and we used the values $\eps_0=5 \cdot 10^{-5}$ and $\eps_1=10^{-3}$ in \eqref{initconds0} and \eqref{bsystem}. We will now go through a detailed error analysis for the example solutions plotted in fig. \ref{fig:expls}.

\begin{figure}[h]
 \begin{center}
  \includegraphics[width=0.4\textwidth,height=170pt]{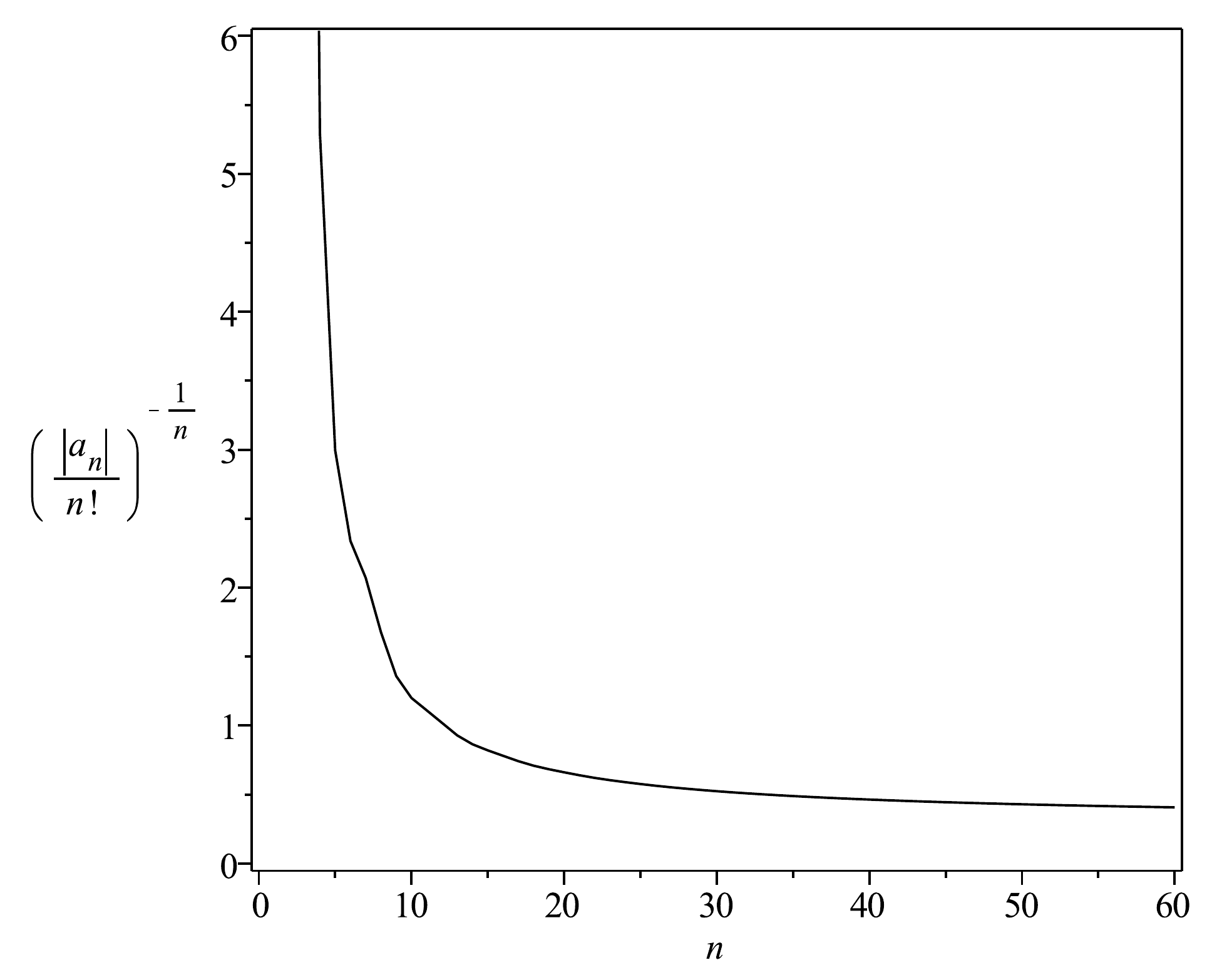}
\hspace{1cm}
  \includegraphics[width=0.425\textwidth,height=170pt]{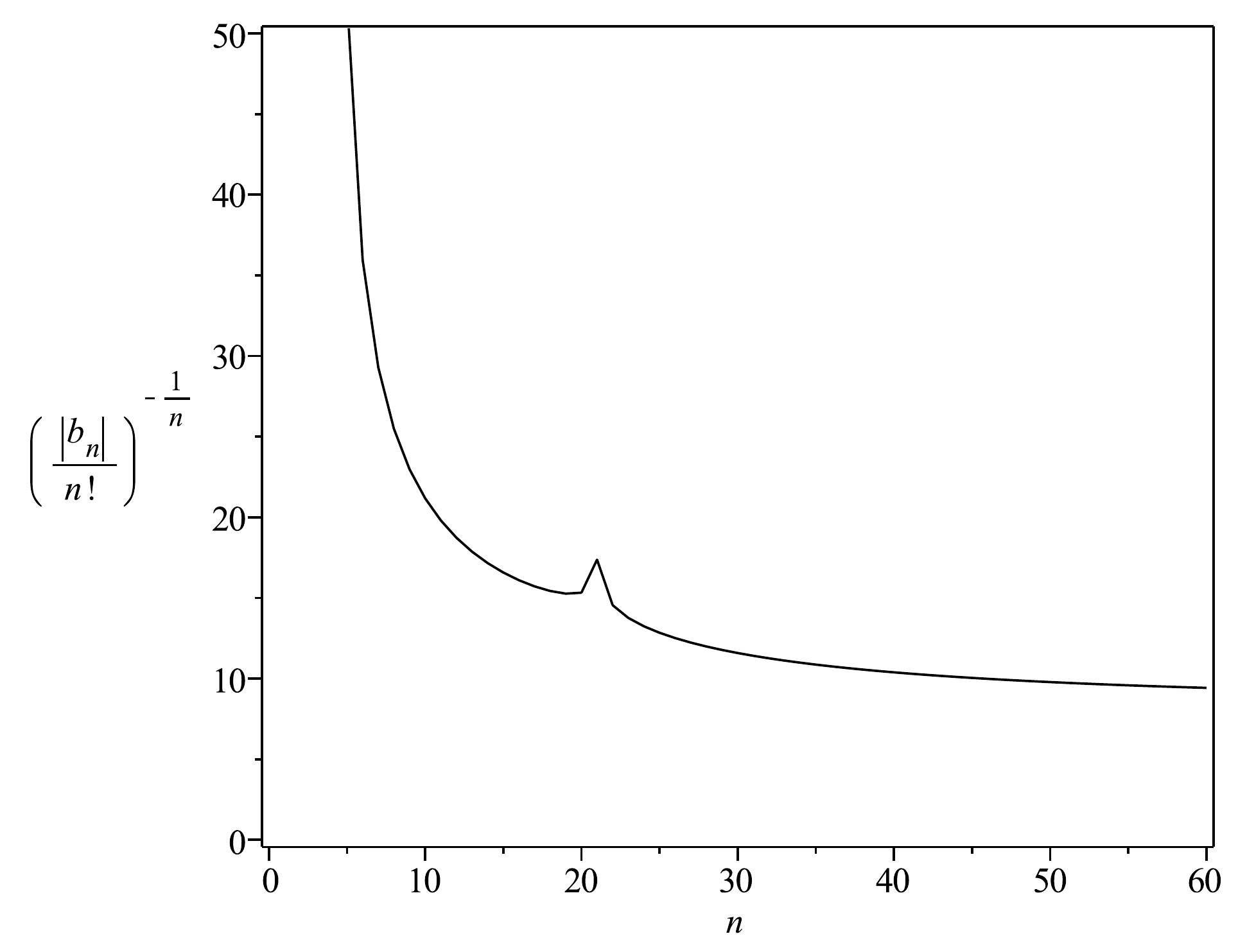}
 \end{center}
\caption[Radius of convergence for Taylor expansions of $f(R)$]{The result of a root test applied to the first $60$ coefficients of \eqref{tayexp0} on the left and of \eqref{tayexpRp} on the right for the green fixed point solution in fig. \ref{fig:expls}.}
\label{fig:roottest}
\end{figure}

The first source of error in our solution strategy comes from the truncated Taylor expansions \eqref{tayexpRp}-\eqref{tayexpRm}. We investigated the radius of convergence of \eqref{tayexp0} and \eqref{tayexpRp} by performing a root test on the first $60$ coefficients for each of the example solutions. The result for the green example solution is shown in fig. \ref{fig:roottest}.
The root test on the left hand side points to a convergence radius of $r_0 \approx 0.4$ for the Taylor series around zero. The plot on the right pertains to a root test for the Taylor series around $R_+$ and seems to indicate that the convergence radius of that series is $r_+ \approx 9$. The two convergence radii have to satisfy the relation $r_+ \leq R_+ + r_0 $. With $R_+ \approx 7.4$ we see that the values for $r_0$ and $r_+$ can be taken as rough estimates.

\begin{figure}[h]
 \begin{center}
  \includegraphics[width=0.5\textwidth,height=170pt]{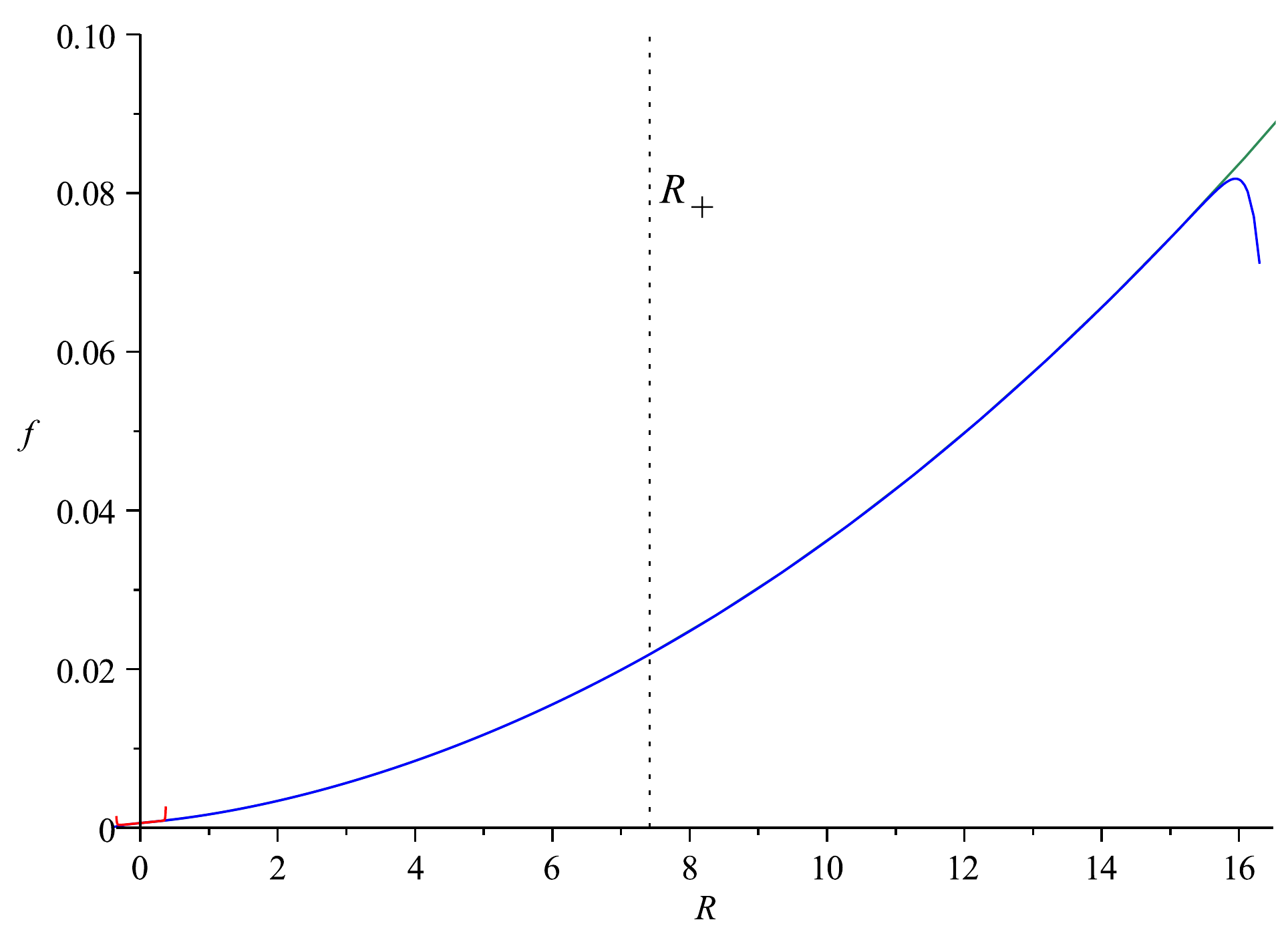}
  \includegraphics[width=0.4\textwidth,height=170pt]{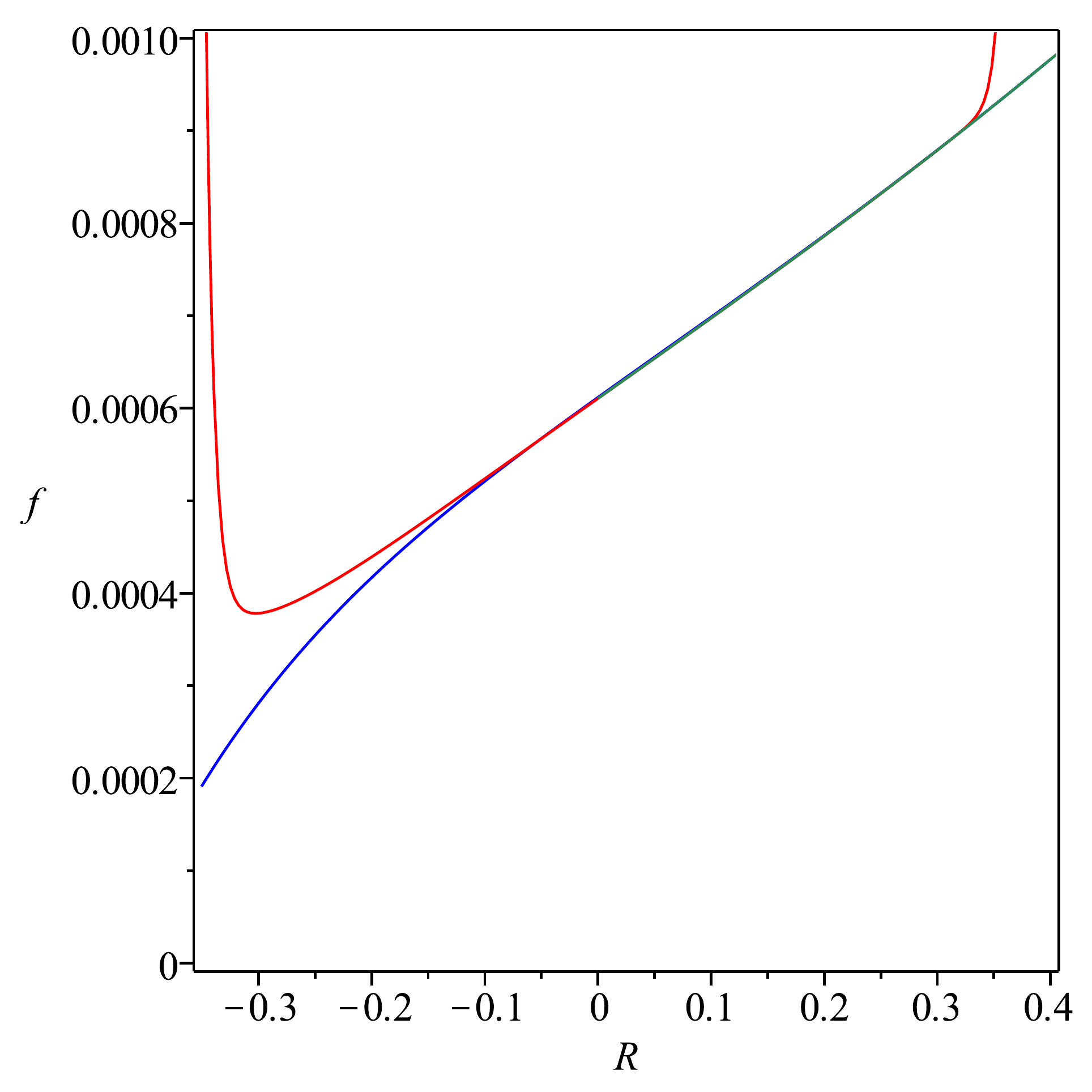}
 \end{center}
\caption[$f(R)$ example solutions with its Taylor approximations]{The green example solution together with its Taylor approximations of order $60$ around $R_+$ and zero in blue and red, respectively. The magnification on the right shows their behaviour round $R=0$.}
\label{fig:TaylorAndSol}
\end{figure}

If we use these Taylor approximations and plot them together with the numerically computed solution we get fig. \ref{fig:TaylorAndSol}. We notice the rapid divergence of the Taylor expansions from the numerical solution towards the end of the respective convergence interval but in between the numerical solution is indistinguishable from the Taylor approximation.

Zooming in on the Taylor approximations in the region around $R=0$, cf. fig. \ref{fig:TaylorAndSol} (right), we can see that for negative values of $R$ the $b$-series approximation in blue starts diverging earlier than the red $a$-series approximation. This plot is consistent with a radius of convergence $r_+ \approx R_+ + r_0$ at $R=R_+$ which would satisfy the aforementioned relation between $r_+$ and $r_0$. 

The analogous plots to fig. \ref{fig:TaylorAndSol} for the first and second derivatives of $f(R)$ for the green solution example show equally good agreement with the only difference that the Taylor expansions start diverging slightly earlier as the number of derivatives increases. Altogether these results present strong evidence that we can trust the Taylor expansions used to bridge the singularities.

With this in mind, the error analysis proceeds along the following lines. We estimate the relative error $\tilde \Delta_{a,s}$ on the initial conditions \eqref{initconds0} from truncating the Taylor series \eqref{tayexp0} by taking the modulus of the last term in the Taylor expansion. Supposing the true initial conditions \eqref{initconds0} are within this error of the ones we actually have, we can determine the amount by which we have to vary the two parameters $(a_0,a_1)$ in order to obtain this variation $\tilde \Delta_{a,s}$ of the initial conditions. We then take the result $\Delta_{a,s}$ as the corresponding error on the pair $(a_0,a_1)$.

We also have to consider the intrinsic error of the solver, i.e. even if we started with exact initial conditions at $\eps_0$, the integrator will induce a small error in each step which leads to a numerical solution different from the true solution. The size of this error $\Delta_{a,n}$ can be determined by using plots like fig. \ref{fig:solcrit} and varying $a_1$ by ever smaller amounts. At some point we will be able to see numerical fluctuations appearing and we have reached the maximum accuracy. These fluctuations limit the accuracy to which we can determine the zero of $\delta f_\text{sol}$ we are looking for. The value of $\Delta_{a,n}$ can then be read off  from the $a_1$-axis of this plot. Since these two errors are uncorrelated we obtain the total error on $(a_0,a_1)$ according to
\be \label{toterra}
\Delta_a=\sqrt{\Delta_{a,s}^2 + \Delta_{a,n}^2}.
\ee
To obtain the resulting error on the corresponding solution point $(b_0,b_1)$ we now evolve this total error $\Delta_a$ with the numerical integrator by starting from the modified pair $\left(1+\Delta_a\right) \cdot(a_0,a_1)$ and compute the error $\Delta_{b,n}$ on the pair $(b_0,b_1)$. We then also have to include the truncation error of the Taylor series at $R_+$, viz. \eqref{tayexpRp}, in a similar manner as for the truncation error at zero, and finally compute the total error $\Delta_b$ on the pair $(b_0,b_1)$ with the corresponding form of \eqref{toterra}.

By similar methods we can assess the error on the asymptotic parameters $A,B,C$. At $R_\infty$ we have two sources of errors. Firstly, we have to take into account that the error on the point in the $b$-plane will have evolved to a new error $\Delta_{\infty,n}$ on the initial conditions at $R_\infty$ and hence on the asymptotic parameters $A,B,C$. Secondly, we also have a truncation error $\Delta_{\infty,s}$ produced by the asymptotic series \eqref{asymp}. Again, we combine these two errors into the total error $\Delta_\infty$ on any of the parameters $A,B,C$ by using the analogous form of \eqref{toterra}. Compared to the errors on solution points in the $a$- and $b$-plane the error on the asymptotic parameters is relatively large. This is to be expected from the fact that we have to integrate over a comparatively long range, namely from $R_+$ to $R_\infty=200$, in order to determine these parameters and thus the error on the solution point in the $b$-plane can evolve to larger values. Additionally, we truncate the asymptotic series at order $R^0$, i.e. we keep only the first three orders, which also contributes to the size of $\Delta_\infty$. Despite the magnitude of $\Delta_{\infty}$, the essential message delivered by the asymptotic parameters, that the constraint \eqref{safedisc} is fulfilled, remains unaltered for all example solutions.

The various values for the errors on the three points in the $a$- and $b$-plane as well as on the asymptotic parameters are summarised in table \ref{tab:errs}.

\begin{table}[h]
\begin{center}
\begin{tabular}{|l|c|c|c|}
 \hline
  FP & green & magenta & black \\ 
  \hline
 \hline $\Delta_{a,s}$ & $4\cdot 10^{-11}$ & $3\cdot 10^{-12}$ &$3\cdot10^{-14} $ \\
 \hline $\Delta_{a,n}$ & $1\cdot 10^{-14}$ & $1\cdot 10^{-14}$ & $1\cdot 10^{-12}$  \\
 \hline $\Delta_a$ & $4\cdot 10^{-11}$ & $3\cdot 10^{-12}$ &  $1 \cdot 10^{-12}$  \\
 \hline \hline $\Delta_{b,s}$ & $2.8\cdot 10^{-7}$ & $1.4\cdot 10^{-7}$ & $1\cdot 10^{-9}$ \\
 \hline $\Delta_{b,n}$ & $3\cdot 10^{-9}$ & $2\cdot 10^{-10}$ & $3\cdot 10^{-10}$ \\
 \hline $\Delta_b$ & $2.8\cdot 10^{-7}$ & $1.4\cdot 10^{-7}$ & $1\cdot 10^{-9}$ \\
 \hline \hline $\Delta_{\infty,s}$ & $3.5\cdot 10^{-5} $ &  $5\cdot 10^{-5}$ & $2\cdot 10^{-5}$  \\
 \hline $\Delta_{\infty,n}$ & $1.8\cdot 10^{-5}$ &  $1.3\cdot 10^{-5}$ &  $3\cdot 10^{-7}$ \\
 \hline $\Delta_{\infty}$ & $4\cdot 10^{-5}$  & $5.2\cdot 10^{-5}$ & $2\cdot 10^{-5}$\\
\hline
\end{tabular}
\end{center}
\caption[Error values for the $f(R)$ example fixed point solutions.]{Error values for the parameters given in table \ref{tab:par} of the fixed point solutions shown in fig. \ref{fig:expls}. $\Delta_{a,s}$, $\Delta_{a,n}$ and $\Delta_a$ are the $a$-series error, the numerical error and the total error on the solution point in the $a$-plane. $\Delta_{b,s}$, $\Delta_{b,n}$ and $\Delta_b$ have the analogous meaning for the solution point in the $b$-plane and $\Delta_{\infty,s}$, $\Delta_{\infty,n}$ and $\Delta_{\infty}$ for the asymptotic parameters $A,B,C$. All errors are relative errors.}
\label{tab:errs}
\end{table}

We have verified that the position of these example solutions in the $a$- and $b$-plane remains unaltered if we decrease the distances $\eps_0$ and $\eps_1$ used to compute the initial conditions close to zero and $R_+$. This can be used to shrink the series errors $\Delta_{a,s}$ and $\Delta_{b,s}$ in table \ref{tab:errs} to smaller values.In fact, the value for $\Delta_{b,s}$ in the case of the black example solution has been computed with $\eps_1=10^{-4}$ instead of the usual value $\eps_1=10^{-3}$ which has been used in all other cases. In this way we can make sure that the different solution lines in the $b$-plane are not an artefact of truncating the $b$-series and would still be present in the exact case despite the fact that they can get very close to each other.

\subsection{Structure of the $a$-plane} \label{sec:aplane}
From the configuration of lines in the $a$-plane on the right hand side in fig. \ref{fig:als} it might seem that the point where the brown and the olive line from $I_-$ intersect the black solution line from $I_+$ on the singular line $\gamma_3$ represents a fixed point function $f(R)$ defined on $[R_-,\infty)$. As we will see now this is not the case. In performing the Taylor expansion \eqref{tayexp0} around zero we obtain expressions $a_n(a_0,a_1)$, $n=2,3,...$ such as \eqref{expra2}, relating the higher coefficients to the first two. As is already the case for $a_2$ these expressions have denominators that vanish along lines in the $a$-plane which are shown as dashed black lines in fig. \ref{fig:as} together with the solution lines. We have plotted the first four singular lines $\gamma_i$ corresponding to singular denominators of the coefficients $a_i$ for $i=2,3,4,5$. For example, the singular curve $\gamma_2$ is given by
\be \label{asing2} 
a_1=\gamma_2( a_0 ) = -\frac{3 a_0\left(64\pi^2 a_0 +1\right)}{192\pi^2 a_0-7}. 
\ee
Each coefficient $a_{n+1}$ receives an additional factor in its denominator compared to $a_n$ and so we get a series of singular curves. The next one is $\gamma_3$,
\be \label{asing3} 
\gamma_3 ( a_0 )= -\frac{3a_0 \left(256 \pi^2 a_0+5\right)}{768\pi^2 a_0-25},
\ee
and similar expressions hold for all other singular curves.

It is indeed straightforward to derive a general expression for the $n$-th singular curve $\gamma_n$. Substituting the Taylor series \eqref{tayexp0} into the normal form of \eqref{fp-Dario} we get $\gamma_n$ by setting the coefficient of $a_n$ at order $R^{n-2}$ to zero, $n=2,3,\dots$. The relevant 
terms can be picked out by hand from the normal form and one obtains the following condition:
\be
\frac{3}{(n-3)!}(a_1+a_0) + \frac{4}{(n-2)!}\left(192\pi^2 (a_1+a_0)a_0 -7a_1 +3a_0 \right)=0
\ee
This expression is also valid for $n=2$ if we adopt the convention to keep only the second term in that case. If we solve this condition for $a_1$ we obtain a general expression for $\gamma_n$.

Let us now carefully look at $\gamma_3$. The constraint equation we get from plugging \eqref{tayexp0} into \eqref{fp-Dario} at first order in $R$ is given by
\be \label{a3cons}
 9\left( \left( -768 \pi^2 a_0 + 25 \right)a_1 -768 \pi^2 a_0^2 -15 a_0 \right) a_3
 + \alpha_{22}a_2^2 + \alpha_{21}a_2 + \alpha_{20}=0,
\ee
where $a_2$ is known from zeroth order as in \eqref{expra2} and the $\alpha_{ij}$ depend on $a_0$ and $a_1$ only:
\begin{align}
 \alpha_{22} &= 54 \left(-128\pi^2 a_0 +3\right)\\
 \alpha_{21} &= 6 \left(-576 \pi^2 a_1^2 + 3\left(-640\pi^2 a_0 +3\right)a_1 -640\pi^2a_0^2 -7a_0\right)  \\
 \alpha_{20} &= 8 \left( -144\pi^2 a_1^3 - \left( 480\pi^2 a_0+1\right) a_1^2 -6\left(48\pi^2 a_0+1\right)a_0a_1\right)
\end{align}
As explained before, the singular curve $\gamma_3$ is then simply given by setting the coefficient of $a_3$ in \eqref{a3cons} to zero. For any solution $f(R)$ represented by a point on $\gamma_3$ we must also have
\be
\alpha_{22}a_2^2 + \alpha_{21}a_2 + \alpha_{20}=0.
\ee
Using \eqref{expra2}, we can determine a discrete set of solutions for $a_0$ and $a_1$ from the last equation together with the condition that we are on $\gamma_3$. Solving for them and calculating $a_2$ gives
\be \label{crosspt}
 p_3 = (a_0,a_1)=\left(2.6589\cdot 10^{-4},1.9684\cdot 10^{-4}\right), \qquad a_2=1.9913\cdot 10^{-4}. 
\ee
There are actually five possible solutions for the present case but we have singled out the unique non-zero solution corresponding to the point $p_3$ where the solution lines cross $\gamma_3$ in fig. \ref{fig:als} (right). All other solutions have $a_0>7\cdot10^{-3}$ and are thus far beyond the range we have developed the solution lines in. This entails that any solution line crossing $\gamma_3$ has to do so at discrete points such as $p_3$. And indeed, this is exactly what we can see in fig. \ref{fig:als}: all three solution lines, the two from $I_-$ and the one from $I_+$ intersect $\gamma_3$ at $p_3$, the only point on $\gamma_3$ in the plot  which can represent a solution. At this stage, the third coefficient $a_3$ is left undetermined at $p_3$, and all higher coefficients $a_4,a_5 \dots$ will be unique functions of $a_3$. The parameter counting arguments now tell us that we expect a discrete set of solutions on $I_+$ and another discrete set of solutions on $I_-$. This is a heuristic argument why we should not expect to find a solution at $p_3$ which is valid on $I_-\cup I_+$ as these two discrete sets will in general not overlap. 

There is a way to determine the unique Taylor expansion around zero of any fixed point solution located on a solution line running through $p_3$ which proceeds without determining the free parameter $a_3$ by numerical integration. We should expect that the initial conditions at zero depend continuously on the initial conditions at $R_+$. We can then exploit \eqref{a3cons} once more to determine $a_3$. The condition of continuity implies that if we solve \eqref{a3cons} for $a_3$ and take the limit as $a_0,a_1$ approach $p_3$ we will get a finite value.\footnote{whereas in the neighbourhood of $p_3$ the same process for any point $p \in \gamma_3$, will produce a divergent expression.} We can then proceed to obtain a unique Taylor expansion at $p_3$ by exploiting the higher order constraints by which the coefficients $a_4,a_5,...$ are fixed. Since we are taking limits of an expression depending on two variables in which both numerator and denominator become zero, it is not surprising to find the limit value depends on the direction along which we approach $p_3$. If $m$ denotes the slope of the straight line along which we take the limit, the function 
\be
m \mapsto \lim_{(a_0,a_1) \rightarrow p_3} a_3(a_0,a_1,m)
\ee
depends smoothly on $m$. Thus there is no reason to expect that the limits along the solution lines from $I_+$  and $I_-$ have to coincide. In fact, computing numerically the values of the slopes in each case and taking the limit, one finds the three different values
\be \label{a3vals2}
a_{3+}=-1.297\cdot 10^{-4}, \quad a_{3-}^b=9.518\cdot 10^{-4} \quad \text{and} \quad a_{3-}^o=7.239\cdot 10^{-4}
\ee
for the solution valid on $I_+$, the brown solution and the olive solution on $I_-$.

What we have illustrated here in the case of the singular curve $\gamma_3$ holds analogously for all the other singular lines as well. This implies that the singular lines furnish a certain structure on the $a$-plane since solution lines are not free to cross them at any point. On each singular line there are only a few points that can be determined numerically where a solution line can go through. In general, the precise Taylor expansion at zero of the solution located at one of these points will depend on the slope with which the solution line crosses the singular line.

\begin{figure}[h]
\begin{center}
 \includegraphics[scale=0.6]{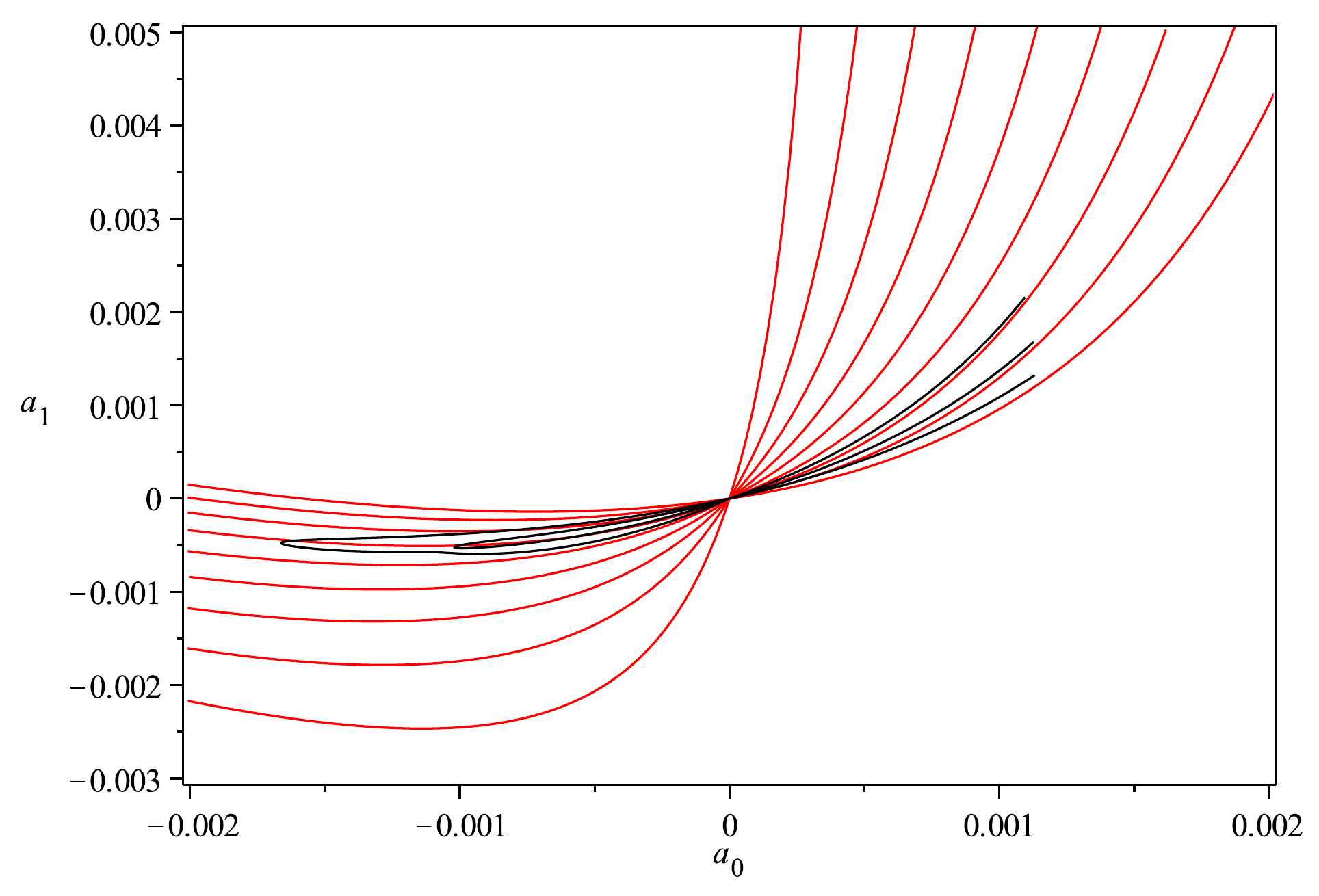}
\end{center}
 \caption[Singular lines in the $a$-plane]{The solution lines (black) of fig. \ref{fig:as} in the $a$-plane together with the singular lines (red). From bottom to top in the range of negative $a_0$ the singular lines are $\gamma_2, \gamma_3,\dots,\gamma_{10}$ stemming from the coefficients $a_2,a_3,\dots,a_{10}$ of the $a$-series \eqref{tayexp0}.}
\label{fig:singlinesbig}
\end{figure}

Let us finally comment on the reason why it is imperative to use \eqref{solcrit} with the behaviour shown in fig. \ref{fig:solcrit} as a criterion to decide if we count a point as a solution point. Consider searching for fixed point solutions on $I_+$ by integrating from $R_+$ with the approach described in sec. \ref{approach}, exploiting the corresponding form of \eqref{solcrit} at zero instead of $R_+$. Furthermore, let us first focus on the range $a_0>0$ of the $a$-plane in fig. \ref{fig:singlinesbig}, where we have plotted the first nine singular lines and the solution lines for $I_+$.

According to its definition the quantity $\delta f_\text{sol}$ can become arbitrarily small as we approach a solution line and shows singular behaviour as we cross a singular line of the $a$-series. More precisely, since the $n$-th singular line is only relevant for coefficients $a_k(a_0,a_1)$ with $k\geq n$ and we are working with an $a$-series of order $5$, $\delta f_\text{sol}$ will only show singular behaviour close to the lines $\gamma_2,\dots,\gamma_5$. Away from these two special cases $\delta f_\text{sol}$ has a typical size that varies across the $a$-plane (as long as we stay above $\gamma_2$ in fig. \ref{fig:singlinesbig} numerical integration from $R_+$ to zero is generically possible despite the threat posed by moveable singularities). Speaking in orders of magnitude, $\delta f_\text{sol}$ starts at $10^{-3}$ between $\gamma_2$ and $\gamma_3$ and drops to $10^{-8}$ in the area between $\gamma_4$ and $\gamma_5$. If we now take into account the next order in the Taylor expansion around zero, i.e. we include the term with coefficient $a_6(a_0,a_1)$, we find that its size is always several orders of magnitude smaller than $\delta f_\text{sol}$ in these regions and thus will have no effect on existing solution lines. This changes once we consider the region above $\gamma_5$ towards $\gamma_6$ and beyond. Here, the critical quantity $\delta f_\text{sol}$ becomes comparable in size to the $6$th order correction of the $a$-series, both evaluating to around $10^{-14}$ as we approach $\gamma_6$.
When searching for solutions in a region above the singular curve $\gamma_n$ it is therefore essential to use a Taylor series at zero of order at least $n+1$. For example, using a Taylor series of order $6$ at zero we found a first solution point of a solution line situated between $\gamma_5$ and $\gamma_6$ in the region of positive $a_0$ and we would expect that more solution lines can be found continuing this pattern.

Coming back to why we should not just require $|\delta f_\text{sol}|$ to be smaller than some upper bound for a point to be regarded a solution point, it is clear from this discussion that there is no sensible way to find such a bound. Moreover, the value of $|\delta f_\text{sol}|$ depends sensitively on $\eps_0$. For example, with a $5$th order Taylor series at zero and $\eps_0=5\cdot10^{-5}$ a scan of the type in fig. \ref{fig:solcrit} between $\gamma_5$ and $\gamma_6$ at $a_0 = 7\cdot 10^{-4}$ produces values of $\delta f_\text{sol}$ between $10^{-12}$ and $10^{-14}$ but $\delta f_\text{sol}$ never changes sign. Decreasing $\eps_0$ in several steps to $\eps_0=10^{-6}$, the same scan shows correspondingly decreasing values of $|\delta f_\text{sol}|$ reaching $10^{-18}$ to $10^{-21}$ in the last step. During none of these steps there is a sign change of $\delta f_\text{sol}$ and thus no evidence for a solution line. This again shows that just imposing an upper bound on $|\delta f_\text{sol}|$ cannot be sufficient. It also confirms the necessity of the $6$th order term in the $a$-series if one intends to find solutions between these two singular lines.

What we have described here in the region $a_0>0$ of the $a$-plane in fig. \ref{fig:singlinesbig} holds analogously in the region of negative $a_0$. Indeed, the bottom part of the large solution line on this side has been developed with a Taylor series of order six at zero from the beginning.

One might also wonder if there are similar singular lines in the $b$-plane arising from the expansion \eqref{tayexpRp}. An investigation into the expressions for its coefficients indeed shows that such singular lines are present. However, the solution lines we have found in the $b$-plane are all a safe distance away from them and therefore are not expected to exert any influence.

\section{Analysis of eigenoperators}
\label{sec:eops}
The result of the previous section that the fixed point equation \eqref{fp-Dario} admits several lines of solutions in the range $0\leq R < \infty$ is qualitatively very different from scalar field theory in the LPA. As mentioned previously, it is nevertheless in principle possible to experimentally measure the location on one of the solutions lines relevant for Nature and this result in itself can therefore still be accommodated in the asymptotic safety scenario. However, we now extend the analysis of the flow equation \eqref{flow-Dario} beyond its fixed point version and investigate perturbations around the fixed points of the previous section, leading to the really problematic result announced at the end of sec. \ref{par-count-Benedetti} of continuous eigenspectra.

Adapted to the present context the equation \eqref{perturbations} for perturbations of the effective action around a fixed point becomes
\begin{equation}
\label{pert-f}
f(R,t) = f(R) + \delta\!f(R) = f(R) + \eps\, \mathrm{e}^{-2\lambda t} \, v(R) \,,
\end{equation}
where we have also substituted $\lambda \mapsto 2\lambda$ due to the classical scaling dimension of $\tilde R$. Here and in the following we keep with the convention to omit the asterisk on the fixed point solution $f(R)$. Using this to perform the linearisation of the flow equation \eqref{flow-Dario} around $f(R)$ leads to the eigen-operator equation
\begin{multline} 
\label{eigen}
-\lambda\left\{ 768\pi^2v +\frac{40v'}{(R-2)f'-2f} + 
 \frac{(R^4-54R^2-54)v''-(R^3+18R^2+12)v'}{9f''-[R-3]f'+2f}\right\}
=\\
768\pi^2\left(Rv'-2v\right)+\frac{40 \left(R v''-4v'\right) - {\tilde\cT}_2\left( [R-2]v'-2v \right)}{(R-2)f'-2f}+\\
\frac{(R^4-54R^2-54)Rv'''-(R^3+18R^2+12)\left(Rv''-v'\right)+18(R^2+2)(v'+6v'')}{9f''-[R-3]f'+2f}\\
-{\tilde\cT}^{\bar h}_0\frac{9v''-[R-3]v'+2v}{9f''-[R-3]f'+2f}
\end{multline}
for $v(R)$. All terms proportional to the eigenvalue $\lambda$ are collected on the left hand side and we have used the expressions \eqref{t2} and \eqref{th} for compactness.  The first contribution on the left and right originates from the left hand side of the flow equation \eqref{flow-Dario}, the second terms on either side come from \eqref{T2} and the remaining pieces stem from \eqref{Tbh}.

\subsection{Asymptotic expansion of eigenoperators}
\label{sec:eopsasy}
In order to apply the parameter counting method to the eigenoperator equation \eqref{eigen} it is useful to first develop the asymptotic behaviour of the eigenoperators $v(R)$ and thereby understand the structure of the parameter space of solutions as $R\to \infty$. This can be done by using the tools developed in sec. \ref{sec:asy-fp} for the asymptotic analysis of the fixed point equation. We therefore write
\be
\label{veps-scaling}
v_\eps(R) = v(R/\eps)
\ee
to implement the change of variables $R\mapsto R/\eps$ and to convert the limit $R\to \infty$ into $\eps\to 0$. The power of $\eps$ as dictated by the leading behaviour of $v(R)$ on the right to enable us to take the limit $\eps \to 0$ has been omitted by linearity of \eqref{eigen}. Analogous to \eqref{fepsexp} we then expand
\be
\label{defvasymp}
v_\eps(R) = v_0(R)+\eps v_1(R) + \eps^2 v_2(R) + \dots
\ee
into the individual contributions $v_n(R)$ that in general implicitly depend on $\eps$ although they are not allowed to vanish faster than $\eps$ or diverge faster than $1/\eps$. The eigenoperator equation \eqref{eigen} is then re-written in terms of $v_\eps$ and $f_\eps$ of  \eqref{fepsexp} by substituting $R\mapsto R/\eps$, $f^{(n)}\mapsto \eps^{n-2} f^{(n)}_\eps$ and $v^{(n)}\mapsto \eps^n v^{(n)}_\eps$. After using the asymptotic expansion of fixed points \eqref{fepsexp} with \eqref{f0}, \eqref{f1}, \eqref{f2} and \eqref{defvasymp} the result is then expanded as a series in the explicit dependence on $\eps$. This gives a hierarchy of differential equations order by order in $\eps$ for the individual contributions $v_n(R)$ in \eqref{defvasymp}.

The lowest order is $\mathcal{O}\!\left(\eps^{-1}\right)$ and contains contributions from the third terms on the left and right of \eqref{eigen} coming originally from \eqref{Tbh}. It reads
\be
\label{v0eq}
R^3 v_0'''-R^2v_0''+6Rv_0'-10v_0 +\lambda\left\{ R^2v_0''-Rv_0'\right\}=0\,.
\ee
As a third order ordinary differential equation this has three linearly independent solutions. For a generic value of $\lambda$ they are
\begin{equation} \label{v0}
v_0(R) = R^2 \qquad \text{and} \qquad v_0(R) = R^p \quad \text{with} \quad p = 1-\frac{\lambda}{2}\pm\frac{1}{2}\sqrt{\lambda^2-4\lambda-16}\,,
\end{equation}
where we have normalised their coefficient to one using linearity of \eqref{v0eq}.

It is worth noting that these power-law solutions are not the ones expected from considering only the left hand side of the flow equation \eqref{flow-Dario} given by the first terms on the left and right of \eqref{eigen}. This would be valid if the quantum corrections on the right hand side of the flow equation could be neglected as $R\to \infty$ and lead to $v_0 = R^{2-\lambda}$ instead. The powers in \eqref{v0eq} are however determined by the physical scalar sector \eqref{Tbh} and the left hand side of the flow equation only becomes relevant at higher orders in $\eps$.

A further interesting aspect of the leading solutions \eqref{v0} is that $v_0(R) = R^2$ is independent of the RG eigenvalue $\lambda$ since this solution separately solves the term proportional to $\lambda$ in \eqref{v0eq}. We will see shortly however that this property does not persist at sub-leading orders with a dependence on $\lambda$ already appearing at the next order in $\eps$.

Proceeding to the next order in \eqref{defvasymp} we find that at $\mathcal{O}\!\left(\eps^{0}\right)$ the third terms on the left and right in \eqref{eigen} contribute with the previous order $v_0(R)$ but now also the first terms on either side enter the calculation. We obtain
\be
\label{v1eq}
R^3 v_1'''-R^2v_1''+6Rv_1'-10v_1 +\lambda\left\{ R^2v_1''-Rv_1'\right\}=
R^{p-1} \left( c_1+c_2\cos\ln R^2+c_3\sin\ln R^2\right)\,,
\ee
where $p$ is as in \eqref{v0} for $v_1(R)$ as the sub-leading term to $v_0(R)=R^p$ and $p=2$ if $v_1(R)$ is the sub-leading term for $v_0(R)=R^2$. The $c_i$ are functions of the RG eigenvalue $\lambda$ and the asymptotic fixed point parameters $A,B,C$. The left hand side of this equation on its own reproduces the solutions of \eqref{v0eq} which are thus discarded. If the special solution is combined with the leading order and $\eps$ set to one we obtain
\begin{multline}
\label{Rp}
v(R) = R^p + R^{p-1}\Big\{ \frac{3\left(384\pi^2[p^2-2p-25]A+p[p+30]\right)}{(p-3)(p^2-p-5)}\\
+\frac{768\pi^2(p^2-2p-5)\left[\gamma(-C,B)\cos\ln R^2+\gamma(B,C)\sin\ln R^2\right]}{(p^2-6p+13)(p^4-2p^3-5p^2+10p+25)}\Big\}+O(R^{p-2})\,,
\end{multline}
where we have abbreviated
\be
\gamma(x,y):=2(2p^2-4p-5)x+(p^3-4p^2-6p+15)y
\ee
and left the $\lambda$-dependence of $p$ implicit. For the leading term $v_0 =R^2$ the result is
\begin{multline}
\label{R2}
v(R)=R^2 + R\left\{ \frac{6(960\pi^2\lambda A-6\lambda+1)}{\lambda+4} \right.	\\
	\left.+\frac{768\pi^2}{5}\left([B-2C]\cos\ln R^2+[C+2B]\sin\ln R^2\right)\right\}+O(1)\,.
\end{multline}
The asymptotic expansion \eqref{defvasymp} can be taken to even higher orders but the result is lengthy and contains integrals that can no longer be done explicitly. The results of this section do not depend on their explicit form. We note however that the denominators in \eqref{eigen} are the same as for the fixed point equation \eqref{fp-Dario} and thus they do not lead to additional factors beyond \eqref{singpoint}. Hence \eqref{safedisc} is the only restriction on the parameters $A,B,C$.

There are several special cases for which the precise form of the asymptotic expansion changes. If $\lambda=-5/2$ then $p=2,5/2$ and the three power-law solutions in \eqref{v0} are no longer linearly independent. Similarly, if $\lambda=2(1\pm\sqrt{5})$ the square root in the definition of $p$ in \eqref{v0} vanishes and the two values for $p$ coincide as $p=\mp \sqrt{5}$ respectively. One finds that the missing third leading solution in these cases is $R^p\ln R^2$, where correspondingly $p=2$ or $p=\mp\sqrt{5}$. Other special cases are given by $\lambda=-4$, when \eqref{R2} becomes invalid, or any value of $\lambda$ for which any of the denominators in \eqref{Rp} vanish. In these cases a contribution $\propto R\ln R^2$ appears instead in the expansions. For all these exceptions higher order terms in the asymptotic series can be computed as in the generic case and in the following we will not take the explicit form of these special cases into account as they do not alter any of the conclusions. The important property we need is that for any $\lambda$ there are three linearly independent asymptotic solutions to the eigenoperator equation \eqref{eigen}.

Using this result, an arbitrary eigenoperator $v(R)$ can therefore be written as a linear combination of the three solutions obtained above,
\be
\label{vgeneral}
v(R) = \alpha_1 w_1 (R)+\alpha_2 w_2(R)+ w_3(R)\,,
\ee
where $\alpha_1,\alpha_2$ are two free parameters and we have normalised the coefficient of one of the solutions, here denoted $w_3$, to one.

As mentioned in sec. \ref{sec:asymptotic-safety} one finds complex eigenvalues in polynomial truncations of $f(R)$. Consequently we will allow $\lambda$ to be complex in the solutions \eqref{Rp} and \eqref{R2} and hence the general eigenoperator solution \eqref{vgeneral} may also be complex valued. In particular, the two coefficients $\alpha_1,\alpha_2$ will be treated as complex parameters. Since the coefficients of $v^{(n)}$ and $\lambda v^{(n)}$ in the eigenoperator equation \eqref{eigen} are real, if $v$ is a solution with eigenvalue $\lambda$ then the pair $\bar \lambda, \bar v$ is also a solution. This is used to obtain real perturbations 
\begin{equation} \label{deltaf}
\delta\!f(R) = \eps\, \mathrm{e}^{-2\lambda t} \, v(R)\,,
\end{equation}
cf. \eqref{pert-f}, by adding the two solutions coming from $\lambda, v$ and $\bar \lambda, \bar v$ together with complex conjugate coefficients $\eps$ and $\bar \eps$. The property of being relevant, marginal or irrelevant for eigenoperators introduced in sec. \ref{sec:non-pert-renorm} is now decided by the real part of $\lambda$ instead.

\subsection{Counting parameters for eigenoperators}
\label{sec:eopsparcount}
Now that we are in possession of the asymptotic behaviour of the eigenoperator solutions, we can apply the parameter counting methods of sec. \ref{par-counting} to gain qualitative insight into the parameter space of solutions.

The general asymptotic solution \eqref{vgeneral} has two parameters $\alpha_1,\alpha_2$. As mentioned at the end of the previous section, we take them to be complex since the solutions $w_i$ are complex in general. To unify the discussion here, we will do so even if $\lambda$ is such that the solutions $w_i(R)$ are real.\footnote{Note that $\lambda$ is real $\Leftrightarrow$ $|p|^2=5$.  Thus \eqref{Rp} can still be complex valued even if $\lambda$ is real.} As we see in the following this leads to the correct way of counting parameters in both cases.

If the eigenoperator equation \eqref{eigen} is expressed in normal form as in \eqref{normal} with $v$ instead of $f$ the only fixed singularities, corresponding to algebraic poles of the right hand side, come from the coefficient of $v'''$ in \eqref{eigen}. This is the same as for $f'''$ in the fixed point equation \eqref{fp-Dario} and thus gives rise to the three fixed singularities $R_c= R_-, 0, R_+$, cf. sec. \ref{par-count-Benedetti}. The corresponding constraints \eqref{pole-const} take the form
\be
\label{constraint}
h_2\, v''(R_c)+h_1\, v'(R_c)+h_0\, v(R_c) =0
\ee
due to linearity of \eqref{eigen}. The coefficients $h_i$ depend linearly on the eigenvalue $\lambda$ as $h_i = h_i^1 \lambda + h_i^0$, with the components $h_i^j$ containing some complicated dependence on the values $f(R_c), f'(R_c)$ and  $f''(R_c)$. These constraints restrict the parameters in the general solution \eqref{vgeneral} such that smoothness of $v$ is retained across $R=R_c$. Note that since we consider $v$ to be complex, equation \eqref{constraint} represents a complex constraint, i.e. one for its real and one for its imaginary part.

If we first focus on the range $R\geq 0$ as the natural domain of validity for \eqref{eigen} due to its derivation, only the two fixed singularities $R=0, R_+$ are operative. As for the fixed point equation we can be confident that the corresponding constraints \eqref{constraint} act independently and therefore determine the two parameters $\alpha_1$ and $\alpha_2$ in the general solution \eqref{vgeneral} to assume values in a discrete set. Note that this is true for any given eigenvalue $\lambda$. As a result we expect a discrete set of eigenoperator solutions for each eigenvalue and the eigen-spectrum is not quantised.

In particular, since $f(R)$ in \eqref{eigen} represents a fixed point solution well-defined for all $R\geq 0$ and the denominators in \eqref{eigen} are the same as in the fixed point equation \eqref{fp-Dario}, there are no moveable singularities in the eigenoperator equation that would further restrict the allowed values for $\lambda$ to one-dimensional subsets of the real line or complex one-dimensional subsets of the complex plane. The expectation from parameter counting is therefore that a discrete set of eigenoperator solutions exists for any value of $\lambda$.

As a result there will be an uncountable number of relevant directions for each of the fixed points found in sec. \ref{numerics}, leading to a complete breakdown of predictivity for any of the quantum field theories one might intend to construct around them.

The only way out of this conclusion in the present context is to extend the range on which the flow equation \eqref{flow-Dario} is considered to hold to the whole real line and take $-\infty<R<\infty$. Now also the third singular point becomes effective and we are provided with three constraints of form \eqref{constraint} at $R=R_c=R_-, 0, R_+$. The additional constraint from $R=R_-$ would then over-constrain the two parameters of the general eigenoperator solution \eqref{vgeneral} and fix $\lambda$ to assume values in a discrete sub-set of the complex plane. Unfortunately, as described in sec. \ref{numerics}, no globally valid fixed point solution that could be used in \eqref{eigen} has been found numerically.

\subsection{Asymptotic properties of finite perturbations}
\label{sec:finiteperts}
Presented with the situation where eigenoperator solutions are expected for every eigenvalue $\lambda$ due to the presence of two free parameters in the general asymptotic solution \eqref{vgeneral}, the question arises whether some of the asymptotic solutions can be rejected for reasons that become apparent once the analysis is extended beyond the strictly infinitesimal level of the previous section.

This is the case in scalar field theory, where eigenoperators that grow faster than a power of the field in the asymptotic regime are rejected for various reasons when they are considered as finite seeds of perturbations \cite{Morris:1998}.

Perturbing around a fixed point with the ansatz \eqref{deltaf} is valid as long as $\eps$ is small enough so that
\be
\label{dfvsf}
\df(R,t)=\eps\, \mathrm{e}^{-2\lambda t} \, v(R)\ll f(R)
\ee
holds. Since the leading asymptotic behaviour of $f(R)$ is $\propto R^2$, cf. \eqref{asymp}, this condition has to be investigated in the light of the potentially faster growing leading term $R^p$ of the asymptotic eigenoperator solution \eqref{Rp}.

In order to do so we need some remarks on the relation between $\lambda$ and $p$. Inverting the expression for $p$ in \eqref{v0} we get
\be
\label{lambda-p}
 \lambda = 2-p-5/p \,.
\ee
Using this we find
\be
\Real\lambda=2-\Real(p)-5\Real(p)/|p|^2
\ee
which entails
\begin{equation} \label{realp}
\Real p > 2 \Leftrightarrow \Real \lambda < -\frac{10}{|p|^2} \qquad \text{and} \qquad
\Real p < 2 \Leftrightarrow \Real \lambda > -\frac{10}{|p|^2}.
\end{equation}
Note that these inequalities are satisfied for both solutions $p$ to \eqref{lambda-p} for a given $\lambda$. We recall, that using these values for $p$ in \eqref{Rp} gives rise to two terms in the general solution \eqref{vgeneral} and the third is provided  by \eqref{R2}.

We first consider (marginally) irrelevant eigenoperators, i.e. we require $\Real \lambda \leq 0$. From \eqref{realp} we see that (marginally) irrelevant eigenoperators may have either $\Real p > 2$ or $\Real p\leq2$. In the latter case therefore the dominant contribution for large $R$ in \eqref{vgeneral} is given by the leading term of \eqref{R2} and is thus $\propto R^2$. Hence, if $\eps$ is small enough such that \eqref{dfvsf} is satisfied for a particular value of $R$ it will be satisfied for all $R\to \infty$ at any given $t$. Lowering the cutoff scale $k$ in $t =\ln k/k_0$ results in a decreasing exponential in \eqref{dfvsf} and the perturbation $\delta\!f$ indeed falls back into the fixed point.

If on the other hand $\Real p >2$, i.e. we have $\Real \lambda < -10/|p|^2$ from \eqref{realp}, the dominant behaviour of \eqref{vgeneral} at large $R$ is $\propto R^p$ as given by the leading term of \eqref{Rp}. Since the fixed point solution behaves as $f(R) \propto R^2$ in this regime, the condition \eqref{dfvsf} will always be violated at some large enough $R$, for any small but finite $\eps$ and at a given value of $t$. Substituting the ansatz $f(R,t)=f(R) + \delta\!f(R,t)$ into the full flow equation \eqref{flow-Dario} and using the fact that $\delta\!f(R,t) \propto R^p$ dominates at large $R$ one finds that the right hand side does not grow faster than $R^2$ and thus $\delta\!f$ has to satisfy the left hand side separately. Hence, $\delta\!f$ again assumes the separable form \eqref{deltaf} but now $\lambda$ is such that $p=2-\lambda$ is the power expected on dimensional grounds. This new eigenvalue $\lambda = 2-p$ is the effective RG eigenvalue for large $R$ and implies that these perturbations are truly irrelevant, i.e. that they too fall back into the fixed point as $t$ is lowered.

Let us now consider the relevant perturbations, leaving aside the marginally relevant ones. Since now $\Real \lambda > 0$, we have $\Real p < 2$ from \eqref{realp} and the dominant behaviour of the general solution \eqref{vgeneral} will be $\propto R^2$ coming from \eqref{R2}. Consequently, if the inequality \eqref{dfvsf} is satisfied at some finite $R$ it holds for all $R\to \infty$ at constant $t$. Together the relevant operators  give rise to perturbations of the asymptotic form $\delta\!f(R,t) = g(t)R^2$, where
\be
\label{relevant-t}
g(t)=\sum_{\lambda_n > 0} \eps_n \exp (-2\lambda_n t)
\ee
is a sum over the corresponding eigenvalues. This expression is obtained by taking the normalised contribution $w_3$ in \eqref{vgeneral} to be the leading solution $\eqref{R2}$. Lowering $t$ results in a growing $g(t)$ and \eqref{dfvsf} will eventually be violated. Once this happens the $t$-dependence of $\delta\!f(R,t)$ is instead determined by the full flow equation \eqref{fp-Dario}.

We therefore see that all of the perturbations found in the previous sections behave in the expected way and therefore there is no reason to discard any particular class of eigenoperators based on this analysis. However, as we discuss in the following chapter, one finds from a very different analysis that there is indeed a fundamental problem that concerns all eigenoperators found here.

%% file: red-ops/red-ops.tex
\chapter{Redundant operators in the $f(R)$ approximation}
\label{sec:red-ops}
The results of the previous chapter in the $f(R)$ truncation of quantum gravity as derived in Benedetti and Caravelli's paper \cite{Benedetti:2012dx} concerning global solutions to the fixed point and eigenoperator equations are surprising in the context of the functional renormalisation group. The expectation is that under normal circumstances the fixed point equations give rise to a discrete number of fixed point solutions whose eigenspectra are quantised. The lines of solutions of the previous chapter, with each fixed point on them supporting a continuous eigenspectrum, are drastically different from this standard picture.

The purpose of the following sections is to show that these unconventional findings for the $f(R)$ truncation\footnote{From here on the phrase '$f(R)$ truncation' always refers to the incarnation \cite{Benedetti:2012dx} of it.} cannot be taken as physical. Instead the observed proliferation of fixed points with non-quantised eigenspectra can be understood to reflect reparametrisation redundancy inherent in the flow equations \eqref{flow-Dario} and manifests itself in the form of redundant operators to which we come now.

\section{Redundant operators}
\label{sec:red-ops-def}
In general, the effective action $\Gamma_k[\phi]$ in the setup of non-perturbative renormalisation outlined in sec. \ref{sec:non-pert-renorm} is a functional of the (set of) field(s) $\phi$. We are free to consider a different parametrisation of this field as expressed by the infinitesimal change of field variable
\begin{equation}
\label{changevar}
\phi(x) \mapsto \phi(x)+\eps\, F[\phi](x)
\end{equation}
with an arbitrary functional $F$ depending on the field $\phi$ and coordinates $x$. Using this in the effective action gives rise to an operator
\begin{equation}\label{opredef}
\mathcal{O}_k= \frac{\delta \Gamma_k}{\delta \phi} \cdot F
\end{equation}
produced by this change of field variable and as a result such operators are clearly non-physical. When the flow equations are linearised to investigate perturbations around a fixed point as described around \eqref{perturbations} it may happen that the eigenoperator equation \eqref{eigenopequ} admits solutions that take the form given above, i.e.
\begin{equation} \label{redop}
\mathcal{O}= \frac{\delta \Gamma_*}{\delta \phi} \cdot F,
\end{equation}
where now the fixed point effective action $\Gamma_*$ is re-parametrised according to \eqref{changevar}. An operator of this form that also appears as a solution to the eigenoperator equation for a given RG eigenvalue $\lambda$ is called a redundant operator and the associated coupling a redundant coupling \cite{WR}.\footnote{Also known in the literature as an inessential operator (coupling).}

It may seem that \eqref{redop} is an eigenoperator with a well-defined RG eigenvalue and thus with an associated coupling with a well-defined mass dimension. However, its non-physical character becomes visible once it is realised that the flow equation for the effective action \eqref{equ:FRGE} is only one particular way of expressing the concept of Wilsonian renormalisation in the continuum. It can then be shown that the RG eigenvalue of a redundant operator actually depends on the particular implementation of the Wilsonian renormalisation group, i.e. transforming the flow \eqref{equ:FRGE} to an equivalent version will in general lead to a different RG eigenvalue for \eqref{redop}, cf. \cite{WR, DietzMorris:2013-2}. In particular, redundant operators cannot be uniquely classified as relevant, marginal or irrelevant. Instead, as their non-physical origin from field reparametrisations dictates, redundant operators have to be discarded and the true physical space of eigenoperators is only obtained after factoring out the subspace of redundant operators.

Note that one requirement for an operator to be redundant is that it is a solution of the eigenoperator equation for a definite RG eigenvalue $\lambda$. There are cases where an operator that comes from a redefinition of the field variable as in \eqref{redop} can be written as a linear combination of eigenoperators with different RG eigenvalues. Suppose for concreteness that it can be written as
\begin{equation}\label{O0}
\mathcal{O}_0=\frac{\delta \Gamma_*}{\delta \phi} \cdot F =\alpha_1 \mathcal{O}_1+ \alpha_2 \mathcal{O}_2,
\end{equation}
where $\mathcal{O}_1, \mathcal{O}_2$ are two eigenoperators associated to different RG eigenvalues $\lambda_1,\lambda_2$ and the coefficients $\alpha_1,\alpha_2$ are non-vanishing. Even though $\mathcal{O}_0$ is equivalent to a reparametrisation of the field and therefore non-physical, it does not have an associated RG eigenvalue $\lambda$ and hence no associated coupling as in \eqref{perturbations}. As a result it cannot be used to eliminate one of the two couplings $g_1(t),g_2(t)$ associated with $\mathcal{O}_1$ or $\mathcal{O}_2$. If \eqref{O0} is used to express $\mathcal{O}_2$ in terms of $\mathcal{O}_1$ and $\mathcal{O}_0$ to get
\begin{equation}
g_1(t)\mathcal{O}_1+g_2(t)\mathcal{O}_2=\left(g_1(t)+\frac{\al_1}{\al_2}\, g_2(t)\right)\mathcal{O}_1 + \frac{g_2(t)}{\al_2}\mathcal{O}_0 \, ,
\end{equation}
the effective coupling $g_1(t)+\al_1/\al_2 g_2(t)$ of $\mathcal{O}_1$ contains the two independent couplings $g_1, g_2$. Setting $g_2=0$ in an attempt to reduce the number of couplings due to the presence of $\mathcal{O}_0$ would correspond to wrongly discarding the coupling $g_2$ associated to $\mathcal{O}_2$.\footnote{It is useful to assume both $\mathcal{O}_1$ and $\mathcal{O}_2$ to be relevant but it is clear that taking one or both of them to be (marginally) irrelevant would not lead to a reduction of couplings either.}

Under normal circumstances in the context of functional renormalisation, redundant operators are rare. Normal circumstances again refers to a situation where the eigenspectrum of fixed points is quantised, and therefore, regarding the RG eigenvalue $\lambda$ as a parameter, only a discrete set of values for $\lambda$ leads to eigenoperator solutions. The relation \eqref{redop} then represents an additional constraint on $\lambda$ and therefore in general can be expected to have no solution. However, the constraint \eqref{redop} can fail to be an independent constraint if there is an underlying symmetry present in the flow equations and examples of this are known in scalar field theory \cite{DietzMorris:2013-2}. As we will see now, this constraint fails spectacularly in the $f(R)$ approximation to quantum gravity for a very different reason.

\section{Breakdown of the $f(R)$ approximation}
\label{sec:breakdown}
In the context of the $f(R)$ truncation, the infinitesimal field reparametrisation \eqref{changevar} becomes
\begin{equation}\label{repf(R)}
\gmn(x) \mapsto \gmn(x) + \eps\, F_{\mu\nu}[g](x)
\end{equation}
and the resulting equation \eqref{redop} for a redundant operator takes the form
\begin{equation}\label{intredopf(R)}
\mathcal{O} = \int d^dx \sqrt{g}\, F_{\mu\nu} \frac{\delta \Gamma_*}{\delta \gmn},
\end{equation}
where we work in $d$ dimensions for the moment. The equations of motion of the fixed point action $\Gamma_*=\int d^dx \sqrt{g}\,f_*(R)$ are given by
\begin{equation}
\label{eomf(R)}
\frac{\delta \Gamma_*}{\delta \gmn} = \frac{1}{2}g^{\mu\nu}f_*-R^{\mu\nu}f_*' +\nabla^\mu\nabla^\nu f_*' +g^{\mu\nu} \Delta f'_*.
\end{equation}
Since the $f(R)$ truncation of the previous chapter was derived on four-spheres, the last two terms in this expression vanish. For the same reason, the reparametrisation in \eqref{repf(R)} takes the form
\begin{equation}
F_{\mu\nu} = \zeta(R) \gmn
\end{equation}
for some function $\zeta(R)$. The integrated eigenoperator on the left in \eqref{intredopf(R)} is
\begin{equation}
\mathcal{O}=\int d^dx \sqrt{g}\, v(R),
\end{equation}
so that a redundant eigenoperator has to satisfy
\begin{equation}
\label{redopf(R)}
v(R) = \zeta(R) E_d(R),
\end{equation}
where
\begin{equation}
\label{eomRconst}
E_d(R) = \frac{d}{2}f_*(R) - Rf_*'(R)
\end{equation}
are the equations of motion of $f(R)$ gravity for constant Ricci curvature $R$ in $d$ dimensions.

Clearly, if a fixed point solution $f_*(R)$ is such that the equations of motion \eqref{eomRconst} never vanish the reparametrisation function $\zeta(R)$ in \eqref{redopf(R)} can simply be defined by dividing through by $E_d(R)$. In this case every eigenoperator $v(R)$ represents a reparametrisation of the metric field and is therefore redundant.

Setting $d=4$ it can be checked numerically that none of the fixed point solutions defined on the range $0\leq R <\infty$ of sec. \ref{numerics} lead to a zero in \eqref{eomRconst}. The inevitable consequence therefore is that none of the eigenoperators found in sec. \ref{sec:eops} are physical and that the physical space of eigenoperators for all fixed points in this version of the $f(R)$ truncation is empty. In fact, the marginal eigenoperator that represents the direction in theory space along the lines of fixed point solutions in fig. \ref{fig:as} is also redundant, implying that after factoring out reparametrisations, the lines of solutions all shrink to points. But even this discrete set of fixed points still has empty spaces of physical eigenoperators and therefore does not have physical relevance. This is a complete breakdown of the $f(R)$ approximation.

In general, if there is a solution to the equations of motion \eqref{eomRconst} the relation \eqref{redopf(R)} immediately requires any redundant eigenoperator to vanish at the same value of $R$. This is the constraint on eigenoperators represented by \eqref{redop} as mentioned at the end of the previous section. Whether such a constraint is operative or not can depend crucially on the range of $R$ for which \eqref{redopf(R)} is required to hold.

Indeed, if \eqref{redopf(R)} is imposed for all $-\infty<R<\infty$ for the $f(R)$ truncation of the previous chapter it represents a true constraint as it can be shown that the equations of motion \eqref{eomRconst} must now have a solution. To see this we note that if the asymptotic expansion of fixed point solutions \eqref{asymp} is substituted into \eqref{eomRconst} we obtain
\begin{equation}
\label{asyeom}
E_4(R) = R\left(\frac{3}{2}A+(B-2C)\cos\ln R^2 +(C+2B)\sin\ln R^2\right) + \mathcal{O}(1).
\end{equation}
This vanishes infinitely often as $R\to \infty$ unless $\frac{9}{20} A^2>B^2+C^2$ which is a stronger condition than the known cone condition \eqref{safedisc} on the asymptotic parameters $A,B,C$. But if this stronger cone condition is satisfied there nevertheless has to be a zero for $E_4(R)$ somewhere on the real line since, as discussed at the end of sec. \ref{par-count-Benedetti}, the parameter $A$ of the leading term has to be the same for the two limits $R\to \pm \infty$ (while $B$ and $C$ may be different).

As discussed in sec. \ref{sec:eops}, if a fixed point solution solution to \eqref{fp-Dario} valid on the whole real line had been found numerically its eigenoperator spectrum would automatically be quantised due to a sufficient number of fixed singularities. Thus, only a discrete and likely even only a finite set of eigenoperators $v(R)$ would be available for the redundancy relation \eqref{redopf(R)}, which as a result of the additional constraint just discussed would generically have no solution. The eigenoperator spaces of such truly global fixed point solutions would therefore certainly contain physical eigenoperators.

%% file: LPA/LPA.tex
\chapter{The LPA and background fields}
\label{ch:LPA}
The breakdown of the $f(R)$ approximation discussed in the previous chapter was traced to the fact that the equations of motion of $f(R)$ gravity on constant curvature spaces do not have a zero for any of the fixed point solutions discovered in chapter \ref{sec:f-of-R}. Given this result, a natural question to ask is whether this collapse of renormalisation group flow indicates a problem intrinsic to the asymptotic safety scenario or it represents instead an artefact created by the approximations adopted in the derivation of the RG flows of the $f(R)$ truncation. A first hint that points towards the latter situation may be taken from the study \cite{Demmel:2014sga}, where it is shown that the $f(R)$ truncation is much better behaved in the context of conformal gravity in three dimensions. Allowing all values $-\infty<R<\infty$ for the Ricci scalar, the authors find a discrete set of globally well-defined fixed point solutions.

Ultimately however, to assess the effects of approximations is certainly difficult in the context of quantum gravity alone, on the one hand since quantum gravity is uncharted territory and on the other hand since the pertaining RG flows are hard to obtain computationally. It can therefore be an enlightening strategy to take one step back by mimicking the same approximations in the familiar context of scalar field theory, where many established results are available to compare against, and where the computational complexities are manageable. This will help to understand the extent to which the approximations implemented in quantum gravity can alter the correct physical picture.

In the following we will carry out this programme in the context of a single scalar field in three dimensions and in the LPA. The focus will be on investigating the effects of the single field approximation discussed below \eqref{truncation-effaction}, which is ubiquitous in asymptotic safety and has been used for the derivation of all three $f(R)$ truncations treated in chapter \ref{sec:f-of-R} in particular.

As mentioned there, by now there are studies going significantly beyond the single field approximation in different ways such as \cite{Becker:2014qya,Manrique:2010am,Manrique:2009uh,Manrique:2010mq,Codello:2013fpa,Christiansen:2014raa,Groh:2010ta,Eichhorn:2010tb,Becker:2014jua}, but none of them allow to draw definite conclusions as to effects of the approximation for the $f(R)$ truncation.

\section{LPA with field dependent cutoff operators}\label{sec:LPA}
In order to study the effects of the single field approximation in scalar field theory, it is necessary to introduce a background field. This is of course an artificial step from the point of view of scalar field theory, but it allows us to copy the approach taken in asymptotic safety for gravity and we will show that the standard results can be recovered in the background field formalism.

\subsection{Setup and derivation of flow equation} \label{sec:LPAsetup}
One of the main reasons for the use of the background field method in gravity is the fact that the cutoff operator $\mathcal{R}_k$ in the cutoff action \eqref{equ:cutoffaction} is then allowed to depend on the background metric, cf. sec. \ref{sub:adaptations-gravity}. This ensures that the cutoff action stays quadratic in the fluctuation field which in turn ensures the general structure of the flow equation \eqref{equ:FRGE} with the inverse propagator modified by the cutoff operator on its right hand side. In order to mimic this approach in scalar field theory, we split the bare quantum field $\phi$ into a fixed background field $\pb$ and a fluctuation field $\vp$ according to 
\begin{equation} \label{bgrscalar}
 \phi = \pb + \vp.
\end{equation}
The path integral \eqref{partfunc} is now shifted so that it is the fluctuation field $\vp$ which is integrated over. As in gravity, the flow equation \eqref{equ:FRGE} still holds even if the cutoff operator depends on the background field $\mathcal{R}_k=\mathcal{R}_k\!\left(-\partial^2,\pb \right)$, provided the cutoff action \eqref{equ:cutoffaction} is quadratic in the fluctuation field. The effective action is now a functional of both the classical fluctuation field $\vp^c$ and the background field, $\Gamma_k = \Gamma_k[\vp^c,\pb]$, and the Hessian on the right hand side of the flow equation \eqref{equ:FRGE} is taken with respect to the classical fluctuation field.

Employing the background field method in this way leads to greater flexibility in the choice of cutoff. We note that this flexibility has very importantly been made use of in the derivation of the $f(R)$ truncation in gravity as discussed around \eqref{Deltas}. However, it comes at the price of having to work with an effective action depending on two arguments: the fluctuation and the background field. The single field approximation remedies that as we now describe in detail for scalar field theory, cf. the analogous discussion in sec. \ref{sub:adaptations-gravity}. Let us decompose
\begin{equation} \label{approx}
 \Gamma_k[\vp^c,\pb]= 
 \Gamma_k[\phi^c]+ \hat \Gamma_k[\vp^c,\pb],
\end{equation}
where we have defined $\phi^c=\pb+\vp^c$ as the classical counterpart of the full quantum field, and $\Gamma_k[\phi^c] = \Gamma_k[0,\phi^c]$ is defined by substituting the total classical field for the background field, cf. \eqref{truncation-effaction}. By definition therefore, the remainder vanishes for vanishing (classical) fluctuation field: $\hat \Gamma_k[0,\pb]=0$. In this sense, $\hat \Gamma_k$ captures the deviation in the effective action from $\phi^c = \pb$. The single field approximation consists in entirely neglecting this second term $\hat \Gamma_k$, setting it to zero from the outset. Terms contained in $\hat \Gamma_k$ are therefore not taken into account in the Hessian on the right hand side of the flow equation \eqref{equ:FRGE}, nor are the corresponding running couplings included on the left hand side. After the Hessian of $\Gamma_k[0,\phi^c]$ in \eqref{equ:FRGE} has been evaluated we are allowed to set $\vp^c=0$ in all subsequent calculations, thereby identifying the total classical field and the background field. The advantage of the single field approximation is that we now have to deal with an effective action depending on only one field, $\phi^c$.

We stress again that our motivation for adopting this approximation arises entirely from our aim to follow the approach adopted in most of the asymptotic safety literature and, as mentioned above, from the underlying role this approximation assumed in the context of chapters \ref{sec:f-of-R} and \ref{sec:red-ops}.

In the single field approximation we can now take the effective action to be of form $\Gamma_k[\phi^c]$. The LPA in scalar field theory is characterised by truncating it to
\begin{equation} \label{equ:LPA}
 \Gamma_k[\phi^c] = \int_x \left( \frac{1}{2} \left(\partial_\mu \phi^c \right)^2 + V(\phi^c) \right),
\end{equation}
discarding all higher derivative terms and setting the wavefunction renormalisation to one (cf. \cite{Bervillier:2013}, however). We retain a general potential $V(\phi^c)$, in particular it need not be symmetric under $\phi^c \mapsto -\phi^c$. For simplicity of notation, the $k$ dependence of the potential is left implicit.
 
Let us now comment on the choice of cutoff which is a crucial feature in this context. We choose to work with the following cutoff operator, a generalisation of the optimised cutoff \cite{Litim:2001}:
\be 
\label{equ:cutoffh}
\mathcal{R}_k\!\left(-\partial^2,\pb(x)\right) = \left(k^2+\partial^2-h(\pb)\right)\theta\!\left(k^2+\partial^2-h(\pb)\right)\,.
\ee
Here $h$ is some general function of the background field $\pb(x)$ and RG time $t$, and it may or may not itself depend on the effective action \eqref{equ:LPA} through $V$. In order to derive the flow equation originating from this form of cutoff we note that within the LPA the field can effectively be treated as spacetime independent  \cite{Morris:1994ie} and thus in momentum space the cutoff operator is diagonal:
\begin{equation*}
 \mathcal{R}_k\!\left(p^2,\pb\right) = \left(k^2 - p^2 -h(\pb)\right) \theta\! \left(k^2 - p^2 -h(\pb)\right)\,.
\end{equation*}
On the right hand side of \eqref{equ:FRGE} we need the time derivative of this, given by
\begin{equation}
 \label{equ:cutoffh-tder}
 \partial_t \mathcal{R}_k = \left(2k^2 -\partial_t h(\pb)\right) \theta \left(k^2 - p^2 - h(\pb)\right).
\end{equation}
Combining this with the second functional derivative of \eqref{equ:LPA}, the evaluation of \eqref{equ:FRGE} leads to
\begin{equation}
 \label{equ:flowLPA}
 \partial_t V - \frac{1}{2}(d-2) \phi V' +dV = 
 \frac{(1-h)^{d/2}}{1-h+V''}\left(1-h-\frac{1}{2}\partial_t h +\frac{1}{4}(d-2)\phi h'\right)\theta(1-h),
\end{equation}
where as usual primes denote derivation with respect to the argument $\phi$. In the process we have absorbed a constant by a field redefinition and adopted scaled variables according to
\begin{equation} \label{scaledvars}
 \tilde \phi = k^{\frac{1}{2}(2-d)} \phi^c, \qquad \tilde V(\tilde \phi) =k^{-d}V(\tilde \phi)
  \qquad \text{and} \qquad \tilde h(\tilde \phi)=k^{-2}h(\tilde \phi),
\end{equation}
followed by an immediate renaming by dropping the tilde.

We remember that in the single field approximation we can set $\vp^c=0$ in $\phi^c=\vp^c+\pb$ once the Hessian of the ansatz \eqref{equ:LPA} has been taken. Consequently, the only field variable appearing in the flow equation \eqref{equ:flowLPA} is the scaled total classical field $\phi$. In particular, since we no longer distinguish between the background and total classical fields, all appearances of $h$ in the flow equation \eqref{equ:flowLPA} have the dependence $h=h(\phi)$.

We will study the consequences of a background field dependent cutoff operator under adoption of the single field approximation for two qualitatively different choices of $h$. As our first choice, we specialise to the case where 
\begin{equation} \label{cutoff1}
h=\alpha k^{4-d}\pb^2,
\end{equation}
$\al$ being a dimensionless free parameter introduced by hand which controls the field dependence of
the cutoff operator. This leads to a cutoff operator with a purely explicit dependence on the background field and we analyse this case in sec. \ref{sec:gencutoff}.

Our second choice will give rise to a cutoff operator with an implicit background field dependence through the potential by setting 
\begin{equation}\label{cutoff2}
h=\al V''(\pb),
\end{equation}
the dimensionless parameter $\al$ playing the same role as in the previous example. Our investigation pertaining to this choice is presented in sec. \ref{sec:LPALitim}.

When the single field approximation is employed it can be necessary to impose an additional constraint on the cutoff operator to ensure that the correct one-loop $\beta$-function is reproduced. For the flow in gauge theory considered in \cite{Gies:2002af} the cutoff operator $\cutoff \!\left(\Gamma^{(2)}\right)$ depends on the two-point function itself, leading to what is known as a spectral flow, and the additional constraint is given by a particular limit of the cutoff operator as $\Gamma^{(2)} \to 0$. It is  therefore interesting to investigate how the perturbative one-loop $\beta$-function of scalar field theory is affected by our choices of cutoff operator in the single field approximation, but we note that the fact that an additional constraint on the cutoff operator has to be implemented is itself a first sign of possible non-physical behaviour.

We first consider the standard case by setting $d=4$ and $h=0$ in \eqref{equ:flowLPA}. The Gaussian fixed point is $V=V_*=1/4$ and the perturbative flow of the $\phi^4$ coupling can be found by setting
\be
\label{perturbative}
V={1\over4}+{\lambda(t)\over4!} \,\phi^4
\ee
and expanding the right hand side of \eqref{equ:flowLPA} in $\lambda$. Matching the coefficients of $\phi^4$ leads to the perturbative $\beta$-function
\be
\label{beta-function}
\beta(\lambda) \equiv \partial_t\lambda= 6\lambda^2\,.
\ee 
This is the standard result after reinstating the constant $1/2(4\pi)^2$ on the right hand side that was absorbed by a field redefinition in \eqref{equ:flowLPA}.

If we now take $h$ of form \eqref{cutoff1} with $\alpha \neq 0$ there is no longer a Gaussian fixed point solution, as we confirm in sec. \ref{sec:gencutoff} and perturbation theory cannot be recovered in this simple way.

On the other hand, if we take the cutoff form \eqref{cutoff2} in the flow equation \eqref{equ:flowLPA} we see that the Gaussian fixed point $V_*=1/4$ is a solution for any value of $\alpha$. Using the ansatz \eqref{perturbative} and expanding the right hand side of \eqref{equ:flowLPA} leads to the standard beta function \eqref{beta-function}, irrespective of the value taken by $\alpha$. This shows that for the spectral flow given by the choice \eqref{cutoff2} there is no additional constraint needed to recover the standard one-loop $\beta$-function. Thus, at this level in perturbation theory, no non-physical effects are visible but as we will see in sec. \ref{sec:LPALitim} there are dramatic non-physical alterations at the non-perturbative level.

In the context of the $f(R)$ truncation \cite{Benedetti:2012dx} leading to a collapse of the space of eigenoperators as discussed in chapter \ref{sec:red-ops}, the cutoff operators also depend on the Laplacian shifted by a field dependent quantity to implement the rule \eqref{typeI} for \eqref{Deltas}. Although the situation there is much more involved in the sense that the field dependence of the cutoff operators is a mixture of the two choices \eqref{cutoff1} and \eqref{cutoff2}, and the other two versions of the $f(R)$ truncation \cite{Codello:2008,Machado:2007} feature a rather different field dependence of the cutoffs,
we expect to capture the essence of implementing field dependent cutoffs and to be able to exemplify possible consequences.

\subsection{The Wilson-Fisher and Gaussian fixed points} \label{sec:WF}
In this section we first recover well-known results about three-dimensional single component scalar field theory  by setting $h=0$ in \eqref{equ:flowLPA}, in order to then investigate what are the effects of using a generalised cutoff operator of type \eqref{cutoff1} or \eqref{cutoff2} with $\al \neq 0$ on both the fixed point structure and the eigenspectra in the following sections. Correspondingly, we will set $d=3$ from now on.

For $h=0$ the cutoff operator \eqref{equ:cutoffh} reduces to the standard version of the optimised cutoff \cite{Litim:2001} which has been used before in the context of three dimensional scalar field theory. We are here going to reproduce results that can be found for example in \cite{Litim:2002cf} but we nevertheless go through their derivation as our approach differs from the one used there and as we will follow the same steps for the more complicated flow equation \eqref{equ:flowLPA} when $h \neq0$ in the next sections.

Eliminating the field dependence in the cutoff operator by setting $h=0$ in \eqref{equ:flowLPA} and focussing on scale invariant fixed point potentials, we obtain the fixed point equation
\begin{equation}
 \label{equ:fpWF}
 3V_*- \frac{1}{2} \phi V'_* = \frac{1}{1+V''_*}.
\end{equation}

We are first interested in the behaviour of the potential at large field values. Assuming that at zero order we  can neglect the quantum corrections on the right hand side of \eqref{equ:fpWF} we find $V_0(\phi)=A\phi^6$ as the solution of the left hand side, $A$ being a real constant. Of course, substituting this into the right hand side, we see that for $A\ne0$ the corrections are indeed subleading. At first order, we use $V_*(\phi)=V_0(\phi)+V_1(\phi)$ on the left hand side of \eqref{equ:fpWF} and equate this to the quantum corrections stemming from $V_0$ only. At this order we keep only the leading term in a Taylor expansion in $1/\phi$ of the right hand side which leads to
\begin{equation}
 \label{equ:Vasy1}
 3V_1- \frac{1}{2} \phi V_1' = \frac{1}{30A\phi^4}\,.
\end{equation}
The general solution of this equation just reproduces the leading term $V_0$ but the special solution supplies us with the subleading contribution $V_1(\phi)=1/(150A\phi^4)$. Continuing with this process the asymptotic series for large $\phi$ can be developed to arbitrary order, the first terms of which are:
\begin{equation}  \label{equ:VWFasy}
V_*(\phi) = A{\phi}^{6}+{\frac {1}{150}}\,{\frac {1}{A{\phi}^{4}}}-{\frac {1}{6300
}}\,{\frac {1}{{A}^{2}{\phi}^{8}}}+{\frac {1}{243000}}\,{\frac {1}{{A}
^{3}{\phi}^{12}}}-{\frac {1}{67500}}\,{\frac {1}{{A}^{3}{\phi}^{14}}} + \dots 
 \end{equation}
We note that this series depends on only one free parameter $A$ and, after mapping onto the corresponding variables, the first two terms of this expansion coincide with ref. \cite{Litim:2002cf}.

\begin{figure}[ht]
  \begin{center}
  $
   \begin{array}{cccc}
     \includegraphics[width=0.45\textwidth,height=0.295\textheight]{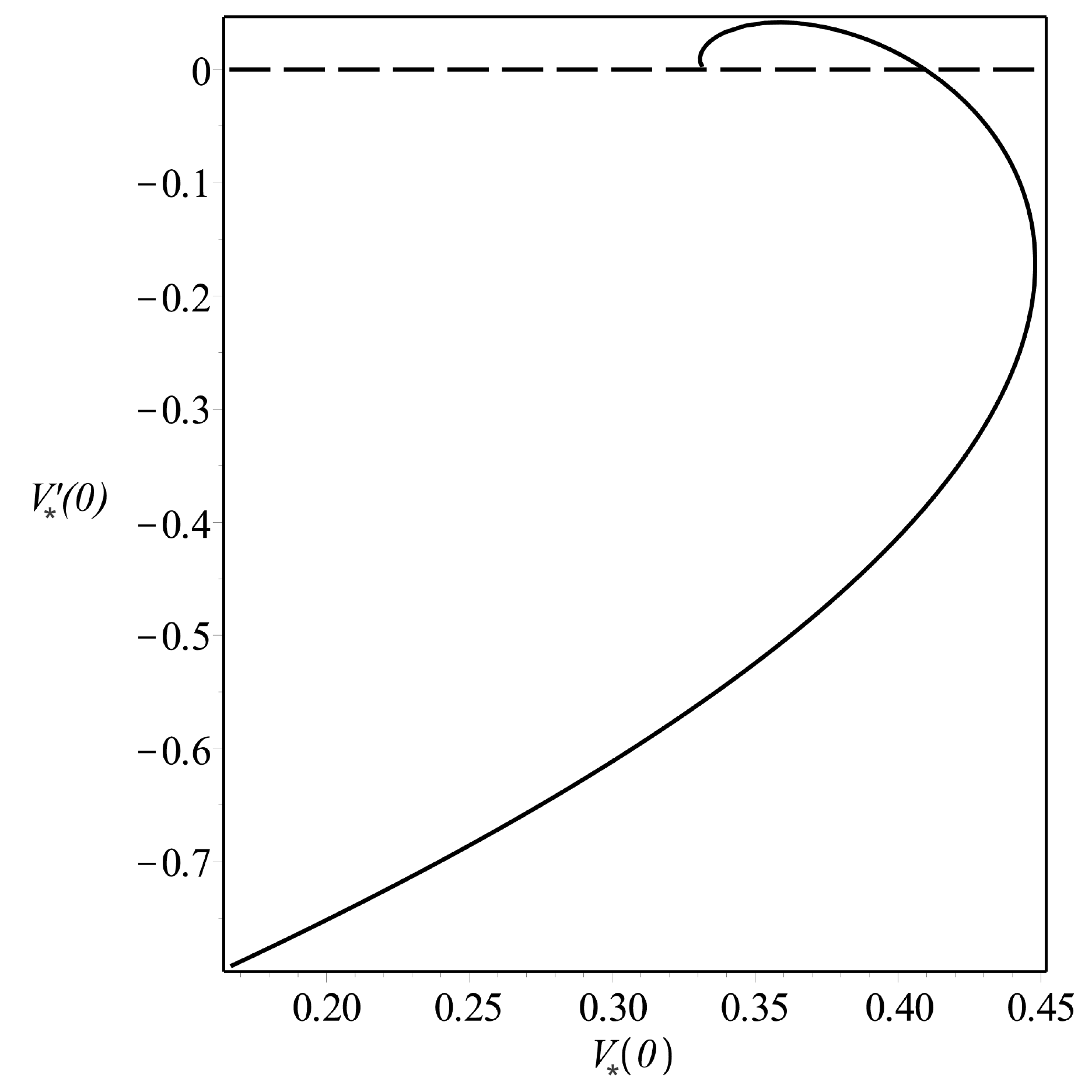} & \hfill &  \hfill &
     \includegraphics[width=0.4\textwidth,height=0.3\textheight]{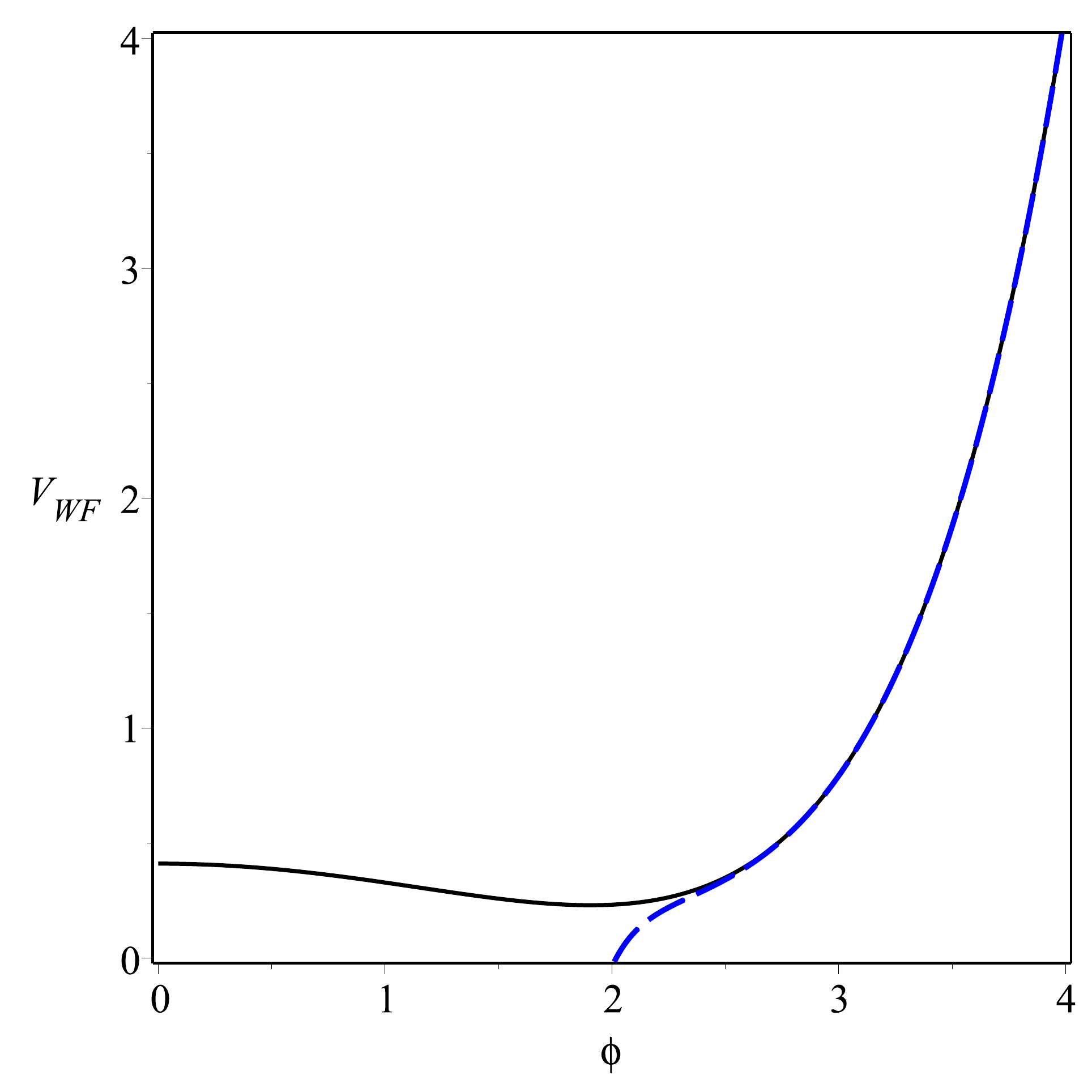} 
   \end{array}$
  \end{center}
   \caption[The Wilson-Fisher fixed point]{Varying the asymptotic parameter $A$ from $A\to0$ at the top to $A=50$ at the bottom we plot $V'_*(0)$ against $V_*(0)$ obtained by numerical integration (left). The Wilson-Fisher fixed point 	potential in black and solid (right) together with the asymptotic approximation in blue and dashed.}
\label{fig:WF}
\end{figure}

The strategy we have pursued is to numerically integrate \eqref{equ:fpWF} starting with initial conditions calculated from the asymptotic series \eqref{equ:VWFasy} at some large enough $\phi=\phi_\infty>0$ such that the series is well convergent.\footnote{As we can see from \eqref{equ:VWFasy} the value of $\phi_\infty$ depends on $A$ but $\phi_\infty=30$ results in accurate enough initial conditions for values of $A$ as small as $A=10^{-5}$.} We then record the values of $V_*(0)$ and $V'_*(0)$ as a function of the asymptotic parameter $A$. The result is shown in fig. \ref{fig:WF}, where we have varied $A$ from $A\to0$ at the top end of the curve to $A=50$ at the bottom end of the curve.

In principle (and later we will see in practice) there can exist fixed point potentials with no symmetry under $\phi\leftrightarrow-\phi$. However, since the fixed point differential equation \eqref{equ:fpWF} is symmetric under $\phi\leftrightarrow-\phi$, there would then automatically be two solutions: $V_*(\phi)$ and $V_*(-\phi)$. Therefore solutions with no symmetry can be recognised by the fact that in a plot of $V'_*(0)$ {\it vs.} $V_*(0)$, there are pairs of points with the same $V_*(0)$ but equal and opposite $V'_*(0)$. Fixed point solutions that are odd under $\phi\leftrightarrow-\phi$ would additionally have $V_*(0)=0$ of course, whilst even potentials are associated to only one point since $V'_*(0)=0$. 

We can see from the plot in  the left hand panel of fig.~\ref{fig:WF}, that there are in fact just two fixed point potentials and both of these are even. The first solution appears at $A=0$ and this corresponds to the Gaussian fixed point whose exact solution is simply $V_\mathrm{G}(\phi)=1/3$.

The second solution is the Wilson-Fisher fixed point with (to 5 significant figures):
\begin{equation}
\label{equ:WFpars}
 A_{\mathrm{WF}}=0.00100  \qquad \text{and} \qquad V_\mathrm{WF}(0)=0.40953
\end{equation}
and according to the plot in fig. \ref{fig:WF} there are no other  fixed point solutions for larger $A$.
The Wilson-Fisher fixed point potential is plotted on the right in fig. \ref{fig:WF} (black and solid line)
together with the asymptotic approximation given by \eqref{equ:VWFasy} with $A=A_\mathrm{WF}$ (blue and dashed).

We now turn our attention to perturbations around the Gaussian and the Wilson-Fisher fixed points by writing
\begin{equation}
 \label{equ:perts}
 V(\phi,t)=V_*(\phi) + \eps v(\phi) \exp(-\lambda t),
\end{equation}
where $V_*$ stands for either of the fixed point potentials and $\eps$ is infinitesimal and dimensionless, cf. the discussion around \eqref{perturbations}. Substituting the ansatz \eqref{equ:perts} into the full flow equation \eqref{equ:flowLPA} with $h=0$ and linearising in $\eps$ we arrive at the eigenoperator equation
\begin{equation}
 \label{equ:WFeops}
 (\lambda-3) v  + \frac{1}{2} \phi \,v' ={\frac {v'' }{ \left( 1+V_*''  \right) ^{2}}}.
\end{equation}
At the Gaussian fixed point, where $V_*''=0$, this eigenoperator equation can be solved exactly and one recovers the marginal operator $v(\phi)=\phi^6 + subleading$ and  the expected six relevant operators with eigenvalues $\lambda=1/2,1,\dots,5/2,3$. In addition to this we find the infinite sequence of irrelevant operators with leading pieces  $\phi^7,\phi^8,\dots$ corresponding to negative half integer values of $\lambda$.

For the Wilson-Fisher fixed point $V_*'' \neq0$ and the eigenoperator equation cannot be solved analytically anymore. In order to find its solutions numerically, we follow a similar strategy as for the fixed point equation. The first step consists in developing the large field behaviour of the eigenoperator $v(\phi)$. In doing so we expand $v(\phi)=v_0(\phi)+v_1(\phi)+\dots$, where $v_0(\phi)$ solves the left hand side of \eqref{equ:WFeops} and $v_n(\phi)$ is obtained by substituting $v(\phi)=v_0(\phi)+v_1(\phi)+\dots+v_n(\phi)$  into the left hand side and the same expansion of one lower order into the right hand side of \eqref{equ:WFeops}, keeping only the leading term of a Taylor expansion in $1/\phi$ of the right hand side and solving the resulting differential equation. Of course, in this process we have to use the asymptotic expansion \eqref{equ:VWFasy} of $V_*(\phi)$ to corresponding order. The first few terms in the resulting asymptotic series for the eigenoperators take the form
\begin{multline}
\label{equ:WFeopsasy}
 v(\phi) = {|\phi|}^{-2\,\lambda+6}
	      -{\frac {1}{4500}}\,{\frac { \left( 2\,\lambda-5\right) \left( 2\,\lambda-6 \right) }{A_\mathrm{WF}^{2}}} {|\phi|}^{-2\,\lambda-4} \\
	      +{\frac {1}{94500}}\,{\frac { \left( 2\,\lambda-5 \right) \left( 2\,\lambda-6 \right)}{A_\mathrm{WF}^{3}}}{|\phi|}^{-2\,\lambda-8} 
	      +\dots
\end{multline}
where we have normalised the eigenoperator such that the first term in this expansion has unit coefficient
and $A_\mathrm{WF}$ is given in \eqref{equ:WFpars}.

Using the asymptotic series \eqref{equ:WFeopsasy} for any given $\lambda$ to calculate the initial conditions at some sufficiently large $\phi_\infty$, we numerically integrate \eqref{equ:WFeops} to determine $v'(0)$. Since the fixed point potential $V_\mathrm{WF}$ is an even function of $\phi$, the eigen-perturbations can be classified as even or odd functions of $\phi$. To begin with we are interested in even eigen-perturbations, so we impose the condition $v'(0)=0$. This will quantise the eigenspectrum leaving us with a discrete set of eigenoperators. Apart from the vacuum energy $v=1$ with $\lambda=3$ which is clearly an exact solution of \eqref{equ:WFeops}, the only relevant eigen-perturbation we find has the eigenvalue
\begin{equation*}
 \lambda_\mathrm{WF}=1.539,
\end{equation*}
corresponding to a critical exponent $\nu=1/\lambda=0.649$ which is within 3\% of the results obtained at 
$O(\partial^2)$ of the derivative expansion and other approximation methods, such as Monte Carlo methods, resummed perturbative calculations or the scheme proposed by Blaizot, M\'endez-Galain and Wschebor \cite{Benitez:2011xx}.

For odd eigen-perturbations the spectrum will instead be quantised by the condition $v(0)=0$. The only two relevant odd eigenoperators are then $v(\phi)=\phi$ with $\lambda=5/2$ which solves the left hand side
of \eqref{equ:WFeops} and $v(\phi)=V_\mathrm{WF}'(\phi)$ with $\lambda=1/2$. The latter however is a redundant
eigenoperator that corresponds to the change of field variable $\phi \mapsto \phi + const.$, cf. sec. \ref{sec:missingvacua}.

\subsection{Cutoff operators with explicit field dependence} \label{sec:gencutoff}
Having confirmed in the previous section that our methods reproduce known results, we are now going to investigate how the fixed point structure and the critical exponents change if we consider the flow equation \eqref{equ:flowLPA} with \eqref{cutoff1} and $\al \neq 0$. Apart from a proliferation of fixed points,
we will also encounter a partial breakdown of the LPA in the sense of sec. \ref{sec:breakdown}.

The fixed point equation we are now interested in solving is obtained from \eqref{equ:flowLPA} by substituting
the dimensionless version of \eqref{cutoff1}, 
\begin{equation} \label{equ:FPgen}
 3V_*- \frac{1}{2} \phi V'_*  = 
 \frac{\left(1-\al\phi^2\right)^{\frac{3}{2}} \left(1- \frac{1}{2}\al \phi^2 \right)}{1-\al \phi^2+V''_*}\,\theta\!\left(1-\al\phi^2\right)
\end{equation}
and we will let $\al$ take positive and negative values in the following. Due to the fractional exponent on the right hand side however, the analysis proceeds along the same lines as in the previous section only for $\al<0$. In particular, the asymptotic expansion for large $|\phi|$ depends on only one free parameter $A$,
\begin{multline}
 V_*(\phi) = A \phi ^{6}+{\frac {1}{150}}\,{\frac {{|\alpha|}^{5/2} \left| \phi \right| }{A}}
 +{\frac {1}{6300}}\,{\frac {{|\alpha|}^{3/2} \left( 105\,A-{\alpha}^{2} \right) }{{A}^{2} \left| \phi \right| }} \\
 +{\frac {1}{486000}}\,{\frac {\sqrt {|\alpha|} \left( 6075\,{
A}^{2}-270\,{\alpha}^{2}A+2\,{\alpha}^{4} \right) }{{A}^{3} 
 \left| \phi \right| ^{3}}} + \dots .
\end{multline}

If $\al>0$ the right hand side of the fixed point equation \eqref{equ:FPgen} becomes zero at $|\phi|=\phi_c=1/\sqrt{\al}$ and vanishes identically for all $|\phi|>\phi_c$. This is a direct consequence of the field dependent version of the optimised cutoff we are using. For $|\phi|>\phi_c$ the solution of \eqref{equ:FPgen} is therefore simply given by $V_*(\phi)=A\phi^6$ and this solution has to be matched onto the solution obtained from the full fixed point equation for $|\phi|<\phi_c$. Due to the fractional power on the right hand side of \eqref{equ:FPgen} it is problematic to start numerical integration directly at the matching point $\phi=\phi_c$. We therefore develop the potential into a generalised Taylor expansion around the matching point according to
\begin{equation} \label{equ:Vtaylor}
 V_*(\phi)= a_0 + a_1\left(\frac{1}{\sqrt{\al}}-\phi \right)^{\frac{1}{2}} +\frac{a_2}{2!}\left(\frac{1}{\sqrt{\al}}-\phi \right)
 + \frac{a_3}{3!}\left(\frac{1}{\sqrt{\al}}-\phi \right)^\frac{3}{2}+ \frac{a_4}{4!}\left(\frac{1}{\sqrt{\al}}-\phi \right)^2 + \dots.
\end{equation}
To do this in practice it is convenient to perform the change of variable $\phi \mapsto 1/\sqrt{\al}-u^2$
before substituting this expansion into \eqref{equ:FPgen}\footnote{The form of the generalised Taylor expansion \eqref{equ:Vtaylor} is dictated by the fractional power in \eqref{equ:FPgen}.}. For the fixed point equation to be satisfied order by order in $u$ we then find that $a_0$ and $a_2$ are free parameters, corresponding to two initial conditions at the matching point, whereas $a_1=a_3=0$ and all higher coefficients $a_4,a_5,\dots$ are functions of $a_0$ and $a_2$. Since we want to match the solution \eqref{equ:Vtaylor} at $\phi=1/\sqrt{\al}$ to the asymptotic solution $V_*(\phi)=A\phi^6$ we choose the initial conditions accordingly:
\begin{equation}\label{equ:a0a2}
a_0=\frac{A}{\al^3} \qquad \text{and} \qquad a_2 = -\frac{12A}{\al^\frac{5}{2}}.
\end{equation}
In this way,\footnote{Actually, it is convenient to make the identification \eqref{equ:a0a2} before substituting \eqref{equ:Vtaylor} into \eqref{equ:FPgen} to avoid the appearance of vanishing denominators at intermediate stages.} the higher coefficients $a_4,a_5,\dots$ become functions of $A$ and $\al$, 
\begin{equation} \label{asyexpcoeffs}
a_{{4}}=360\,{\frac {A}{{\alpha}^{2}}}, \qquad a_{{5}}={\frac {16\sqrt {2}}{5}}\,\frac{{
\alpha}^{{\frac {13}{4}}}}{{A}}, \qquad a_{{6}}=-{\frac {8}{75}}\,{
\frac {135000\,{A}^{4}+{\alpha}^{10}}{{\alpha}^{3/2}{A}^{3}}}, \dots
\end{equation}
We note that this form of $a_4$ leads to automatic matching of the second derivative $V''_*(\phi)$ at $\phi=\phi_c$ but higher derivatives of the potential will diverge as we approach the matching point from the left. This is clearly an artefact of the cutoff choice and does not have any physical significance. We will see now that despite this non-smoothness of the potential the overall picture emerging from our analysis is consistent.

\begin{figure}[ht]
  \begin{center}
  $
   \begin{array}{ccccc}
     \includegraphics[width=0.3\textwidth,height=0.225\textheight]{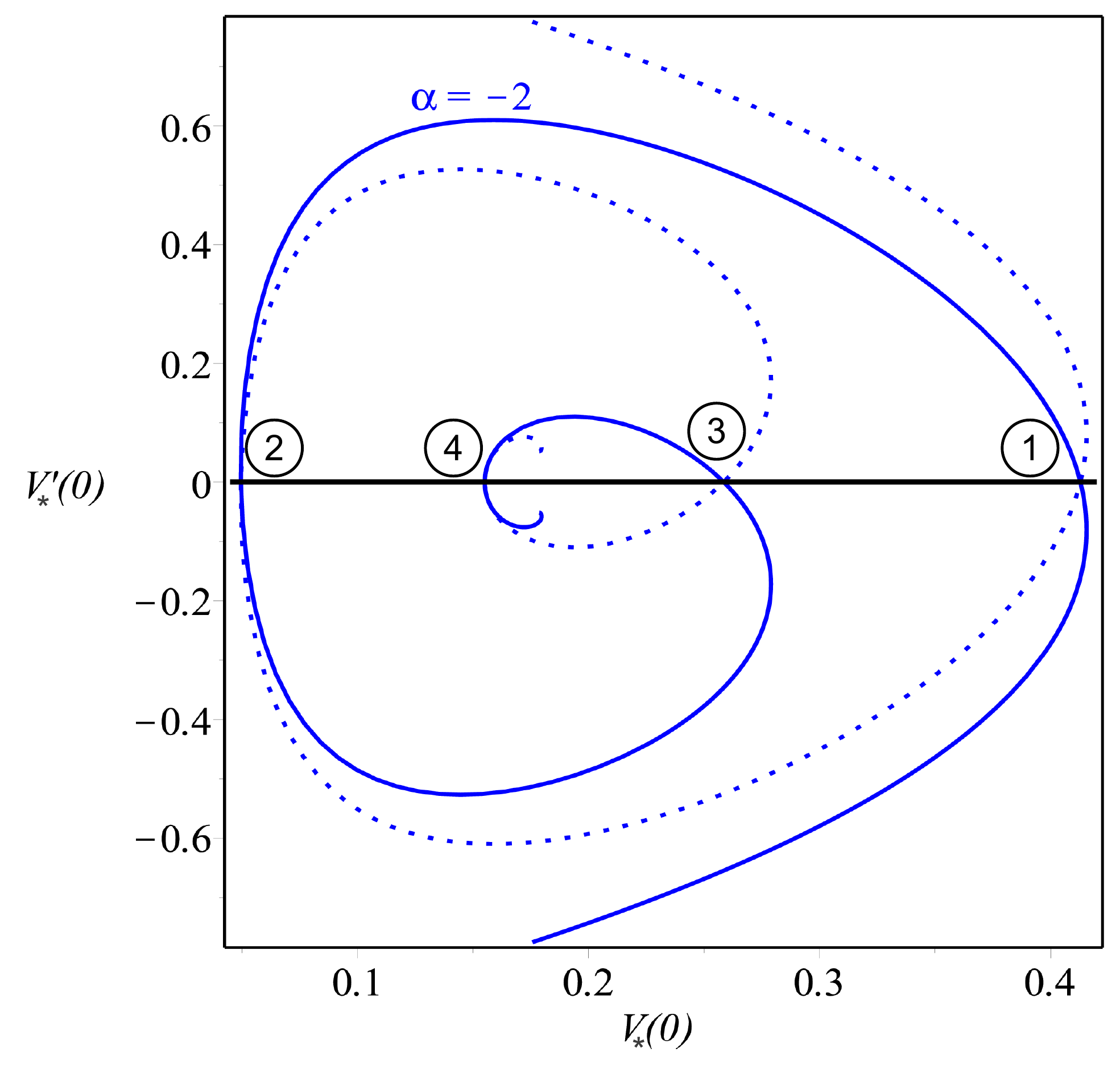}  & &
     \includegraphics[width=0.3\textwidth,height=0.225\textheight]{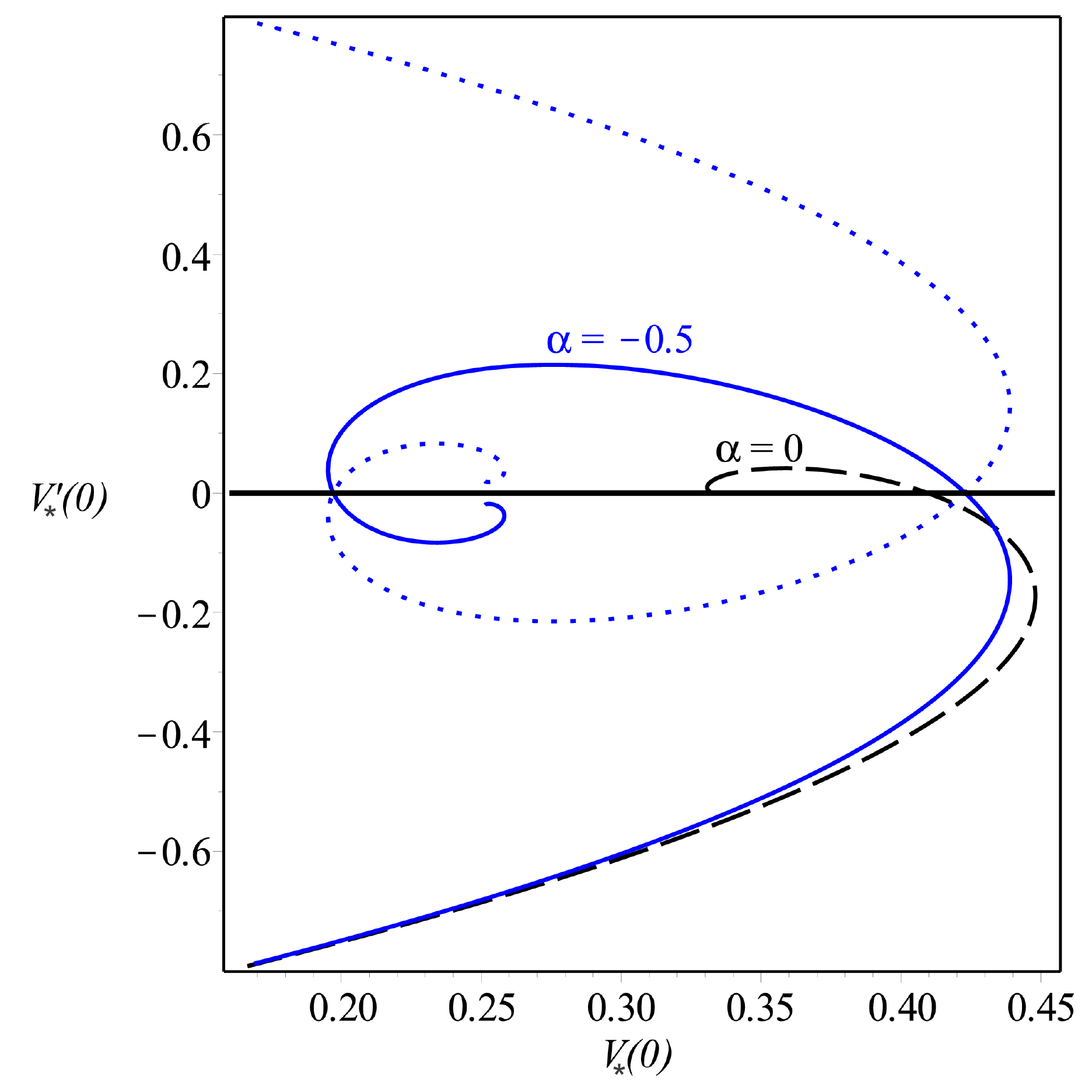} && 
     \includegraphics[width=0.3\textwidth,height=0.225\textheight]{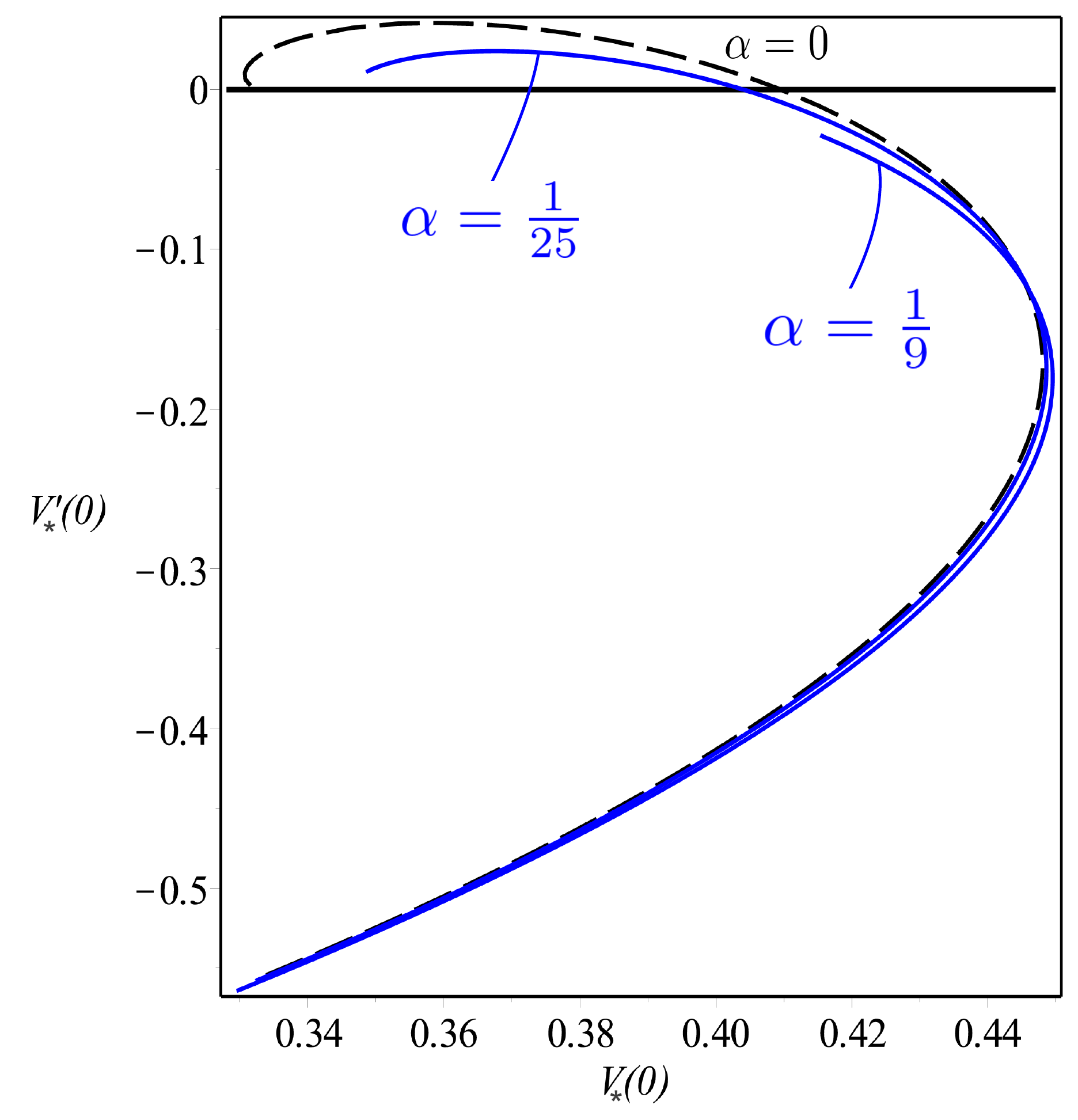}
   \end{array}$
  \end{center}
   \caption[Finding scalar field fixed point potentials with field dependent cutoffs] {Varying the asymptotic parameter $A$ for different values of $\al$ leads to these spirals in $V_*(0),V'_*(0)$ - space. In dashed and black we have reproduced the case $\al=0$ from fig. \ref{fig:WF} whereas the blue, solid curves belong to various values of $\al \neq0$. The blue dotted curves correspond to the blue curves reflected in the $V_*(0)$ axis. From the discussion below \eqref{equ:VWFasy}, we know that fixed point solutions correspond to the points where the blue curves and dotted-blue curves cross. }
\label{fig:gen}
\end{figure}

After these preparations and just like in the case of $h=0$ in the previous section we now vary the asymptotic parameter $A$ and numerically integrate the fixed point equation \eqref{equ:FPgen} to find $V_*(0)$ and $V'_*(0)$ as functions of $A$. The result is shown in fig. \ref{fig:gen} for various positive and negative values of $\al$. 

Let us first discuss the two plots pertaining to $\al<0$. As soon as $\al$ becomes negative, what was the Gaussian fixed point at the top end of the curve corresponding to $A=0$, drops below the $V_*(0)$ - axis and thus it is no longer present. Instead a new fixed point with non-constant potential comes into being which for $\al=-0.5$ is located at $V_*(0)\approx0.2$. Close to where the Wilson-Fisher fixed point was for $\al=0$ we still find a second fixed point and we will see that it still has only one relevant even eigenperturbation
(ignoring the vacuum energy)  like the original Wilson-Fisher fixed point.

As we decrease $\al$ further, more and more fixed points are created. On the left in fig. \ref{fig:gen} we see easily that for $\al=-2$ there are already four fixed points, the one on the far right corresponding to the Wilson-Fisher fixed point. These fixed points are labelled \tc1 to \tc4 in the direction of decreasing $A$.

Less easy to see, even given the dotted blue curve to guide the eye, is the pair of points $V_*(0),\pm V'_*(0)$ close to \tc4, corresponding to the first non-symmetric fixed point pair. As $\al$ is decreased still further these non-symmetric solutions become more visible and are joined by both more symmetric fixed point solutions and other pairs of non-symmetric fixed point solutions.

On the other hand, if we go to $\al>0$ the spiral from $\al=0$ progressively unwinds. We immediately lose the Gaussian fixed point but retain the Wilson-Fisher fixed point. Increasing $\al$ further however leaves us with no fixed points at all.

In order to understand the nature of the new fixed points and to justify that, provided $\al$ is not too large and positive, an image of the Wilson-Fisher fixed point is still present, we have to investigate their eigenspectra. The eigenoperator equation takes the form
\begin{equation}
 \label{equ:eopsgen}
  (\lambda-3) v  + \frac{1}{2} \phi \,v' =  
  \frac{\left(1-\al\phi^2\right)^{\frac{3}{2}} \left(1- \frac{1}{2}\al \phi^2 \right)}{\left(1-\al \phi^2+V_*''\right)^2}v''(\phi)
  \, \theta \! \left(1-\al \phi^2 \right).
\end{equation}
For each fixed point potential $V_*$ this equation can be analysed with the same methods as in sec. \ref{sec:WF} provided $\al<0$. For $\al>0$ the right hand side of \eqref{equ:eopsgen} again  vanishes for $|\phi| \geq \phi_c=1/\sqrt{\al}$ and we have to match the solution in the range $\phi<\phi_c$ to the solution $v(\phi) = \phi^{6-2\lambda}$ of the left hand side at $\phi=\phi_c$. This is done along the same lines as for the fixed point equation \eqref{equ:FPgen} via a generalised Taylor expansion of the eigenoperator $v$ at the matching point.

In the following we will focus our attention on the even fixed points in fig. \ref{fig:gen}, leaving aside the two non-symmetric solutions close to fixed point \tc4. This in turn allows us to search for even eigenoperator solutions of \eqref{equ:eopsgen} with the aim of comparing the results to the single even eigenperturbation found for the Wilson-Fisher fixed point in the previous section.

As labelled in fig.~\ref{fig:gen}, we then find that the $n$-th fixed point has $n$ relevant even eigenoperators if we do not take into account the vacuum energy. In particular we see that the first fixed point always has one relevant even eigendirection which justifies regarding it as the Wilson-Fisher fixed point. Of course, the eigenvalue associated to this relevant direction depends on $\al$ as can be seen from table \ref{tab:evWF}. It decreases along with $\al$ and we have checked that it is still relevant at $\al=-8$ which suggests that it may slowly tend to zero from the right as $\al \rightarrow -\infty$. In table \ref{tab:evs} we give the eigenvalues pertaining to the relevant even eigenoperators of the additional new fixed points labelled in fig. \ref{fig:gen} on the left.
\begin{table}[h]
\parbox{0.45\textwidth}{
\centering
  \begin{tabular}{|c|c|c|c|c|}
   \hline $\al$ & $1/25$ & 0 & $-0.5$ & $-2$ \\
   \hline $\lambda_\mathrm{WF}$ & 1.62 & 1.54 & 1.17 & 0.89 \\
   \hline
  \end{tabular}
\caption[Dependence of the Wilson-Fisher scaling exponent on $\al$]{
The eigenvalue of the single relevant even eigendirection of the Wilson-Fisher fixed point 
for the values of $\al$ used in fig. \ref{fig:gen}.}
\label{tab:evWF}
  }
  \hspace{1cm}
 \parbox{0.45\textwidth}{
 \centering
\begin{tabular}{|c|c|c|c|c|}
  \hline
   FP & $\lambda_1$ & $\lambda_2$ & $\lambda_3$ & $\lambda_4$ \\
  \hline 2 & 2.35 & 0.76 & - & - \\
  \hline 3 & 2.02 & 1.43 & 0.60 & - \\
  \hline 4 & 2.10 & 1.69 & 1.08 & 0.39 \\
  \hline
  \end{tabular}
\caption[New relevant eigen-directions for field dependent cutoffs]{Each new fixed point possesses one more relevant even eigendirection. These are the eigenvalues for the labelled additional fixed points in the plot on the left in fig. \ref{fig:gen}.}
\label{tab:evs}
}
\end{table}

Overall, we conclude that as soon as $\al \neq 0$ we encounter significant differences to the standard picture of the fixed point structure for a three-dimensional single scalar field. For appropriate values of $\al$ spurious fixed points can be created, previous fixed points can be lost and the number of possible relevant even eigendirections changes with $\al$. More than that, these are not the only problems as we will see now.

\subsubsection{Missing vacua and redundant operators}
\label{sec:missingvacua}
We now come back to the topic of redundant operators as introduced in sec. \ref{sec:red-ops-def} and the possibility that a given approximation of the effective action can break down as described for the $f(R)$ truncation in sec. \ref{sec:breakdown}. In the following we show that a version of this pathology also occurs in the present context when $\al \neq 0$. To this end we remark that even though we have so far been concerned with even eigenoperators only, we now include odd eigenoperators in the discussion below.

At the level of the LPA in scalar field theory, the relation \eqref{redop} for an eigenoperator $v$ to be redundant assumes the form
\begin{equation} \label{equ:redcond}
v(\phi)=\zeta(\phi) V_*'(\phi),
\end{equation}
where we are free to choose the function $\zeta$, corresponding to $F$ in \eqref{redop}, with the only requirement that it has to be non-singular on its domain of definition $\phi \in (-\infty , \infty)$. This implies that turning points of the fixed point potential $V_*$ are zeros of the redundant eigenoperator.

\begin{figure}[ht]
\centering
 \includegraphics[width=0.5\textwidth,height=0.35\textheight]{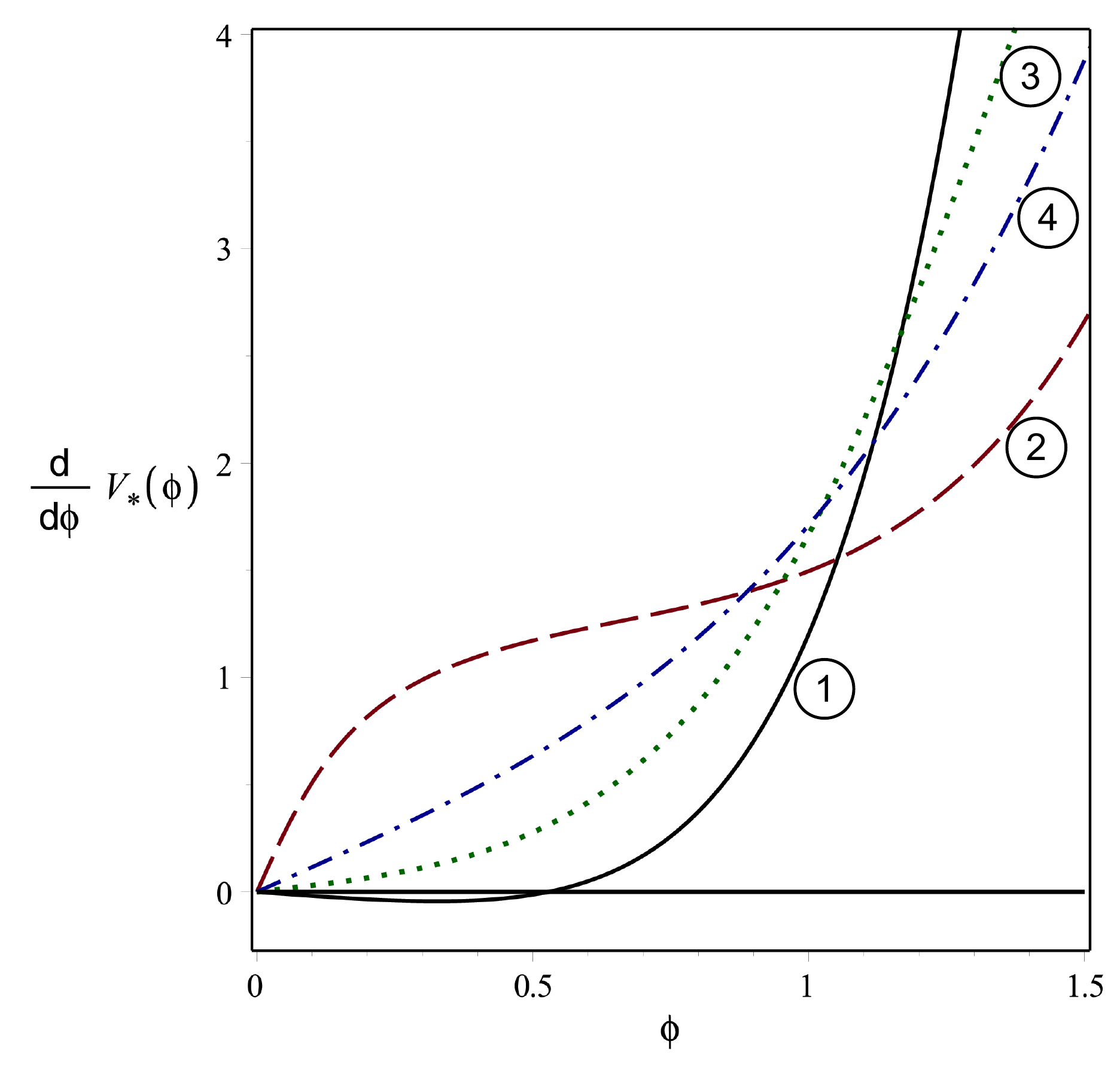}
 \caption[Absence of turning points of fixed point potentials]{The derivative of all four fixed point potentials from fig. \ref{fig:gen} (left).
 Only the Wilson-Fisher fixed point features a turning point at non-zero field.}
 \label{fig:vps}
\end{figure}

Considering the Wilson-Fisher fixed point (with arbitrary $\al$) we always find a non-trivial turning point $V_\mathrm{WF}'(\pm \phi_0)=0$ at some $\phi_0 > 0$. For $\al=0$ this can be seen from fig. \ref{fig:WF} (right) and for $\al=-2$ we have plotted $V_\mathrm{WF}'$ in fig. \ref{fig:vps}. Given that the eigenoperator equation \eqref{equ:eopsgen} is already constrained such that it gives rise to a quantised eigenspectrum, the additional condition $v(\phi_0)=0$ implied by \eqref{equ:redcond} overconstrains the system and we will find no solutions at all. In the case of the Wilson-Fisher fixed point we have explicitly checked for the values of $\al\neq0$ given in fig. \ref{fig:gen} that $v(\phi_0)\neq 0$ is indeed the case for all relevant eigenoperators, both odd and even. Hence \eqref{equ:redcond} cannot hold for a non-singular function $\zeta$ and none of its eigenoperators are redundant. For $\al=0$ there is one exception, given by the redundant odd eigenoperator $v=V_\mathrm{WF}'$ with $\lambda=1/2$. This happens only for $\alpha=0$ since in this case the cutoff operator is field independent which restores the shift symmetry $\phi(x) \mapsto \phi(x) +\delta$ of the unscaled equation \eqref{equ:FRGE}. Upon expressing this symmetry in terms of dimensionless variables as in \eqref{scaledvars} it can also be found in \eqref{equ:flowLPA} with $h=0$, but since the dimensionless version involves the RG scale $k$ it is of course broken in the fixed point equation \eqref{equ:fpWF}. It may be worth emphasising that this shift symmetry is a symmetry of the flow equation, meaning that if $V$ is a solution then the transformed $V$ is also a solution. This should be contrasted to symmetries of a given solution $V$ of the flow equation, under $\phi \mapsto -\phi$ say. The fact that this redundant operator originates from a symmetry inherent in the flow equation is expected, as discussed at the end of sec. \eqref{sec:red-ops-def}.

The state of affairs is different for all the other labelled fixed points in fig. \ref{fig:gen} (left). The only turning point for these fixed points is at $\phi=0$ which is a consequence of their $Z_2$-symmetry, cf. fig. \ref{fig:vps}. Even eigen-perturbations, i.e. eigen-perturbations satisfying $v'(0)=0$, are still not redundant as they must have $v(0) \neq 0$ if they are not to vanish identically. However, since odd eigenoperators vanish at $\phi=0$ they satisfy $v'(0)\neq0$ and hence would be redundant if the fixed point potential does not vanish faster than linearly, i.e. if $V_*''(0)\neq 0$. From the fixed point equation \eqref{equ:FPgen} we see that $V_*''(0)= 0$ implies $V_*(0)=1/3$. Together with $V'_*(0)=0$ these unique initial conditions will only define a fixed point solution for a discrete set of values for $\al$ (e.g. if $\al=0$ they describe the Gaussian fixed point) but can be excluded otherwise. We therefore see that generically all odd eigen-perturbations of all fixed points other than the Wilson-Fisher fixed point are redundant if $\al<0$. The non-existence of a vacuum solution $V'(\phi)=0$ for $\phi >0$ leading to this redundancy for these additional fixed points is clearly an artefact incurred by the single field approximation in combination with a field dependent cutoff operator.

\subsection{A cutoff operator depending on the potential}\label{sec:LPALitim}
The results of the previous section were obtained with the purely explicit field dependent choice \eqref{cutoff1} for the function $h$ in the cutoff operator \eqref{equ:cutoffh}. As a second example, we now consider the choice \eqref{cutoff2} which, contrary to \eqref{cutoff1}, has occurred previously in the literature \cite{Litim:2002hj}. It carries intrinsic scale dependence from the potential $V(\phi)$ of \eqref{equ:LPA} and for $\alpha=1$ leads to a spectral flow in the sense of \cite{Gies:2002af}. We have seen at the end of sec. \ref{sec:LPAsetup} that the correct one-loop $\beta$ function in four dimensions is then recovered for any $\alpha$. One might hope, with the background field dependence of the cutoff being thus adapted to the structure of \eqref{equ:FRGE} that at least qualitatively non-physical behaviour would now be avoided. However we will see in this section that this hope is unfounded.

Using \eqref{cutoff2} in the general expression \eqref{equ:flowLPA}, the flow equation in dimensionless
variables becomes
\begin{equation}
\label{equ:V''Flow}
\partial_{t}V-\frac{1}{2}(d-2)\phi V'+dV=\dfrac{\big(1-\al(1+\frac{1}{2}\partial_{t})V''+\frac{1}{4}\al(d-2)
\phi V'''\big)\big(1-\al V''\big)^{d/2}}{1+(1-\al)V''}\,\theta(1-\al V'').
\end{equation}
We emphasise again that the background field from \eqref{equ:cutoffh} has disappeared from the flow equation \eqref{equ:V''Flow} due to the single field approximation as described below \eqref{approx}. In \cite{Litim:2002hj} the choice $\al =1$ was considered and we see that this leads to a significant simplification in the flow equation as the denominator on the right hand side becomes equal to one.

The fixed point equation in three dimensions that we wish to solve is now given by:
\begin{equation}
\label{equ:V''Fixed}
3V_{*}-\frac{1}{2}\phi V_{*}'=\dfrac{(1-\al V_{*}''+\frac{1}{4}\al\phi V_{*}''')(1-\al V_{*}'')^{3/2}}{(1+(1-\al)V_{*}'')}\,\theta(1-\al V_{*}'').
\end{equation}
The analysis of this equation presents many similarities to the treatment of \eqref{equ:FPgen} in the previous
section. At the same time an important subtlety arises from the appearance of the potential as the function
that is being solved for in the step function on the right hand side. In the following we will concentrate on the case $\alpha>0$.

As before, solving the left hand side gives again the asymptotic solution \ensuremath{V_{*}(\phi)=A\phi^{6}}. Since $\alpha>0$ this is indeed an exact solution of \eqref{equ:V''Fixed} since the right hand side vanishes above the critical field value \ensuremath{V_{*}''(\phi_c)=1/\al}, where we can always assume $\phi_c>0$ due to the symmetry $\phi \mapsto -\phi$ of the fixed point equation. This solution has to be matched to the solution obtained from the full fixed point equation for all \ensuremath{\lvert\phi\rvert\leq\phi_{c}}, cf. sec. \ref{sec:gencutoff}. This is achieved via a Taylor expansion around $\phi=\phi_c$,
\begin{equation}
\label{equ:V''Taylor}
V_{*}(\phi)=b_0 + b_1\left(\phi_c-\phi \right)
 +\frac{b_2}{2!}\left(\phi_c-\phi \right)^2+ F(\phi_c-\phi),
\end{equation}
where the relevant side captured by this expansion is $\phi\leq\phi_c$ and we have allowed for a general remainder function $F$, for reasons that will become clear shortly. We emphasise here that the basic requirement of finite initial conditions $V_*(\phi_c),V_*'(\phi_c),V_*''(\phi_c)$ for the fixed point equation \eqref{equ:V''Fixed} at the matching point $\phi_c$ dictates the form of the first three terms in the expansion \eqref{equ:V''Taylor} with finite $b_i$. If we choose $b_i=V^{(i)}(\phi_c)$ for $i=0,1,2$ then this furthermore implies $F^{(i)}(\phi_c)=0$ for $i=0,1,2$ but a priori does not impose any conditions on higher derivatives of $F$ at the matching point.

From the above exact asymptotic solution $V_*(\phi)=A\phi^6$ we find that the $\theta$-function in \eqref{equ:V''Fixed} determines the matching point to be \ensuremath{\phi_c=(30A\al)^{-1/4}}. Choosing the initial conditions at $\phi_c$ such that matching to the asymptotic solution is achieved up to the second derivative results in
\begin{equation}\label{bs}
b_{0}=A\phi_{c}^{6}=A(30A\al)^{-3/2},\qquad b_{1}=-6A\phi_{c}^{5}=-6A(30A\al)^{-5/4},\qquad b_{2}=30A\phi_{c}^{4}=\frac{1}{\al}.
\end{equation}
Upon substituting the expansion \eqref{equ:V''Taylor} into the fixed point equation \eqref{equ:V''Fixed}, it turns out that the left and right hand sides cannot be made to correspond to each other if the remainder $F$ just continues \eqref{equ:V''Taylor} as a standard Taylor expansion. We find that the two sides can be compared order by order in $\phi_c-\phi$ if the the remainder instead takes the form
\begin{equation}
\label{equ:F}
F(\phi_c-\phi)=(\phi_c-\phi)^{\gamma}\, \bar F(\phi_c-\phi),
\end{equation}
where $\gamma = 16/5$ and the function $\bar F(\phi_c-\phi)$ assumes the form of a generalised Taylor series
\begin{equation}
\label{equ:rtaylor}
\bar F(\phi_c-\phi)=c_{0}+c_{1}(\phi_c-\phi)^{1/5}+c_{2}(\phi_c-\phi)^{2/5}+c_{3}(\phi_c-\phi)^{3/5}+\dots,
\end{equation}
similar to \eqref{equ:Vtaylor}. As expected, the coefficients $c_i$ are then functions of the initial conditions \eqref{bs} and $\alpha$. Expressed in terms of $\phi_c$, the first few are
\begin{multline}
\label{cs}
c_{0}=-\dfrac{25}{88}\left(\dfrac{25}{72}\alpha^{-4}\phi_c^{-2}\right)^{\frac{1}{5}},
\quad c_{1}=\dfrac{125}{5984}\left(\dfrac{625}{162}\alpha^{-13}\phi_{c}\right)^{\frac{1}{5}},\\ 
c_{2}=-\dfrac{71875}{61611264}\left(\dfrac{5}{48} \alpha^{-17}\phi_{c}^{4}\right)^{\frac{1}{5}}.
\end{multline}
Since $\gamma>3$ in \eqref{equ:F}, the matching thus obtained will cause \ensuremath{V_{*}'''(\phi_c)=0} but all higher derivatives diverge as the matching point is approached from the left. As we noted with equation \eqref{equ:Vtaylor}, this is an artefact of the type of cutoff chosen and has no physical significance. 

Similar to the strategy in the previous section, we now use the expansion \eqref{equ:V''Taylor} to determine initial conditions slightly to the left of $\phi_c$ as a function of the asymptotic parameter $A$ and numerically integrate the fixed point equation \eqref{equ:V''Fixed} down to $\phi=0$ where we record $V_{*}(0)$ and $V'_{*}(0)$. In principle, we would also have to record $V''_*(0)$ in order to have a complete set of initial conditions. As discussed in sec. \ref{sec:WF}, a global solution would then be given by two sets of such initial conditions, where the components $V_*(0)$ and $V_*''(0)$ are equal in each set, whilst $V_*'(0)$ is equal and opposite. As we will see in the following, we can safely disregard the $V_*''(0)$-component in this procedure as we do not find any two pairs $\left(V_*(0),V_*'(0)\right)$ with the same $V_*(0)$ and equal and opposite $V_*'(0) \neq0$. If that was the case, it would be necessary to check that also the second derivative at $\phi=0$ agrees for these pairs. Since we only find solutions satisfying $V_*'(0)=0$, implying that they are symmetric under $\phi \mapsto -\phi$, the same value for the asymptotic parameter $A$ can be used to describe the behaviour at $\phi \rightarrow \pm\infty$.

\begin{figure}[ht]
  \begin{center}
  $
   \begin{array}{ccccc}
     \includegraphics[width=0.3\textwidth,height=0.225\textheight]{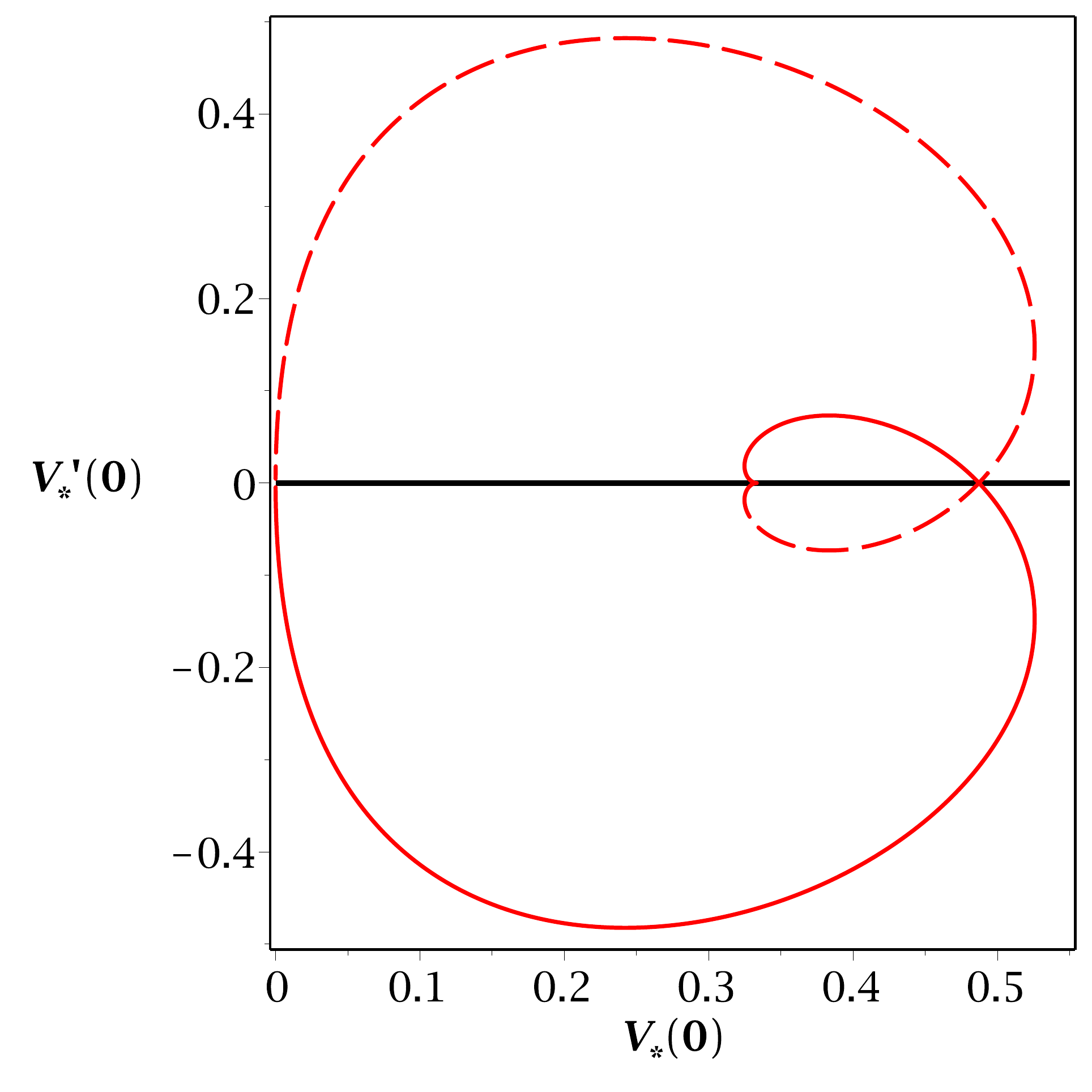}  & &
     \includegraphics[width=0.3\textwidth,height=0.225\textheight]{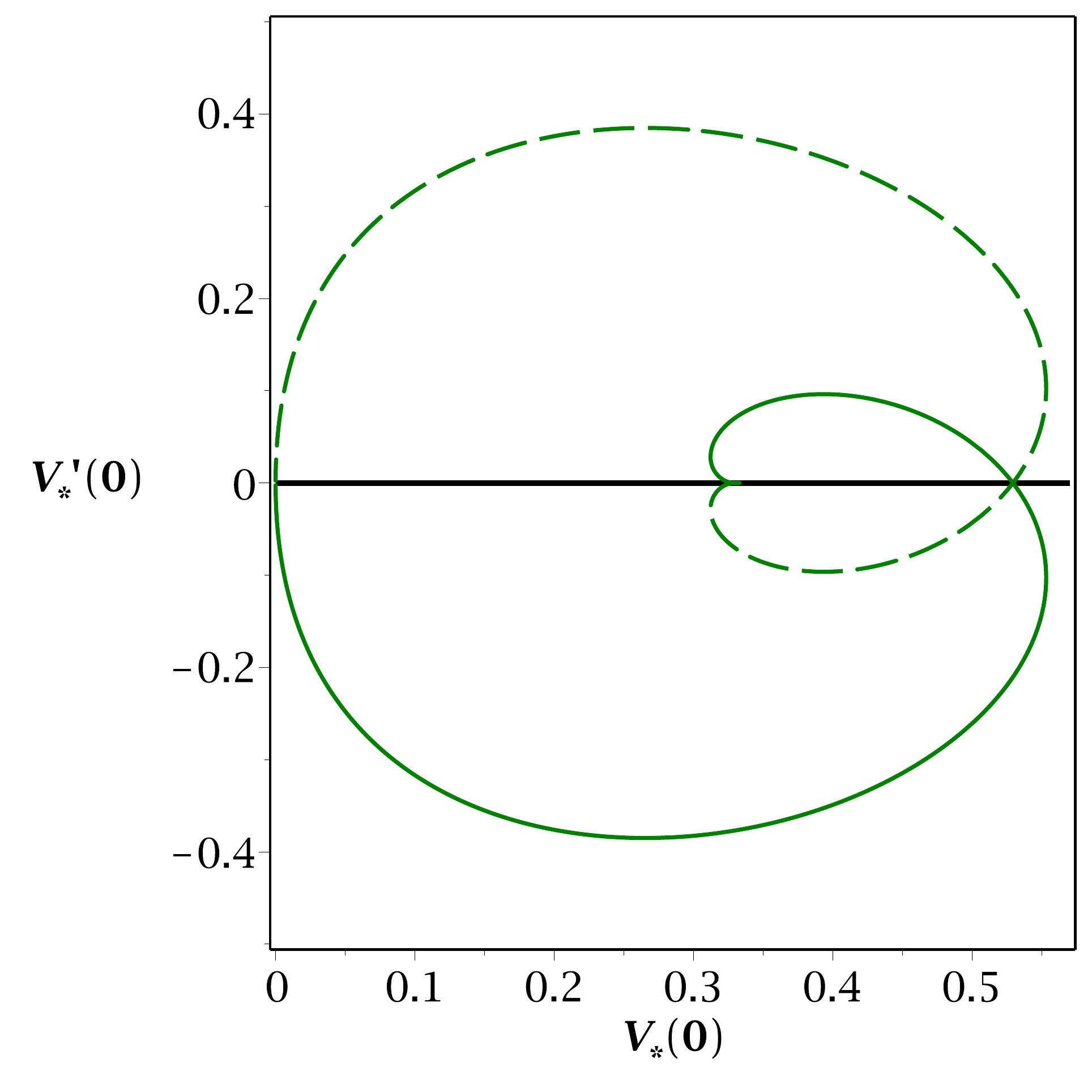} && 
     \includegraphics[width=0.3\textwidth,height=0.225\textheight]{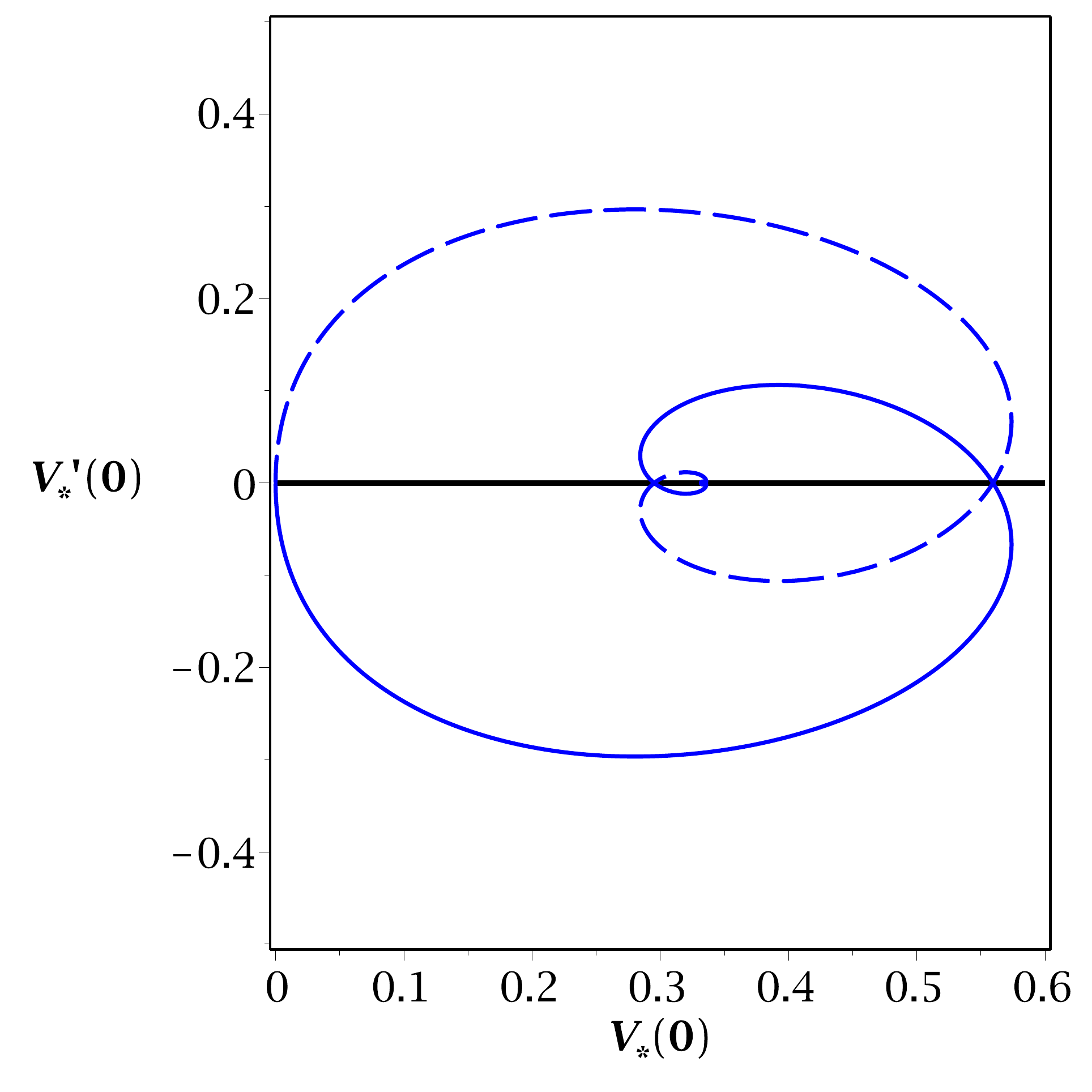}
   \end{array}$
  \end{center}
   \caption[Fixed point structure for a cutoff depending of the potential]{Varying the asymptotic parameter $A$ from $A=0$ to $A\to\infty$, for different values of $\al$    leads to these spirals in $V_*(0),V'_*(0)$ - space, where  $A=0$ corresponds to the point \ensuremath{(V'_{*}(0)=0, V_{*}(0)=1/3)}, and $A\to\infty$ tends to the origin. From left to right the curves show \ensuremath{\al=0.5\text{ (red), }\al=1\text{ (green) and } \al=2} (blue). The dashed curves correspond to the solid curves reflected in the $V_*(0)$ axis. From the discussion below \eqref{equ:VWFasy}, we know that fixed point solutions correspond to the points where the solid and dashed curves cross.}
\label{fig:V''FixedPoints}
\end{figure}

\begin{figure}[ht]
  \begin{center}
  $
   \begin{array}{ccccc}
     \includegraphics[width=0.3\textwidth,height=0.225\textheight]{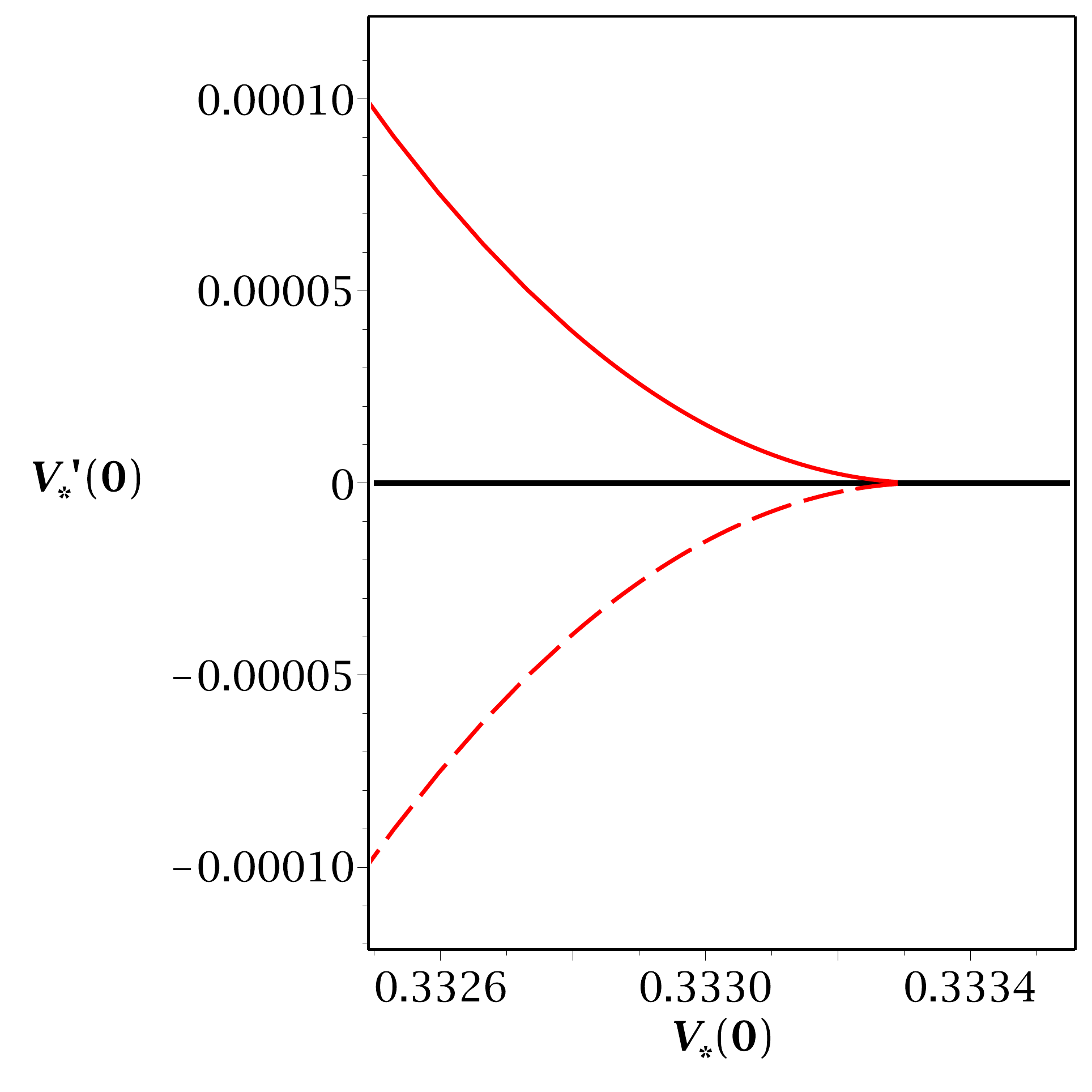}  & &
     \includegraphics[width=0.3\textwidth,height=0.225\textheight]{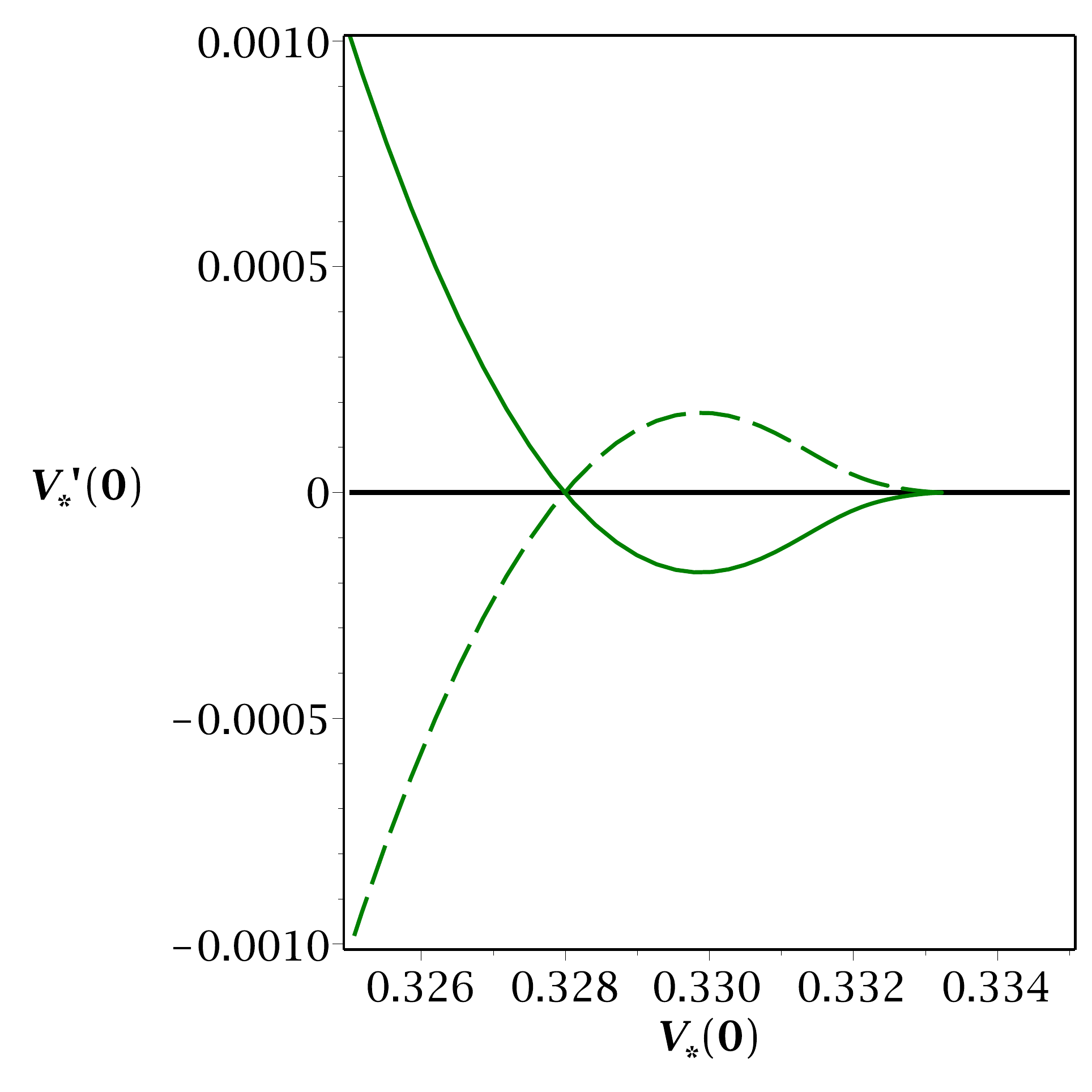} && 
     \includegraphics[width=0.3\textwidth,height=0.225\textheight]{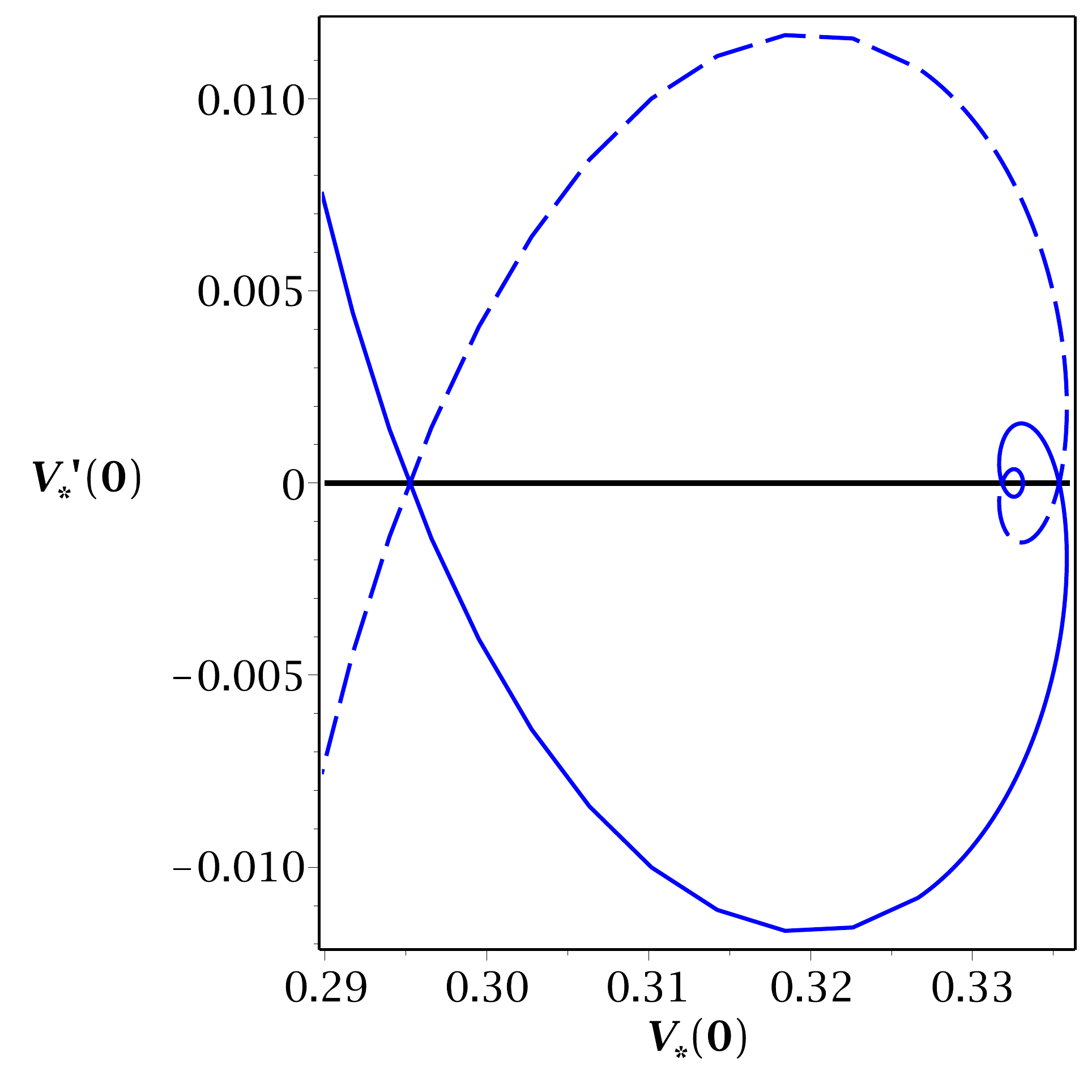}
   \end{array}$
  \end{center}
   \caption[Zoom into fig. \ref{fig:V''FixedPoints}]{Low values of asymptotic parameter $A$ detail from figure \ref{fig:V''FixedPoints}. Each plot corresponds to the equivalent values of \ensuremath{\al} in figure \ref{fig:V''FixedPoints}. For large \ensuremath{\al} more fixed points appear near the Gaussian fixed point at \ensuremath{V_{*}(0)=1/3}.}
\label{fig:V''Zoom}
\end{figure}

The results of our analysis are shown in fig. \ref{fig:V''FixedPoints} and \ref{fig:V''Zoom} for the cases \ensuremath{\al=0.5,\ 1 \text{ and } 2}. 
If we first discuss the three plots of the full parameter space given by \ensuremath{0<A<\infty} showing
\ensuremath{V_{*}(0) \text{ vs. } V'_{*}(0)} for \ensuremath{\al=0.5,1\ \&\ 2} in fig. \ref{fig:V''FixedPoints}, we find that for all values of \ensuremath{\al} there exist at least two fixed point solutions. As \ensuremath{A\to 0} we find the Gaussian fixed point at \ensuremath{V_{*}=1/3} as already mentioned towards the end of sec. \ref{sec:LPAsetup}. Furthermore, for all three values of $\al$ there is a second fixed point solution characterised by $V_*(0)>0.4$ which we identify with the Wilson-Fisher fixed point in the present context. This identification is made by analogy in shape of the spiral in fig. \ref{fig:V''FixedPoints} to fig. \ref{fig:WF} (left). As \ensuremath{A\to \infty} there appears to be another fixed point solution at \ensuremath{V_{*}(0)=V'_{*}(0)=0}, however this point represents the limit $A\to \infty$ and does not have sensible asymptotic behaviour.

From fig. \ref{fig:V''FixedPoints} (right) we can already see that depending on $\al$ there can be more than these two standard fixed points.
This is confirmed by the plots in fig. \ref{fig:V''Zoom} which show magnifications corresponding to the plots
in fig. \ref{fig:V''FixedPoints}. As we vary \ensuremath{\al} we find that for \ensuremath{\al\gtrapprox 0.85} a first additional non-trivial fixed point appears which, as can be seen from fig. \ref{fig:V''Zoom} (middle), is still present in the important case $\al=1$ considered in \cite{Litim:2002hj}. Increasing $\al$ to above one results in even more fixed points, e.g. for $\al =2$ there are three additional fixed points, cf. fig. \ref{fig:V''Zoom} (right). The number of these non-trivial fixed points that appear near \ensuremath{A= 0} increases dramatically as we increase \ensuremath{\al}, e.g. for \ensuremath{\al=4} there are in excess of 20 new fixed points. As we increase \ensuremath{\al} the spiral in \ensuremath{V_{*},V'_{*}} space flattens towards the \ensuremath{V_{*}} axis and each non-trivial fixed point moves away from the Gaussian fixed point on the \ensuremath{V_{*}} axis. 

Since we know that the correct description of a single scalar field in three dimensions includes only one non-trivial fixed point we conclude already at this stage that the fixed point equation \eqref{equ:V''Fixed} is invalidated by these findings.

This conclusion is further supported by an investigation of the eigenspectra of the fixed point solutions described in this section. Interestingly, creating a plot analogous to fig. \ref{fig:vps} for the non-trivial fixed point solutions in fig. \ref{fig:V''Zoom} shows that each of these displays a minimum at non-zero field, i.e. a decisive obstruction to redundant operators as expressed in \eqref{equ:redcond}. However, as we will see now, we encounter a very different problem.

Linearising the flow equation \eqref{equ:V''Flow} in three dimensions and assuming separation of variables as in \eqref{equ:perts} leads to the following linear differential equation of third order for eigenoperators $v(\phi)$:
\begin{multline}
\label{eopsVpp}
(3-\lambda)v -\frac{1}{2}\phi\, v'  =\left\{-\frac{3}{2}\frac{
\left( 1-\al V''+\frac{1}{4}\alpha \phi V''' \right) \sqrt {1-
\alpha V''} \al \, v''}{ 1+ (1-\alpha) V'' } + \Bigg[\frac{\alpha \left( \frac{\lambda}{2} -1\right)v'' +\frac{1}{4}\alpha \phi \, v'''}{ 1+ (1-\alpha) V'' } \Bigg. \right. \\
 \left.\Bigg.+\frac{1}{4}\frac { \left( -\alpha \phi V''' +4 \alpha V'' -4 \right)  
 \left( 1-\alpha \right) }{(1+ (1-\alpha) V'')^2} \,v'' \Bigg] \left( 1-\alpha V''\right) ^{3/2}
 \right\}\theta\!\left(1-\al V''\right)
\end{multline}
Primes denote differentiation with respect to $\phi$ and for typographical clarity we have omitted the asterisk to denote a fixed point solution for the potential.

Since the step function from the right hand side of the fixed point equation \eqref{equ:V''Fixed} appears in the same form on the right hand side of the eigenoperator equation, the asymptotic solution $v(\phi)=\phi^{6-2\lambda}$, given by solving its left hand side, has to be matched onto the solution of the full equation \eqref{eopsVpp} at the same field value $\phi_c=(30A\al)^{-1/4}$ as for the fixed point equation.

Analogous to \eqref{equ:V''Taylor}, we set up an expansion of $v(\phi)$ valid to the left of the matching point which by matching derivatives up to the second with the asymptotic solution, takes the following form:
\begin{multline}
\label{vexp}
 v(u) = (30\al A)^{\frac{1}{2}(\lambda-3)} + 2(\lambda-3)(30\al A)^{\frac{1}{4}(2\lambda-5)} u \\ + (2\lambda-5)(\lambda-3)(30\al A)^{\frac{1}{2}(\lambda-2)} u^2 + H(u).
\end{multline}
For brevity we use the variable $u = \phi_c -\phi$ in this expression. We have $u\geq0$ and $H(u)$ is a remainder that has to vanish faster than $u^2$ in order to not affect the required matching of $v$, $v'$ and $v''$ to the asymptotic solution at $u=0$.

We now use the behaviour \eqref{equ:V''Taylor} with \eqref{bs}, and \eqref{equ:F} with \eqref{equ:rtaylor}, and \eqref{cs} of the fixed point solutions in the limit $u\to 0$ in the eigenoperator equation \eqref{eopsVpp} together with \eqref{vexp}. It is straightforward to check that the first ratio on the right hand side of \eqref{eopsVpp} is the dominant contribution as $u\to 0$, taking the form
\begin{equation}\label{orderrhs}
\propto\, \al^{-\frac{3}{10}+\frac{\lambda}{2}} A^{-\frac{11}{10}+\frac{\lambda}{2}}
		(\lambda-3)\left(\lambda-\frac{5}{2}\right)\, u^{\frac{4}{5}}\,,
\end{equation}
while the other two are of higher order in $u$. On the other hand, since the coefficients in \eqref{vexp} are fixed such that matching is achieved with the solution of the left hand side of \eqref{eopsVpp}, this side of the eigenoperator equation behaves as $\mathcal{O}\!\left(u^2\right)$. We note that these considerations hold for any allowed remainder $H$ in \eqref{vexp}.

Thus we see that the eigenoperator equation cannot be satisfied at the matching point $\phi=\phi_c$ unless $\lambda=3,5/2$. The first eigenvalue corresponds to the vacuum energy, and the second to the odd eigenoperator $v(\phi)=\phi$, both of them solving the left and right hand sides of \eqref{eopsVpp} separately. For the fixed point solutions with the behaviour \eqref{equ:V''Taylor} and \eqref{equ:F}, the eigenoperator equation admits no other global solutions. In particular, with the present set-up, none of the non-trivial fixed point solutions found above possess a relevant eigenoperator resembling the one needed for the Wilson-Fisher fixed point.

We note that these conclusions are valid for the non-trivial fixed points of this section since for the Gaussian fixed point $A=0$ and \eqref{orderrhs} is no longer valid. At the Gaussian fixed point $V_*=1/3$ the right hand side of the eigenoperator equation does no longer require a matching procedure since the step function is identically equal to one. An analysis of the resulting eigenoperator equation confirms the standard spectrum of relevant eigenoperators of the Gaussian fixed point, cf. below \eqref{equ:WFeops}, for all three values of $\al$ considered in fig. \ref{fig:V''FixedPoints}.

We therefore see that the flow equation \eqref{equ:V''Flow} for $\al \neq 0$ gives qualitatively very different results for both fixed points and eigenoperators compared to the correct description of sec. \ref{sec:WF}.  Nevertheless, we will show in sec. \ref{sec:sWI-Vpp} that also for the cutoff choice \eqref{cutoff2} in \eqref{equ:cutoffh} the correct description of single component scalar field theory in three dimensions can be recovered with the help of the modified split Ward identity.

\section{Including the modified split Ward identity} \label{sec:backgroundfield}
In the previous sections we have seen how the use of cutoffs of form \eqref{equ:cutoffh} with \eqref{cutoff1} or \eqref{cutoff2} and $\al\neq0$ together with the single field approximation has led to a rather severe distortion of the established fixed point structure for the theory of a single scalar field in $d=3$ dimensions. As we show now, it is however possible to recover the correct results as presented in sec. \ref{sec:WF} by appropriately taking into account the dependence of the effective action on the background field through the modified split Ward identity.

\subsection{Derivation and interpretation} \label{sec:sWI-derivation}
In order to derive the modified split Ward identity, we apply the background field split \eqref{bgrscalar} to the partition function \eqref{partfunc} to obtain
\begin{equation} \label{partfuncsca}
 Z[J,\pb]=\int \mathcal{D}\vp \exp\left(-S[\vp+\pb]-S_k[\vp,\pb]+ J\cdot \vp\right)
\end{equation}
The cutoff action $S_k$ now becomes a functional of both fields and takes the form $S_k[\vp,\pb] = \frac{1}{2}\vp \cdot \mathcal{R}_k[\pb] \cdot\vp $, satisfying the requirement that it has to be quadratic in the fluctuation field. We note that the classical action $S$ is invariant under the combined shift 
\begin{equation} 
 \label{shifts}
 \pb \mapsto \pb + \eps(x) \qquad \text{and} \qquad \vp \mapsto \vp - \eps(x),
\end{equation}
whereas the cutoff action and the source term break this invariance. Hence we can write
\begin{equation*}
 Z[J,\pb+\eps]=\int \mathcal{D}\vp \exp\left(-S[\vp+\pb]-S_k[\vp-\eps,\pb+\eps]+ J \cdot (\vp-\eps)\right)
\end{equation*}
and, taking $\eps$ to be infinitesimal, the resulting variation of the functional $W=\ln Z$ becomes 
\begin{equation*}
 \frac{\delta W}{\delta \pb}\cdot\eps = \eps\cdot \mathcal{R}_k \cdot \frac{\delta W}{\delta J}
 -\frac{1}{2}\frac{\delta W}{\delta J} \cdot \left(\frac{\delta \mathcal{R}_k}{\delta \pb_a}\, \eps_a\right) \cdot \frac{\delta W}{\delta J}
 -\frac{1}{2} \mathrm{Tr} \left [\left(\frac{\delta \mathcal{R}_k}{\delta \pb_a}\, \eps_a \right) \cdot \frac{\delta^2 W}{\delta J \delta J}\right] - J \cdot \eps,
\end{equation*} where, as defined below \eqref{partfunc}, the dot indicates the contraction of the corresponding free indices. We now express this equation in terms of the effective action
\begin{equation} \label{Legendre-trafo}
 \Gamma_k[\vp^c,\pb] = J \cdot \vp^c - W[J,\pb] - S_k[\vp^c,\pb]
\end{equation}
by taking into account the identities
\begin{equation*}
 \vp^c = \frac{\delta W}{\delta J}, \qquad \frac{\delta \Gamma_k}{\delta \pb}=-\frac{\delta W}{\delta \pb}-\frac{\delta S_k}{\delta \pb},
 \qquad \frac{\delta^2 W}{\delta J \delta J} = \left[\frac{\delta^2 \Gamma_k}{\delta \vp^c \delta \vp^c}+\mathcal{R}_k\right]^{-1}
\end{equation*}
and arrive at
\begin{equation*}
 \left(\frac{\delta \Gamma_k}{\delta \pb}-\frac{\delta \Gamma_k}{\delta \vp^c}\right)\cdot \eps
 = \frac{1}{2} \mathrm{Tr} \left [\left(\frac{\delta^2 \Gamma_k}{\delta \vp^c \delta \vp^c}+\mathcal{R}_k\right)^{-1}\cdot \frac{\delta \mathcal{R}_k}{\delta \pb_a}\, \eps_a\right].
\end{equation*}
Since $\eps$ is arbitrary, this identity still holds after dropping it:
\begin{equation}
 \label{sWI}
  \frac{\delta \Gamma_k}{\delta \pb_a}-\frac{\delta \Gamma_k}{\delta \vp^c_a} = \frac{1}{2} \mathrm{Tr} \left [\left( \frac{\delta^2 \Gamma_k}
 {\delta \vp^c \delta \vp^c} + \mathcal{R}_k \right)^{-1} \frac{\delta \mathcal{R}_k}{\delta \pb_a}\right].
\end{equation}
This is the (modified) split Ward identity (sWI) capturing the dependence of the effective action on the background field as introduced by the cutoff action breaking the symmetry \eqref{shifts} of the classical action.

In the context of scalar field theory the sWI has appeared previously in \cite{Litim:2002hj},
expressed in terms of $\tilde \Gamma_k[\phi^c,\pb] = \Gamma_k[\phi^c-\pb,\pb]$. An analogous sWI for Yang-Mills theory can be found in \cite{Reuter:1993kw,Litim:1998nf,Reuter:1997gx,Litim:2002ce} and for scalar QED in \cite{Reuter:1994sg}. An appropriate version of the sWI for conformally reduced gravity has been used in \cite{Manrique:2009uh}.

The structure of the sWI is very close to that of the flow equation \eqref{equ:FRGE}, in particular note that the trace is over the same indices, which is a result of the following alternative way of deriving it. Taking the partition function \eqref{partfuncsca} and rewriting it as
\begin{equation*}
 Z[J,\pb]=\int \mathcal{D}\phi \exp\left(-S[\phi]-S_k[\phi-\pb,\pb]+ J\cdot (\phi-\pb)\right),
\end{equation*}
where $\phi=\vp+\pb$ is the total quantum field, it suffices to take a functional derivative with respect to 
the background field and follow a similar procedure as in the derivation of the flow equation \eqref{equ:FRGE} to arrive at the sWI \eqref{sWI}. This more direct derivation is the one followed in refs. \cite{Reuter:1993kw,Litim:2002ce,Reuter:1994sg,Reuter:1997gx,Manrique:2009uh}. The derivation above however displays more clearly its interpretation as a modified Ward identity expressing the breaking of the shift symmetry \eqref{shifts}. 

In this context it is important to stress that the sWI \eqref{sWI} represents an additional constraint on the effective action and thus needs to be investigated in its own right. If $\Gamma_k[\vp^c,\pb]$ is a solution of the flow equation \eqref{equ:FRGE} at the exact level and we modify the effective action by an arbitrary $k$-independent  functional depending only on the background field, $\Gamma'_k[\vp^c,\pb]=\Gamma_k[\vp^c,\pb]+F[\pb]$, we obtain a new effective action which is still a solution of the flow equation. On the other hand,
if $\Gamma_k[\vp^c,\pb]$ satisfies the sWI \eqref{sWI} the same will no longer be true for $\Gamma'_k[\vp^c,\pb]$, except in the case where $F$ is just a constant.

We can see this also by noting that through \eqref{Legendre-trafo} the effective action $\Gamma'_k$ corresponds to a different bare action than the one that gives $\Gamma_k$. If the latter is associated to the partition function \eqref{partfuncsca} the former corresponds to the same partition function with the replacement $S[\vp+\pb] \rightarrow S[\vp+\pb] - F[\pb]$. According to the derivation of the sWI above, this modification of the path integral would lead to an identity different from \eqref{sWI} due to the fact that the bare action in this path integral does no longer respect the shift symmetry \eqref{shifts}. In other words, the flow equation \eqref{equ:FRGE} does not include the information that in the path integral it is derived from the bare action $S[\vp+\pb]$ depends only on the total field instead of on both fields separately. Indeed, treating the general case now, we note that using a bare action $S[\vp,\pb]$ depending separately on the fluctuation field and the background field in the partition function \eqref{partfuncsca} will lead to the same flow equation \eqref{equ:FRGE} but of course the sWI \eqref{sWI} would no longer hold.

We thus see that as soon as the background field method is employed and the effective action is a functional of both fields separately, the flow equation \eqref{equ:FRGE} will generically admit additional solutions as it alone is not enough to determine the background field dependence of the effective action.

These arguments show that the sWI \eqref{sWI} is not necessarily satisfied for an effective action with background field dependence that solves the flow equation \eqref{equ:FRGE}, but as we see now it suffices to satisfy the sWI at one particular point $k=k_0$ along the flow to ensure that it holds for any scale $k$. If we denote the right hand side of \eqref{Legendre-trafo} by $\mathcal{F}[\vp^c,\pb,k]$ the flow equation is just $\partial_t \Gamma_k[\vp^c,\pb] = \partial_t \mathcal{F}[\vp^c,\pb,k]$ and similarly the sWI becomes $\frac{\delta}{\delta \pb}\left\{\Gamma_k[\vp^c,\pb] - \mathcal{F}[\vp^c,\pb,k]\right\}=0$. Given that the flow equation is satisfied, we see that the time derivative of the sWI automatically vanishes.	

A second reason for imposing the sWI on the effective action independently of the flow equation \eqref{equ:FRGE} has its roots in the fact that the separate background field dependence of $\Gamma_k[\vp^c,\pb]$ comes from the cutoff operator $\mathcal{R}_k[\pb]$. A priori, there is a great deal of freedom in the choice of cutoff operator as far as its dependence on the background field is concerned, and each choice might lead to a different set of solutions to the flow equation. The sWI \eqref{sWI} however captures the dependence of the effective action on the background field as introduced by the cutoff operator and thus has the potential of re-adjusting the structure of the solution space of the flow equation accordingly. In sec. \ref{sec:sWI-LPA} we will see an explicit example of this.

So far we have assumed that the effective action has not been truncated to lie in a subspace of the full theory space. In practice this is however always necessary, in which case $\Gamma_k[\vp^c,\pb]$ only contains operators of a certain specified type. In such a case the arguments given above, showing how the sWI enforces that the bare action depends only on the total field $\vp+\pb$, are no longer applicable, as we cannot make the connection back to a partition function that gives rise to the truncated effective action. In the same way, we cannot generally prove for truncations that the sWI will be satisfied along the whole flow given that it holds at one particular point of it. Nevertheless, judging from the discussion above it would seem that for a truncation which is sufficiently ``close'' to an exact solution of the flow equation, explicitly imposing the sWI will increase the reliability of the truncation.

\subsection{The LPA with the modified split Ward identity}
\label{sec:sWI-LPA}
As argued in the previous section, the sWI \eqref{sWI} has to be considered in conjunction with the flow equation \eqref{equ:FRGE}. By following this procedure we will now see how this allows us to overcome the difficulties encountered in sec. \ref{sec:LPA} in the presence of a field dependent cutoff operator and, in particular, recover the standard fixed point structure of 3-dimensional single component scalar field theory.

At the level of the LPA the ansatz we make for the effective action is
\begin{equation} \label{effact2}
 \Gamma_k[\vp,\pb] = \int_x \left\{ \frac{1}{2}\left(\partial_\mu \vp\right)^2 \right.
 \left. +\frac{1}{2}\left(\partial_\mu \pb\right)^2 + \gamma\partial_\mu\vp \partial^\mu\pb + V(\vp,\pb)\right\},
\end{equation}
where we have canonically normalised the first two derivative terms, included a relative normalisation constant $\gamma$ in the third term and re-named $\vp^c \rightarrow \vp$. This truncation represents the generalised LPA in the context of an effective action which separately depends on the background field. If we wish to impose $Z_2$-symmetry on one or separately on both of the arguments of the effective action the constant $\gamma$ in this ansatz would have to be set to zero.

In what follows we will use the general form \eqref{equ:cutoffh} of the cutoff operator, which we reproduce here for convenience:
\begin{equation} \label{equ:cutoffh1}
 \mathcal{R}_k\left(-\partial^2,\pb\right) =\left(k^2+\partial^2-h(\pb)\right) \theta \! \left(k^2+\partial^2-h(\pb)\right),
\end{equation}
We recall that $h$ is an arbitrary function of the background field and renormalisation group time. Even though we will keep the discussion at the level of a general $h$, we remark that  the reader may specialise to either \eqref{cutoff1} or the appropriate generalisation of \eqref{cutoff2} at any stage. Inserting this and \eqref{effact2} into the sWI \eqref{sWI}, performing the trace and adopting dimensionless variables after absorbing a constant into the definition of the potential and the fields, we are led to the sWI in the LPA which is the first line in the following system of equations. A similar calculation for the flow equation \eqref{equ:FRGE} leads to the second line:
\begin{subequations}
\label{system}
\begin{align}
 \partial_\vp V -\partial_{\pb} V &= \frac{h'}{2} \frac{(1-h)^{d/2}}{1-h+\partial^2_{\vp}V}\,\theta(1-h), \label{sWI-LPA} \\
  \partial_t V - \frac{1}{2}(d-2)\left(\vp \partial_\vp V +\pb \partial_{\pb}V\right) +dV 
 &=  \label{flow2} \\
  & \mkern-40mu \frac{(1-h)^{d/2}}{1-h+\partial^2_{\vp}V} \left(1-h-\frac{1}{2}\partial_t h
 		 +\frac{1}{4}(d-2)\pb h'\right)\theta(1-h).\notag
\end{align}
\end{subequations}
In these equations we have re-instated the space dimension $d$. We thus obtain a coupled system of non-linear partial differential equations governing the dependence of the potential $V$ on the background field and the classical field. 

It is instructive to consider the case \eqref{cutoff1} and explicitly set $h(\pb)=\al \pb^2$, $d=3$ in \eqref{flow2}. Restricting ourselves to fixed points, the result is 
\begin{equation}
 \label{flow2-special}
 3V_* -\frac{1}{2}\left(\vp \partial_\vp V_* +\pb \partial_{\pb}V_*\right) 
  =\frac{\left(1-\al \pb^2\right)^{3/2}}{1-\al \pb^2 +\partial_\vp^2 V_*}\left(1-\frac{1}{2}\al\pb^2\right)\theta\left(1-\al\pb^2\right),
\end{equation}
which we compare to the corresponding fixed point equation \eqref{equ:FPgen} obtained in the single field approximation. As described below \eqref{approx} this approximation is obtained by neglecting the remainder $\hat \Gamma_k$ in \eqref{approx} from the outset and identifying the background field with the total field
$\pb = \phi$ after the evaluation of the Hessian in \eqref{equ:FRGE}. If we make this identification in \eqref{flow2-special}, which also implies $\vp=0$, we recover \eqref{equ:FPgen}. 

Being able to contrast \eqref{flow2-special} with its approximated version \eqref{equ:FPgen} makes the 
consequences of the single field approximation stand out. The right hand side of \eqref{flow2-special} depends on the fluctuation field only through the second derivative of the potential whereas on the right hand side of the approximated equation \eqref{equ:FPgen} every single term contributes. It is therefore to be expected that the solutions to the two equations will show strong qualitative differences and we will confirm this in the following.

We first come back to an analysis of the full coupled system \eqref{system}. In order to gain insight into the combined solution space, we consider the following change of variables
\begin{equation} \label{changevars}
 V = (1-h)^{d/2}\hat V, \qquad \vp = (1-h)^\frac{d-2}{4} \hat \vp -\pb, \qquad t = \hat t -\ln \sqrt{1-h}
\end{equation}
in the region $h(\pb)<1$.
In terms of the running scale $k$, the last equality translates into $\hat k =k\sqrt{1-h}$ which is precisely the replacement that is needed in \eqref{equ:cutoffh1} in order to transform this cutoff operator into the standard field independent cutoff operator where $h=0$.\footnote{To see this, we should keep in mind that $[h]=2$ in \eqref{equ:cutoffh1} whereas in \eqref{changevars} we use the rescaled version of $h$ with $[h]=0$.} 
Since we are working with dimensionless variables, this transformation of the renormalisation group scale $k$ induces corresponding transformations on the field and the potential as expressed in the first two equations in \eqref{changevars}.

Applying the change of variables \eqref{changevars} to \eqref{system} leads to the equivalent system
\begin{subequations}
\label{systemt}
\begin{align}
 \partial_{\pb} \Vh &= \frac{h'}{2(1-h)}\, \mathcal{W}, \label{sWI-LPAt} \\
  \frac{1}{2}(d-2)\pb \partial_{\pb}\Vh &= \frac{1}{1-h}\left(1-h-\frac{1}{2}\partial_t h +\frac{1}{4}(d-2)\pb h'\right)  \mathcal{W}, \label{flow2t}
\end{align}
\end{subequations}
where we have abbreviated
\begin{equation*} 
 \mathcal{W}=\partial_{\hat t} \hat V+ d\hat V-\frac{1}{2}(d-2)\hat\vp \partial_{\hat \vp} \hat V - \frac{1}{1+\partial^2_{\hat \vp} \hat V}.
\end{equation*}
By combining the two equations in \eqref{systemt}, we see immediately that this system is in turn equivalent to
\begin{equation} \label{systemfinal}
\partial_{\pb} \Vh 	=0, \qquad \mathcal{W}			=0
\end{equation}
as long as $h$ does not satisfy $1-h-\frac{1}{2}\partial_t h =0$. However, the last equation has the solution $h(\pb)=1+c(\pb)\exp(-2t)$, where $c$ is an arbitrary function of the background field. If we substitute this form of $h$ into the cutoff operator \eqref{equ:cutoffh1} we find that it is actually independent of the scale $k$, i.e. it does not lead to a renormalisation group flow and thus we can safely discard this special case.

From now on we implement the condition $\partial_{\pb} \Vh=0$ in \eqref{systemfinal} by dropping the dependence of $\Vh$ on the background field and tacitly taking $\Vh$ to depend only on the transformed field $\ph$ and renormalisation group time $\that$. We can therefore conclude that the original system \eqref{system} is fully equivalent to $\mathcal{W}=0$, i.e.
\begin{equation} \label{reduced}
 \partial_{\hat t} \hat V+ d\hat V-\frac{1}{2}(d-2)\hat\vp \partial_{\hat \vp} \hat V = \frac{1}{1+\partial^2_{\hat \vp} \hat V},
\end{equation}
in the sense that the change of variables \eqref{changevars} supplies a one to one mapping between the solutions $V_t(\vp,\pb)$ of \eqref{system} and the background field independent solutions $\hat V_{\hat t}(\hat \vp)$ of the reduced flow equation \eqref{reduced}. This reduced flow equation however is precisely the flow equation we would have obtained with a standard background field independent cutoff operator, given
by \eqref{equ:flowLPA} with $h=0$. 

As we can see from the simplified system \eqref{systemt}, it is only through the requirement that the sWI \eqref{sWI-LPAt} hold that the solutions of \eqref{reduced} are actually independent of the background field.
This is the additional crucial ingredient supplied by the sWI which is not contained in the flow equation \eqref{flow2t} alone.

The possibility to reduce the renormalisation group flow \eqref{flow2} for any cutoff operator of form \eqref{equ:cutoffh1} to the standard flow \eqref{reduced} is a striking manifestation of universality. Even though the flows \eqref{flow2} may potentially look very different from the standard flow \eqref{reduced} depending on the choice of $h$, exploiting the sWI paves the way to recognising their equivalence.

In order to flesh out the details of this equivalence of renormalisation group flows, let us specialise to the investigation of their fixed points. At a fixed point we drop all $t$-dependence in the flow equation \eqref{flow2} and thus the dimensionless potential has to be renormalisation group time independent, $\partial_t V=0$. Accordingly, we also require $\partial_t h=0$ at a fixed point and denote the corresponding $t$-independent function by $h_*$. Using \eqref{changevars} and the chain rule, we see that setting $\partial_t V =0$ and $\partial_th =0$ in \eqref{flow2} is equivalent to setting $\partial_{\hat t} \hat V=0$ in the reduced flow \eqref{reduced}. As expected, we therefore find a one to one mapping of the fixed points of the two renormalisation group flows which is given by the fixed point version of the change of variables \eqref{changevars}, i.e.
\begin{equation}
 \label{FPs-map}
 V_*(\vp,\pb)=(1-h_*(\pb))^{d/2}\,\hat V_*\!\!\left((1-h_*(\pb))^{\frac{2-d}{4}}(\vp+\pb)\right).
\end{equation}
It now becomes apparent how the inaccurate fixed point structure of the renormalisation group flow in sec. \ref{sec:gencutoff} is avoided by the approach described in this section. Specialising the fixed point version of \eqref{reduced} to $d=3$ we recover the fixed point equation \eqref{equ:fpWF}, accurately describing the fixed point structure of 3-dimensional single component scalar field theory. Through \eqref{FPs-map} we conclude that even the full background field dependent flow \eqref{flow2} admits only the Gaussian and Wilson-Fisher fixed points originating from \eqref{equ:fpWF}, provided we also impose the pertaining sWI \eqref{sWI-LPA}. A similar resolution of the inaccurate fixed point structure of sec. \ref{sec:LPALitim} is only partly
achieved by the map \eqref{FPs-map} as in this case this relation between fixed point solutions still takes the form of a differential equation. We will come back to this and complete the equivalence in sec. \ref{sec:sWI-Vpp}.

Taking it one step further, the equivalence of \eqref{system} to \eqref{reduced} also manifests itself at the level of the eigenspectra and eigenoperators of these flows. Proceeding in the usual way, we write
\begin{equation}\label{linVt}
 V_t(\vp,\pb)=V_*(\vp,\pb) + \eps\, v(\vp,\pb)  \exp(-\lambda t)
\end{equation}
and require that the original system \eqref{system} be satisfied up to linear order in $\eps$. Similarly, for
\begin{equation} \label{linVht}
 \hat V_{\hat t}(\hat \vp) = \hat V_*(\hat \vp) + \eps \, \hat v(\hat \vp) \exp(-\lambda \hat t)
\end{equation}
we impose that \eqref{reduced} hold up to linear order in $\eps$. Our goal is to show by using the correspondence \eqref{changevars} that for every eigenoperator $v$ with associated eigenvalue $\lambda$ we find a corresponding eigenoperator solution $\vh$ of the linearisation of \eqref{reduced} with the same eigenvalue $\lambda$ and vice versa, thereby proving equality of eigenspectra. In order to linearise the flow equation \eqref{flow2} after substituting \eqref{linVt} we also need to perturb the function $h$ according to
\begin{equation}
\label{linh1}
 h_t(\pb) = h_*(\pb) + \eps \, \delta h(t,\pb).
\end{equation}
With this and \eqref{linVt}, the change of variables \eqref{changevars} becomes at order $\eps$,
\begin{equation} \label{dh-tdep}
 v \exp(-\lambda t) = (1-h_*)^\frac{d-\lambda}{2}\vh \exp(-\lambda t)
		      -\frac{\delta h}{2}\frac{(1-h_*)^{\frac{d}{2}-1}}{1+\partial_{\ph}^2\Vh_*}, 
\end{equation}
where the variables are now related by the fixed point version of the change of variables \eqref{changevars},
\begin{equation}
\label{linchangevars}
 V_* = (1-h_*)^{d/2}\hat V_*, \qquad \vp = (1-h_*)^\frac{d-2}{4} \hat \vp -\pb, \qquad t = \hat t -\ln \sqrt{1-h_*}.
\end{equation}
We then recognise from \eqref{dh-tdep} that $\delta h \exp(\lambda t)$ has to be $t$-independent and thus we can write 
\begin{equation}
 \label{linh2}
   h_t(\pb) = h_*(\pb) + \eps \, \kappa(\pb)\exp(-\lambda t).
\end{equation}
This separable form of the perturbation $\delta h$ in \eqref{linh1} is a necessary requirement if the two systems \eqref{system} and \eqref{reduced} are to possess the same eigenspectra and is tied to the separability used in \eqref{linVt} and \eqref{linVht}. In particular, it is applicable if the function $h$ depends on renormalisation group time $t$ only through the potential $V$ as in \eqref{cutoff2}. Applying this to \eqref{dh-tdep}, we obtain the relation between the original eigenoperator $v(\vp,\pb)$ and the eigenoperator $\vh(\ph)$,
\begin{equation} \label{rel-eops}
 v = (1-h_*)^\frac{d-\lambda}{2}\vh
		      -\frac{\kappa}{2}\frac{(1-h_*)^{\frac{d}{2}-1}}{1+\partial_{\ph}^2\Vh_*}. \end{equation}
Using \eqref{linVt} and \eqref{linh2} in the original system \eqref{system}, we are led to the following linearised equations for the eigenoperator $v$:
\begin{subequations}
\label{linsystem}
\begin{align}
 \partial_\vp v - \partial_{\pb} v = \frac{1}{2}\frac{(1-h_*)^{d/2}}{1-h_*+\partial_\vp^2 V_*}\left(\kappa'-\frac{d}{2}
				      \frac{\kappa}{1-h_*}h_*'
				      -\frac{\partial_\vp^2 v-\kappa}{1-h_*+\partial_\vp^2V_*} h_*'\right),
				      \label{sWI-lin} \\
(d-\lambda)v-\frac{1}{2}(d-2)(\vp+\pb)\partial_\vp v = \frac{(1-h_*)^{d/2}}{1-h_*+\partial_\vp^2 V_*}
				      \left((\lambda-d-2)\frac{\kappa}{2}
				      -\frac{(1-h_*)(\partial_\vp^2 v - \kappa)}{1-h_*+\partial_\vp^2 V_*} \right). \label{flow2-lin}
\end{align}
\end{subequations}
The first equation is the linearised sWI and the second equation comes from linearising the flow equation \eqref{flow2} and subtracting the linearised sWI multiplied by the factor $\frac{1}{2}(d-2)\pb$.

It is then a straightforward calculation to confirm that the linearised system \eqref{linsystem} is fully
equivalent to the linearisation of the reduced flow equation \eqref{reduced} according to \eqref{linVht}, given by
\begin{equation}
 \label{linreduced}
 (\lambda-d)\vh + \frac{1}{2}(d-2)\ph \vh' = \frac{\vh''}{\left(1+ \Vh_*''\right)^2}.
\end{equation}
As expected from the equivalence of the full renormalisation group flows \eqref{system} and \eqref{reduced} described by the change of variables \eqref{changevars}, the equivalence at the linearised level is given by the adapted change of variables \eqref{linchangevars} and the relation between the eigenoperators \eqref{rel-eops}, mapping an eigenoperator solution $v(\vp,\pb)$ of \eqref{linsystem} with eigenvalue $\lambda$ onto the eigenoperator solution $\vh(\ph)$ of \eqref{linreduced} with the same eigenvalue $\lambda$. Hence the eigenspectrum of the flow equation \eqref{flow2} complemented by the sWI \eqref{sWI-LPA} is identical to the eigenspectrum of the reduced flow \eqref{reduced}. 

Specialising to $d=3$ in \eqref{linreduced}, we recover the eigenoperator equation \eqref{equ:WFeops} and thus we conclude that also at the linearised level, the inaccurate description of 3-dimensional single component scalar field theory described in sec. \ref{sec:gencutoff} can be rectified completely by keeping track of the background field through the sWI \eqref{sWI-LPA}.

We remark that the discussion so far is restricted to the range $h(\pb)<1$ where the change of variables \eqref{changevars} is valid. For $h(\pb)>1$ the right hand sides of the system \eqref{system} vanish and we obtain a different solution which then has to be matched onto the previous solution at $h(\pb)=1$. Since we are mostly interested in $d=3$ dimensions let us briefly illustrate how this matching can be done for the fixed point and eigenoperator solutions in this case.

Focussing on fixed point solutions first, the solution of \eqref{sWI-LPA} and the fixed point version of \eqref{flow2} for $h_*(\pb)>1$ is $V_*(\vp,\pb)=A(\vp+\pb)^6$, in analogy to the solution of the left hand side of \eqref{equ:FPgen}. This solution needs to be matched onto the solution \eqref{FPs-map} at $h_*(\pb) \rightarrow 1$ for both possibilities, the Wilson-Fisher fixed point and the Gaussian fixed point. 
Concentrating on the Wilson-Fisher fixed point, we see that the argument of $\hat V_*$ in \eqref{FPs-map} diverges in this limit since $d>2$.\footnote{Depending on the precise form of $h_*$ there could be an exception to this if $\vp+\pb \rightarrow 0$ at the same time. In what follows we will therefore assume that $\vp \neq -\pb$ in the limit $h_*(\pb)\rightarrow 1$.} This implies that we can make use of \eqref{equ:VWFasy} describing the asymptotic behaviour of $\hat V_\mathrm{WF}$ and obtain
\begin{equation} \label{potsolasy}
V_\mathrm{WF}(\vp,\pb) = A_\mathrm{WF}(\vp+\pb)^6 + \frac{(1-h_*)^{5/2}}{150A_\mathrm{WF}(\vp+\pb)^4} + \dots
\end{equation}
in the limit $h_*(\pb)\rightarrow 1$. All subleading terms in this expansion vanish individually in the limit we are considering. Assuming that therefore the remainder vanishes as well, we find perfect matching between the two solutions at $h_*(\pb)=1$. From \eqref{potsolasy} we can also appreciate that this global solution is smooth in $\vp$ at $h_*(\pb)=1$ and therefore smooth everywhere in $\vp$. On the other hand, we can see that the third and all higher derivatives of $V_\mathrm{WF}$ with respect to the background field $\pb$ will diverge as $h_*(\pb)\rightarrow 1$ as caused by the subleading terms in \eqref{potsolasy}. This coincides with the observation we made just below \eqref{asyexpcoeffs} and the same conclusion can also be drawn from the sWI \eqref{sWI-LPA} itself, once we know that $V_\mathrm{WF}$ is smooth in $\vp$. As before, these divergences of derivatives of the fixed point potential with respect to the background field are a direct consequence of the cutoff choice \eqref{equ:cutoffh1}. However we see that as a result of a full treatment of the background field dependence as dictated by the sWI they do not occur for the fluctuation field $\vp$. The same statements are of course true for the Gaussian fixed point whose global form we can write down explicitly:
\begin{equation*}
 V_\mathrm{G}(\vp,\pb)=\frac{1}{3}(1-h_*(\pb))^{3/2} \theta(1-h_*(\pb)).
\end{equation*}
Note that this potential is in fact only a function of the background field.

In order to investigate a similar extension of the eigenoperator solutions $v(\vp,\pb)$ into the range $h(\pb)>1$, we first note that also at the linearised level the condition $h(\pb)>1$ is equivalent to $h_*(\pb)>1$ as the derivative of the step function does not lead to a contribution to the linearised equations. If we want to match the solution \eqref{rel-eops} valid in the range $h_*	(\pb)<1$ to the solution $v(\vp,\pb)=|\vp+\pb|^{-2\lambda+6}$ of the left hand sides of \eqref{linsystem} we can proceed in a similar way as for the fixed point solution. For the Wilson-Fisher fixed point we can make use of the asymptotic expansion \eqref{equ:WFeopsasy} of the eigenoperator solutions, the argument of $\hat v$ in \eqref{rel-eops} being divergent in the limit $h_*(\pb)\rightarrow 1$. With the second term in \eqref{rel-eops} vanishing in this limit, we find that
\begin{equation*}
v(\vp,\pb) = |\vp+\pb|^{-2\lambda+6} -\frac{1}{4500}\frac{(2\lambda-5)(2\lambda-6)}{A_\mathrm{WF}}
\frac{(1-h_*)^{5/2}}{|\vp+\pb|^{2\lambda+4}} + \dots
\end{equation*}
and by the same arguments as below \eqref{potsolasy} we see that $v$ is a globally defined function which is smooth in $\vp$ and twice differentiable in $\pb$. As a consequence of the step function in our choice of cutoff all higher derivatives of $v$ with respect to the background field diverge as $h_*(\pb)\rightarrow 1$.

\subsection{Dealing with a cutoff depending on the potential} \label{sec:sWI-Vpp}
 A crucial step in the process of showing how the sWI reduces the background field dependent flow  \eqref{flow2} to the standard flow of a single scalar field in the LPA \eqref{reduced} is represented by the change of variables \eqref{changevars}. This is a well-defined change of variables if the cutoff modification $h$ is an explicit function of the background field but it does not supply us with a direct definition of $V(\vp,\pb)$ if $h$ is a function of the potential itself. For example, this is the case for 
 \begin{equation}
 \label{hVpp}
  h_t(\pb)= \left. \partial^2_\vp V_t(\vp,\pb)\right|_{\vp=\pb},
 \end{equation}
where in this instance we have emphasised the $t$-dependence that is usually left implicit. This is the appropriate generalisation of \eqref{cutoff2} in the context of an effective action fully depending on the background and fluctuation fields at the level of the LPA. We will now show how the steps of the previous section can be adapted for this particular modification of the cutoff operator \eqref{equ:cutoffh1} in order to illustrate how the sWI again leads to the correct description of single component scalar field theory in $d=3$, thereby comprehensively rectifying the inaccurate results found in sec. \ref{sec:LPALitim}.

We start by implicitly defining a new field variable $\tilde \vp$ through
\begin{equation}
 \label{tphi}
 \pb = \frac{1}{2}\frac{\tilde \vp}{\left(1+\Vh''(\tilde \vp)\right)^{1/4}},
\end{equation}
where as before $\Vh$ is a solution of the reduced flow equation \eqref{reduced}. This is trivially possible at the Gaussian fixed point but also at the Wilson-Fisher fixed point as plotting the right hand side of \eqref{tphi} proves it to be a strictly monotonically increasing function. The same will be true for small $t$-dependent perturbations around either of these fixed points and the following steps hold whenever this relation can be inverted to obtain $\tilde \vp$ as a function of the background field and renormalisation group time.
In general, the allowed range of $\pb$ in $\tilde \vp(\pb)$ is bounded, e.g. at the Wilson-Fisher fixed point 
we have
\begin{equation*}
 \lim_{\tilde \vp \rightarrow -\infty} \pb(\tilde \vp) = -\frac{1}{2\left(30 A_\mathrm{WF}\right)^{1/4}}
 <  \pb(\tilde \vp) < \frac{1}{2\left(30 A_\mathrm{WF}\right)^{1/4}} = \lim_{\tilde \vp \rightarrow \infty} \pb(\tilde \vp)
\end{equation*}
as can be deduced from its asymptotic form \eqref{equ:VWFasy}.We then define the following change of variables
\begin{equation}
\label{changevarsVpp}
\begin{gathered}
 \ph = \left(1+\Vh''(\tilde \vp)\right)^{1/4} (\vp + \pb), \qquad V(\vp,\pb) = \left(1+\Vh''(\tilde \vp)\right)^{-3/2} \Vh(\ph), \\
  t = \hat t + \ln \sqrt{1+\Vh''(\tilde \vp)},
\end{gathered}
\end{equation}
where we always regard $\tilde \vp$ as a function of $\pb$. A short calculation then shows
\begin{equation}
\label{hVpp-relation}
 1-h(\pb)=1-\left. \partial^2_\vp V(\vp,\pb)\right|_{\vp=\pb}=\frac{1}{1+\Vh''(\tilde \vp)}.
\end{equation}
Using this identity, the change of variables \eqref{changevarsVpp} assumes the form of the change of variables
\eqref{changevars} of the previous section, implying that the results between the equivalence of the complicated system \eqref{system} with \eqref{hVpp} and the standard flow \eqref{reduced} carry over to the present case.

In particular, the relation \eqref{hVpp-relation} gives 
\begin{equation} \label{solh}
 h_*(\pb) = \left. \partial^2_\vp V_*(\vp,\pb)\right|_{\vp=\pb} = \frac{\Vh_*''(\tilde \vp)}{1+\Vh_*''(\tilde \vp)},
\end{equation}
through which the fixed point solutions are given by \eqref{FPs-map} with $d=3$, $\Vh_*$ being either the Gaussian fixed point or the Wilson-Fisher fixed point of the standard flow \eqref{reduced}. In the case of the Gaussian fixed point we simply have $h_*=0$ and $V_*=1/3$.

Furthermore, \eqref{hVpp-relation} also implies that in \eqref{linh2} we have
\begin{equation*}
\kappa(\pb) =  \left. \partial^2_\vp v(\vp,\pb)\right|_{\vp=\pb} = \frac{\hat v''(\tilde \vp)}{\left(1+
\Vh_*''(\tilde \vp)\right)^2}
\end{equation*}
which together with \eqref{solh} defines the eigenoperator solutions via \eqref{rel-eops} and as an immediate consequence  the eigenspectra of the standard flow \eqref{reduced} in $d=3$ are again shown to be equal to the ones of the corresponding fixed points of the system \eqref{system}. For example, at the Gaussian fixed point $V_*=1/3$ the eigenoperator solutions of \eqref{linsystem} are
\begin{equation*}
v(\vp,\pb)=\vh(\vp+\pb)-\frac{1}{2}\vh''(2\pb),
\end{equation*}
where $\vh$ is any eigenoperator solution of \eqref{linreduced} at the Gaussian fixed point $\Vh_*=1/3$.

This completes our discussion of how the inaccurate results of sec. \ref{sec:gencutoff} and \ref{sec:LPALitim}
in the context of the single field approximation with field dependent cutoff operators can be understood and corrected by the use of the modified split Ward identity \eqref{sWI}.

%% file: conformal/conformal.tex
\chapter[Conformal gravity]{Background independence and conformal gravity}
\label{sec:conformal}

The background field formalism in the context of functional renormalisation leads to an effective action that depends separately on the background field, as we have seen in the previous chapter, and this dependence is induced notably by a background field dependent cutoff operator. The effective action therefore lives in an appropriately enlarged theory space, compared to when only the total field is present. We have also seen that a common approach to deal with this extended theory space is to project the effective action onto the subspace where the total field and the background field are identified, i.e. to make use of the single field approximation. The essential insight gained from the previous chapter is that for RG flows in this approximation the flow equation \eqref{equ:FRGE} is very likely not sufficient to avoid non-physical results. But if the single field approximation is discarded, the RG flow described by \eqref{equ:FRGE} evolves in the full theory space spanned by operators involving both the total field and the background field. In a sense, the change from the original theory space to this enlarged theory space happened almost by hand, namely by manually introducing a background field, and hence these additional directions in theory space have to be constrained by an appropriate identity involving the background field, i.e. the split Ward identity \eqref{sWI}. We have emphasised in sec. \ref{sec:sWI-derivation} that the flow equation alone is not powerful enough to eliminate the additional freedom gained by introducing a background field. As the results of the previous chapter demonstrate, the sWI can be used to fully control the background field dependence of the effective action and restore the physically correct RG flow.

While the previous chapter dealt with scalar field theory, it is desirable to follow the same strategy in asymptotic safety for gravity. One may hope that combining the flow equation with the split Ward identity is a way forward for resolving the serious problems of the $f(R)$ truncation described in chapters \ref{sec:f-of-R} and \ref{sec:red-ops}. However, extending the single field $f(R)$ truncation to a fully-grown bi-field truncation is computationally extremely challenging. In this chapter we will therefore implement the sWI alongside the flow equation in the less involved setting of conformally reduced gravity and we will see that this strategy is again sufficient to fully control the background field dependence of the effective action.

A related second motivation for implementing the sWI along with the flow equation has to do with background independence. This is the simple statement that a background field split as in \eqref{background-split} must not affect physical observables in the sense that they have to be fully independent of the background metric $\bgmn$. The background field may (and is bound to) appear at intermediate stages of the calculation but is required to drop out at the very end, e.g. in the evaluation of Green's functions. The setup and the ensuing flow equation for the effective action \eqref{effacgrav} of sec. \ref{sub:adaptations-gravity} never requires the background field $\bgmn$ to be specified \cite{Reuter:1996}, implying that no background configuration assumes a distinguished role in this approach of quantising gravity and that effectively quantisation proceeds on all backgrounds simultaneously. But making sure that the formalism does not rely on a particular background field is only the first step towards background independence of physical observables since it could still be the case that each background field configuration leads to \emph{different} observables due to the additional background field dependence in the effective action. Instead, background independence is a further condition on the effective action that needs to be imposed separately, and in addition to the flow equation. We will see in this chapter that by taking care of the background field dependence of the effective action, the sWI also implements exact background field independence to the maximum extent possible for the setup of the following sections.

For more details on the first step towards background independence by quantisation on all backgrounds simultaneously see also \cite{Reuter:2008wj,Reuter:2008qx,Manrique:2010am,Manrique:2009uh}. Background independence as a condition in the above described strict sense has recently been implemented in \cite{Becker:2014qya}, and will certainly play an important role as the field of asymptotic safety progresses into more advanced stages of its development. 

Asymptotic safety for conformal gravity has been investigated previously in single field truncations in \cite{Reuter:2008wj, Reuter:2008qx, Bonanno:2012dg,Machado:2009ph} and \cite{Demmel:2014sga,Demmel:2013myx,Demmel:2012ub}, where the latter three studies focus on three-dimensional conformal $f(R)$ truncations. Work beyond the single field approximation in conformal gravity was first carried out in \cite{Manrique:2009uh} in the bi-field LPA, where it was found that if the sWI is imposed alongside the flow equation, no non-perturbative fixed points exist. As we will see in sec. \ref{sec:confLPA} the contrary is true in the LPA for our study. 

\section{The conformal truncation}
\label{sec:conftrunc}
We truncate to the conformal degree of freedom of the metric \eqref{background-split} by writing
\be
\label{equ:confpar}
\gmn = f(\chi + \vp)\, \hat g_{\mu\nu} \qquad \text{and} \qquad \bgmn = f(\chi)\, \hat g_{\mu\nu},
\ee
where we consider only metrics that are conformally equivalent to the fixed reference metric $\hat g_{\mu\nu}$. For the moment we have left unspecified the way in which the conformal factor is parametrised as encoded in the function $f$. In order to represent a valid parametrisation of the conformal factor,  it needs to be positive. Previously used choices include $f(\phi) = \exp(2\phi)$, e.g. \cite{Machado:2009ph}, or $f(\phi)=\phi^2$, e.g. \cite{Bonanno:2012dg}. The definitions \eqref{equ:confpar} imply that we can view $\chi(x)$ as the background conformal factor and $\vp(x)$ as the quantum conformal fluctuation field, making up the total quantum conformal factor $\phi(x) = \chi(x) + \vp(x)$. A path integral quantisation of $\vp$ will then lead to a classical fluctuation field $\vp^c = \langle \vp \rangle$ and total classical field $\phi^c = \langle \phi \rangle = \chi + \vp^c$. The quantum fluctuation field of the metric $\hmn = \gmn-\bgmn$ is given by $\hmn = \left[f(\phi)-f(\chi)\right]\hat g_{\mu\nu}$. Note that the difference on the right hand side may no longer be positive, corresponding to the fact that $\hmn$ need not be a metric field.

Although the field dependence of the parametrisation function in \eqref{equ:confpar} is left arbitrary for now, we will assume in the following that $f$ does not depend on the RG scale $k$, cf. the discussion related to this point in chapter \ref{sec:conclusions}.

In this conformal truncation, the effective action takes the form $\Gamma_k[\vp^c,\chi]$, where we have omitted a parametric dependence on the fixed reference metric $\hat g$, and it satisfies the flow equation 
\be
\label{equ:flowGamma}
\partial_t \Gamma_k[\vp^c,\chi] = \frac{1}{2}\mathrm{Tr}\left[\frac{1}{\sqrt{\bar g}\sqrt{\bar g}}\frac{\delta^2\Gamma_k}
				  {\delta \vp^c \delta \vp^c}+ \cutoff[\chi]\right]^{-1} \partial_t \cutoff[\chi].
\ee
This version of the general flow equation \eqref{equ:FRGE} is adapted to the present setup and will be derived in the next section. We emphasise again that the background field dependence of the cutoff operator in this expression is a consequence of the background field method as discussed in sec. \ref{sub:adaptations-gravity} and leads to RG flows that in the single field approximation would therefore be prone to the type of non-physical behaviour discussed in the previous chapter.

As announced before, here we keep track of the background field dependence through the sWI \eqref{sWI} which in the present context assumes the following form,
\be
\label{equ:sWiGamma}
\frac{1}{\sqrt{\bar g}}\left(\frac{\delta\Gamma_k}{\delta \chi}-\frac{\delta \Gamma_k}{\delta \vp^c}\right)
      =\frac{1}{2}\mathrm{Tr}\left[\frac{1}{\sqrt{\bar g}\sqrt{\bar g}}\frac{\delta^2\Gamma_k}
				  {\delta \vp^c \delta \vp^c}+ \cutoff[\chi]\right]^{-1} 
				  \left(\frac{1}{\sqrt{\bar g}}\frac{\delta \cutoff[\chi] }{\delta \chi}+Y_f \cutoff[\chi]\right).
\ee
Exact background independence as described in the introduction of this chapter would be realised if the right hand side of the sWI was zero, implying that the effective action is only a functional of the total field $\phi^c = \chi + \vp^c$. The presence of the cutoff operator however causes the right hand side to be non-vanishing in general. It is only in the limit $k\rightarrow0$ that the cutoff operator drops out and background independence can be restored exactly. We note therefore that imposing the sWI in addition to the flow equation \eqref{equ:flowGamma} automatically ensures exact background independence in the limit $k\rightarrow0$. The function $Y_f$ in \eqref{equ:sWiGamma} is connected to the parametrisation of the conformal factor in \eqref{equ:confpar} and will be determined explicitly in sec. \ref{sec:confsWI-derivation}.

When all degrees of freedom of the metric are considered in full gravity, the sWI \eqref{equ:sWiGamma} receives additional terms on its right hand side. Beyond the analogue of \eqref{equ:sWiGamma} for the classical metric fluctuation field $\hmn^c$ it contains terms originating from cutoff terms in the ghost sector, the ghost action itself and the gauge fixing term. In the limit $k\rightarrow 0$ all cutoff terms are required to vanish and exact background independence can in principle be restored by noting that the gauge fixing and ghost terms on the right hand side of the sWI are expected to drop out upon going on-shell.

For the truncation considered in this chapter we will show in sec. \ref{sec:combine} that the sWI can be imposed alongside the flow equation for any value of $k$, leading to a reduced set of flow equations in a manner similar to scalar field theory in sec. \ref{sec:sWI-LPA}. This transformed system of flow equations is independent of the parametrisation function $f$ of \eqref{equ:confpar} and therefore the RG flow it describes is also independent of the conformal parametrisation, as one may expect on physical grounds. This exemplifies how the sWI can take appropriate care of the separate background field dependence of $\Gamma_k[\vp^c,\chi]$ all along the renormalisation group flow and that in particular in the limit $k\rightarrow 0$ it implements exact background independence.

However, the transformed system is expressed in new variables and it is formulated in terms of a new RG scale $\hat k$. In contrast to sec. \ref{sec:sWI-LPA}, it turns out that fixed points with respect to the original RG scale $k$ are not in general related to fixed points with respect to the transformed scale $\hat k$. Instead, $k$-independent solutions of the original system are given by $\hat k$-dependent solutions of the transformed system and vice versa, and thus the two RG scales can in principle lead to different fixed point structures. But this is only the case if $f$ is left completely general. For parametrisations in \eqref{equ:confpar} of power law type we show in sec. \ref{sec:fpmap} that this distinction is absent and the fixed point structures of the two RG flows can be mapped onto each other. In this way, requiring fixed points of the original flow to correspond to fixed points of the reduced flow can be viewed as a selection criterion for parametrisation functions $f$. This is the approach we will take from sec. \ref{sec:fpmap} onwards.

\section{The flow equation for the conformal factor}
\label{sec:flowequ-derivation}
Truncating theory space to contain functionals of the conformal factor only by writing \eqref{equ:confpar} leaves us to consider effective actions that depend on the two scalar fields $\vp^c$ and $\chi$. This simplifies the flow equation for $\Gamma_k$ as we no longer have to take into account any gauge fixing or ghost terms, where the choice of a fixed reference metric $\hat g$ in \eqref{equ:confpar} can be viewed as effectively fixing the gauge. Nevertheless, the flow equation does not simply reduce to that of scalar field theory in the presence of a background field but still contains important features indicating its origin from a flow equation on a truncated theory space of quantum gravity. In order to spell out these differences we go through the derivation of \eqref{equ:flowGamma} from the path integral in some detail.

The starting point is the Euclidean path integral over the fluctuation part of the conformal factor,
\be
\label{equ:pathint-conffact}
\exp(W_k) = \int  \mathcal{D}\vp \, \exp\left(-S[\chi + \vp] - S_k[\vp,\bar g] + S_\mathrm{src}[\vp,\bar g]\right).
\ee
In this equation and in what follows we will keep the background metric $\bar g$ for the sake of generality and make the identification as in the second equation of \eqref{equ:confpar} only when necessary at a later stage. Under the conformal truncation \eqref{equ:confpar}, the bare action for gravity turns into the bare action for the total conformal factor, 
\begin{equation} \label{equ:confBareAction}
S[\bar g + h] \rightarrow S[\chi +\vp].
\end{equation}
The cutoff action $S_k$ has to suppress the momentum modes of the dynamical field $\vp$ and takes the form
\be
\label{equ:cutoff-action}
S_k[\vp,\bar g] = \frac{1}{2}\int d^dx \sqrt{\bar g(x)}\, \vp(x) \cutoff\![\bar g]\, \vp(x),
\ee
where we are operating in $d$ dimensions and have explicitly kept the Riemannian measure of the background field in the same way as it appears in the cutoff action $S_k[\hmn,\bgmn]$ of the full gravitational path integral, cf. \cite{Reuter:1996}. We have also made explicit the dependence of the cutoff operator on the background metric. Again bearing the gravitational setting in mind, the source term is
\be
\label{equ:source}
S_\mathrm{src}[\vp,\bar g] = \int d^dx\sqrt{\bar g(x)}\, \vp(x)J(x).
\ee
Taking a $t$-derivative of \eqref{equ:pathint-conffact} leads to
\be
\label{equ:dtWk}
\partial_t W_k = -\frac{1}{2}\int d^dx\sqrt{\bar g(x)}\, d^dy\sqrt{\bar g(y)}\, \partial_t\! \cutoff[\bar g](x,y) 
		  \langle \vp(x) \vp(y)\rangle,
\ee
where we have rewritten the cutoff operator using its kernel according to
\be
\label{equ:kernel-notation}
\cutoff\!(x,y) = \mathcal{R}_{k,x}\, \frac{\delta(x-y)}{\sqrt{\bar g(y)}},
\ee
with the subscript in $\mathcal{R}_{k,x}$ indicating that this differential operator acts on $x$-dependence. We then proceed in the usual way via a Legendre transformation with respect to the source,
\be
\label{equ:LegendreTrans}
\tilde \Gamma_k[\vp^c,\bar g] = \int d^dx\sqrt{\bar g(x)} \, J(x)\vp^c(x) - W_k[J,\bar g],\qquad \text{with} \qquad
	\vp^c =\frac{1}{\sqrt{\bar g}}\frac{\delta W_k}{\delta J}
\ee
so that using the well-known identity
\be
\label{equ:rel-expval}
\langle\vp(x) \vp(y)\rangle = 
\left(\frac{1}{\sqrt{\bar g(x)}\sqrt{\bar g(y)}}\frac{\delta^2\tilde \Gamma_k}{\delta \vp^c(x)\delta\vp^c(y)}\right)^{-1} + \vp^c(x)\vp^c(y)
\ee
together with the redefinition $\Gamma_k[\vp^c,\bar g] = \tilde \Gamma_k[\vp^c,\bar g] - S_k[\vp^c,\bar g]$ turns \eqref{equ:dtWk} into
\be
\label{equ:flowGamma-detail}
\partial_t\Gamma_k[\vp^c,\bar g] = \frac{1}{2}\int_x \int_y
				  \left[\Gamma^{(2)}(x,y) + \cutoff\!(x,y)\right]^{-1} \, \partial_t \cutoff(y,x).
\ee
Here, we have generalised the convention \eqref{dotconvention} to $\int_x \equiv \int d^dx\sqrt{\bar g(x)}$ and written
\begin{equation}
\Gamma^{(2)}(x,y)=\frac{1}{\sqrt{\bar g(x)}\sqrt{\bar g(y)}}\frac{\delta^2 \Gamma_k}{\delta \vp^c(x)\delta\vp^c(y)}
\end{equation}
for the Hessian of the effective action. We have further exploited the fact that the kernel $\cutoff\!(x,y)$ is symmetric. The flow equation \eqref{equ:flowGamma-detail} is the detailed form of \eqref{equ:flowGamma} and it contains the appropriate factors of the background field in places where they are to be expected in a truncation of a gravitational effective action.

We note at this point that because of the replacement \eqref{equ:confBareAction} in the path integral \eqref{equ:pathint-conffact} using \eqref{equ:confpar}, the effective action in \eqref{equ:flowGamma-detail} has already a dependence on $\chi$ that should appear as a third argument. Later on we will make the identification $\bgmn = f(\chi)\, \hat g_{\mu\nu}$ so that the second argument of the effective action $\Gamma_k[\vp^c,\bar g]$ turns into the background conformal factor $\chi$, accounting for all dependence on  $\chi$.

Connected with the conformal truncation \eqref{equ:confBareAction}, there is another issue worth mentioning. Any truncation ansatz for the effective action $\Gamma_k[\vp^c,\bar g]$ in \eqref{equ:flowGamma-detail} that reflects its gravitational nature will contain the parametrisation function $f$ entering through the second equation in \eqref{equ:confpar}, cf. \eqref{equ:ansatzGamma}. As mentioned previously, we will show in sec. \ref{sec:combine} that this dependence on the way the conformal factor is parametrised can be eliminated through the use of the sWI of the following section. But there is a more subtle way in which the parametrisation function enters the current setup. Starting from a bare action of the metric field $S\!\left[\gmn\right]=S[\bgmn + \hmn]$ and applying the conformal truncation \eqref{equ:confpar} leads to a bare action $S_f[\chi + \vp]$ that depends on the parametrisation function $f$. In principle this $f$-dependence carries through to the effective action in the flow equation \eqref{equ:flowGamma-detail} and this would seem to be an additional dependence on $f$ that is not explicitly represented in any ansatz one might propose for $\Gamma_k[\vp^c,\bar g]$. However, if we formally imagine the effective action to be expanded in terms of all allowed operators,
\begin{equation} \label{equ:GammaOpExp}
\Gamma_k[\vp^c,\bar g] = \sum_n g_n(k)\, \mathcal O_n[\vp^c,\bar g]
\end{equation}
with renormalisation group scale dependent couplings $g_n(k)$, there will be a map between the couplings of this expansion and the bare couplings contained in the bare action $S[\gmn] = S[f(\chi+\vp)\hat g_{\mu\nu}]$ as given by the Legendre transform \eqref{equ:LegendreTrans}. Adopting a different choice of parametrisation function $f$ in \eqref{equ:confpar} will only lead to a correspondingly modified map between the bare couplings and the couplings in the effective action \eqref{equ:GammaOpExp}. In other words, if we can find a satisfactory solution of the flow equation \eqref{equ:flowGamma-detail} different parametrisations $f$ simply correspond to different ways of mapping the renormalised couplings of the effective action onto the bare couplings of the bare action in the path integral approach.\footnote{This is similar to the concept of universality, where different bare actions lead to the same renormalised effective action.} Given that we can derive any physical observable from the effective action, we may regard any $f$-dependence that affects only the bare couplings in this way as physically irrelevant.

\section{The split Ward identity for the conformal factor}
\label{sec:confsWI-derivation}
The derivation of the sWI \eqref{equ:sWiGamma} proceeds along the same lines as in scalar field theory, cf. sec. \ref{sec:sWI-derivation}, with the necessary adaptations being made explicit in the following.

For the sake of clarity it will be useful in this section to make the identification 
\begin{equation}\label{equ:choiceBackg}
\bgmn=f(\chi)\,\delta_{\mu\nu}
\end{equation}
of \eqref{equ:confpar} with the fixed reference metric taken to be flat. This choice will be used in actual computations in later sections.

The path integral for the conformal factor \eqref{equ:pathint-conffact} then becomes
\be
\label{equ:pathint-conffact-sWI}
\exp(W_k) = \int  \mathcal{D}\vp \, \exp(-S[\chi + \vp] - S_k[\vp,\chi] + S_\mathrm{src}[\vp,\chi]).
\ee
The split symmetry of the bare action analogous to \eqref{shifts} now reads
\be
\label{equ:split-symmetry}
\vp(x) \mapsto \vp(x) + \eps(x) \qquad \chi(x) \mapsto \chi(x) -\eps(x),
\ee
but, as before, the cutoff action breaks this invariance,\footnote{The source term also breaks split invariance but does not contribute to the separate background field dependence in $\Gamma_k[\vp^c,\chi]$, as the ensuing calculation shows.}  cf. \cite{Becker:2014qya,Reuter:2008qx}. It is the violation of split symmetry that signals the loss of background independence, both at the level of the path integral and at the level of the effective action.

Applying the shifts \eqref{equ:split-symmetry} with an infinitesimal $\eps(x)$ to the path integral \eqref{equ:pathint-conffact-sWI} with $\sqrt{\bar g}=f(\chi)^{d/2}$ in the cutoff action \eqref{equ:cutoff-action} and the source term \eqref{equ:source} leads to
\be
\label{equ:varW-sWI}
-\int_w f^{-\frac{d}{2}}\frac{\delta W_k}{\delta \chi} \,\eps = \int_w \left[
		J -\frac{d}{2}\dclnf \langle \vp\rangle J 
		-\cutoff\! \langle \vp\rangle +\frac{d}{4}\dclnf\langle \vp \,\cutoff \vp
		\rangle + \frac{1}{2}\langle \vp \,\partial_\chi\! \cutoff \vp\rangle\right]\eps.
\ee
The Legendre transformation \eqref{equ:LegendreTrans} with the shifts \eqref{equ:split-symmetry} gives
\be
\int_w f^{-\frac{d}{2}}\frac{\delta W_k}{\delta \chi} \, \eps = 
		\int_w \left[\frac{d}{2}\dclnf J\vp \,- \right.
		\left. f^{-\frac{d}{2}}\frac{\delta \tilde \Gamma_k}{\delta \chi} \right] \eps,
\ee
which we use in \eqref{equ:varW-sWI}. Performing similar steps as in the previous section, 
exploiting \eqref{equ:rel-expval} and redefining $\Gamma_k[\vp^c,\chi] = \tilde \Gamma_k[\vp^c,\chi] - S_k[\vp^c,\chi]$ then results in
\begin{equation}
\label{equ:sWI-detail-eps}
\int_w f^{-\frac{d}{2}}\left(\frac{\delta \Gamma_k}{\delta \chi}-\frac{\delta \Gamma_k}{\delta \vp^c}\right)\eps =
 \frac{1}{2}\int_x \int_y
 \Delta(x,y)\left[\frac{d}{2}\,\partial_\chi\! \ln\! f(y) \,\cutoff\!(y,x)+\partial_\chi\! \cutoff\!(y,x)\right]
 \eps(y).
\end{equation}
Here we have again used the kernel notation \eqref{equ:kernel-notation}, the integral notation introduced below \eqref{equ:flowGamma-detail} and abbreviated
\be
\label{equ:Delta}
\Delta(x,y) = \langle \vp(x) \vp(y)\rangle - \vp^c(x)\vp^c(y)
	    = \left[f(x)^{-\frac{d}{2}}f(y)^{-\frac{d}{2}}\frac{\delta^2 \Gamma_k}{\delta\vp^c(x)\delta\vp^c(y)}+\cutoff\!(x,y)\right]^{-1}.
\ee
The parametrisation function $f$ inherits its coordinate dependence from the background field $\chi$ and it is therefore useful to write $f(x) \equiv f(\chi(x))$, where appropriate. The sWI is obtained by dropping the arbitrary choice of $\eps(w)$,
\begin{multline}
\label{equ:sWI-detail}
f(w)^{-\frac{d}{2}}\left(\frac{\delta \Gamma_k}{\delta \chi(w)}-\frac{\delta \Gamma_k}{\delta \vp^c(w)}\right) =\\
 \frac{1}{2}\int_x \int_y
 \Delta(x,y)\left[\frac{d}{2}\,\partial_\chi\! \ln\! f(y) \,\cutoff\!(y,x)+\partial_\chi \cutoff\!(y,x)\right]
 \frac{\delta(y-w)}{f(w)^{\frac{d}{2}}}.
\end{multline}
We see that \eqref{equ:sWI-detail} is the written out form of \eqref{equ:sWiGamma}, where 
\be
Y_f(x,w) = \frac{d}{2}\dclnf\frac{\delta(x-w)}{f(w)^{\frac{d}{2}}}
\ee
is an additional contribution, beyond the background field dependence of the cutoff operator itself, that originates from the background field contained in the measure of the cutoff action \eqref{equ:cutoff-action}. We will see in sec. \ref{sec:combine} that both contributions are crucial for the implementation of the sWI \eqref{equ:sWI-detail} on the entire renormalisation group flow.

\section{Beyond the local potential approximation}
\label{sec:BeyondLPA}
In order to truncate the effective action of the conformal factor we make use of the derivative expansion. Since we will no longer need the quantum field from here on we rename the classical field $\vp^c \to \vp$ and then make the ansatz
\be
\label{equ:ansatzGamma}
\Gamma_k[\vp,\chi] = \int d^dx f(\chi)^\frac{d}{2}\left(-\frac{1}{2}K(\vp,\chi)\left(\bar \nabla_\mu \vp\right)^2
		      +V(\vp,\chi)\right)
\ee
in which we keep a general potential $V(\vp,\chi)$ at zeroth order of the derivative expansion and a general function $K(\vp,\chi)$ at $\mathcal{O}\big(\partial^2\big)$ for the fluctuation field $\vp$. It is understood that both of them depend on renormalisation group time $t$. Since this effective action represents an approximation in the conformal sector of quantum gravity we have explicitly kept the appropriate measure factor. Similarly, the kinetic term in \eqref{equ:ansatzGamma} has been formulated in terms of the covariant derivative with respect to the background field $\bgmn = f(\chi)\hat g_{\mu\nu}=f(\chi)\delta_{\mu\nu}$ in order to keep as close as possible to the formulation in full quantum gravity, cf. sec. \ref{sub:adaptations-gravity}.

A further conspicuous feature of \eqref{equ:ansatzGamma} is the minus sign in front of the kinetic term of $\vp$, reflecting the negative sign of the kinetic term of the conformal factor as obtained from the Einstein-Hilbert action that is otherwise known as the the conformal factor problem. Note that as far as the effective action is concerned, this sign can be accommodated in a natural way as already discussed in \cite{Reuter:1996}. In the context of conformal gravity it has also been discussed in \cite{Manrique:2009uh,Reuter:2008qx,Reuter:2008wj}.

A full treatment of the effective action at $\mathcal{O}\big(\partial^2\big)$ of the derivative expansion would also include an analogous kinetic term for the background field as well as a possible mixed term. We have omitted them in our ansatz \eqref{equ:ansatzGamma} since we will restrict our analysis to slowly varying background fields $\chi$. This means that from now on we will neglect any derivatives $\partial_\mu \chi(x)$ compared to derivatives of the fluctuation field $\partial_\mu \vp(x)$.

\subsection{Strategy for the evaluation of the traces}
\label{sec:strategy-traces}
An essential step in deriving the flow equation and the sWI for the ansatz \eqref{equ:ansatzGamma} is the evaluation of the traces in \eqref{equ:flowGamma-detail} and \eqref{equ:sWI-detail}. In the following we will derive a general framework that can be used to perform this task in the present context of conformal gravity and for a derivative expansion of the effective action with a flat reference metric $\hat g_{\mu\nu}$ in \eqref{equ:confpar}.

We are interested in the expression
\be
\label{equ:gen-trace}
T_\mathcal{K} = \int d^dxd^dy \sqrt{\bar g(x)}\sqrt{\bar g(y)}\, \Delta(x,y) \mathcal{K}(y,x)
\ee
for a generic kernel $\mathcal{K}$ that can be chosen correspondingly for the flow equation and the sWI, and $\Delta(x,y)$ has been introduced in \eqref{equ:Delta}.

For any function $u(x)$, the kernel $\mathcal{K}(x,y)$ satisfies the relation
\be
\int d^dy \sqrt{\bar g(y)}\,\mathcal{K}(x,y)u(y) = \mathcal{K}_x u(x),
\ee
where $\mathcal{K}_x$ is the associated differential operator acting on $x$-dependence. This relation between the kernel and its differential operator is satisfied if they obey \eqref{equ:kernel-notation} with $\mathcal{K}$ in place of $\cutoff$. Using this in \eqref{equ:gen-trace}, we find
\begin{equation}
\label{temp1}
T_\mathcal{K} = \int d^dx\, d^dy \, \left[\Delta_x\delta(x-y)\right]\left[\mathcal{K}_y \delta(y-x)\right]	
	      = \int \left. d^dx \,\Delta_x \, \mathcal{K}_x \,\delta(x-x')\right|_{x'=x},
\end{equation}
where in the last step the differential operators act on the delta function before setting $x'=x$. For the flow equation \eqref{equ:flowGamma-detail} we simply have $\mathcal{K}_x =  \partial_t \mathcal{R}_{k,x}$ while for the sWI \eqref{equ:sWI-detail} the operator $\mathcal{K}_x$ takes the form
\be
\label{equ:KsWI}
\mathcal{K}_x = \frac{\delta(x-w)}{f(w)^\frac{d}{2}}\left\{\frac{d}{2}\dclnf \mathcal{R}_{k,x} 
+\partial_\chi R_{k,x}\right\}.
\ee
In this last expression, the differential operator $\mathcal{R}_{k,x}$ acts to its right, leaving the delta function untouched. It is a function of the covariant background field Laplacian,
\be
\mathcal{R}_{k,x} = \cutoff\!\!\left(-\bar \nabla^2_x\right) = \cutoff\!\!\left(-f(\chi)^{-1}\partial^2_x \right)
\ee
where we have made use of the conformal approximation \eqref{equ:confpar} and neglected any derivatives of $\chi$. Using this, and by expressing $\delta(x-x')$ in momentum space, \eqref{temp1} can be manipulated as follows,
\begin{align}
T_\mathcal{K} &= \int  d^dx \frac{d^dp}{(2\pi)^d} 
			\left. e^{-i p\cdot x'}\Delta_x \,
			 \mathcal{K}\left(-f^{-1}\partial_x^2\right)e^{ip\cdot x} \right|_{x=x'}	\\
			 &= \int  d^dx \frac{d^dp}{(2\pi)^d}
			 	\left\{e^{-i p\cdot x}\Delta_x e^{i p\cdot x}\right\} \mathcal{K}\left(f^{-1} p^2\right).
\end{align}
In the last step we have again neglected all $\partial_\mu \chi$-terms. Since we make use of this approximation of slow background fields the operator $\mathcal{K}$ in this calculation effectively has no separate dependence on $w$ or $x$ through the background field $\chi$. As shown in sec. \ref{sec:evalsWI}, this is also true for the delta function appearing in \eqref{equ:KsWI}. Abbreviating 
\be
Q(p,x)=e^{-i p\cdot x}\Delta_x e^{i p\cdot x}
\ee
we can thus finally write
\be
\label{equ:traces}
T_\mathcal{K} = \int  d^dx \frac{d^dp}{(2\pi)^d} \,
			 	Q(p,x) \, \mathcal{K}\left(f^{-1} p^2\right).
\ee
The function $Q(p,x)$ will still be $x$-dependent in general, as it may contain instances of the fluctuation field $\vp(x)$ and its derivatives. In the next section we will perform a derivative expansion of $Q(p,x)$ which can then be used in the last expression to evaluate both the flow equation and the sWI.

\subsection{Derivative expansion of $Q(p,x)$}
When the ansatz \eqref{equ:ansatzGamma} for the effective action is substituted into the left hand side of the flow equation \eqref{equ:flowGamma-detail} or the sWI \eqref{equ:sWI-detail}, the resulting expression already has the form of a derivative expansion. On their right hand sides however, we encounter the term
\be
\label{equ:Q}
Q(p,x) = e^{-ip\cdot x}\Delta_x e^{i p\cdot x}=e^{-ip\cdot x}\left[\hess+\mathcal{R}\right]^{-1}e^{i p\cdot x}
\ee
as it appears in \eqref{equ:traces}. Since it will be more convenient in the following, we have expressed \eqref{equ:Delta} in the more compact and equivalent differential operator notation, suppressing $k$-dependence to avoid clutter. With an inverse of an operator containing the fluctuation field $\vp(x)$ and its derivatives, this term in its current form is not presented in a derivative expansion. In order to be able to compare the left hand sides with the corresponding right hands sides order by order in the derivative expansion we proceed as follows.

First, the definition \eqref{equ:Q} can be rewritten as
\be
\label{equ:temp2}
\mathcal{R}\left\{\eipx Q\right\} = \eipx\left(1-\emipx \hess\left\{\eipx Q\right\}\right).
\ee
A derivative expansion of the left hand side can be achieved by using
\begin{multline}
\left(-\partial^2\right)^n\left\{\eipx Q\right\} = \eipx\left[p^{2n}Q -2in p^{2n-2} p^\mu\partial_\mu Q 
	  -n p^{2n-2}\partial^2 Q \right. \\ \left. -2n(n-1)p^{2n-4}p^\mu p^\nu \partial_\mu \partial_\nu Q + \mathcal{O}\big(\partial^3\big)\right]
\end{multline}
in the Taylor expansion
\be
\mathcal{R}(-\partial^2,\chi)\left\{\eipx Q\right\} = \sum_{n=0}^\infty \frac{\mathcal{R}^{(n)}(0,\chi)}{n!}\left(-\partial^2\right)^n\left\{\eipx Q\right\}.
\ee
Summing this series leaves us with the expansion
\begin{multline}
\mathcal{R}(-\partial^2,\chi)\left\{\eipx Q\right\} = \eipx \Big[ \mathcal{R}(p^2,\chi)Q -2i\partial_{p^2} \mathcal{R}(p^2,\chi)p^\mu\partial_\mu Q
		-\partial_{p^2}\mathcal{R}(p^2,\chi) \partial^2 Q \partial_{p^2}^2 
		\\ 
		\left. -2\partial_{p^2}^2 \mathcal{R}(p^2,\chi) p^\mu p^\nu \partial_\mu
		\partial_\nu Q + \mathcal{O}\big(\partial^3\big)\right].
\end{multline}
By substituting this result on the left hand side of \eqref{equ:temp2}, we arrive at the following implicit equation for $Q$,
\begin{multline}
\label{equ:Qimp}
Q = \mathcal{R}\left(p^2,\chi\right)^{-1}\left[1-\emipx \hess\left\{\eipx Q\right\} +2i\partial_{p^2} \mathcal{R}(p^2,\chi)p^\mu\partial_\mu Q \right. \\
    +\left. \partial_{p^2}\mathcal{R}(p^2,\chi) \partial^2 Q + 2\partial_{p^2}^2 \mathcal{R}(p^2,\chi) p^\mu p^\nu \partial_\mu
		\partial_\nu Q + \mathcal{O}\big(\partial^3\big)\right].
\end{multline}
The Hessian operator associated with the ansatz \eqref{equ:ansatzGamma} can be separated into contributions according to the number of derivatives of $\vp$ they contain,
\be
\label{equ:expHess}
\hess = \hess_0 + \hess_1 + \hess_2
\ee
with the individual terms given by
\be
\hess_0 = \partial_\vp^2 V + \frac{K}{f}\partial^2, \qquad \hess_1 = \frac{\partial_\vp K}{f}\partial_\mu\vp \partial^\mu, \qquad
	  \hess_2 = \frac{1}{2}\frac{\partial_\vp^2 K}{f}\left(\partial_\mu \vp\right)^2 +  \frac{\partial_\vp K}{f}\partial^2\vp.
\ee
In the same vein, we also expand
\be
\label{equ:Qsum}
Q = Q_0 + Q_1 + Q_2 + \dots,
\ee
collecting terms with $n$ derivatives of $\vp$ into $Q_n$. Derivatives $\partial_\mu Q_n$ therefore are allocated to $Q_{n+1}$ since we infer from the definition \eqref{equ:Q} that $Q$ depends on $x$ only through the fluctuation field $\vp(x)$ and its derivatives as they appear in the Hessian $\hess$ (as well as the background field $\chi(x)$ whose derivatives we neglect however). Using this expansion and \eqref{equ:expHess} in \eqref{equ:Qimp} leads to implicit equations for the terms $Q_n, \, n=0,1,2$ that are easily solved to give
\begin{align}
\label{equ:Q0}
Q_0 =& \left[\partial_\vp^2 V - \frac{K}{f} p^2 + \mathcal{R} \right]^{-1}  \\
\notag
Q_1 =& -i p^\mu\left[\frac{\partial_\vp K}{f}Q_0 \partial_\mu \vp + 2 \left(\frac{K}{f}+ \partial_{p^2}\mathcal{R}\right)\partial_\mu Q_0\right]Q_0 \\ \notag
Q_2 =& -\left[i p^\mu\left\{\frac{\partial_\vp K}{f}Q_1 \partial_\mu \vp + 2 \left(\frac{K}{f}+ \partial_{p^2}\mathcal{R}\right)\partial_\mu Q_1\right\} + \left(\frac{K}{f} + \partial_{p^2}\mathcal{R} \right)\partial^2 Q_0
\right. \\ \label{equ:Q2} 
& \left. + \frac{\partial_\vp K}{f}\left(\partial_\mu \vp \partial^\mu Q_0 + Q_0 \partial^2 \vp\right)
+\frac{1}{2}\frac{\partial_\vp^2 K}{f}Q_0 \left(\partial_\mu \vp\right)^2
+ 2\partial_{p^2}^2\mathcal{R}\, p^\mu p^\nu\partial_\mu\partial_\nu Q_0\right]Q_0.
\end{align}
In these expressions $\mathcal{R}=\mathcal{R}(p^2,\chi)$, and they can now be used directly in \eqref{equ:traces} to obtain both the flow equation \eqref{equ:flowGamma-detail} and the sWI \eqref{equ:sWI-detail}. Out of these, $Q_1$ is an odd function of momentum and therefore integrates to zero in \eqref{equ:traces}. This is expected since there are no terms in the effective action with only one derivative of $\vp$, but we display $Q_1$ since it appears in the $\mathcal{O}\big(\partial^2\big)$-term $Q_2$.

\subsection{Evaluation of the flow equation}
We are now in a position to evaluate the flow equation \eqref{equ:flowGamma-detail} for the truncation ansatz \eqref{equ:ansatzGamma} using the result \eqref{equ:traces} with $\mathcal{K}(p^2/f) = \dot{\mathcal{R}}(p^2/f)$. Here and in the following the dot notation refers to a derivative with respect to renormalisation group time $t$. Making use of \eqref{equ:Qsum} and \eqref{equ:Q0} we immediately obtain
\be
\label{equ:flowV}
\partial_t V(\vp,\chi) = \frac{\tilde \Omega_{d-1}}{2}f(\chi)^{-\frac{d}{2}}\int dp \,p^{d-1}\frac{\dot{ \mathcal{R}}(p^2/f)}{\partial_\vp^2 V - K p^2/f+\mathcal{R}(p^2/f)}.
\ee
This equation involves the constant $\tilde \Omega_{d-1}= \Omega_{d-1}/(2\pi)^d$, where $\Omega_{d-1}$ is the surface area of the $(d-1)$-dimensional sphere, and is expressed as an integral over the modulus $p=|p^\mu|$. Here and in the following we will omit the obvious bounds for the integral on the right until they become important in sec. \ref{sec:confLPA}. The overall background field factor in front of the integral has its origin in the left hand side of the flow equation and comes from the measure factor contained in the ansatz \eqref{equ:ansatzGamma}.

A similar computation at $\mathcal{O}\big(\partial^2\big)$ of the derivative expansion involving \eqref{equ:Q2} leads to
\be
\label{equ:flowK}
f(\chi)^{-1}\partial_t K(\vp,\chi) = \tilde \Omega_{d-1}\, f(\chi)^{-\frac{d}{2}}\int dp\, p^{d-1}\, P\!\left(p^2,\vp,\chi\right) \dot{ \mathcal{R}}\!\left(p^2/f\right),
\ee
where
\begin{align}
\label{equ:P}
 P =& -\frac{1}{2}\frac{\partial_\vp^2 K}{f}Q_0^2 + \frac{\partial_\vp K}{f}\left(2\partial_\vp^3 V-\frac{2d+1}{d}\frac{\partial_\vp K}{f}p^2\right) Q_0^3 \notag \\ 
			  &  -\left[\left\{\frac{4+d}{d}\frac{\partial_\vp K}{f}p^2-\partial_\vp^3V\right\}\left(\partial_{p^2}\mathcal{R}-\frac{K}{f}\right) \right. \\ \notag
			  & \left. \mkern150mu + \frac{2}{d}p^2\partial_{p^2}^2 \mathcal{R}\left(\frac{\partial_\vp K}{f}p^2-\partial_\vp^3 V\right)\right]\left(\partial_\vp^3 V -\frac{\partial_\vp K}{f}p^2\right) Q_0^4 \\ \notag
 &-\frac{4}{d}p^2\left(\partial_{p^2} \mathcal{R}-\frac{K}{f}\right)^2\left(\partial_\vp^3V-\frac{\partial_\vp K}{f}p^2\right)^2 Q_0^5
\end{align}
is obtained from \eqref{equ:Q2} by integration by parts under the integral in \eqref{equ:flowGamma-detail}. We have chosen to arrange this expression by orders of powers of the $\mathcal{O}\big(\partial^0\big)$-term $Q_0$, \eqref{equ:Q0}.

The flow equations \eqref{equ:flowV} and \eqref{equ:flowK} are still rather general as we have not specified the form of the cutoff operator $\mathcal{R}$. We will come back to this in sec. \ref{sec:combine}.

\subsection{Evaluation of the split Ward identity}
\label{sec:evalsWI}
In principle the sWI \eqref{equ:sWI-detail} can be obtained along the same lines as the flow equation in the previous section. Due to the type of truncation we made for the effective action, \eqref{equ:ansatzGamma}, there is however an additional small but important subtlety that needs to be taken care of. As emphasised just before sec. \ref{sec:strategy-traces} we consider slowly varying background fields $\chi(x)$ so that we can take functional derivatives as in the left hand side of the sWI \eqref{equ:sWI-detail} but neglect all derivatives $\partial_\mu \chi$ compared to derivatives of the fluctuation field $\partial_\mu \vp$. While the derivation of the sWI in sec. \ref{sec:confsWI-derivation} is completely general, we have to neglect derivatives $\partial_\mu \eps(x)$ of the shifts in \eqref{equ:split-symmetry} for our truncation of the effective action in order to remain in the regime of slowly varying background fields $\chi(x)$. To do this it is convenient to start from the version \eqref{equ:sWI-detail-eps} of the sWI as it explicitly contains the shifts $\eps$. With the ansatz \eqref{equ:ansatzGamma} for the effective action, the integrand on its left hand side becomes
\begin{multline}
\label{equ:lhs-sWI}
f^{-\frac{d}{2}}\left(\frac{\delta \Gamma_k}{\delta \chi}-\frac{\delta \Gamma_k}{\delta \vp}\right)
=-\frac{1}{2 f}\left\{\partial_\chi K - \partial_\vp K +\frac{d-2}{2}\dclnf K\right\}
\left(\partial_\mu \vp\right)^2 \\
+\frac{d}{2}\dclnf V+\partial_\chi V- \partial_\vp V + I,
\end{multline}
where, writing $K(x) = K(\vp(x),\chi(x))$, the last term is
\be
\label{equ:I}
I(w) = f(w)^{-\frac{d}{2}}\int d^dx f(x)^\frac{d}{2}\frac{K(x)}{f(x)}\partial_\mu \delta(x-w)\partial^\mu \vp(x).
\ee
Hence, for slowly varying $\eps$ we have
\be
\int_w I(w)\,\eps(w) = 0,
\ee
causing this term to drop out on the left hand side of the sWI \eqref{equ:sWI-detail-eps}. We note here that we would arrive at the same result for the left hand side of \eqref{equ:sWI-detail} if we consider the delta function in \eqref{equ:I} to be slowly varying itself, $\partial_\mu \delta(x-w) = 0$.

The right hand side of the sWI \eqref{equ:sWI-detail-eps} can be evaluated using the techniques of sec. \ref{sec:strategy-traces}. Instead of \eqref{equ:KsWI} we use 
\be
\label{equ:op-sWI}
\mathcal{K}_x = \eps(x)\left\{\frac{d}{2}\,\partial_\chi\! \ln\! f(x) \, \mathcal{R}_{k,x} + \partial_\chi \mathcal{R}_{k,x}\right\}
\ee
resulting in the appropriately modified version of \eqref{equ:traces},
\be
\label{equ:rhs-sWI}
T_\mathcal{K} = \int d^dx \frac{d^dp}{(2\pi)^d}\,Q(p,x)\left[\frac{d}{2}\dclnf \cutoff\!(p^2/f) + \partial_\chi\! \cutoff\!(p^2/f)\right]\eps(x).
\ee
At order $\mathcal{O}\big(\partial^0\big)$ of the derivative expansion \eqref{equ:Qsum} with \eqref{equ:Q0}, this last expression can be used directly with the corresponding part of \eqref{equ:lhs-sWI} in \eqref{equ:sWI-detail-eps} to get the sWI for the potential,
\be 
\label{equ:sWI-V}
\partial_\chi V- \partial_\vp V+\frac{d}{2}\dclnf V
= \frac{\tilde \Omega_{d-1}}{2}f(\chi)^{-\frac{d}{2}}\int dp\, p^{d-1} \,\frac{\partial_\chi \mathcal{R}(p^2/f)+\frac{d}{2}\dclnf \mathcal{R}(p^2/f)}{\partial^2_\vp V -Kp^2/f +\mathcal{R}(p^2/f)}.
\ee
The last term on the left hand side stems from the measure factor in the ansatz \eqref{equ:ansatzGamma} for the effective action. In the same way,  the second term in the numerator on the right hand side has its origin in the measure factor of the cutoff action \eqref{equ:cutoff-action}, whereas the first term is a result of the fact that the cutoff is imposed with respect to the eigenmodes of the background field Laplacian, $-\bar \nabla^2$.

In order to evaluate \eqref{equ:rhs-sWI} at second order of the derivative expansion using \eqref{equ:Q2}, we integrate by parts under the spatial integral in \eqref{equ:rhs-sWI}, just as for the flow equation. The advantage of the representation \eqref{equ:rhs-sWI} is that it makes it apparent that this integration by parts does not pick up additional terms from the operator \eqref{equ:op-sWI} as long as we neglect all derivatives of the background field and the shift function $\eps(x)$. Thus, collecting the $\left(\partial_\mu \vp\right)^2$-terms from \eqref{equ:lhs-sWI} for the left hand side of the sWI \eqref{equ:sWI-detail-eps}, the sWI for the kinetic term assumes the form
\begin{multline}
\label{equ:sWI-K}
\frac{1}{f}\left\{\partial_\chi K - \partial_\vp K +\frac{d-2}{2}\dclnf K\right\}
= \\
\tilde \Omega_{d-1}f^{-\frac{d}{2}}\int dp\, p^{d-1} \,P(p^2,\vp,\chi)\left[\partial_\chi \mathcal{R}+\frac{d}{2}\dclnf \mathcal{R}\right].
\end{multline}
Here, $P$ is the same as for the flow equation, \eqref{equ:P}, and from now on the argument of the cutoff is understood to be $\mathcal{R}=\mathcal{R}(p^2/f)$. On the right hand side we find the two contributions analogous to the right hand side of the sWI for the potential, \eqref{equ:sWI-V}, and on the left hand side the factor $d/2 -1$ comprises the contribution coming from the measure factor and from the background covariant derivative of the kinetic term in the ansatz \eqref{equ:ansatzGamma}.

We also note that for the right hand side of \eqref{equ:sWI-K} too, we could have instead started from \eqref{equ:sWI-detail} with \eqref{equ:KsWI} as long as we regard the delta function $\delta(x-w)$ to be slowly varying, allowing us to neglect its derivatives. However, the justification for this can only be understood when using the form \eqref{equ:sWI-detail-eps} of the sWI together with a correct treatment of the shift function $\eps(x)$ as above.

\subsection{Combining the flow equation with the sWI \\ and background independence}
\label{sec:combine}
Considering the flow equations for the potential \eqref{equ:flowV} and for the kinetic term \eqref{equ:flowK}, we find the background field entering through the parametrisation function $f$ of \eqref{equ:confpar} as a consequence of the background field method in asymptotic safety. A noteworthy observation however is that there appear no derivatives with respect to the background field in either of the flow equations, a direct consequence of the general structure of the flow equation of the effective action \eqref{equ:flowGamma}. For the purpose of solving differential equations, this means that the background field can be regarded as a parameter instead of as an independent variable. Hence we can choose a specific value $\chi=\chi_0$ for the background field in the flow equations \eqref{equ:flowV}, \eqref{equ:flowK} and then try to solve them as a partial differential equation in $t$ and $\vp$ to obtain a solution $\left(V_t(\vp,\chi_0), K_t(\vp,\chi_0)\right)$. By varying the parameter $\chi_0$ and repeating this process we can then build the solution $\left(V_t(\vp,\chi), K_t(\vp,\chi)\right)$. But the background field dependence of any such solution remains undetermined as becomes clear from the fact that with $V_t(\vp,\chi)$ also $V_t(\vp,\chi)+F(\chi)$ is a solution of \eqref{equ:flowV} and \eqref{equ:flowK}, where $F$ is an arbitrary $t$-independent function of the background field, cf. the corresponding discussion in sec. \ref{sec:sWI-derivation}.

This illustrates with a specific example that from an investigation of the flow equation alone we cannot control the background field dependence of the renormalisation group solutions and it is thus by no means possible to establish background independence. As discussed in the introduction to this chapter, the fact that it is possible to successfully quantise for all background configurations simultaneously does not guarantee background independence, but instead the result can be a different effective action for each background field configuration.

The split Ward identity, given by \eqref{equ:sWI-V} for the potential and by \eqref{equ:sWI-K} for the kinetic term, involves the background field as an independent variable in a differential equation, rather than as a parameter. The expectation is that the sWI, when used in conjunction with the flow equation, will eliminate the additional freedom represented by the background field as described here. The aim of this section is to show that this can be achieved by appropriately combining the sWI with the flow equation.

We start by redefining
\begin{equation}
 V \mapsto \frac{\tilde \Omega_{d-1}}{2}V, \qquad K \mapsto \frac{\tilde \Omega_{d-1}}{2}K, \qquad \mathcal{R}\mapsto \frac{\tilde \Omega_{d-1}}{2}\mathcal{R}
\end{equation}
to convert the flow equation \eqref{equ:flowV} and the sWI \eqref{equ:sWI-V} for the potential into
\begin{subequations}
 \label{equ:sysV-resc}
\begin{align}
 \partial_t V(\vp,\chi) &= f(\chi)^{-\frac{d}{2}}\int dp \,p^{d-1}\,Q_0\,\dot{\mathcal{R}}, \label{equ:flowV-resc}\\
 \partial_\chi V- \partial_\vp V+\frac{d}{2}\dclnf V
&= f(\chi)^{-\frac{d}{2}}\int dp\, p^{d-1} \,Q_0\left[\partial_\chi \mathcal{R}+\frac{d}{2}\dclnf \mathcal{R}\right], \label{equ:sWIV-resc}
\end{align}
\end{subequations}
where we have made use of the zeroth order of the derivative expansion $Q_0$ given in \eqref{equ:Q0}. Rescaling the cutoff operator by a positive constant as we have done here is allowed since it does not affect the suppression of modes it is responsible for. In the same way, the flow equation \eqref{equ:flowK} and the sWI \eqref{equ:sWI-K} for the kinetic term are transformed into
\begin{subequations}
\label{equ:sysK-resc}
\begin{align}
 f^{-1}\partial_t K(\vp,\chi) &= 2 f^{-\frac{d}{2}}\int dp\, p^{d-1}\, P\left(p^2,\vp,\chi\right) \dot{\mathcal{R}}, \label{equ:flowK-resc}\\
 f^{-1}\left\{\partial_\chi K - \partial_\vp K +\frac{d-2}{2}\dclnf K\right\}
&= 2f^{-\frac{d}{2}}\int dp\, p^{d-1} \,P(p^2,\vp,\chi)\left[\partial_\chi \mathcal{R}+\frac{d}{2}\dclnf \mathcal{R}\right]. \label{equ:sWIK-resc}
\end{align}
\end{subequations}
Before combining these two pairs of equations it is convenient to discuss the dimensions of the fields. In general, requiring that the line element $ds^2 = \gmn dx^{\mu}dx^{\nu}$ has mass dimension $-2$ leaves us with the freedom of assigning the mass dimension of coordinates to be $\left[x^\mu\right] = c$ so that $\left[\gmn\right]=-2(1+c)$. In the framework of quantum gravity, where we would like to consider arbitrary diffeomorphisms as functions of the coordinates, a natural choice for the mass dimension of $\left[x^\mu\right]$ would be to set $c=0$ and shift the mass dimension of $ds^2$ into the metric field variable $\left[\gmn\right] = -2$. After all, the physical concept of length is encoded in the line element $ds^2$ and the coordinates represent a choice made to describe a given patch of spacetime. 

Independently of the value of $c$, if we insist that also the reference metric $\hat g_{\mu\nu}$ in the conformal truncation \eqref{equ:confpar} is a true metric, consistency would impose $\left[f(\phi)\right] = 0$ so that the mass dimension of the conformal factor would naturally be vanishing, $\left[\phi\right]=[\vp]=[\chi]=0$. At the Gaussian fixed point the dimension of the conformal factor will remain zero even after quantisation but at a non-perturbative fixed point we have to account for in principle different  anomalous dimensions for the fluctuation field $\vp$ and the background field $\chi$. In the following however, we make the approximation in which these two anomalous dimensions are assumed equal so that we have anomalous scaling according to $\left[\phi\right] = [\chi]= \eta/2 \neq 0$ at a non-Gaussian fixed point. As a consequence of a non-zero anomalous dimension for the conformal factor we would then also find $\left[f(\phi)\right]\neq 0$. \footnote{This would lead to $\left[ ds^2\right] = -2 + \left[f(\phi)\right]$, a quantity that depends on our choice of parametrisation. But the line element is a \emph{classical} concept and becomes invalid at a non-perturbative fixed point.} Consequently, we will account for a dimensionful parametrisation function $f$ and keep the anomalous dimension $\eta$ as a parameter.

Coming back to the constant $c$ describing the dimension of coordinates, it is clear that the quantum structure of gravity should not depend on the value we choose for it. Since the conventional choice in quantum field theory on Minkowski space is $c=-1$, ascribing the mass dimension of the line element to the coordinates, we will therefore make the "flat-space" choice $c=-1$ in the following.

We now come to the specification of the cutoff operator $\cutoff$. Its dimension is prescribed by its place in the cutoff action \eqref{equ:cutoff-action} and thus it takes the general form
\begin{equation}
\label{equ:cutoff}
 \mathcal{R}\!\left(p^2/f\right) = - k^{d-\eta-\frac{d}{2}d_f}\,r\!\left(\frac{p^2}{k^{2-d_f}f}\right),
\end{equation}
where we have defined $d_f := [f]$ and the profile function $r(z)$ is a dimensionless function of a dimensionless argument. The crucial ingredient here is the minus sign on the right hand side, reflecting the minus sign of the conformal factor in quantum gravity as mentioned at the beginning of sec. \ref{sec:BeyondLPA}. At a practical level, this sign ensures that singularities are avoided in the modified propagator $\left[\Gamma_k^{(2)}+\cutoff\right]^{-1}$ in \eqref{equ:flowGamma}, cf. \cite{Reuter:1996}.

The cutoff operator performs the suppression of modes correctly if the profile function $r(z)$ satisfies the requirements $\lim_{z\to0} r(z) >0$ and $\lim_{z\to\infty} r(z) = 0$, cf. \eqref{cutoff-props}. For our present purpose of combining the flow equation with the sWI it turns out that choosing a power-law behaviour for the profile function is most convenient. Hence, we set
\begin{equation}
 \label{equ:cutoff-pwrlw}
 r(z) = \frac{1}{z^n}, \qquad \text{implying} \qquad zr'(z) = -nr(z).
\end{equation}
To ensure finiteness of the integrals on the right hand sides of \eqref{equ:sysV-resc} and \eqref{equ:sysK-resc}, the exponent $n$ has to be chosen such that $n>d/2-1$, cf. \cite{Morris:1994ie}. This choice of cutoff leads to the two identities
\begin{subequations}
\begin{align}
 \partial_t \mathcal{R} &= - k^{d-\eta-\frac{d}{2}d_f}\left(d-\eta+2n-\frac{d+2n}{2}d_f\right)r\!\left(\frac{p^2}{k^{2-d_f}f}\right),  \\
 \partial_\chi \mathcal{R}+\frac{d}{2}\dclnf \mathcal{R} &= - k^{d-\eta-\frac{d}{2}d_f}\,\frac{d+2n}{2}\dclnf r\!\left(\frac{p^2}{k^{2-d_f}f}\right).
\end{align}
\end{subequations}
Exploiting this allows us to combine the flow equation \eqref{equ:flowV-resc} with the sWI \eqref{equ:sWIV-resc} into a single equation,
\begin{equation}
\label{equ:combV}
 \dclnf \partial_t V = \alpha \left(\frac{d}{2}\dclnf V + \partial_\chi V-\partial_\vp V\right),
\end{equation}
where the constant $\alpha$ is given by
\begin{equation}
 \alpha = \frac{2(d-\eta)+4n-(d+2n)d_f}{d+2n}.
\end{equation}
The flow equation for the kinetic term \eqref{equ:flowK-resc} and its sWI \eqref{equ:sWIK-resc} can be combined to give the similar equation
\begin{equation}
\label{equ:combK}
 \dclnf \partial_t K = \alpha\left(\partial_\chi K - \partial_\vp K +\frac{d-2}{2}\dclnf K\right).
\end{equation}
The two combined equations \eqref{equ:combV} and \eqref{equ:combK} are solved by the change of variables\footnote{We see from \eqref{equ:combV} and \eqref{equ:combK} that $\alpha=0$ does not lead to an actual RG flow. Hence we exclude this case here and in the following.}
\begin{equation}
 \label{equ:chvars}
 V(k,\vp,\chi) = f(\chi)^{-\frac{d}{2}}\tilde V(\tilde k,\phi), \qquad K(k,\vp,\chi) = f(\chi)^{-\frac{d}{2}+1}\tilde K(\tilde k,\phi), \qquad \tilde k = k f(\chi)^\frac{1}{\alpha},
\end{equation}
where we have highlighted the dependence on the original and the transformed renormalisation group scales and used the total field $\phi = \chi + \vp$. Under this change of variables the systems \eqref{equ:sysV-resc} for the potential and \eqref{equ:sysK-resc} for the kinetic term are respectively equivalent to the first and second equation in the following system,
\begin{subequations}
 \label{equ:sys-red}
 \begin{align}
  \partial_{\tilde t}\tilde V &= -\frac{1}{2}(2n+d)\alpha\,\tilde k^{\alpha(n+d/2)}\int dp\, p^{d-1}\, \tilde Q_0\,r\!\left(p^2\right),   \\
  \partial_{\tilde t}\tilde K &= -(2n+d)\alpha\,\tilde k^{\alpha(n+d/2)}\int dp\, p^{d-1}\, \tilde P\,r\!\left(p^2\right).
 \end{align}
\end{subequations}
Here $\partial_{\tilde t} \equiv \tilde k \partial_{\tilde k}$, and $\tilde Q_0$ and $\tilde P$ are given by applying the change of variables \eqref{equ:chvars} to \eqref{equ:Q0} and \eqref{equ:P} and pulling out a factor of $f^{d/2}$. 

The next step consists in adopting dimensionless variables.  We remark that the mass dimension of the transformed renormalisation group scale $\tilde k$ may deviate from one, $\big[\tilde k \big]=1+d_f/\alpha$, but it turns out that formulating the flow equations in terms of dimensionless variables naturally leads to a new renormalisation group scale given by
\begin{equation}
\label{equ:kbar}
 \kh = \tilde k^\frac{\alpha}{\alpha+d_f}.
\end{equation}
The mass dimension of this is $\big[\kh\big]=1$ and the dimensionless quantities are given by the following redefinitions,
\begin{equation}
\label{equ:dimless}
 \tilde V = \kh^{d}\Vh, \qquad \tilde K = \kh^{d-2-\eta}\Kh, \qquad \phi = \kh^{\frac{\eta}{2}} \phih, \qquad p = \kh\,\hat p.
\end{equation}
We note in passing that $\kh$ is distinguished by transforming the cutoff operator \eqref{equ:cutoff} with \eqref{equ:cutoff-pwrlw} as follows
\begin{equation}
\label{cutoff-kbar}
 \mathcal{R}\big(p^2/f\big) = -f^{-\frac{d}{2}} \, \kh^{d-\eta}\, r\big(p^2/\kh^2\big),
\end{equation}
and thus represents a renormalisation group scale that converts the argument of the profile function $r$ into a form reminiscent of scalar field theories, e.g. \cite{Litim:2002cf}.

Expressing the system \eqref{equ:sys-red} in terms of the dimensionless variables then finally leads to
\begin{subequations}
\label{equ:sys-red-final}
\begin{align}
 \kh\partial_{\kh} \Vh +d\Vh-\frac{\eta}{2}\phih \Vh' &= -\left(d-\eta+2n\right)\int d\hat p\, \hat p^{d-1} \,\hat Q_0 \,r\big(\hat p^2\big), \label{equ:flowV-final} \\
 \kh\partial_{\kh} \Kh +(d-2-\eta)\Kh-\frac{\eta}{2}\phih \Kh' &= -2(d-\eta+2n)\int d\hat p\, \hat p^{d-1} \,\hat P\big(\hat p^2,\phih\big) \,r\big(\hat p^2\big). \label{equ:flowK-final}
\end{align}
\end{subequations}
where now
\begin{subequations}
\label{Q0Pfinal}
\begin{align}
 \hat Q_0 = &\left[\Vh''-\Kh \hat p^2-r\big(\hat p^2\big)\right]^{-1} \label{Q0final} \\ 
 \hat P = &-\frac{1}{2}\Kh'' \hat Q_0^2 + \Kh'\left(2\Vh'''-\frac{2d+1}{d}\Kh' 
 			\hat p^2\right)\hat Q_0^3    \label{Pfinal} \\
    & +\left[\left\{\frac{4+d}{d}\Kh' \hat p^2 - \Vh'''\right\}\left(r'(\hat p^2)+\Kh\right)+\frac{2}{d}\hat p^2 r''\big(\hat p^2\big)\left(\Kh' \hat p^2-\Vh'''\right)\right]\!\left(\Vh'''-\Kh' \hat p^2\right)\!\hat Q_0^4
     \notag \\
    & -\frac{4}{d}\hat p^2 \left(r'\big(\hat p^2\big)+\Kh\right)^2 \left(\Vh'''-\Kh' \hat p^2\right)^2 \hat Q_0^5 \, , \notag 
\end{align}
\end{subequations}
and primes denote derivatives with respect to $\phih$. We will refer to the two equations \eqref{equ:sys-red-final} as the reduced system. It consists of two partial differential equations for $\Vh_{\kh}(\phih)$ and $\Kh_{\kh}(\phih)$ with respect to the renormalisation group scale $\kh$ and the total conformal factor field $\phih$.

The solutions of the original systems \eqref{equ:sysV-resc} and \eqref{equ:sysK-resc} comprising the flow equation and sWI for the potential and the kinetic term are mapped onto the solutions of the reduced system \eqref{equ:sys-red-final} in a one-to-one fashion by the changes of variables \eqref{equ:chvars}, \eqref{equ:kbar} and \eqref{equ:dimless}. In this way, the renormalisation group flow of the ansatz \eqref{equ:ansatzGamma} for the effective action is fully captured by a set of differential equations that do no longer depend on the background field $\chi$ or the parametrisation function $f$ in \eqref{equ:confpar}. The sWI \eqref{equ:sWiGamma} has made it possible to gain complete control over the dependence of the potential and the kinetic term on the background field and has led to the significant simplifcation as embodied by the reduced system \eqref{equ:sys-red-final} compared to the original systems \eqref{equ:sysV-resc}, \eqref{equ:sysK-resc}.

As mentioned at the end of sec. \ref{sec:conftrunc} there is a caveat to this statement which is related to the two RG scales $k$ and $\kh$. Although the reduced system \eqref{equ:sys-red-final} has been obtained by a change of variables from the original RG flow this does not guarantee that specific RG related quantities, such as fixed point solutions, of the original system are given by the corresponding quantities when the reduced system \eqref{equ:sys-red-final} is treated as describing an RG flow in its own right. We come back to this question in the next section.

The change of variables \eqref{equ:chvars} together with \eqref{equ:kbar} turns the effective action \eqref{equ:ansatzGamma} into
\begin{equation}
\label{equ:Gamma-hat}
 \Gamma_k[\vp,\chi] = \int d^dx \left(-\frac{1}{2} \tilde K_{\kh}(\phi) \left(\partial_\mu \phi\right)^2+ \tilde V_{\kh}(\phi)\right),
\end{equation}
where the two renormalisation group scales are related by
\begin{equation}
\label{equ:kbar-k}
 \kh = k^\frac{\alpha}{\alpha+d_f}f(\chi)^\frac{1}{\alpha+d_f}
\end{equation}
and we have made use of the approximation of slowly varying background fields. Hence the solutions are such that the measure factor is absent in \eqref{equ:Gamma-hat} but the right hand side still depends on the background field through $\kh$. This is expected since the right hand side of the sWI \eqref{equ:sWiGamma} contains the background field dependent cutoff operator which is an intrinsic feature of the RG flow. As mentioned in sec. \ref{sec:conftrunc}, the cutoff operator drops out in the limit $k\to0$ and the effective action should become a function of the total field $\phi$ only. As we can see from \eqref{equ:kbar-k}, this is indeed the case. If the exponent of $k$ is positive the limit $k\to 0$ implies $\kh \to 0$ and the right hand side of \eqref{equ:Gamma-hat} becomes a function of $\phi$ only. This remains true if the exponent of $k$ in \eqref{equ:kbar-k} is negative, where now $k\to 0$ corresponds to $\kh \to \infty$ and the right hand side of \eqref{equ:Gamma-hat} would approach an ultraviolet fixed point of \eqref{equ:sys-red-final} (the actual limit $\kh \to \infty$ has to be taken in dimensionless variables as in \eqref{equ:dimless}). Thus, in either case this leads to a background independent effective action in the limit $k\to 0$ as a result of combining the sWI with the flow equation.

\subsection{Equivalence of the RG flows}
\label{sec:fpmap}
As emphasised in the previous section, the reduced system \eqref{equ:sys-red-final} is equivalent to the original flow equations \eqref{equ:flowV} and \eqref{equ:flowK} when also the equations of the sWI \eqref{equ:sWI-V} and \eqref{equ:sWI-K} are imposed. However, this is an equivalence between differential equations. By contrast, we now ask if the RG flow described by the reduced system \eqref{equ:sys-red-final} expressed in terms of the RG scale $\kh$ is equivalent to the RG flow of the original equations \eqref{equ:flowV} and \eqref{equ:flowK}. Such an equivalence of RG flows would require in particular that fixed points of one system correspond to fixed points of the other, and we will see now that whether this is true in the present case depends on the parametrisation function $f$.

Let us first concentrate on fixed point solutions for the potential. Denoting quantities that have been made dimensionless using the original RG scale $k$ with a bar, we write
\begin{equation}\label{Vh}
V(\vp,\chi) = k^{d-\frac{d}{2}d_f} \bar V(\bp, \cb), \qquad \vp = k^{\eta/2}\bp, \qquad \chi = k^{\eta/2}\cb
\end{equation}
for the original potential of \eqref{equ:flowV}, resulting in fixed point solutions satisfying $\partial_t \bar V =0$. On the other hand we have from the change of variables \eqref{equ:chvars} and the corresponding relation in \eqref{equ:dimless} that
\begin{equation}\label{Vt}
V(\vp,\chi) = \kh^d f^{-d/2} \Vh(\phih), \qquad \phih = \left(k/\kh\right)^{\eta/2}(\cb+\bp) \,.
\end{equation}
Using the relation \eqref{equ:kbar-k} between the two RG scales, the last two equations combine to give
\begin{equation}
\bar V(k,\bp,\cb) = \left(k^{-d_f} f\right)^\frac{d\eta}{2(d-\eta+2n)} \Vh(\kh,\phih).
\end{equation}
Bearing in mind that the partial derivative in the fixed point condition $\partial_t \bar V=0$ does not act on the dimensionless arguments $\bp$ and $\cb$, the last relation leads to
\begin{align}
\label{fpcondV}
\partial_t \bar V = \left(k^{-d_f} f\right)^\frac{d\eta}{2(d-\eta+2n)} & \left\{ \frac{1}{\al +d_f}
				\left(\al+ \frac{\eta}{2}\chi \dclnf\right) \partial_{\hat t} \Vh \right. \notag \\
				&\left.
				+ \left(\frac{\eta}{2}\chi \dclnf -d_f\right)\left(\frac{d\eta}{2(d-\eta+2n)}\Vh
				-\frac{\eta}{2(\al+d_f)}\phih \Vh'\right)\right\}.
\end{align}
The kinetic term $K$ can be dealt with similarly. Changing to the dimensionless version using $k$ gives the analogue of \eqref{Vh},
\begin{equation}
K(\vp,\chi) = k^{d-2-\eta-\left(\frac{d}{2}-1\right)d_f} \bar K(\bp,\cb),
\end{equation}
and the analogue of \eqref{Vt} is
\begin{equation}
K(\vp,\chi) = \kh^{d-2-\eta}f^{\frac{d}{2}+1} \Kh(\phih).
\end{equation}
Combining these results in
\begin{equation}
\bar K(\bp,\cb) = \left(k^{-d_f}f\right)^{-\frac{(n+1)\eta}{d-\eta+2n}} \Kh(\phih)
\end{equation}
which we use to calculate
\begin{align}
\label{fpcondK}
\partial_t \bar K = \left(k^{-d_f} f\right)^{-\frac{(n+1)\eta}{d-\eta+2n}} & \left\{ \frac{1}{\al +d_f}
				\left(\al+ \frac{\eta}{2}\chi \dclnf\right) \partial_{\hat t} \Kh \right. \notag \\
				&\left.
				- \left(\frac{\eta}{2}\chi \dclnf -d_f\right)\left(\frac{(n+1)\eta}{d-\eta+2n}\Kh
				+\frac{\eta}{2(\al+d_f)}\phih \Kh'\right)\right\}.
\end{align}
From \eqref{fpcondV} and the last equation we see that in general the fixed point condition $\partial_t \bar V = \partial_t \bar K =0$ of the original flow is not equivalent to the fixed point condition $\partial_{\hat t} \Vh=\partial_{\hat t} \Kh=0$ of the reduced flow \eqref{equ:sys-red-final}. Instead, requiring one of these conditions to hold will in general lead to RG time dependent solutions for the other flow.

This distinction disappears for parametrisation functions of power-law type, $f(\phi) = \phi^{2m}$, where the exponent is required to be an even integer for positivity. For this class of parametrisation functions the combination $\frac{\eta}{2}\chi \dclnf -d_f$ vanishes and we see from \eqref{fpcondV} and \eqref{fpcondK} that now the two fixed point conditions are in fact equivalent:
\begin{equation}
\partial_t \bar V = \partial_t \bar K =0 \qquad \Leftrightarrow \qquad \partial_{\hat t} \Vh=\partial_{\hat t} \Kh=0.
\end{equation}
In this way, by adopting the view that the reduced system of flow equations is required to be equivalent to the original system of flow equations at the level of the RG flows they describe, the parametrisation function in \eqref{equ:confpar} is restricted to be of power-law type. Note that using the sWI has led to the reduced system \eqref{equ:sys-red-final} in the first place and thus it crucially plays into this restriction on the form of parametrisation.

\subsection{Comparison to scalar field theory}
\label{sec:comp-sft}
To compare the present setup to standard scalar field theory, we note that if we absorb all factors containing the parametrisation function $f$ in \eqref{equ:ansatzGamma} into $K$ and $V$ and let $V \mapsto -V$ the result is
\begin{equation} \label{Gamma-sf}
\Gamma \mapsto -\Gamma_{\mathrm{sf}} = - \int d^dx \left(\frac{1}{2}K\left(\partial_\mu \vp\right)^2
		      +V\right),
\end{equation}
where $\Gamma_\mathrm{sf}$ is the ansatz for the effective action one would write down in the derivative expansion up to $\mathcal{O}\!\left(\partial^2\right)$ of standard scalar field theory. Strictly speaking this is true up to the fact that here $K$ and $V$ also depend on the background field, but as far as the flow equation \eqref{equ:FRGE} is concerned the background field just appears as a parameter and does not affect the conclusions we will come to here. There is also the additional difference given by the choice of dimensions as reflected in \eqref{Vh} which will be appropriately taken care of in a moment. Motivated by the overall minus sign included in the cutoff \eqref{equ:cutoff} reflecting the conformal factor sign in quantum gravity, we also make the replacement $\cutoff \mapsto -\cutoff$. Taken together these changes convert the flow equation \eqref{equ:FRGE} into
\begin{equation}\label{flow-comp}
\frac{\partial}{\partial t}\Gamma_\mathrm{sf} = -\frac{1}{2} \, \mathrm{Tr} \left[ \left(\Gamma^{(2)}_\mathrm{sf} + \cutoff\right)^{-1} \frac{\partial}{\partial t} \cutoff \right],
\end{equation}
which is the flow equation for the standard scalar field theory effective action $\Gamma_\mathrm{sf}$ of \eqref{Gamma-sf} with an additional minus sign on its right hand side.

Let us pause here to see how these alterations can be connected with the use of the sWI that led to the change of variables \eqref{equ:chvars}. As mentioned before, the effect of this change of variables is to convert the ansatz \eqref{equ:ansatzGamma} for the effective action into \eqref{equ:Gamma-hat}, which is precisely of form \eqref{Gamma-sf} if we also replace $V \mapsto -V$.  Moreover, we have seen earlier that the new RG scale $\kh$ defined by \eqref{equ:kbar} and the last equation in \eqref{equ:chvars} converts the cutoff operator \eqref{equ:cutoff} into the simpler version \eqref{cutoff-kbar}. The $f^{-d/2}$ factor in this expression cancels the measure factor in \eqref{equ:cutoff-action}, leaving a cutoff action that takes the form used in standard scalar field theory up to its overall sign and the replacement $\eta \mapsto d-2+\eta$. Substituting for the anomalous dimension according to this rule is necessary to pass from the dimension of the conformal factor field to the dimension of a standard scalar field. Hence, after implementing the replacement rules 
\begin{equation}\label{replsf}
V \mapsto -V \qquad \text{and} \qquad \eta \mapsto d-2+\eta
\end{equation}
in \eqref{equ:sys-red-final}, we obtain the same flow equations as for standard scalar field theory with the replacements $\Gamma_\mathrm{sf}\mapsto -\Gamma_\mathrm{sf}$ and $\cutoff \mapsto -\cutoff$, i.e. the flow \eqref{flow-comp}.

This relation to scalar field theory can be exemplified in $d=3$ dimensions. Applying \eqref{replsf} to the reduced flow \eqref{equ:sys-red-final} and evaluating the integrals reproduces the scalar field theory flow equations of ref. \cite{Morris:1994ie} with an additional minus sign on their right hand sides as in \eqref{flow-comp}.\footnote{When the equations are compared it should be noted that the RG time used in \cite{Morris:1994ie} corresponds to $-\hat t$ in the present context and the right hand sides are rescaled with the factor $1/(2\pi)$ compared to \eqref{equ:sys-red-final}.}

The second replacement rule in \eqref{replsf} simply amounts to a redefinition of a parameter and does not represent a structural difference to scalar field theory. Similarly, at a mathematical level, changing the sign of the potential is only a simple change of variables for the equations \eqref{equ:sys-red-final}. Hence, as far as the analysis of the differential equations \eqref{equ:sys-red-final} is concerned, we conclude that the only relevant difference to scalar field theory is a relative minus sign between its left and right hand sides. This close connection to scalar field theory will be used in sec. \ref{sec:fpbeyondLPA} to find an ansatz for the asymptotic behaviour of fixed point solutions.

\section{Fixed point analysis}
Building on the result of sec. \ref{sec:fpmap}, that fixed points of the original flow \eqref{equ:flowV} and \eqref{equ:flowK} correspond to fixed points of the reduced flow \eqref{equ:sys-red-final} for power-law parametrisations of the conformal factor, we first analyse the corresponding LPA in the next section before taking the first steps of an analysis of the full fixed point system in sec. \ref{sec:fpbeyondLPA}.

\subsection{Local potential approximation for the conformal factor}
\label{sec:confLPA}
We therefore first specialise to the lowest order of the derivative expansion by setting $\hat K=1$ and discarding the second equation in \eqref{equ:sys-red-final}. Corresponding to this order of the derivative expansion we also take the anomalous dimension to vanish, $\eta=0$. Untying the changes of variables \eqref{equ:dimless} and \eqref{equ:chvars} with \eqref{equ:kbar-k} shows that this corresponds to setting $K=k^{d-2}$ in the original effective action \eqref{equ:ansatzGamma}, i.e. we indeed obtain the LPA as characterised by a field independent coefficient of the kinetic term. In contrast to scalar field theory this coefficient is here the appropriate power of the RG scale due to the vanishing classical scaling dimension of the conformal factor field.

\begin{figure}[ht]
\centering
\includegraphics[scale=0.65]{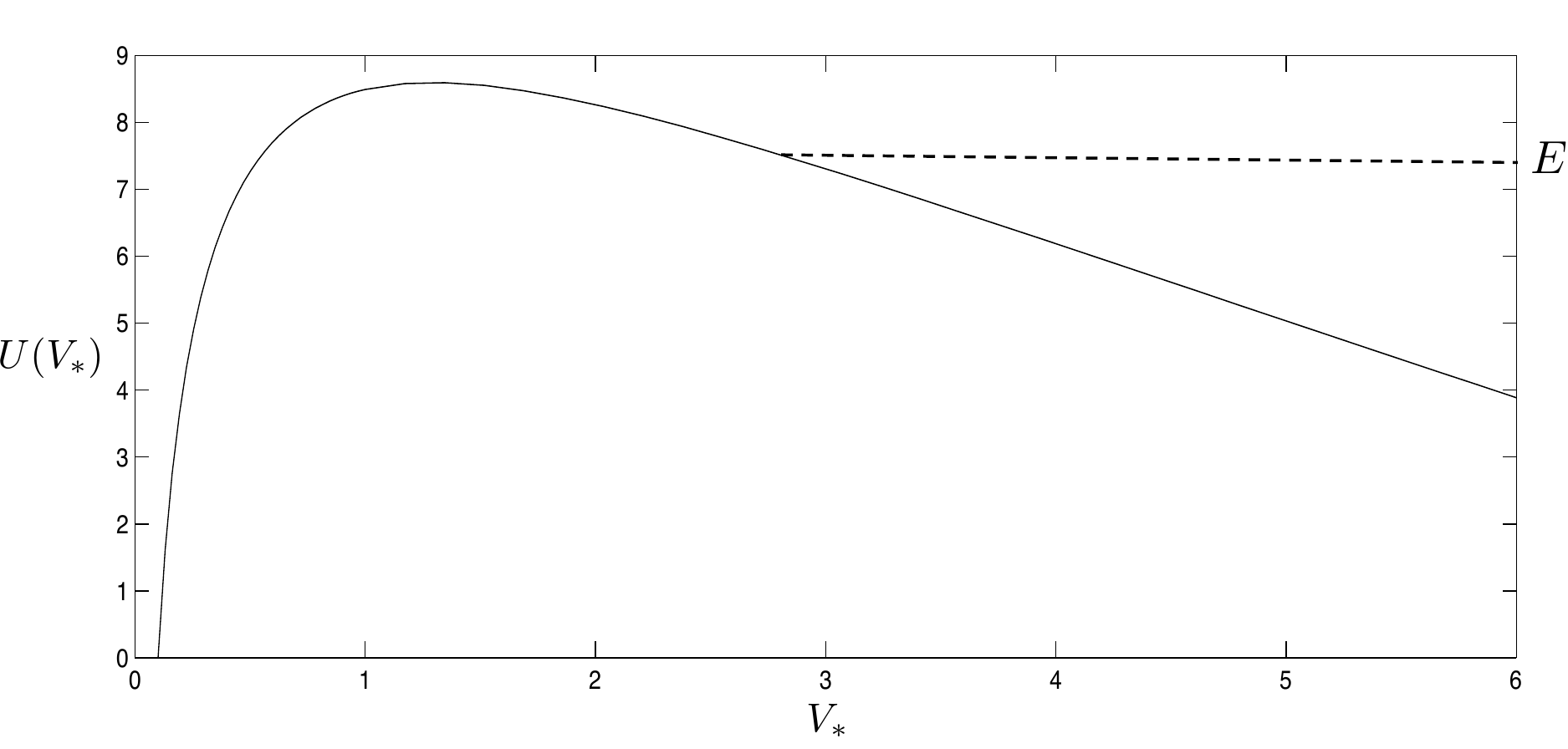}
\caption[The Newtonian potential used in solving the LPA for the conformal factor.]{Characterising the solutions of \eqref{FPLPAconf} with the Newtonian potential $U(V_*)$.}
\label{fig:NewPot}
\end{figure}

With these provisions the fixed point version of \eqref{equ:flowV-final} in $d=4$ dimensions becomes
\begin{equation}
\label{FPLPAconf}
V_* = -2 \int_0^\infty dp \,\frac{p^3}{V_*''p^4-p^6-1}=: \mathcal{F}\!\left(V_*''\right),
\end{equation}
where we have omitted the hat on the fixed point potential for typographical clarity and with $n=2$ made the smallest possible choice for the exponent in the power law cutoff to ensure convergence, cf. below \eqref{equ:cutoff-pwrlw}. One finds that the right hand side $\mathcal{F}$ is a monotonically increasing function, where the allowed range for its argument is (re-naming $\phih \to \phi$)
\begin{equation} \label{convbound}
-\infty<V_*''(\phi)<3\cdot 2^{-2/3}
\end{equation}
to guarantee finiteness of the integral. In fact, any solution $V_*$ for which $V_*''$ reaches the upper bound in this inequality on a finite range for $\phi$, ends in a moveable singularity. We can now proceed as for Newton's equation in classical mechanics and find a first integral by inverting the function $\mathcal{F}$ and solving $dU/dV_* = -\mathcal{F}^{-1}\!\left(V_*\right)$ so that the solutions are labelled by one parameter $E$ and satisfy $E=1/2 (V_*')^2 + U(V_*)$. The "Newtonian potential" $U$ can be determined numerically and takes the form shown in fig. \ref{fig:NewPot}.
Globally valid solutions are obtained only if $E$ is not greater than the maximum displayed by $U$ and if $V_*$ accordingly takes values corresponding to the region to the right of the maximum and above the lower bound provided by $U$ in fig. \ref{fig:NewPot}. All other solutions end at the singularity $V_* \to 0 \Leftrightarrow V_*'' \to -\infty$ at some finite field value $\phi=\phi_c$. If a solution is globally defined we see that $V_*(\phi)\to \infty$ asymptotically which entails $V_*''(\phi)\to 3\cdot 2^{-2/3}$, i.e. $V_*''$ is asymptotically approaching the upper limit of the convergence range \eqref{convbound}. From the second derivative tending to a constant one would infer a two parameter set of solutions. The additional parameter besides $E$ is slightly hidden in the approach using the first integral $U$ here but it can be recovered by exploiting translation symmetry $V_*(\phi) \mapsto V_*(\phi +c)$ for any solution at fixed $E$ of the fixed point equation \eqref{FPLPAconf}. This symmetry can be exploited to implement $\phi \mapsto -\phi$ symmetry of the solutions $V_*(\phi)$, corresponding to time reflection symmetry in the Newtonian analogy.\footnote{Note here the distinction between a symmetry of the fixed point equation and a symmetry of its solutions.}

Hence, the fixed point equation \eqref{FPLPAconf} has a two parameter set of solutions that can be thought of as parametrised by $E$ and $V_*'(0)$, where $E$ is restricted to not lie above a maximum value. If we choose to implement the condition $V_*'(0)=0$ to obtain even solutions, this set reduces to a single ray as given by $E\leq E_\mathrm{max}$.

Coming back to the relation to standard scalar field theory discussed in the previous section, we remark that to obtain the correct fixed point equation of standard four-dimensional scalar field theory, the second replacement in \eqref{replsf} has to be made before setting $\eta=0$. This would lead to an additional term $-\phi V_*'$ on the left hand side and as a result \eqref{FPLPAconf} is structurally different from scalar field theory in a way that goes beyond the modifications of sec. \ref{sec:comp-sft}. Instead, the picture here is reminiscent of standard scalar field theory in the LPA in two dimensions studied in \cite{Morris:1994jc}, where a similar Newtonian potential leads to a semi-infinite line of oscillating solutions for the potential $V_*$. In fact, if we replace $V_*\mapsto -V_*$ in \eqref{FPLPAconf} and change the sign of the right hand side as discussed in the previous section one obtains the Newtonian potential as in fig. \ref{fig:NewPot} reflected about the $V_*$ - axis, now describing periodic fixed point solutions.

In connection with the results for the LPA of this section, the question arises how they compare to the study \cite{Manrique:2009uh}, where the RG flow of the conformal factor has been investigated in a similar bi-field LPA. The ansatz for the effective action in \cite{Manrique:2009uh} reads
\begin{equation}
  \label{equ:Gamma-Reuter}
  \Gamma_k[\phi,\chi] = -\frac{3}{4\pi}\int d^4x\left\{\frac{1}{2G_k}\left(\partial_\mu \phi\right)^2+ \frac{1}{2G_k^B}\left(\partial_\mu \chi\right)^2+ W_k(\phi,\chi)\right\}.
 \end{equation}
It is expressed as a functional of the total conformal factor field $\phi = \chi+\vp$, where $\vp$ and $\chi$ refer to the variables used here in the original flow equations \eqref{equ:flowV} and \eqref{equ:flowK}, and $W(\phi,\chi)$ is a general scale dependent potential. It also contains the scale dependent analogue of Newton's constant $G_k$ for the total field as well as for the background field, $G_k^B$. The reference metric of \eqref{equ:confpar} is also chosen to be $\hat g_{\mu\nu} = \delta_{\mu\nu}$ in \cite{Manrique:2009uh} and we have implemented this in \eqref{equ:Gamma-Reuter} as well as transcribed into the present notation. After deriving the flow equation and the sWI for the potential $W$, it is then found that they cannot be satisfied simultaneously at a non-Gaussian fixed point. Importantly, as in \eqref{equ:cutoff}, the cutoff operator in \cite{Manrique:2009uh} is also implemented with respect to the covariant background field Laplacian. However, a crucial difference between \eqref{equ:Gamma-Reuter} and our ansatz for the effective action \eqref{equ:ansatzGamma} is that the kinetic terms in \eqref{equ:Gamma-Reuter} are not formulated in terms of the background field covariant derivative, nor is the appropriate measure factor containing the background field included in \eqref{equ:Gamma-Reuter}. If the background field is varied, the operator describing high and low momentum modes of the conformal factor in \eqref{equ:ansatzGamma} changes with it, as the gravitational setting dictates, whereas the high and low momentum modes of the ansatz \eqref{equ:Gamma-Reuter} remain the same. Due to this difference, a direct comparison of results is not possible.

\subsection{Fixed points beyond the LPA}
\label{sec:fpbeyondLPA}
The fixed point equations pertaining to the full system \eqref{equ:sys-red-final} in $d=4$ dimensions and with $n=2$ in the cutoff \eqref{equ:cutoff} are
\begin{subequations}
\label{equ:sysfp}
\begin{align}
 4V_*-\frac{\eta}{2}\phi V_*' &= -\left(8-\eta\right)\int_0^\infty \frac{dp}{p}\,Q_0 , \label{equ:fpV} \\
(2-\eta)K_*-\frac{\eta}{2}\phi K_*' &= -2(8-\eta)\int_0^\infty \frac{dp}{p}\,P\big(p^2,\phi\big) , \label{equ:fpK}
\end{align}
\end{subequations}
where we continue to omit the hats and $Q_0$ and $P$ are now given by the corresponding version of \eqref{Q0Pfinal},
\begin{subequations}
\label{Q0Pfp}
\begin{align}
 Q_0 = &\left[V_*''-K_*p^2-\frac{1}{p^4}\right]^{-1} \label{Q0fp} \\
 P = &-\frac{1}{2}K_*'' Q_0^2 + K_*'\left(2V_*'''-\frac{9}{4} K_*' p^2\right)Q_0^3  \notag \\
    & +\left[\left\{2K_*' p^2 - V_*'''\right\}\left(K_*-\frac{2}{p^6}\right)+ \frac{3}{p^6}\left(K_*' p^2-V_*'''\right)\right]\left(V_*'''-K_*' p^2\right) Q_0^4 \label{Pfp} \\
    & -p^2 \left(K_*-\frac{2}{p^6}\right)^2 \left(V_*'''-K_*' p^2\right)^2 Q_0^5, \notag
\end{align}
\end{subequations}
and we now allow for a non-vanishing anomalous dimension.

As to the general structure of the system of fixed point equations, differentiating \eqref{equ:fpV} once, solving for the third derivative $V_*'''$ and substituting the result into \eqref{equ:fpK} reveals that \eqref{equ:sysfp} is of second order in both $V_*$ and $K_*$ and therefore admits a four dimensional space of local solutions around any generic initial value $\phi=\phi_0$, cf. sec. \ref{par-counting}. For the present purpose of finding global solutions valid on the whole real line $-\infty<\phi<\infty$, we can take $\phi_0=0$ and start with the local parameter space spanned by $V_*(0), V_*'(0), K_*(0), K_*'(0)$. Since there are no explicit appearances of the field $\phi$ in \eqref{Q0Pfp} the fixed point equations do not feature any fixed singularities. However, the generalisation of \eqref{convbound} now takes the form,
\begin{equation}
\label{convboundfull}
V_*''(\phi) < 3\left(\frac{K_*(\phi)}{2}\right)^{2/3} \qquad \text{and} \qquad K_*(\phi)>0\, ,
\end{equation}
and any violation of these inequalities on a finite range for $\phi$ will lead to a moveable singularity, thus placing a restriction on parameter space.

Furthermore, a second virtue of the power-law cutoff \eqref{equ:cutoff-pwrlw} besides facilitating the combination of the flow equation with the sWI is that the fixed point equations \eqref{equ:sysfp} enjoy a (non-physical) scaling symmetry as characterised by the following scaling dimensions
\begin{equation}\label{scalings}
[V_*]=4, \qquad [K_*]=-6, \qquad [\phi]=4, \qquad [p]=1.
\end{equation}
Rescaling all quantities in \eqref{equ:sysfp} with the power of a real number as given here leaves the fixed point equations unchanged. This can be exploited to eliminate one parameter of solution space. Note however that the scaling prescriptions do not allow to change the sign of either $V_*$ or $K_*$. From the inequalities in \eqref{convboundfull} it is therefore convenient to eliminate the parameter $K_*(0)$ by fixing it to $K_*(0)=2$.

Finally, since the fixed point equations \eqref{equ:sysfp} are symmetric under $\phi \mapsto -\phi$ one may choose to impose $V_*'(0)=K_*'(0)=0$ to restrict to even fixed point solutions. It has to be emphasised however that at this point requiring either $V_*$ or $K_*$ or both to be even is an additional assumption.

Following this route and regarding the anomalous dimension as just an additional parameter, we so far find from parameter counting that we are left with only the two parameters $V_*(0)$ and  $\eta$. However, as we have seen in sec. \ref{par-counting} and \ref{par-count-Benedetti} in a different context, an asymptotic analysis of the fixed point equations \eqref{equ:sysfp} is needed to capture possible constraints on parameter space as $\phi \to \infty$ and to arrive at conclusive results for parameter counting.

From the structural similarity of \eqref{equ:sysfp} to standard scalar field theory, cf. sec. \ref{sec:comp-sft}, one may be led to investigating the corresponding asymptotic behaviour given to leading order by solving the left hand sides of the fixed point equations, 
\begin{equation}
\label{sfasy}
V_*(\phi)=A\phi^{8/\eta} + \dots \qquad \text{and} \qquad K_*(\phi) = B\phi^{4/\eta-2} +\dots \, ,
\end{equation}
for constants $A,B$. From this one finds that for $0<\eta<8$ the dominant term at large field in the first inequality in \eqref{convboundfull} is $V_*''$ and the only way to avoid a moveable singularity is therefore to have $A<0$, leading to a potential unbounded from below. On the other hand, we can expand \eqref{Q0fp},
\begin{equation}
\label{expQ0}
Q_0 = -\frac{p^4}{K_* p^6+1}-\frac{p^8}{(K_*p^6+1)^2}\,V_*''-\frac{p^{12}}{(K_*p^6+1)^3}(V_*'')^2-\dots \,,
\end{equation}
whenever this series is meaningful either as a convergent or an asymptotic series. The fixed point equation \eqref{equ:fpV} for the potential then evaluates to
\begin{equation}\label{odeVexp}
4V_*-\frac{\eta}{2}\phi V_*' = \frac{(8-\eta)\pi}{3\sqrt{3}}\left(\frac{1}{K_*^{2/3}}
+\frac{1}{3}\frac{V_*''}{K_*^{4/3}}+ \frac{\sqrt{3}}{4\pi}\frac{(V_*'')^2}{K_*^2}+ \dots \right).
\end{equation}
We see a posteriori that this expansion is useful as long as $V_*''/K_*^{2/3}$ is decreasing. With the assumed asymptotic form \eqref{sfasy}, this is the case precisely for $\eta<0$ or $\eta>8$. A brief calculation shows however, that for these ranges of $\eta$ the right hand side in \eqref{odeVexp} cannot be neglected compared to the left hand side and thus that the asymptotic behaviour \eqref{sfasy} is not consistent. Up to the two special cases $\eta=0$ and $\eta=8$, for which the fixed point equations \eqref{equ:sysfp} change structurally and need separate investigation, standard scalar field theory asymptotic behaviour is therefore excluded unless we accept fixed point potentials that are unbounded from below.

In fact, at the present stage, one may consider not only fixed point potentials that are unbounded from above but also fixed point potentials that are unbounded from below. One could argue that the consequences of the wrong sign kinetic term in \eqref{equ:ansatzGamma} may be cured by higher orders in the derivative expansion, thus favouring the former type of fixed point potentials. On the other hand, if higher derivative terms fail to cure the negative sign effects of the kinetic term one may naturally be led to consider fixed point potentials that are unbounded from below so that $-\Gamma_k$ would become bounded from below.

Nevertheless, if we decide to exclude fixed point potentials that are unbounded from below, we learn from the above that the leading asymptotic behaviour of solutions to \eqref{equ:sysfp} is not determined by scaling dimensions. Instead, the quantum corrections on the right hand side of \eqref{equ:sysfp} cannot be neglected in the large field regime. While this is surprising from the point of view of scalar field theory, the same situation was encountered in sec. \ref{sec:asy-fp} for the asymptotic behaviour in the $f(R)$ truncation as encoded in the fixed point equation \eqref{fp-Dario}. A much more comprehensive asymptotic analysis may therefore be required in the present case and although we will discuss some aspects of it in chapter \ref{sec:conclusions}, this is outside the scope of this work.

\begin{figure}[h]
\begin{center}
$
\begin{array}{cc}
\includegraphics[width=0.45\textwidth]{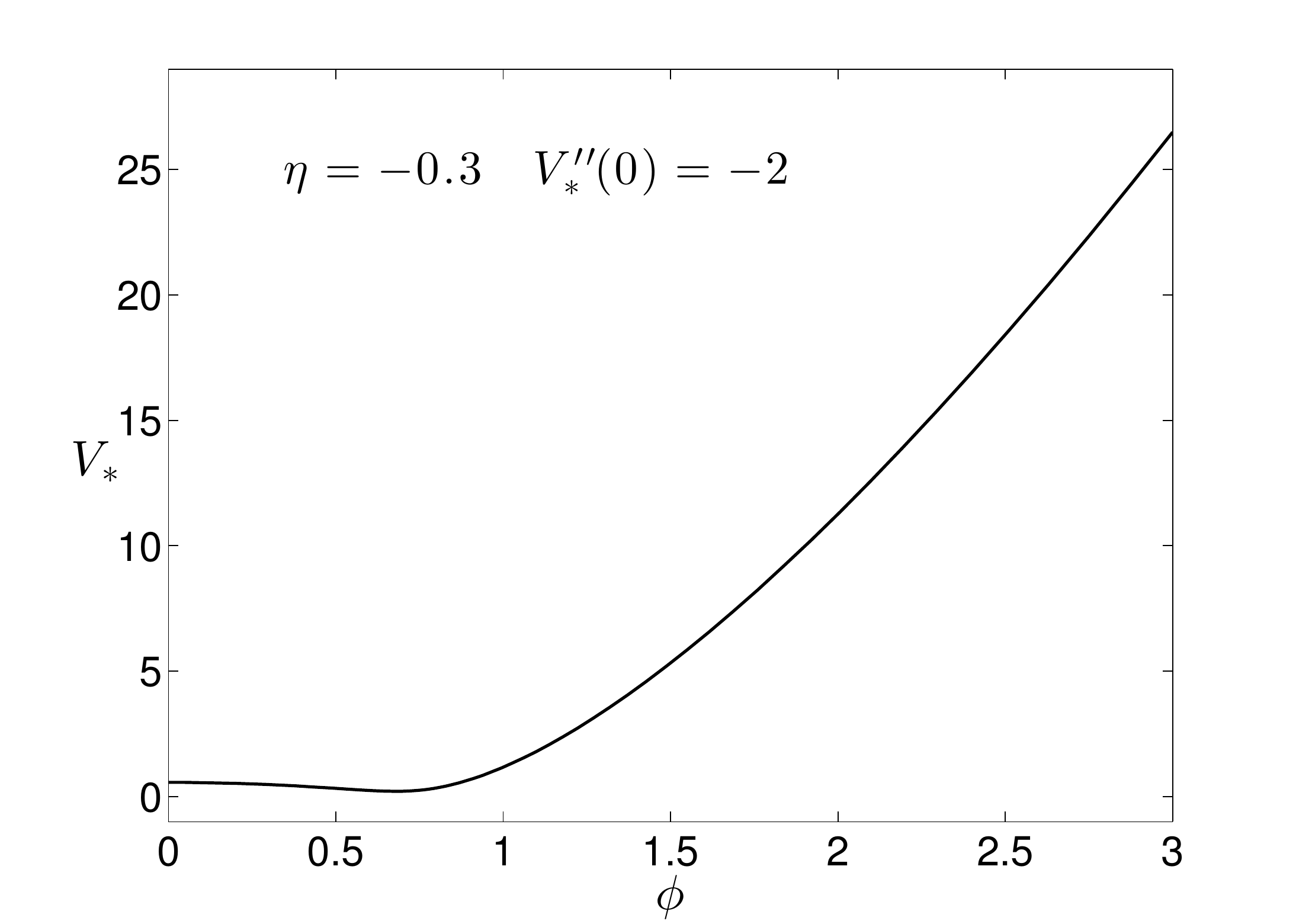} &
\includegraphics[width=0.45\textwidth]{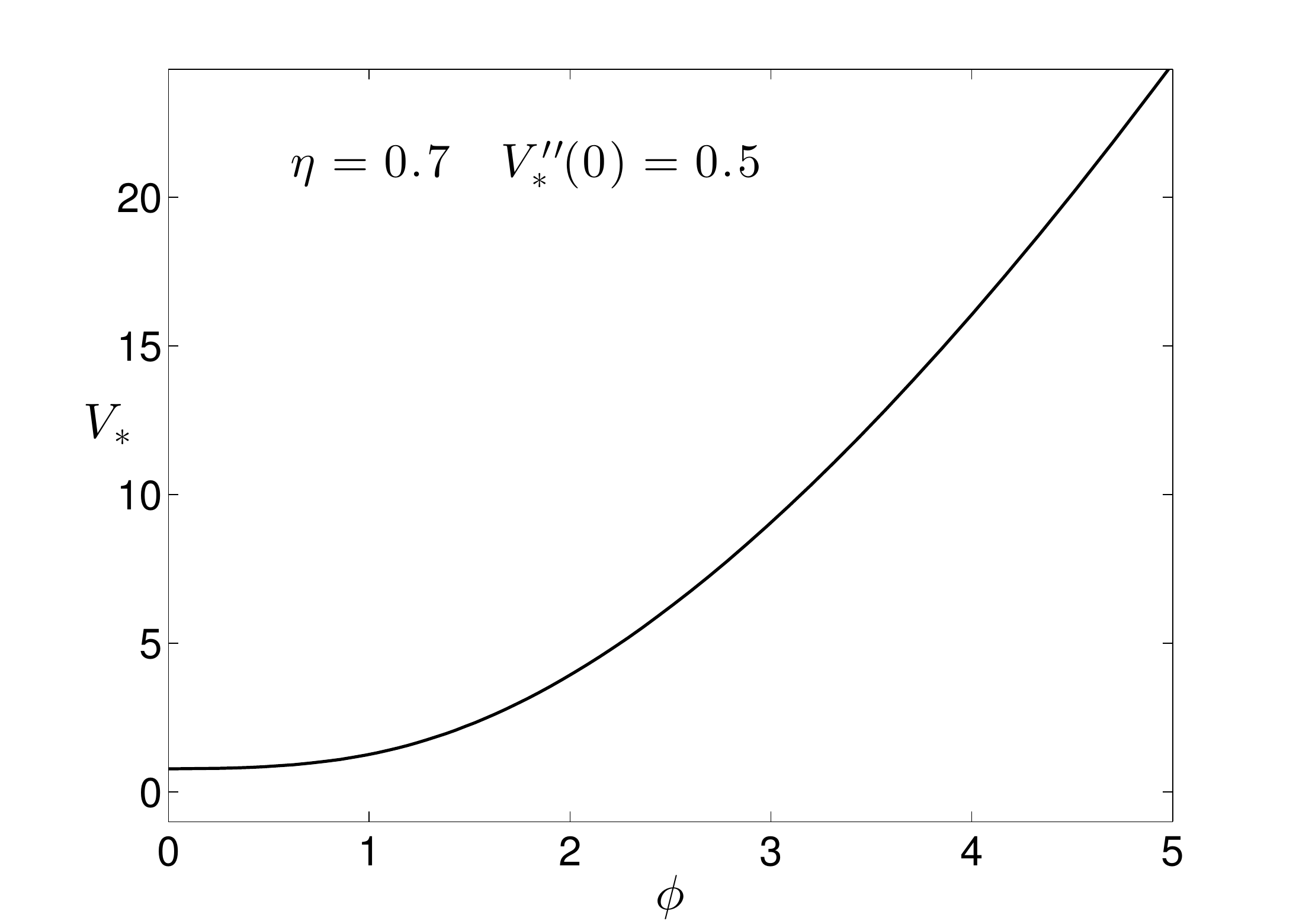} \\
\includegraphics[width=0.45\textwidth]{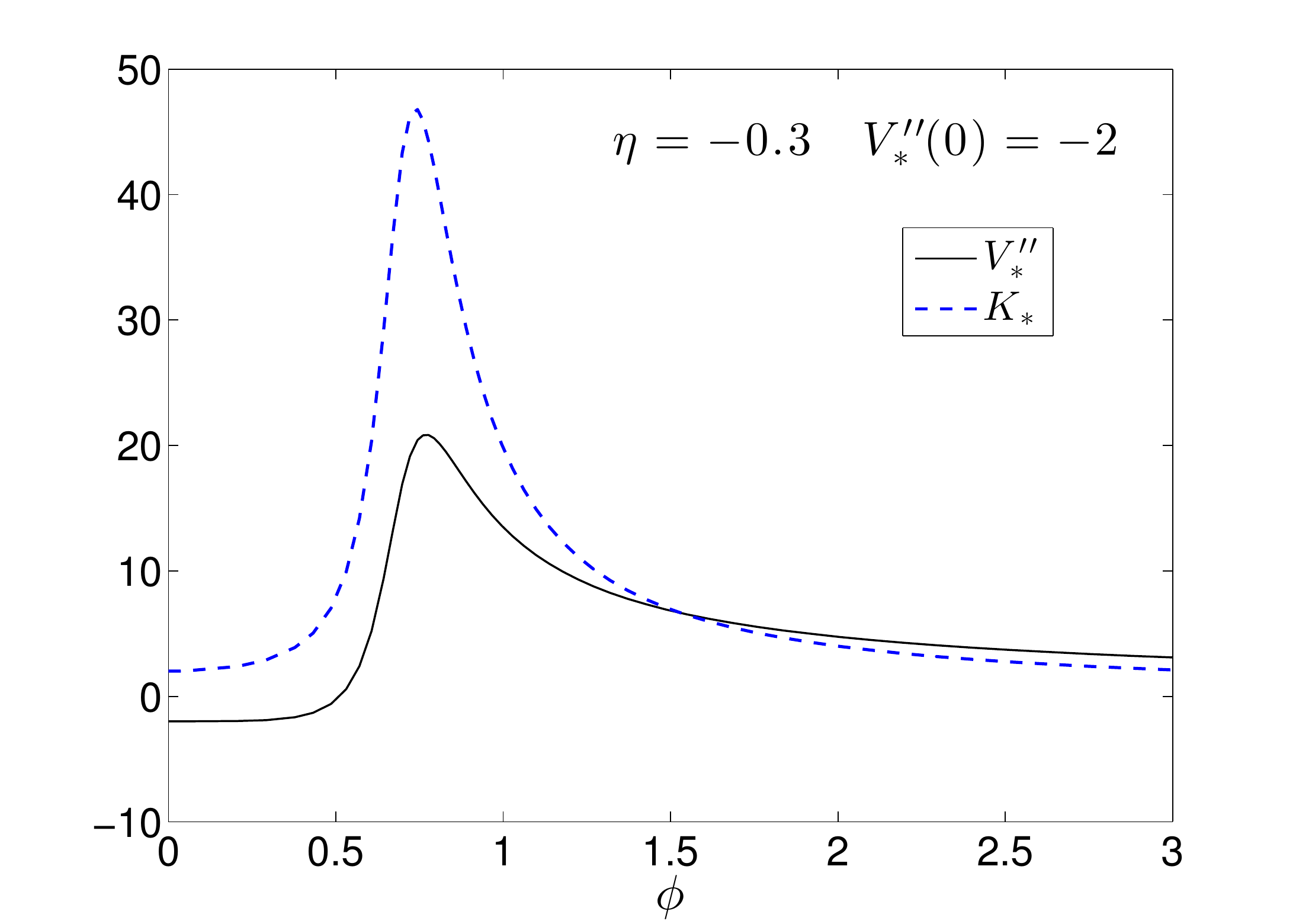} &
\includegraphics[width=0.45\textwidth]{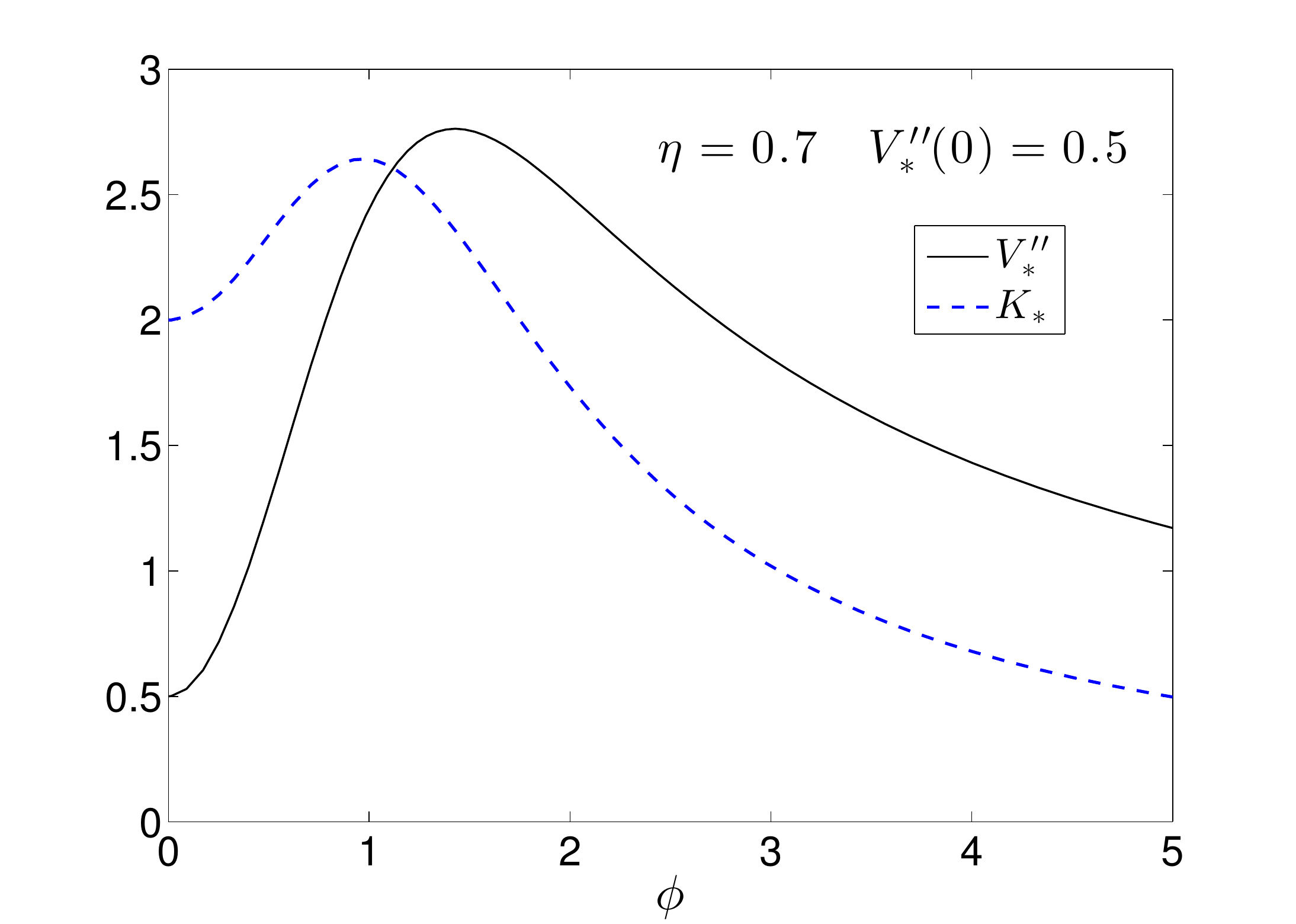}
\end{array}
$
\end{center}
\caption[Numerical integration of conformal fixed point equations]{Numerical integration of \eqref{equ:sysfp} for $\eta=-0.3, \,V_*''(0)=-2$ in the left panel and $\eta=0.7,\, V_*''(0)=0.5$ on the right. The other initial conditions have been fixed as discussed in the text.}
\label{fig:confexsols}
\end{figure}

In principle, the integrals on the right in \eqref{equ:sysfp} can be evaluated using contour integration in the complex plane. The length of the resulting expressions is however such that they become unmanageable. For integrating \eqref{equ:sysfp} numerically, it is therefore advisable to also perform a numerical evaluation of the integrals at each step of the solver. To bring the system \eqref{equ:sysfp} into normal form for actual computations, we solve the differentiated version of \eqref{equ:fpV} for $V_*'''$ and trade the initial condition $V_*(0)$ for $V_*''(0)$, while the \eqref{equ:fpK} is easily solved for the highest derivative $K_*''$. Fig. \ref{fig:confexsols} shows one example integration for negative anomalous dimension on the left and a second for positive anomalous dimension on the right.
In both cases the numerical integration can be carried out to arbitrarily large field, limited only by the efficiency of the solver, and it is interesting to note that the constraints \eqref{convboundfull} seem to be saturated asymptotically.

By varying the two parameters $\eta$ and $V_*''(0)$ one finds that solving the fixed point equations \eqref{equ:sysfp} numerically is in general not hampered by the appearance of moveable singularities as caused by violation of \eqref{convboundfull} at finite field. However, as for parameter counting, an effective and comprehensive numerical analysis of the system \eqref{equ:sysfp} has to build on a thorough understanding of the fixed point solutions at large field and thus has to be deferred to future work. 

%% file: conclusions/conclusions.tex
\chapter{Discussion and Outlook}
\label{sec:conclusions}

A large part of all evidence gathered so far in the literature in favour of asymptotic safety for gravity has been obtained in finite dimensional truncations of the effective action and in the single field approximation.  We have touched on this in sec. \ref{sub:adaptations-gravity}, where we have also mentioned that work has begun to successfully probe asymptotic safety beyond the single field approximation. All studies of full gravity to date that do this and hence keep separate dependence on the background field in the effective action have retained only a finite number of couplings. The step towards infinite dimensional truncations is technically difficult when the single field approximation is adopted, and will thus be even more challenging when this approximation is dropped. An idea of the efforts needed to analyse such a functional truncation in the simpler case where the background and total field are identified can be gained from chapter \ref{sec:f-of-R}.

The basic tenet, on which all further analysis of the differential equations in a functional truncation of the effective action has to build, is the parameter counting method. Requiring solutions to exist globally, i.e. on the whole range of definition of the differential equation, places constraints on parameter space that originate from fixed and moveable singularities as well as from the asymptotic behaviour of solutions. Taking these constraints together, the parameter counting method leads to crucial insight into the structure of the parameter space of solutions. For each independent constraint supplied by fixed singularities or the asymptotic solution of the differential equation, the dimension of parameter space is reduced accordingly. The remaining set of parameters may then be further restricted to possibly disconnected pieces of the same dimension by moveable singularities.

Exploiting this understanding one can then set up an efficient framework for a quantitative analysis of the differential equation that combines analytical and numerical methods. In the search for fixed points of the $f(R)$ truncation in chapter \ref{sec:f-of-R} the analytical part was provided by the asymptotic expansion of fixed point solutions as well as Taylor expansions around the fixed singular points. Matching a numerical solution to the asymptotic expansion with an admissible set of coefficients guarantees the validity of the solution for arbitrarily large $R$. Taylor expanding around fixed singular points allows to implement exactly the corresponding condition on parameter space and leads to a numerically clean way of building a solution across fixed singularities. As described in sec. \ref{numerics}, a careful numerical treatment of the fixed point equation, that often involves very high numerical accuracy, then leads to a confirmation of the results expected from parameter counting.

This overall strategy for tackling functional truncations has proved successful in the context of scalar field theory, e.g. \cite{Morris:1994ie,Morris:1994jc}, but also in conformal $f(R)$ gravity \cite{Demmel:2014sga,Demmel:2013myx,Demmel:2012ub}, and indeed we expect it to be a fruitful approach for any functional truncation in the framework of the functional renormalisation group.

The outcomes of chapters \ref{sec:f-of-R} and \ref{sec:red-ops} concerning the $f(R)$ truncation show that verification of the asymptotic safety scenario in a functional truncation may require an extremely careful derivation of the associated flow equations. This is clear from the result of chapter \ref{sec:f-of-R} that the fixed point equations of two versions of the $f(R)$ truncation do not admit global solutions while the third version leads to continuous sets of fixed point solutions. While the results of chapter \ref{sec:red-ops} show that in the latter case the space of fixed point solutions that are physically sensible is again empty, it nevertheless points towards the possibility that the flow equations of functional truncations are very sensitive to the way they are set up. The single field approximation is one point in deriving the flow equations that has the potential to lead to problematic results, as investigated in detail in sec. \ref{sec:LPA}, but since all three $f(R)$ truncations have exploited the single field approximation it cannot be the only reason for the different solution spaces. Instead, the details of how the cutoff is implemented for the suppression of low momentum modes as well as the chosen strategy for the evaluation of the traces in the flow equation \eqref{equ:FRGE} can both have an impact on the resulting structure of the differential equations. As reviewed in sec. \ref{sub:flows-f(R)}, for the fixed point equations of refs. \cite{Machado:2007, Codello:2008} whose space of global solutions is empty, the cutoff has been implemented with respect to the covariant background field Laplacian according to the rule \eqref{typeI} and the traces  have been evaluated with an asymptotic heat kernel expansion, whereas for the cutoff in \cite{Benedetti:2012dx} the rule \eqref{typeI} is instead used for the three operators \eqref{Deltas} and the traces are evaluated by spectral summation. We have discussed some of the relevant effects these different strategies have on the resulting flow equations in sec. \ref{sub:flows-f(R)}, most notably with respect to fixed singularities.

This sensitivity to details of the implementation is corroborated by the following point, that was already alluded to in sec. \ref{sec:beyondpoly}. Strong evidence in favour of asymptotic safety is found in the studies \cite{Falls:2013,Falls:2014tra} with polynomial truncations to high orders of the function $f(R)$ using the same flow equation as in \cite{Codello:2008}. Taking this evidence seriously by assuming that asymptotic safety is indeed realised in Nature, we should thus be able to verify it in a functional $f(R)$ truncation. As recapitulated above, it was shown in sec. \ref{sec:fpanaMachado} that no global solution to the fixed point equation of \cite{Codello:2008} exists, which may mean that finite dimensional truncations create more favourable conditions than functional truncations with respect to the approximations used, and the freedom available, in setting up the RG flow.

Focussing on the single field approximation, the effects it can cause in the presence of a background field have been discussed extensively in chapter \ref{sec:LPA} in the less involved context of scalar field theory. The overall message from these investigations is that non-physical behaviour in one way or other is likely to be the consequence if background field dependent cutoff operators are used in combination with the single field approximation. This is the case even if the cutoff operator depends on the background field in such a way that the one-loop beta function remains correct in perturbation theory, cf. sec. \ref{sec:LPAsetup}.

Abandoning the single field approximation however means an enlargement of theory space, as additional operators involving the background field now have to be taken into account. At the same time, the underlying number of physical couplings of the theory does not change by introducing a background field split as in \eqref{background-split}, which makes it necessary to keep control of the background field dependence of the effective action and constrain it with the split Ward identity, cf. the discussion at the beginning of chapter \ref{sec:conformal}. That this approach can be sufficient for recovering the correct physical content of RG flows is demonstrated in sec. \ref{sec:sWI-LPA} for the LPA in scalar field theory.

In light of the underlying theme of this thesis, the interplay between functional truncations and the single field approximation, it may be interesting to investigate if the non-physical behaviour found in sec. \ref{sec:gencutoff} and \ref{sec:LPALitim} is also visible in polynomial truncations of the potential. If not, it would be another example of the robustness associated with finite dimensional truncations with respect to the approximations used in implementing the RG flow alluded to above. Drawing the analogy to gravitational flows, this is in fact what should be expected if the results of finite dimensional truncations in gravity in the single field approximation are taken to be trustworthy.

One may also speculate what the consequences would be, if the single field approximation is dropped but one nevertheless decides to not enforce the sWI to hold alongside the flow equation. As before, this could also be investigated in polynomial truncations of the potential $V(\vp,\chi)$  in \eqref{flow2}. At the level of functional truncations however, this is certainly a problematic approach since without the sWI there will not be a unique solution to the flow equation, as emphasised in sec. \ref{sec:sWI-derivation} and at the beginning of sec. \ref{sec:combine}. The success of combining the sWI with the flow equation in sec. \ref{sec:LPA} is such that one could take the point of view that if a truncation of the effective action leads to solutions as long as the flow equation is considered on its own, but no solutions exist if the sWI is imposed alongside the flow equation, the derivation of the corresponding RG flows has to be revised or the possibility has to be entertained that there are no physically trustworthy solutions for that truncation.

A first result for implementing the sWI in conjunction with the flow equation in a gravitational context and for a functional truncation has been obtained in chapter \ref{sec:conformal}. Truncating to the conformal degree of freedom of the metric, the sWI has been combined with the flow equation with an ansatz for the effective action that goes beyond the LPA in the fluctuation field, cf. \eqref{equ:ansatzGamma}. The constraining power of the sWI has led to a set of reduced flow equations which no longer contain explicit background field dependence. For this result to be possible it was important to reflect the gravitational context with appropriate modifications in both the flow equation \eqref{equ:flowGamma} and the sWI \eqref{equ:sWiGamma}, and the ansatz \eqref{equ:ansatzGamma} similarly had to contain the correct background field factors as appearing in the measure and the covariant derivative of the kinetic term.

While the LPA case of this truncation is treated in sec. \ref{sec:confLPA}, it turns out that the full system of flow equations beyond the LPA \eqref{equ:sys-red-final} is very challenging to analyse comprehensively. We have outlined a number of important aspects such an analysis has to include for the corresponding fixed point equations \eqref{equ:sysfp} in sec. \ref{sec:fpbeyondLPA} but the main bottleneck is to gain a detailed understanding of the associated asymptotic behaviour of solutions. While an expansion such as \eqref{expQ0} leading to \eqref{odeVexp} and a similar expansion for the equation \eqref{equ:fpK} is certainly a good starting point for developing this understanding, it is restricted to behaviours where $V_*''/K_*^{2/3}$ is decreasing for large field. Note for example that an attempt to expand $Q_0$ in \eqref{Q0fp} in terms of $K_*$ analogously to the expansion \eqref{expQ0} in $V_*''$ leads to divergent integrals on the right of \eqref{equ:sysfp}. This happens since in the denominator of $Q_0$ in \eqref{Q0fp} the term $K_*p^2$ cannot be neglected for large $p$, no matter how small $K_*$ may be for large field.

A second route to a better understanding of the asymptotic behaviour of fixed point solutions originates from the insight gained numerically that over significant ranges of the parameters $V_*''(0)$ and $\eta$ the constraints \eqref{convboundfull} are saturated as $\phi\to\infty$. This can be exploited to introduce a small variable $u(\phi)$ capturing this behaviour for large field and expanding the integrands on the right in \eqref{equ:sysfp} accordingly. The integrals themselves can then be computed as an expansion in $u(\phi)$ using contour integration in the complex plane. However, this will again provide only a partial understanding of asymptotic behaviour as there are other ranges for the parameters $V_*''(0)$ and $\eta$ that do not lead to a saturation of \eqref{convboundfull} in the large field regime.

Nevertheless, future work may reveal that these approaches can be complemented with additional techniques for treating the cases not covered by the above so that all possible asymptotic behaviour is accounted for. Once this has been achieved, the parameter counting method can be fully applied to gain insight into the parameter space of global solutions. The expectation to be confirmed or refuted is that sufficient constraints are built into the system \eqref{equ:sysfp} to allow for only a discrete set of fixed point solutions. In particular, it should be possible to determine the anomalous dimension $\eta$ from such an analysis of global solutions in which it is expected that the scaling symmetry \eqref{scalings} will play an important role, cf. the studies \cite{Morris:1996xq,Morris:1994jc}.

We also note that the flow equations \eqref{equ:sys-red-final} may be easily transformed into the flow equations of standard scalar field theory at $\mathcal{O}\!\left(\partial^2\right)$ of the derivative expansion using the rules described in sec. \ref{sec:comp-sft}. They may then be used in $d=4$ dimensions to investigate triviality of scalar field theory beyond the LPA, the expectation being that they should not lead to physically sensible non-perturbative fixed points, see e.g. \cite{Rosten:2008ts,Hasenfratz:1985dm}.

Coming back to the gravitational context of chapter \ref{sec:conformal}, there are two important directions for generalisation of the work presented there. Firstly, we have made the approximation of identifying the anomalous dimension of the conformal background field with the anomalous dimension of the fluctuation field. A priori, these two quantities are different and it would be interesting to investigate the consequences of implementing the sWI in this context. Secondly, and perhaps more importantly, the parametrisation function $f$ in \eqref{equ:confpar} may be allowed to depend on the RG scale $k$. This is analogous to the $t$-dependence of the function $h$ in sec. \ref{sec:LPAsetup} used to introduce background field dependent cutoffs in \eqref{equ:cutoffh} and is indeed natural in the context of functional renormalisation. The flow equation \eqref{equ:flowGamma} will then pick up this $t$-dependence and will thereby be linked more directly than before to the background field dependent parametrisation function.

Completing this programme will certainly lead to further insights as to how the sWI places the necessary restrictions on the solutions allowed by the flow equation. In the context of the bi-metric Einstein-Hilbert truncation in full gravity, the sWI has been implemented recently in \cite{Becker:2014qya} for $k\to 0$ by insisting on background independence in this limit. As discussed in sec. \ref{sec:conftrunc}, imposing the sWI and requiring background independence are equivalent to each other for $k\to 0$, as well as to the statement that split symmetry as in \eqref{equ:split-symmetry} is restored and the effective action is thus a functional of the total field only. For $k>0$ this is no longer true and it was found in \cite{Becker:2014qya} that split symmetry cannot be restored for non-vanishing $k$. Note however that it is in general nevertheless possible for the sWI to be imposed for $k>0$. This is certainly the case for the setup of chapter \ref{sec:conformal}, where the implementation of the sWI for any $k$ leads to the reduced system \eqref{equ:sys-red-final} whose solutions are indeed split symmetric as they depend on the total field only. However, the solutions of the original flow \eqref{equ:flowV} and \eqref{equ:flowK} are of course not split symmetric for $k>0$ as follows from the change of variables \eqref{equ:chvars} connecting them. As in \cite{Becker:2014qya}, split symmetry and thus background independence is restored in the limit $k\to 0$ as discussed below \eqref{equ:Gamma-hat}. 

Hence, implementing the sWI automatically takes care of ensuring background independence and restoration of split symmetry when it can be achieved at all, i.e. in the limit $k\to 0$, as we have discussed in sec. \ref{sec:conftrunc}, but it equally ensures that the crucial constraints on the RG flow it embodies are met for $k>0$. This is perhaps one of the most relevant conceptual insights for asymptotic safety that can be gained from this work.

The possibility that asymptotic safety is realised in Nature is certainly very compelling. It would be a triumph for quantum field theory in general and certainly shift the quantum field theoretical spotlight a little away from perturbative and closer to non-perturbative quantum field theory. One may argue that the validity of asymptotic safety for gravity is firmly established only once it has been verified in functional truncations. As the work presented here shows, this will either be in an infinite dimensional truncation beyond the single field approximation or it would have to be accompanied by more insight into how the interplay between the single field approximation and functional truncations lends physical credibility to the results. 